\documentclass[12pt]{article}
\usepackage[utf8]{inputenc}

\usepackage{graphicx}
\usepackage{natbib}
\usepackage{amsmath}
\usepackage{amsfonts}
\usepackage{url}

\newcounter{ctr}

\newcounter{ctr1}

\newcounter{ctr2}

\newcounter{ctr3}

\newenvironment{theorem*}[1]{{\bf Theorem #1} \begin{itshape}}{\end{itshape}}

\newenvironment{corollary*}[1]{{\bf Corollary #1} \begin{itshape}}{\end{itshape}}

\newenvironment{proposition*}[1]{{\bf Proposition #1} \begin{itshape}}{\end{itshape}}

\newcommand{\ud}{\, {\rm d} \kern-.015em }

\newcommand{\modulus}[1]{\left| \kern.05em #1 \kern.05em \right|}
\newcommand{\norm}[1]{\left\| \kern.05em #1 \kern.05em \right\|}
\newcommand{\inner}[1]{\left\langle \kern.05em #1 \kern.05em \right\rangle }

\newcommand{\pick}[2]{\renewcommand{\arraystretch}{0.6}
\left( \kern-.4em \begin{array}{c} #1 \\ #2 \end{array} \kern-.4em \right) }

\usepackage{xr}
\makeatletter

\usepackage{amssymb}
\usepackage{amsmath}
\usepackage{subcaption}
\usepackage{graphicx}
\usepackage{geometry}
\usepackage[utf8]{inputenc}
\usepackage{natbib}
\usepackage{multibib}
\usepackage{xcolor}
\usepackage[official]{eurosym}
\usepackage{wrapfig}
\usepackage{booktabs}
\usepackage{graphics}
\usepackage[bookmarks=false]{hyperref}
\usepackage[graphicx]{realboxes}
\usepackage{bm}
\numberwithin{equation}{section}
\usepackage{siunitx}
\usepackage{gensymb}
\usepackage{makecell}
\usepackage[T1]{fontenc}
\usepackage{titling}
\usepackage{authblk}
\usepackage{appendix}
\usepackage{enumitem}   
\usepackage{multibib}
\newcites{SM}{References}

\title{Accounting for Seasonality in Extreme Sea Level Estimation}
\author[1,*]{Eleanor D'Arcy}
\author[1]{Jonathan A. Tawn}
\author[2]{Am\'elie Joly}
\author[2]{Dafni E. Sifnioti}
\affil[1]{\small STOR-i Centre for Doctoral Training, Department of Mathematics and Statistics, Lancaster University, LA1 4YR, UK}
\affil[2]{\small EDF Energy R\&D UK Centre, Croydon, CR0 2AJ, UK}
\affil[*]{\footnotesize{Correspondence to: e.darcy@lancaster.ac.uk}}

\date{\today}

\begin{document}

\maketitle

\begin{abstract}
Reliable estimates of sea level return levels are crucial for coastal flooding risk assessments and for coastal flood defence design. We describe a novel method for estimating extreme sea levels that is the first to capture seasonality, interannual variations and longer term changes. We use a joint probabilities method, with skew surge and peak tide as two sea level components. The tidal regime is predictable but skew surges are stochastic. We present a statistical model for skew surges, where the main body of the distribution is modelled empirically whilst a non-stationary generalised Pareto distribution (GPD) is used for the upper tail. We capture within-year seasonality by introducing a daily covariate to the GPD model and allowing the distribution of peak tides to change over months and years. Skew surge-peak tide dependence is accounted for via a tidal covariate in the GPD model and we adjust for skew surge temporal dependence through the subasymptotic extremal index. We incorporate spatial prior information in our GPD model to reduce the uncertainty associated with the highest return level estimates. Our results are an improvement on current return level estimates, with previous methods typically underestimating. We illustrate our method at four UK tide gauges.   
\end{abstract}


\newpage

\section{Introduction}\label{Intro}
Extreme sea levels pose an increasing risk to coastline communities. In the absence of any mitigation, the impacts can be severe: fatality, infrastructure damage and habitat destruction. Estimates of sea level return levels are fundamental for various purposes, including coastal flood defence design and flood risk assessments. A return level is the value we expect the annual maximum sea level to exceed with probability $p$. For a stationary series, this corresponds to a value exceeded once every $1/p$ years, on average. We are particularly interested in rare events, where $p\in[10^{-4},10^{-1}]$ to cover sea levels that are important to a range of industries affected by coastal flooding. For example, we obtain data from the Heysham tide gauge; Heysham is a coastal town in north-west England that is home to two nuclear power stations. Nuclear regulators require accurate return level estimates for $p=10^{-4}$. We also consider a further three gauges at Lowestoft, Newlyn and Sheerness; Sheerness is particularly important since it is located on the river Thames estuary where extreme sea levels can propagate down the river towards London. The annual return level, or equivalently its exceedance probability, changes with year if the series is non-stationary and discussions of appropriate design levels involve the intended lifespan of the defence~\citep{rootzen2013}. We focus on a fixed $p$ and recognise that the return level varies across years.

Since the UK is regularly subject to coastal flooding, estimating extreme sea levels is crucial. The 1953 North Sea flood is the worst on record for the 20\textsuperscript{th} century in the UK. Coastal defences were breached in 1,200 places leading to the evacuation of 30,000 people, damage to 24,000 properties and a death toll of 307 in England alone. The damage was estimated at £1.2 billion in 2014~\citep{wadey2015}. Following this event, coastal flood defences were upgraded around most of the UK. Recently, there has been growing concern regarding anthropogenic sea level rise due to climate change. Rises in the mean sea level, coupled with changes in storm frequency and size, can increase the likelihood of coastal flooding. Therefore, it is increasingly important to accurately estimate sea level return levels so that coastline communities are protected against events such as that in 1953. The latest best estimates for extreme sea levels can be found in~\cite{CFB2018}.

\begin{figure}
    \centering
    \includegraphics[width=0.5\textwidth]{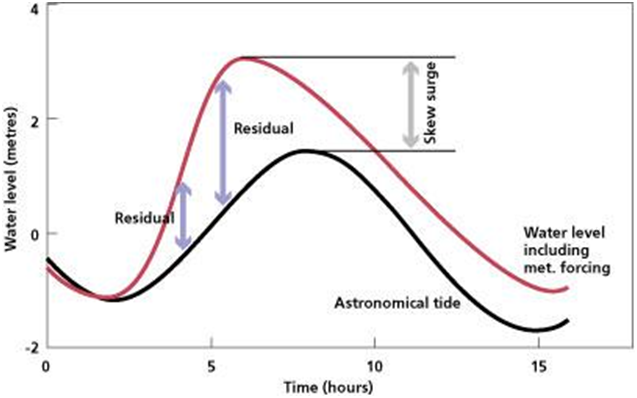}
    \caption{Sea levels during the passage of a surge for a single tidal cycle ~\citep{nationaltidegaugenetwork}.}
    \label{ssexplan}
\end{figure}

\cite{pughwoodworth2014} give a comprehensive overview of sea level processes. Sea levels are a combination of mean sea level, tide, surge and waves. We consider still water level (with waves filtered out), for simplicity we refer to this as sea level. The mean sea level trend is removed so that tide and surge are the only components. Tides are the regular and predictable changes in sea levels driven astronomically; these are well understood and perfectly forecast~\citep{egbert2017}. Surges define any departure from the predicted tidal regime, often resulting from meteorological forces such as storms and are hence sometimes called storm surges or non-tidal residual (as in Figure~\ref{ssexplan}); these are stochastic. However, surges can also include gauge recording errors and tidal prediction errors, but these errors are typically negligible at well maintained gauges, such as those on the UK National Tide Gauge Network (see Section~\ref{data}). Surges can also be influenced by tide-surge interaction, this is the change in the distribution of surge that is dependent on the tidal level, because the surge is essentially a wave which is influenced by the water depth. This interaction is more prominent in shallow water areas, with the largest surges typically occurring mid-tide on the rising tide and the smallest at high tide~\citep{PrandleWolf1978,Tawn1992}.

Early methods to estimate extreme sea levels modelled the observed sea levels directly, ignoring the known tidal component. More recent approaches focus on the convolution of the surge and tide distributions. Due to the complex dependence between surge and tide, we instead consider skew surge and peak tide; skew surge is the difference between the maximum observed sea level and the maximum predicted tide (peak tide) within a tidal cycle, regardless of their timing (see Figure~\ref{ssexplan}). There is a single skew surge value associated with each peak tide value (every 12 hours 26 minutes), as opposed to hourly or 15 minute interval observations of surge and tide. Even though there are fewer skew surge and peak tide observations, these components are preferred because they have much weaker dependence;~\cite{Williams2016} demonstrate that it is reasonable to assume these are independent at most UK sites.

We build on the skew surge joint probabilities method (SSJPM) of~\cite{Batstone2013}, which assumes peak tide and skew surge are independent and that they are stationary processes within and across years, with extreme skew surges modelled by a generalised Pareto distribution (GPD). In our approach, we account for seasonality in the skew surges by adding a daily covariate to the rate of threshold exceedance and scale parameter of the GPD model. We also introduce a tidal covariate to capture skew surge-peak dependence. Skew surges also exhibit temporal dependence, we account for this using the extremal index~\citep{Tawn1992} but in a different way to previous analyses. Since tides are deterministic, we choose our tidal samples so that they account for monthly and interannual variations. We estimate return levels by deriving distributions for the annual maxima sea levels. We obtain the estimates using our proposed method and for the SSJPM, as well as some intermediate methods, at four sites on the UK coastline, finding that the SSJPM tends to underestimate return levels.

It is fundamental to recognise the uncertainty associated with return level estimates of the highest order to add value to the point estimates. We construct confidence intervals for our estimates using a stationary bootstrap procedure~\citep{Politis1994}. This preserves the realism of the sea level processes. Our approach is the first to use a bootstrap procedure for uncertainty quantification on extreme sea level estimates. We add a prior distribution to the GPD shape parameter based on spatial information~\citep{CFB2018}; this significantly reduces the uncertainty associated with our return level estimates. 

We discuss the relevant extreme value theory and existing methodology for extreme sea level estimation in Section~\ref{background}. In Section~\ref{exploratory_analysis} we introduce the data and explore the seasonality of each component as well as the dependence between them. Section~\ref{methodology} describes the methodology for deriving the annual maxima distribution, starting with an idealised solution derived under simplified assumptions that are then relaxed. Our return level estimates are compared with previous methods in Section~\ref{results}. Additional supporting material is presented in \cite{DarcySM}; when cross referencing, the figure, table or equation number is preceded by `S'.

\section{Background to Method}\label{background}

\subsection{Relevant Extreme Value Methods}\label{evmethods}
When deriving a model for extremes, it is natural to first consider the maximum $M_n$ of a sequence of independent and identically distributed (iid) continuous random variables $Z_1,\ldots,Z_n$, i.e., $M_n=\max\{Z_1,\ldots,Z_n\}$. This sequence has marginal distribution function $F$ and upper end point $z^F$. If there exists sequences of constants $\{a_n>0\}$ and $\{b_n\}$, so that the rescaled block maximum $(M_n-b_n)/a_n$ has a nondegenerate limiting distribution as $n\rightarrow\infty$, then this limit must have the form
\begin{equation}
    G(z)=\exp\bigg\{-\bigg[1+\xi\bigg(\frac{z-\mu}{\sigma}\bigg)\bigg]^{-1/\xi}_{+}\bigg\},\label{GEV_form}
\end{equation}
where $x_+=\max\{x,0\}$ for parameters ($\mu,\sigma,\xi)\in\mathbb{R}\times\mathbb{R}_{+}\times\mathbb{R}$ representing the location, scale and shape, respectively~\citep{Coles2001}. This is the generalised extreme value distribution (GEV) that encompasses three families: for $\xi>0$, this corresponds to the Fr\'echet distribution, $\xi<0$ the Weibull and $\xi=0$ the Gumbel. Note $\xi=0$ should be interpreted as the limit as $\xi\rightarrow 0$. This result provides asymptotic motivation for using the GEV as a parametric model for observed block maxima and thus a basis to estimate and extrapolate to high return levels. However, this assumes an underlying iid process, which is unrealistic. 

Now we relax the independence assumption, so that $Z_1,\ldots,Z_n$ is a stationary sequence with the same marginal distribution function $F$. This corresponds to a series whose variables may be mutually dependent, but whose statistical properties are homogeneous through time. The limiting distribution of the rescaled block maxima of a stationary process satisfying a long-range dependence condition, which ensures events long apart in time are near independent, is $G^{\theta}(z)$ with $G(z)$ as in equation~\eqref{GEV_form} and $\theta\in(0,1]$ the extremal index~\citep{Leadbetter1983}. For an independent series $\theta=1$, but the converse is not true.

When a process exhibits extremal dependence, groups of extreme events form above high thresholds; these groups are called clusters. We define different clusters as those separated by some number of non-extreme values (i.e., below a high threshold $z$), this is called the run length $r$. Within a cluster, extreme events are considered as dependent whilst exceedances in different clusters are assumed to be independent. The extremal index tells us about clusters since it can be estimated empirically as the reciprocal of the mean cluster size of exceedances of $z$. This is known as the runs method~\citep{SmithWeissman1994}. Clusters are identified using an arbitrary choice of run length; this is the main pitfall with the approach.~\cite{FerroSegers2003} propose the intervals estimator, based on the limiting distribution of normalised times between exceedances of $z$. This distribution is exponential for independent random variables, otherwise it is a mixture distribution of an exponential with mean $\theta^{-1}$ and a degenerate probability distribution at zero, with probabilities $\theta$ and $1-\theta$, respectively.~\cite{FerroSegers2003} also propose an automatic declustering scheme using the intervals estimate. Both methods only estimate $\theta$ in the observed range of $z$. In practice the runs and intervals estimators of $\theta$ are actually estimators of the subasymptotic extremal index, for threshold level $z$ and run length $r$, defined by~\cite{LedfordTawn2003} as \begin{equation}
    \theta(z,r)=\mathbb{P}(\max\{ Z_2,\ldots, Z_r\}<z| Z_1>z).\label{subas_exi}
\end{equation}Then the extremal index is the limit of expression~\eqref{subas_exi} as $z\rightarrow z^F$ and $r\rightarrow\infty$, with $z$ and $r$ tending to their respective limits in a related fashion.

We can also define extremes as exceedances of a high threshold $u$. If expression $\eqref{GEV_form}$ holds, then for an arbitrary term $Z$ in the sequence $Z_1,\ldots,Z_n$,\begin{equation}
    \mathbb{P}(Z>b_n+a_n z\;|\;Z>a_n+b_n u)\rightarrow H_u(z) \quad \text{ where } \quad
    H_u(z)=\bigg[1+\xi\bigg(\frac{z-u}{\sigma_u}\bigg)\bigg]_+^{-1/\xi}\label{GPD_form}
\end{equation} for $z>u$ as $n\rightarrow\infty$, with $a_n$, $b_n$ as previously and $(\sigma_u,\xi)\in\mathbb{R}_{+}\times\mathbb{R}$ the scale and shape parameters, respectively~\citep{Coles2001}. The shape parameter is the same as that for the GEV, whilst the scale is threshold dependent since $\sigma_u=\sigma+\xi(u-\mu)$ for $\mu$ and $\sigma$ the GEV parameters. This is the generalised Pareto distribution (GPD). If $Z_1,\ldots,Z_n$ are iid, then exceedances of a high threshold $u$ are iid and have limiting GPD tail model \begin{equation*}
    \mathbb{P}(Z>z)=\lambda_u\bigg[1+\xi\bigg(\frac{z-u}{\sigma_u}\bigg)\bigg]_+^{-1/\xi}
\end{equation*} for $z>u$ where $\lambda_u=\mathbb{P}(Z>u)$. Again, if $Z_1,\ldots,Z_n$ are stationary, a common approach is to identify clusters and decluster them to yield an approximately independent sequence of cluster maxima for which the GPD remains a valid model \citep{FawcettWalshaw2007}.

For non-stationary processes, it is common to allow the parameters of a stationary statistical model to vary with time or another covariate. In a block maxima framework, observations in a block are assumed to be iid, so covariates in the GEV parameters cannot change within a block. In contrast, the GPD allows the covariates to vary uncontrolled over consecutive observations. A range of methods can be adopted to incorporate covariates in the model parameters $\lambda_u, \sigma_u, \xi$, including via harmonics~\citep{ColesTawnSmith1994}, splines~\citep{Jonathan2014} or generalised additive models~\citep{ChavesDavison2005}.

\subsection{Existing Methodology}\label{existingmethods}
The earliest methods to estimate sea level return levels fit a GEV to the annual maxima~\citep{graff1978concerning,ColesTawn1990} or the annual $r$-largest observed sea levels~\citep{tawn1988sealevels}. These direct approaches ignore the known tidal component even though it induces non-stationarity into the sea level series, so that the GEV limit may not be a good approximation.~\cite{DixonTawn1999} show that these direct methods underestimate extreme sea levels, especially at longer return levels for tidally dominant sites.

\cite{PughVassie1978} were the first to exploit the  decomposition of sea levels into tide and surge for extreme sea level estimation. This method is known as the joint probabilities method (JPM), where the probability distribution of extreme sea levels is derived by convolution of the distributions of these two components. The JPM gives consideration to all surge values in the data regardless of when they occurred relative to high tide. This approach forms the basis of the subsequent methods, but has some restrictive and unrealistic assumptions. Since the empirical surge distribution is used in the JPM, return level estimates are constrained by the sum of the highest predicted tide and the highest observed surge. Hourly surge observations are assumed to be independent, this is unrealistic as surge exhibits strong temporal dependence~\citep{TawnVassie1989}.~\cite{PughVassie1978} do account for dependence between tide and surge by dividing the tidal range into ordered bands of equal probability and estimating the surge distribution for each band, but this gives results which are sensitive to the choice of the number of bands and their boundary levels.

The revised joint probabilities method (RJPM) of~\cite{Tawn1992} attempts to address these limitations. For the upper tail of surges, an extreme value distribution is used to allow extrapolation beyond what has been observed, hence improving return level estimates. To account for temporal dependence, the extremal index is used.~\cite{Tawn1992} and~\cite{dixontawnvassie1998} model the surge-tide dependence by allowing parameters of the GEV to be functions of the tidal level. This is a difficult task because the relationship is complex~\citep{PrandleWolf1978}.~\cite{dixontawn1994} and~\cite{haigh2010} find that the RJPM is the best performing of these methods for observed and simulated data.

\cite{Batstone2013} propose the skew surge joint probabilities method (SSJPM) to overcome the requirement of modelling the dependence of surge on tide. Their approach assumes that skew surge and peak tide are independent.~\cite{Williams2016} demonstrate this is a good approximation both physically and empirically, for many sites in Europe and the USA. The SSJPM fits a GPD to the upper tail of skew surges, whilst the empirical distribution is used for the main body of the distribution. The extremal index is used to measure dependence for sea levels in adjacent tidal peaks. They find $\theta\approx{1}$ for all sites, suggesting no evidence of dependence in the upper tail.~\cite{baranes2020} adapt the SSJPM to account for interannual variations in the tidal regime, considering summer and winter separately and assuming peak tides and skew surge are stationary within each season.


 
All the methods discussed so far estimate extreme sea levels at a single site, but this gives uncertain return level estimates for sites with shorter observation periods. Spatial pooling can improve estimates at sites with limited or no data. \cite{bernardara2011} use regional frequency analysis to estimate extreme surges. This involves grouping statistically similar sites into homogeneous regions, then fitting an extreme value model with a constant shape parameter over all sites~\citep{HoskingWallis1997}. More recent approaches account for uncertainty in the region selection~\citep{Asadi2018,Rohrbeck2020}. \cite{Batstone2013} use hindcast sea level data to interpolate SSJPM estimates along the UK coast. Other methods allow parameters of the extreme value distribution to vary with spatial covariates.~\cite{ColesTawn1990} do this, allowing the location parameter of the GEV for sea level annual maxima to depend on the harmonic tidal constituents. Whereas,~\cite{dixontawnvassie1998} spatially smooth parameters of the extreme value model for surges used in the RJPM, using predictable tidal variations along the UK coastline.

\section{Exploratory Analysis}\label{exploratory_analysis}
\subsection{Data}\label{data}
We use data from the UK National Tide Gauge Network obtained from the British Oceanographic Data Centre; these undergo rigorous quality control before release. Sea level elevations are recorded at 44 sites along the UK coastline. We consider data from Heysham, Lowestoft, Newlyn and Sheerness only. Heysham is located on the west coast of England and has records from 1964-2016, with 17\% missing. Lowestoft is on the east coast of England, with data available from 1964-2020 (4\% missing). Sheerness is at the Thames Estuary, also on the east coast, with data available from 1980-2016 and 9\% missing. Newlyn is located on the south coast of England and has records from 1915-2020, with 17\% missing. The data are in metres above chart datum.

To compare the relative importance of skew surge and peak tide across sites, we define a surge-tide index as the observed range of skew surges divided by the range of peak tides. Heysham and Newlyn are tidally dominant with indices 0.65 and 0.70, respectively, whilst Lowestoft and Sheerness are surge dominant with indices 3.36 and 1.66, respectively. We choose to study these sites because they are typically affected by different storms and all have a long observational duration. Heysham and Lowestoft are of particular interest because of their widely different surge-tide indices. Newlyn has the longest study period in the tide gauge network.~\cite{howardWilliams2021} use climate model simulations to conclude that skew surge is dependent on peak tide at Sheerness, so we also study this site. The highest astronomical tide (HAT) observed is 10.72m, 2.92m, 6.10m and 6.26m for Heysham, Lowestoft, Newlyn and Sheerness, respectively. The data were preprocessed in an attempt to remove the linear non-stationary effect of sea level rise on skew surges caused by climate change and isostatic rebound, see~\cite{CFB2018} for details. Therefore all results are presented relative to the mean sea level in 2017.

\label{lineartrend}

\subsection{Seasonality and Temporal Dependence of Skew Surge and Peak Tide}\label{ss_seasonality}
\begin{figure}
    \centering
    \includegraphics[width=0.49\textwidth]{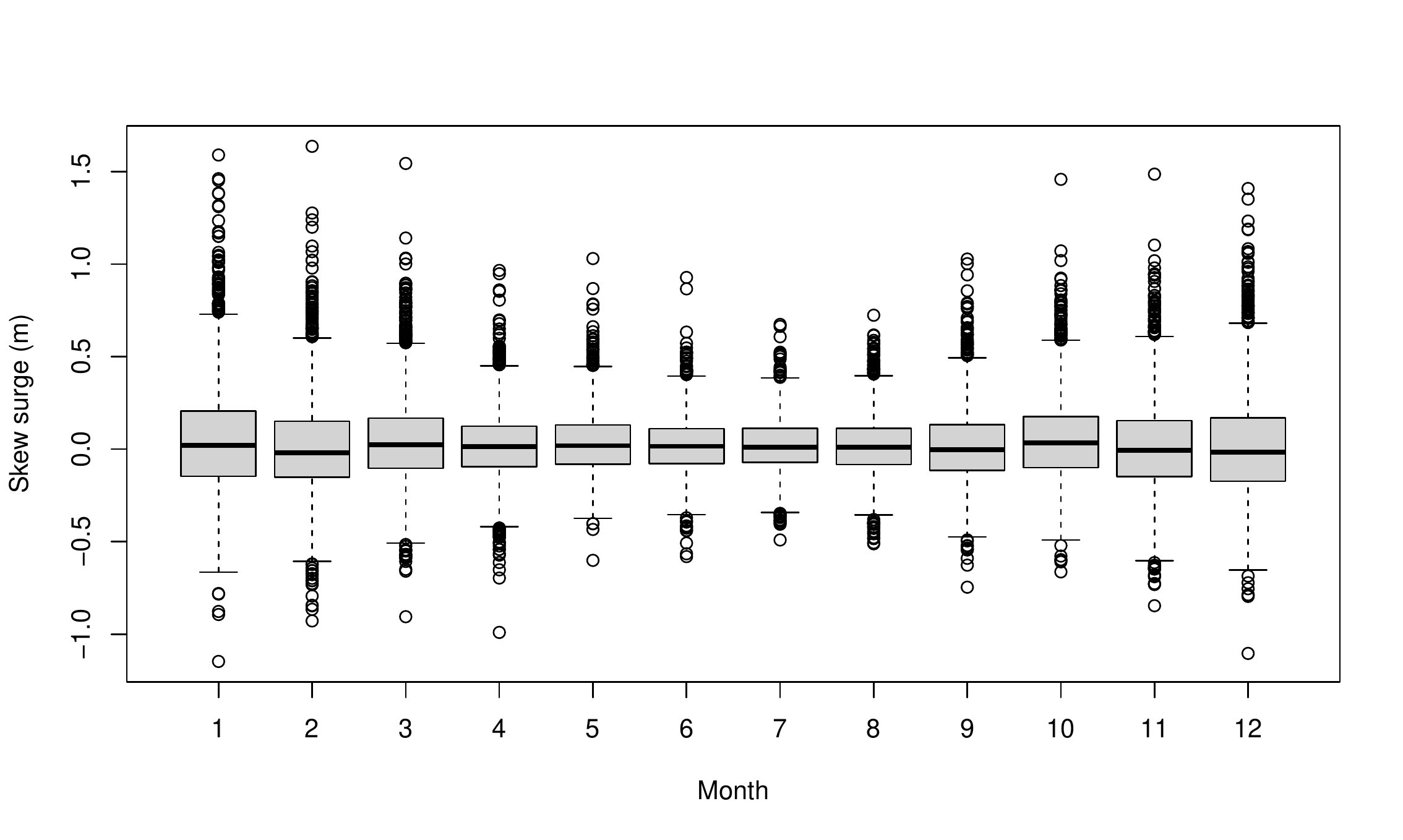}    \includegraphics[width=0.49\textwidth]{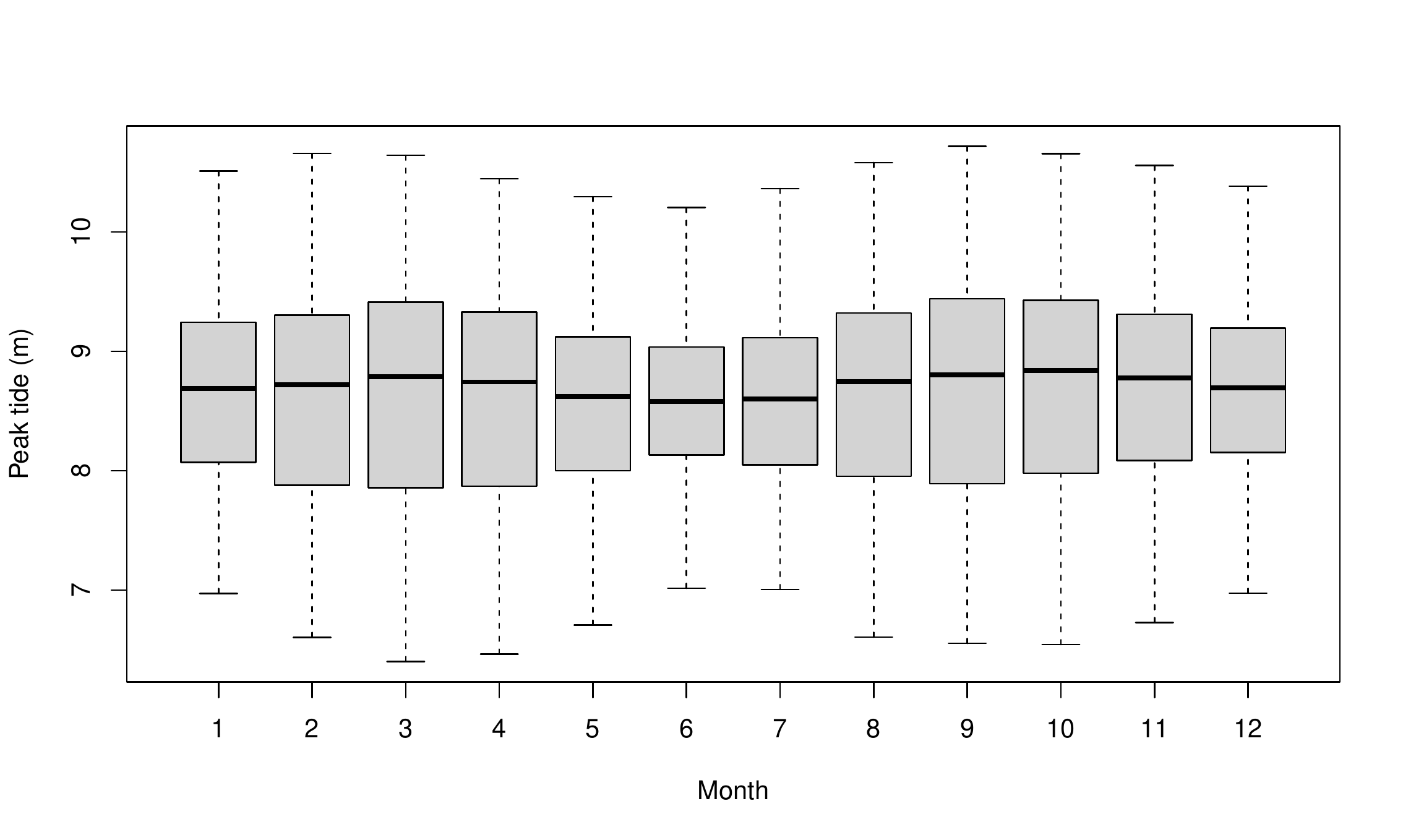}
    \caption{Monthly boxplots of skew surge (left) and peak tide (right) at Heysham.}
    \label{ss_mt_bps_HEY}
\end{figure}

Figures~\ref{ss_mt_bps_HEY} and~\ref{ss_mt_bps_all} show monthly boxplots of skew surge and peak tide observations at Heysham and the remaining sites, respectively. The median of the monthly skew surges remain relatively constant across months at Heysham, Newlyn and Sheerness but there is some variation at Lowestoft. The skew surge range varies across the year, exhibiting clear seasonality. Extreme skew surges and extreme peak tides typically occur in winter and at the equinoxes, respectively. Tides also exhibit interannual variability, including the 18.6 year lunar nodal cycle and the 8.85 year cycle of lunar perigee~\citep{pughwoodworth2014}.

To further assess seasonality, we estimate the probability that a randomly selected peak tide $X$ is from month $j$ where $j=1-12$, given it is higher than some value $x_q$, the $q$th quantile of the distribution of peak tides, with $q\in[0,1]$. This is defined as \begin{equation}
    \tilde P_{X}(j;x_q)=\hat{\mathbb{P}}(m(X)=j\;|\;X>x_q),\label{prob}
\end{equation} 
where $m(X)$ denotes the month of occurrence of $X$ and $\hat{\mathbb{P}}(\cdot)$ is calculated empirically. If peak tides are identically distributed over the year and all months have equal duration, then $\tilde P_{X}(j;x_q)=1/12$ for all $j$ and $q$. Of course, months vary in length, but a significant departure from $\tilde P_{X}(j;x_q)=1/12$ indicates that peak tides are not identically distributed over a year. Months with the largest peak tides will have higher values for $\tilde P_{X}(j;x_q)$ for large $q$. 

\begin{figure}
    \centering
    \includegraphics[width=0.49\textwidth]{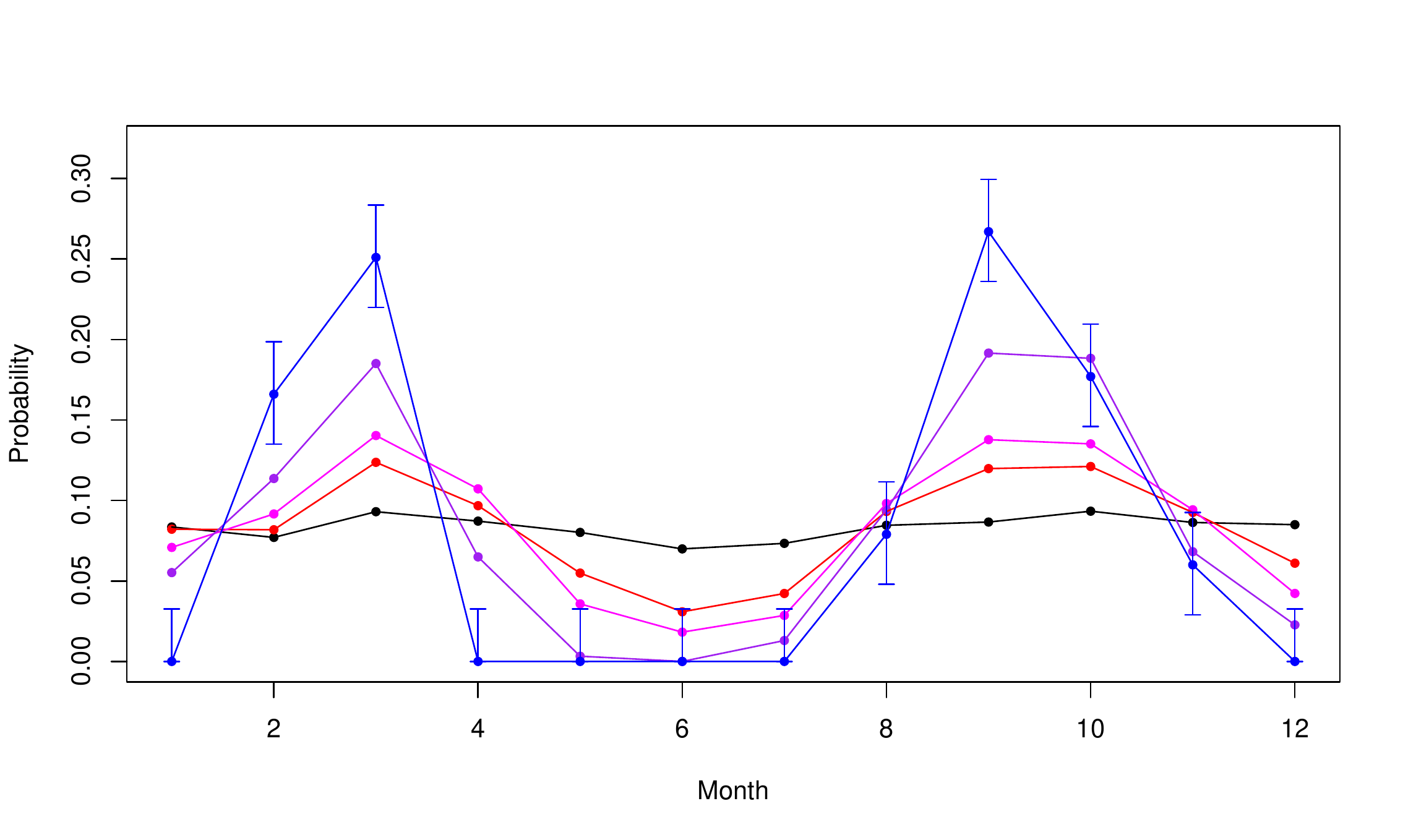}
    \caption{Estimates $\tilde P_{X}(j;x_{q})$ for $j=1-12$ and $q=0.5$ (black), $0.9$ (red), 0.95 (magenta), 0.99 (purple), 0.999 (blue) at Heysham, with 95\% confidence intervals for $q=0.99$.}
    \label{mt_seasonality_plot_HEY}
\end{figure}

Figure~\ref{mt_seasonality_plot_HEY} shows the estimates $\tilde P_{X}(j;x_q)$ for a range of $q$ at Heysham (see Figure~\ref{mt_seasonality_plot} for the remaining sites). These show that $\tilde P_{X}(j;x_{0.5})\approx1/12$ for all $j=1-12$ at Heysham, Newlyn and Sheerness. At Lowestoft this varies, agreeing with Figure \ref{ss_mt_bps_all}. For all $q\geq 0.9$, there is clear evidence that $\tilde P_{X}(j;x_q)$ is largest in months nearest the equinoxes at all sites. At Heysham, $\tilde P_{X}(j;x_{0.999})>0$ for $j=2,3,8,9,10$; we find the same results at Newlyn but $\tilde P_{X}(1;x_{0.999})>0$ also. However, at Lowestoft and Sheerness, $\tilde P_{X}(j;x_{0.999})=0$ in months close to the spring equinox. Confidence intervals on these estimates are constructed by exploiting the property that the number of times that an event $X>x_q$ occurs in month $j$ follows a multinomial distribution, with probabilities $\tilde P_{X}(j;x_{q})$. Figures~\ref{mt_seasonality_plot_HEY} and~\ref{mt_seasonality_plot} also show 95\% confidence intervals for $\tilde P_{X}(j;x_{0.999})$ at each site. These indicate that the differences in $\tilde P_{X}(j;x_{0.999})$ discussed above for each month are statistically significant.

Skew surge and peak tide both exhibit temporal dependence; we are only interested in modelling the former since peak tides are deterministic. Figure \ref{fig::acf_ss} shows the autocorrelation function (acf) plots for skew surge at each site. The correlation is stronger at lower lags, as expected, and tends to zero at higher lags; although this doesn't always reach zero due to seasonality. At Heysham and Newlyn, the correlation is significantly higher for all lags above 1, than at Lowestoft and Sheerness. In Section~\ref{sec::suppmat_exi} of~\cite{DarcySM} we explore temporal dependence in the skew surges further by looking at the two mean measures of extremal dependence $\chi$ and $\bar\chi$~\citep{Coles1999} at different high thresholds and lags. We find similar patterns over lags, but weaker dependence, for the extreme values.

\subsection{Skew Surge and Peak Tide Dependence}\label{ss_mt_indep}

\cite{PrandleWolf1978} examine the tide-surge interaction, both empirically and physically, for sites in the southern North Sea, finding that the most extreme surges occur on rising tides. This complex dependence structure motivates the use of skew surge and peak tide in the SSJPM~\citep{Batstone2013}.~\cite{Williams2016} conclude that the assumption of skew surge-peak tide independence is broadly well supported empirically and based on valid physical reasoning. However,~\cite{CFB2018} and~\cite{Williams2016} identify a weak correlation at Sheerness.~\cite{howardWilliams2021} use oceanographic numerical model simulations to show that the highest skew surges tend to occur on lower peak tides at Sheerness. These findings raise questions about the validity of the independence assumption in the SSJPM. Ignoring the observed dependence would result in overestimates of return levels. We use a range of exploratory data analyses to investigate these claims in Section~\ref{ss_mt_indep_supp} of~\cite{DarcySM}. We find that it is reasonable to assume skew surge-peak tide independence at Heysham, Lowestoft and Newlyn, but not at Sheerness. Our results suggest that this dependence structure is changing throughout the year, with the strongest dependence found in May.

\section{Novel Methodology}\label{methodology}
\subsection{Introduction}
We begin by deriving the distribution of both the sea level monthly maxima $M_{j}$ for $j=1-12$ and annual maxima $M$ under several simplifying assumptions that we subsequently relax. We use the fact that the peak sea level $Z_i$ in tidal cycle $i$ can be written as the sum of skew surge $Y_i$ and peak tide $X_i$ in that tidal cycle, for all $i=1,\ldots,T$ where $T$ is the total number of tidal cycles. In Section~\ref{simple_method} we develop a simplified model, assuming skew surges are iid and independent of peak tide. We relax these assumptions to develop a novel method that reflects the realism of the sea level process; accounting for skew surge seasonality (Section~\ref{acc_ss_seasonality}), skew surge-peak tide dependence (Section~\ref{acc_ssmt_dep}) and temporal skew surge dependence (Section~\ref{ss_dep}). Unless stated otherwise, inference is conducted in a likelihood framework. We provide 95\% confidence intervals for parameter estimates based on the Hessian; this assumes extreme skew surges are independent, which is reasonable for model selection but in Section~\ref{rl_final} we use a bootstrap procedure for uncertainty quantification.

Extreme sea levels up to the $\sim20$ year return period can occur with various combinations of skew surge and peak tides, e.g., a typical skew surge value combined with the largest peak tide, through to an extreme skew surge with a typical peak tide. Since peak tides are bounded above by HAT, return periods over 20 years can only be achieved with extreme skew surges, with the very largest return periods requiring skew surges bigger than already observed. We are interested in return levels corresponding to return periods of 1 year and above, so we require an estimate for the distribution of all possible skew surges, though we are particularly interested in modelling the upper tail for inference at high return periods. Thus in each variant of our method, we develop a model for the whole skew surge distribution under the associated assumptions about the skew surge process.

\subsection{Idealised Case}\label{simple_method}
Under the assumption that the skew surges are identically distributed, we estimate the distribution of skew surges below a threshold $u$ using the empirical distribution $\tilde F_Y$. This is adequate because tide gauges on the UK National Tide Gauge Network typically have long observation lengths (>20 years) so the empirical distribution describes the main body of the data well. To enable extrapolation in the tail we use the GPD model~\eqref{GPD_form}, with constant parameters. Then, our model is
\begin{align}
    F_{Y}(y)=\begin{cases} \tilde{F}_Y(y) \quad\quad &\text{if }y\leq u\\ \\ 1-\lambda_u\big[1+\xi\big(\frac{y-u}{\sigma_u}\big)\big]_+^{-1/\xi}  &\text{if }y> u \end{cases}\label{ss_nonstat_model}
\end{align} 
where $\lambda_u=1-\tilde{F}_Y(u)$ and $\sigma_u,\xi$ are the parameters of the GPD. We take $\lambda_u=0.05$. To simplify notation, we subsequently drop the $u$ subscript on the scale $\sigma$ and rate $\lambda$ parameters.
 
We also assume that skew surge is independent of peak tide, and as peak tides are deterministic, knowing the tidal cycle $i$ determines the peak tide $X_i$, so we have \begin{equation*}
    \mathbb{P}(Z_i\leq z)=\mathbb{P}(X_i+Y_i\leq z)
=\mathbb{P}(Y_i\leq z-X_i)=F_{Y}(z-X_i), \mbox{ for }-\infty<z<\infty.
\end{equation*}

Let $T_j$ denote the number of tidal cycles in month $j$. We use sequential monthly peak tide samples $\{X_{j_i}; j=1-12,i=1,\ldots,T_j\}$ where $j_i$ denotes the $i$th peak tide in month $j$. Then, if skew surges are independent and peak tides repeat exactly on an annual cycle, the distribution of month $j$ maximum sea level, $M_j$, is 
\begin{equation}
    \mathbb{P}(M_j\leq z)=\prod\limits_{i=1}^{T_j}\mathbb{P}(Y_i\leq z- X_{j_i})
    =\prod\limits_{i=1}^{T_j}F_{Y}(z- X_{j_i}).
\label{monmax_noseasonality}
\end{equation}
Then the annual maxima skew surge distribution is given by,
\begin{equation}
    \mathbb{P}(M\leq z)  = \prod\limits_{j=1}^{12 } \prod\limits_{i=1}^{T_j}F_{Y}(z- X_{j_i}).
\label{annmax_noseasonality}
\end{equation} 
Since peak tides are temporally dependent, the tidal samples $\{X_{j_i}\}$ used in expressions~\eqref{monmax_noseasonality} and \eqref{annmax_noseasonality} must be from contiguous peak tides. 

Although cycles of periods up to a year dominate, peak tides have longer term periodicities. Previous methods, such as~\cite{Tawn1992}, assumed peak tides are stationary within and across years so that the annual maxima distribution is given by \begin{equation}
        P(M\leq z)=\bigg(\prod\limits_{i=1}^{T} F_Y(z-X_{i})\bigg)^{1/K},\label{eqn::RJPM_ss}
\end{equation} for $T$ the total number of observations, $K$ the number of years of observation and $F_Y$ the stationary skew surge distribution~\eqref{annmax_noseasonality}. We incorporate interannual peak tide variations by taking the average of the yearly patterns over $K$ years, where $K\geq 19$ to address all nodal cycle variations. We denote peak tide on the $i$th tidal cycle in month $j$ of year $k$ by $X_{j_i}^{(k)}$ for $j=1-12,\;i=1,\ldots,T_j^{(k)}$ and $k=1,\ldots,K$. Here $T_j^{(k)}$ is the number of tidal cycles in month $j$ and year $k$, which varies over $k$ because the timing of the first cycle in the month can vary annually. Then the monthly and annual maxima sea level distributions are
\begin{equation}
    \mathbb{P}(M_j\leq z)=\frac{1}{K}\sum_{k=1}^K\prod\limits_{i=1}^{T_j^{(k)}}F_{Y}(z- X^{(k)}_{j_i}), 
\mbox{ and }
    \mathbb{P}(M\leq z)  =
\frac{1}{K}\sum\limits_{k=1}^{K}\prod\limits_{j=1}^{12}\prod\limits_{i=1}^{T_j^{(k)}}F_{Y}(z- X_{j_i}^{(k)}),
\label{aver_max}
\end{equation}
respectively. This no longer has the property that the distribution of the annual maxima is the product of the monthly maxima distribution, as the monthly maxima are now dependent due to associations in peak tides in different months across years. 

\subsection{Skew Surge Seasonality}\label{acc_ss_seasonality}
As discussed in Section \ref{ss_seasonality}, skew surges are not identically distributed across a year. Here we describe how we capture seasonality in three ways: for values above the threshold, below the threshold and in the exceedance probability. We define extreme skew surges using a month-specific threshold $u_j$, defined as a quantile of each monthly distribution. We use the 0.95 monthly empirical quantiles. In Section~\ref{discussion} we discuss the merits of this approach relative to taking a threshold that varies on a daily scale.

Firstly we look at exceedances of the monthly thresholds and model these using the GPD as the asymptotic justification, given in Section~\ref{evmethods}, is still likely to hold at a monthly level. To account for seasonality we allow the parameters to change over time in a periodic fashion, as discussed in Section~\ref{evmethods}. We consider four models to describe how these parameters change with time, expressed in terms of a daily or monthly covariate. As a first approach to account for seasonality in extremes, \cite{carter1981} suggested allowing each month to have a separate distribution. To do so we fit a GPD with a monthly covariate on the scale $\sigma_j$ and shape $\xi_j$ parameters; we refer to this as Model~$S0$. We use this basic approach for comparison only. To obtain a more parsimonious model, we fix the shape parameter to be the same value over all months but keep a monthly covariate on the scale parameter; we call this Model~$S1$. These models give discontinuities at transitions between months, they assume skew surges are identically distributed within a month and have a large number of parameters. Therefore we introduce a covariate $d=1-365$ that denotes the day in year to capture within-year variations smoothly. For notational simplicity, we assume $d$ is entirely defined using the tidal cycle index $i$ and month $j$, so that there exists a function $h$ where $d=h(j,i)$. We consider two harmonic parameterisations of the scale parameter with a daily covariate. Model~$S2$ uses a single harmonic defined by \begin{equation}
    \sigma_d=\alpha_\sigma+\beta_\sigma\sin\bigg(\frac{2\pi}{f}(d-\phi_\sigma)\bigg),\label{sinescale}
\end{equation} for parameters $\alpha_\sigma>\beta_\sigma>0$, $\phi_\sigma\in[0,365)$ and periodicity $f=365$. Model~$S3$ uses two harmonics with periodicities of $f$ and $f/2$. The shape parameter is equal across months for both of these models, so that Models~$S2$ and $S3$ have 4 and 6 parameters, respectively.

\begin{figure}
    \centering
    \includegraphics[width=0.5\textwidth]{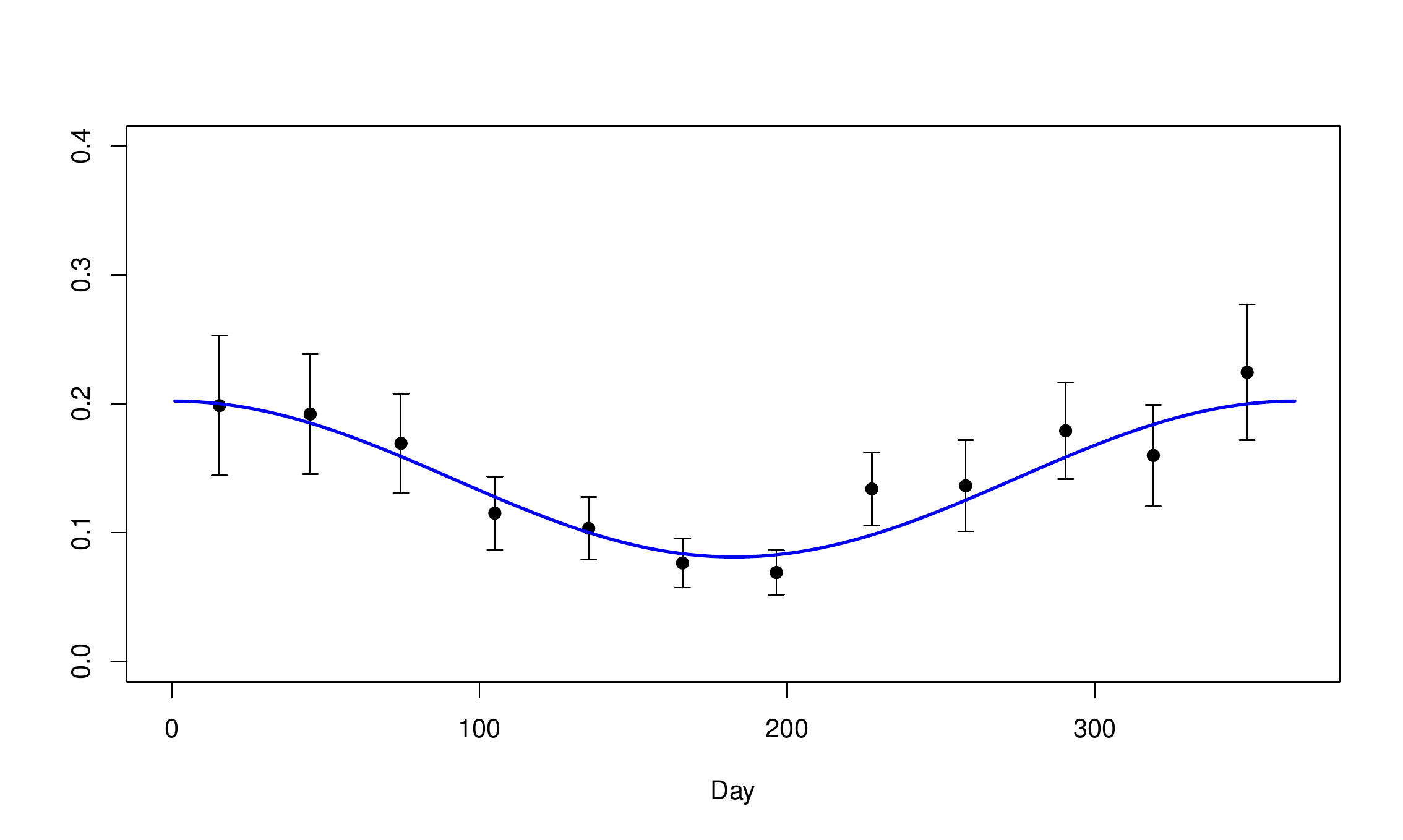}\includegraphics[width=0.5\textwidth]{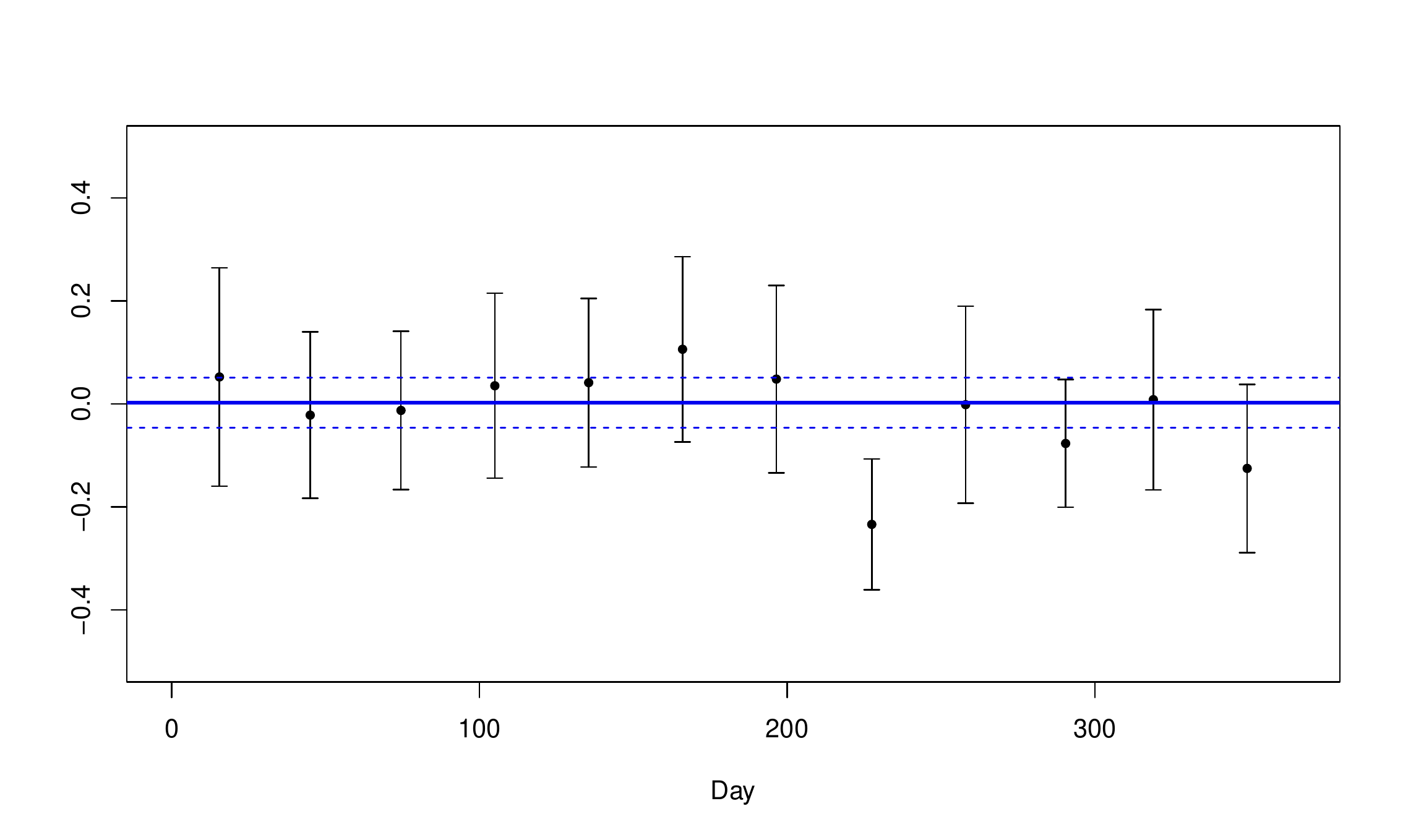}
    \caption{Scale (left) and shape (right) parameter estimates for Model~$S2$ (blue) and Model~$S0$ (black) at Heysham. 95\% confidence intervals are added to Model~$S0$ parameter estimates (shown by black error bars) and to the shape parameter estimate of Model $S2$ (blue dashed lines).}
    \label{ss_param_ests_fig_HEY}
\end{figure}

We now fit the models outlined above to skew surge threshold exceedances at each site. Figures~\ref{ss_param_ests_fig_HEY} and~\ref{ss_param_ests_fig_all} show Model~$S0$ parameter estimates for Heysham and the remaining sites, respectively. Since the 95\% confidence intervals for the monthly shape parameters have considerable overlap for all sites, it is reasonable to have a common value across months as in Models~$S1$, $S2$ and $S3$. Table \ref{AICBIC_ss} reports the AIC and BIC scores for each model at all sites. These scores suggest that restricting the shape parameter to be an unknown constant, over months, improves model fit compared to Model~$S0$ in all cases. Model~$S2$ is selected as our final model for extreme skew surges as the evidence across all sites shows this is reasonable. Table~\ref{tbl:param_ests} gives the parameter estimates for Model~$S2$ at each site. Notice that the shape parameter estimates have 95\% confidence intervals that encompass zero at each site. Figures \ref{ss_param_ests_fig_HEY} and \ref{ss_param_ests_fig_all} show the scale and shape parameter estimates graphically; $\hat\sigma_d$ is higher in the winter and lower in the summer, as expected. The estimated scale parameters of Model~$S2$, when averaged over each month, do not differ significantly from the estimates of Model~$S0$ but do capture a smooth transition within each month.

We also considered using a log link function for the scale parameter instead of the identity link function for Models~$S2$ and $S3$ but found that the identity link yields a better model fit. Furthermore the identity link function has the advantage of preserving the threshold stability property when covariates are included \citep{EastoeTawn2009}.

\begin{table}[]
    \centering
    \resizebox{\textwidth}{!}{
    \begin{tabular}{lccccccccc}
        \hline
         \multicolumn{2}{c}{} & \multicolumn{2}{c}{Heysham} & \multicolumn{2}{c}{Lowestoft} & \multicolumn{2}{c}{Newlyn} & \multicolumn{2}{c}{Sheerness}  \\
         Model & No. Parameters & AIC & BIC & AIC & BIC & AIC & BIC & AIC & BIC \\ \hline
           
         $S1$ & 13 & -66.66 & -119.52 & -57.40 & -118.72 & -77.36 & -143.83 & -67.51 & -123.52 \\
         $S2$ & 4 &  -69.5 & -176.61 & -60.47 & -171.93 & -88.89 & -209.74 & -76.18 & -178.03\\
         $S3$ & 6 &  -65.65 & -161.98 & -56.53 & -156.85 & -86.06 & -194.82 & -75.06 & -166.73 \\ 
         $S4$ & 5 & 2.99 & 8.42 & 1.82 & 7.39 & 5.77 & 11.82 & -181.57 & 3.53 
    \end{tabular}
    }
    \caption{AIC and BIC scores at each site for each skew surge model relative to Model~$S0$ scores, except Model~$S4$ which is measured relative to Model~$S2$.}
    \label{AICBIC_ss}
\end{table}

Another fundamental part of capturing within year seasonality is through the rate parameter $\lambda$. So far, we have assumed this is constant as the threshold has been set at the same quantile across months. Here, we add a daily covariate $d$ to capture smooth changes in $\lambda$ throughout each month, whist recognising that the average exceedance rate across a month is equal for all months. Let $V_d$ be a binary random variable representing whether a skew surge value exceeds its month-specific threshold $u_j$ or not, so that $V_d\sim\text{Bernoulli}(\lambda_d)$. Therefore, we use a logit link function $g(\cdot)$ to model $\lambda_d$ and capture daily changes using a generalised linear model (GLM). To account for within month variations, we relate $g(\lambda_d)$ to the day in month $d_j\in[1,31]$, standardised by the monthly mean day $\bar d_j$, so that $(d_j-\bar d_j)\in[-15,15]$ approximately. We parametrise the gradient using the day in year $d$ to account for different gradients across the year using a harmonic with periodicity $f=365.$ We refer to this model as Model $R0$ and this is given is given by
\begin{equation}
    g(\lambda_d)=g(\lambda)+(d_j-\bar d_j)\beta_\lambda\sin\bigg(\frac{2\pi}{f}(d-\phi_\lambda)\bigg),\label{rate_day}
\end{equation} for $\beta_\lambda>0$, $\phi_\lambda\in[0,365)$ which are parameters to be estimated and $\lambda$ the exceedance rate in a month (here $\lambda=0.05$). Fitting this model to our data demonstrates there is more variation in exceedance probabilities during spring and autumn compared to summer and winter. At all sites, the greatest range in $\lambda_d$ was $\sim 0.03$ in April and October. This agrees with our scale parameter model where the steepest gradient is found in spring and autumn (see Figures \ref{ss_param_ests_fig_HEY} and \ref{ss_param_ests_fig_all}). At all sites, the fitted model for $\lambda_d$ has negative gradient in months at the beginning of the year, so that the exceedance probability is higher earlier in the month (closer to winter), whereas the slope is positive later in the year.

Finally, we use a month-specific empirical distribution $\tilde F_Y^{(j)}$ for skew surges below the threshold. Bringing this together with the parameterisations of $\sigma_d$ and $\lambda_d$ (equations~\eqref{sinescale} and \eqref{rate_day}, respectively) for the upper tail model, the final full skew surge distribution $F_Y$ is dependent on month $j$ and day of the year $d$, and is given by \begin{align}
    F^{(d,j)}_{Y}(y)=\begin{cases} \tilde{F}^{(j)}_Y(y) &\text{if }y\leq u_j\\ \\ 1-\lambda_d\big[1+\xi\big(\frac{y-u_j}{\sigma_d}\big)\big]_+^{-1/\xi} &\text{if }y> u_j. \end{cases}\label{final_ss_model}
\end{align} As $F^{(d,j)}_{Y}(y)$ changes with day $d$ and $d=h(j,i)$, it also changes for every tidal cycle $i$. Consequently the estimated monthly and annual maxima distributions of sea levels are \begin{align}
    &\mathbb{P}(M_j\leq{z})=\frac{1}{K}\sum\limits_{k=1}^{K}\prod\limits_{{i}=1}^{T_j^{(k)}}F_Y^{(d,j)}({z- X_{j_i}^{(k)}}),\label{sldist_ss_stat} \\
    &\mathbb{P}(M\leq{z})=\frac{1}{K}\sum\limits_{k=1}^{K}\prod\limits_{{j}=1}^{12}\prod\limits_{{i}=1}^{T_j^{(k)}}F_Y^{(d,j)}({z- X_{j_i}^{(k)}}).\label{sldist_ss_stat2}
\end{align}

\begin{table}[]
    \centering
    \resizebox{\textwidth}{!}{
    \begin{tabular}{lcccc}
        \hline
         & Heysham & Lowestoft & Newlyn & Sheerness\\ \hline
         \multicolumn{5}{l}{Model $S2$} \\ \hline
         $\alpha_\sigma$ & 0.14 (0.13, 0.15) & 0.15 (0.14, 0.16) & 0.076 (0.073, 0.080) & 0.11 (0.10, 0.12)\\
         $\beta_\sigma$ & 0.060 (0.050, 0.070) & 0.080 (0.070, 0.090) & 0.024 (0.020, 0.028) & 0.052 (0.043, 0.061) \\
         $\phi_\sigma$ & 271.51 (262.77, 280.23) & 266.32 (260.01, 272.63) & 273.58 (265.06, 282.10) & 272.11 (262.66, 281.56) \\
         $\xi$ & 0.002 (-0.042, 0.051) & 0.024 (-0.023, 0.071) & -0.040 (-0.074, 0.006) & 0.037 (-0.029, 0.10)\\ \hline
         \multicolumn{5}{l}{Model $S4$} \\ \hline
         $\gamma_\sigma^{(x)}$ & 0.002 (-0.005, 0.009) & -0.0051 (-0.040, 0.030) & 0.0048 (-0.00048, 0.010) & -0.012 (-0.026, 0.0011)\\\hline
         \multicolumn{5}{l}{Model $R1$} \\ \hline
         $\beta_\lambda$ & 0.0087 (0.0004, 0.017) & 0.022 (0.015, 0.030) & 0.024 (0.018, 0.030) & 0.022 (0.014,0.032) \\
         $\phi_\lambda$ & 155.66 (100.74, 210.59) & 175.16 (155.86, 194.46) & 209.50 (195.48, 223.52) & 184.31 (160.94,207.69) \\
         $\alpha_\lambda^{(x)}$ & -0.13 (-0.18, -0.079) & -0.055 (-0.101, 0.009) & -0.063 (-0.099, 0.108) & -0.32 (-0.37,-0.26) \\
         $\beta_\lambda^{(x)}$ & 0.14 (0.068, 0.21) & -0.016 (-0.084, 0.051) & 0.061 (0.014, 0.108) & 0.23 (0.14, 0.31) \\
         $\phi_\lambda^{(x)}$ & 311.86 (281.78, 341.94) & 359.95 (265.77, 454.15) & 352.38 (299.32, 405.44) & 278.54 (260.44, 293.63)\\ 
    \end{tabular}
    }
    \caption{Parameter estimates for the scale parameter for Models $S2$ and $S4$, and the rate parameter for Model $R1$ with 95\% confidence intervals, at each site.}
    \label{tbl:param_ests}
\end{table}

\subsection{Skew Surge Dependence on Peak Tide}\label{acc_ssmt_dep}
In Section~\ref{ss_mt_indep} we conclude that skew surge-peak tide independence is a reasonable assumption at Heysham, Lowestoft and Newlyn, but not at Sheerness. Here, we describe how we account for this dependence in our skew surge model. We do this for the upper tail by adding a tidal covariate to the GPD scale and rate parameters, but not to the shape parameter $\xi$ to avoid additional uncertainty. For values below the threshold, we use different empirical distributions of skew surges depending on their associated peak tide band, a similar approach to \cite{PughVassie1978}.

We first consider how the threshold exceedance probability varies with peak tide. Here, we extend the GLM parametrisation of expression~\eqref{rate_day} by adding a peak tidal covariate $x$ to the rate parameter $\lambda_d$. We linearly standardise peak tide via $(x-\bar x)/s_x$, where $\bar x$  is the mean and $s_x$ the standard deviation of all peak tide observations at each site. We parametrise the gradient in the same was as expression~\eqref{rate_day}, using the day in year $d$ and a harmonic with periodicity $f=365$ to capture smooth daily changes within a year. This captures the time varying dependence structure between skew surge and peak tide that we found in Section~\ref{ss_mt_indep_supp} of~\cite{DarcySM}. We use the notation $\lambda_{d,x}$ and the following model, denoted $R1$,\begin{equation}
    g(\lambda_{d,x})=g(\lambda_d)+\bigg(\frac{x-\bar x}{s_x}\bigg)\bigg[\alpha_\lambda^{(x)}+\beta_\lambda^{(x)}\sin\bigg(\frac{2\pi}{f}(d-\phi_\lambda^{(x)})\bigg)\bigg], \label{eqn::ss_rate_tide}
\end{equation}
for $g(\cdot)$ the logit link function, $g(\lambda_d)$ defined by expression~\eqref{rate_day} and $\alpha_\lambda^{(x)}\in\mathbb{R}$, $\beta_\lambda^{(x)}>0$ and $\phi_\lambda^{(x)}\in[0,365)$ being parameters to be estimated. The full model for $\lambda_{d,x}$ has 5 parameters.

\begin{figure}
    \centering
    \includegraphics[width=0.9\textwidth]{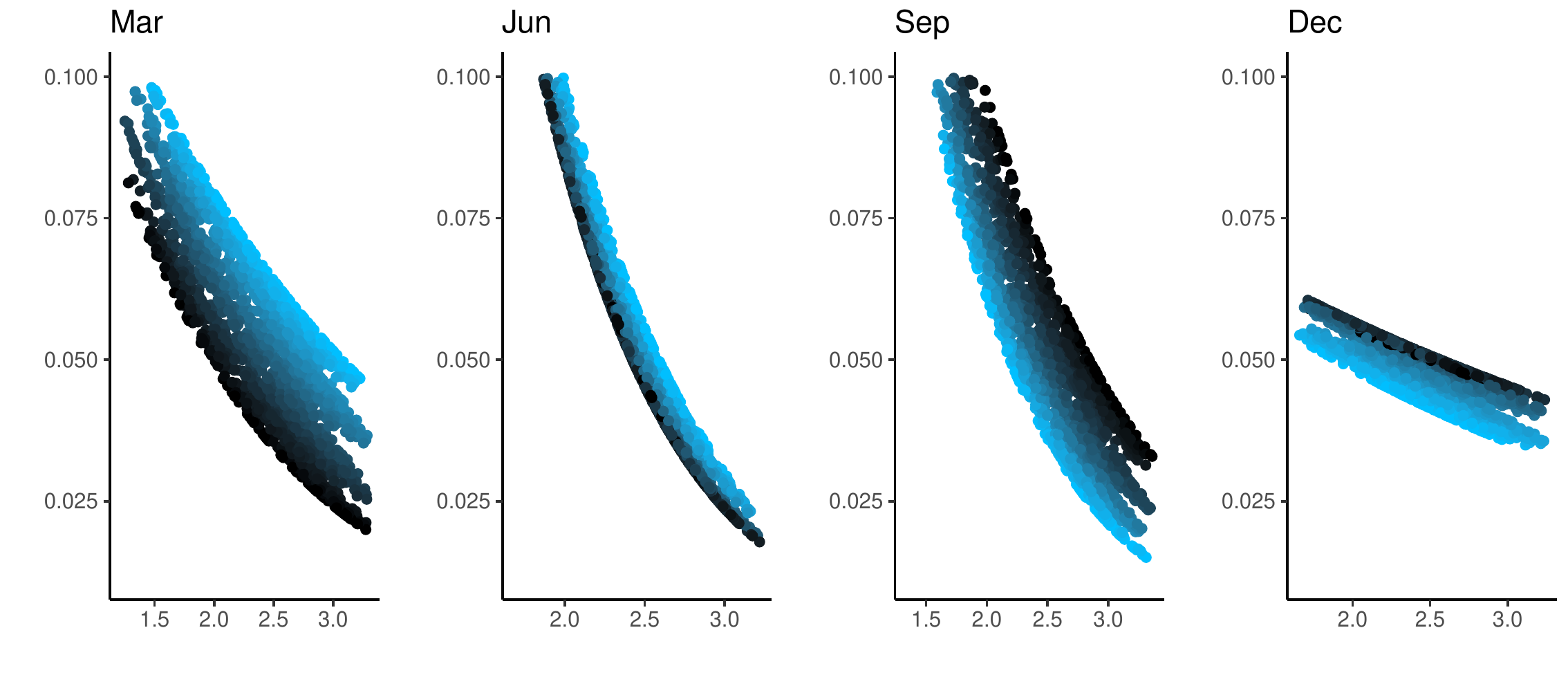}
    \caption{Estimated exceedance probability $\lambda_{d,x}$ under Model $R1$ in March, June, September and December with respect to $x$ being peak tide (in metres) and $d_j$ being day in month at Sheerness. Darker (lighter) points represent days later (earlier) in the month.}
    \label{fig::exprob_tideday_SHE}
\end{figure}

We fit this GLM to model the exceedance probabilities at each site and give parameter estimates in Table~\ref{tbl:param_ests}. Figure \ref{fig::exprob_tideday_SHE} shows these results graphically at Sheerness for March, June, September and December (see Figure~\ref{fig::exprobtide_MJSD_all} for the other sites). As expected the estimated exceedance probability is lower at higher tides, this result is more significant in months where we found skew surge-peak tide dependence to be stronger (April - September). As in Section~\ref{acc_ss_seasonality}, $\lambda_{d,x}$ changes most with day in spring and autumn corresponding to the greatest range of skew surge within a month. Figure~\ref{fig::exprobtide_av_all} shows the gradient of the tidal covariate for each month, when averaged over days. At Heysham, Newlyn and Sheerness, the gradients are greatest near the equinoxes; at Lowestoft there is little variation across months. 

We compare the Model~$R1$~\eqref{eqn::ss_rate_tide} with Model~$R0$~\eqref{rate_day} using AIC and BIC scores. Model~$R1$ minimises the AIC score at Heysham, Newlyn and Sheerness; this is most notable at Sheerness with a reduction of 139. We obtain almost identical AIC scores at Lowestoft. BIC is also minimised by Model~$R1$, at Heysham and Sheerness, but not at Lowestoft and Newlyn. We also compare Model~$R1$ with the model in expression~\eqref{eqn::ss_rate_tide} but with $\beta_\lambda^{(x)}=0$, so that the gradient term for tide does not vary with day; we found that AIC favoured Model~$R1$ at all sites indicating that the dependence structure varies seasonally.

We also investigate adding a tidal covariate $x$ to the scale parameter of the GPD for extreme skew surges. We build on the existing parametrisation of Model $S2$ in equation \eqref{sinescale} to give Model $S4$, \begin{equation}
    \sigma_{d,x}=\alpha_\sigma+\beta_\sigma\sin\bigg(\frac{2\pi}{f}(d-\phi_\sigma)\bigg)+\gamma_\sigma^{(x)} x,\label{eqn::ss_scale_tide1}
\end{equation} for $\alpha_\sigma>\beta_\sigma>0$ , $\phi_\sigma\in[0,365)$, $\gamma_\sigma^{(x)}\in\mathbb{R}$ parameters to be estimated, $f=365$ the periodicity and $d$ the day in year. Since we found evidence that the dependence of skew surge on peak tide changes within a year, we also considered parameterisations that allow the tidal effect to vary with day or month but found no improvement in fit relative to the loss of parsimony.

Table~\ref{tbl:param_ests} gives estimates of $\hat \gamma_\sigma^{(x)}$ at each site; this tells us how $\sigma_{d,x}$ changes with tide (see Table~\ref{tbl:SMparam_ests} for all Model~$S4$ parameter estimates). At Lowestoft and Sheerness $\hat\gamma_\sigma^{(x)}<0$ which agrees with our results of Section~\ref{ss_mt_indep} that extreme skew surges tend to occur on lower tides, although their 95\% confidence intervals contain 0. To formally compare the fit of Model~$S4$ with~$S2$ we use AIC and BIC scores (see Table~\ref{AICBIC_ss}). Model~$S4$ is only favoured by the AIC at Sheerness, this suggests the simpler model without a tidal covariate is preferred elsewhere. Given that our findings of skew surge-peak tide dependence were different at Sheerness in comparison to the other sites, we examined this further to check it wasn't an artifact of the data measurement or tide extraction processes. We explored the fits of Models~$R1$ and~$S4$ on a 483-year data set from a hydrodynamical model driven by a regional climate model~\citep{howardWilliams2021}. We find that adding peak tide covariates to the scale and rate parameters improves fit, with similar estimates to the observations. See Section~\ref{suppmat_climmodel} of~\cite{DarcySM} for details. Therefore, we proceed with Model~$S4$ for the scale parameter at all sites.

Below the threshold, we found it sufficient to capture the dependence by splitting the empirical distribution of skew surges into three associated peak tide bands, i.e., \begin{equation}
    \tilde F_{j,x}(y)=\tilde F_{j}^{(1)}(y) \; \text{ if } x\leq x_{0.33}^{(j)}, \text{ or } \tilde F_{j}^{(2)}(y) \; \text{ if } x_{0.33}^{(j)}<x\leq x_{0.67}^{(j)}, \text{ or }
            \tilde F_{j}^{(3)}(y) \; \text{ if } x>x_{0.67}^{(j)},\label{eqn::ss_emp_tide}
\end{equation} where $x_q^{(j)}$ is the $q$ quantile of the peak tide distribution in month $j$ and $\tilde F_{j}^{(l)}$ for $l=1,2,3$ is the empirical distribution of skew surges in month $j$ which are associated with the lowest ($l=1$), medium ($l=2$) and highest ($l=3$) band of peak tides. The choice of 3 tidal bands is somewhat arbitrary, but appears sufficient given the weak dependence on peak tide. We could have used a kernel smoother to ensure continuity across bands, as in \cite{bashtannyk2001}. Since our interest lies with the extreme values, we did not explore this further as this would not have made a practical difference to our return level estimates.

So our skew surge model, that is dependent on peak tide, is given by \begin{align}
    F_Y^{(d,j,x)}(y)=\begin{cases}
            \tilde F_{j,x}(y) \quad &\text{if } y\leq u_j\\
            \\
            1-\lambda_{d,x}\big[1+\xi\big(\frac{y-u_j}{\sigma_{d,x}}\big)\big]_+^{-1/\xi} \quad &\text{if } y>u_j,
    \end{cases}\label{ss_tidedep_model}
\end{align} with $\tilde F_{j,x}(\cdot)$, $\sigma_{d,x}$ and $\lambda_{d,x}$ defined in expressions \eqref{eqn::ss_emp_tide}, \eqref{eqn::ss_scale_tide1} and \eqref{eqn::ss_rate_tide}, respectively.



\subsection{Skew Surge Temporal Dependence}\label{ss_dep}
\begin{figure}
    \centering
    \includegraphics[width=0.4\textwidth]{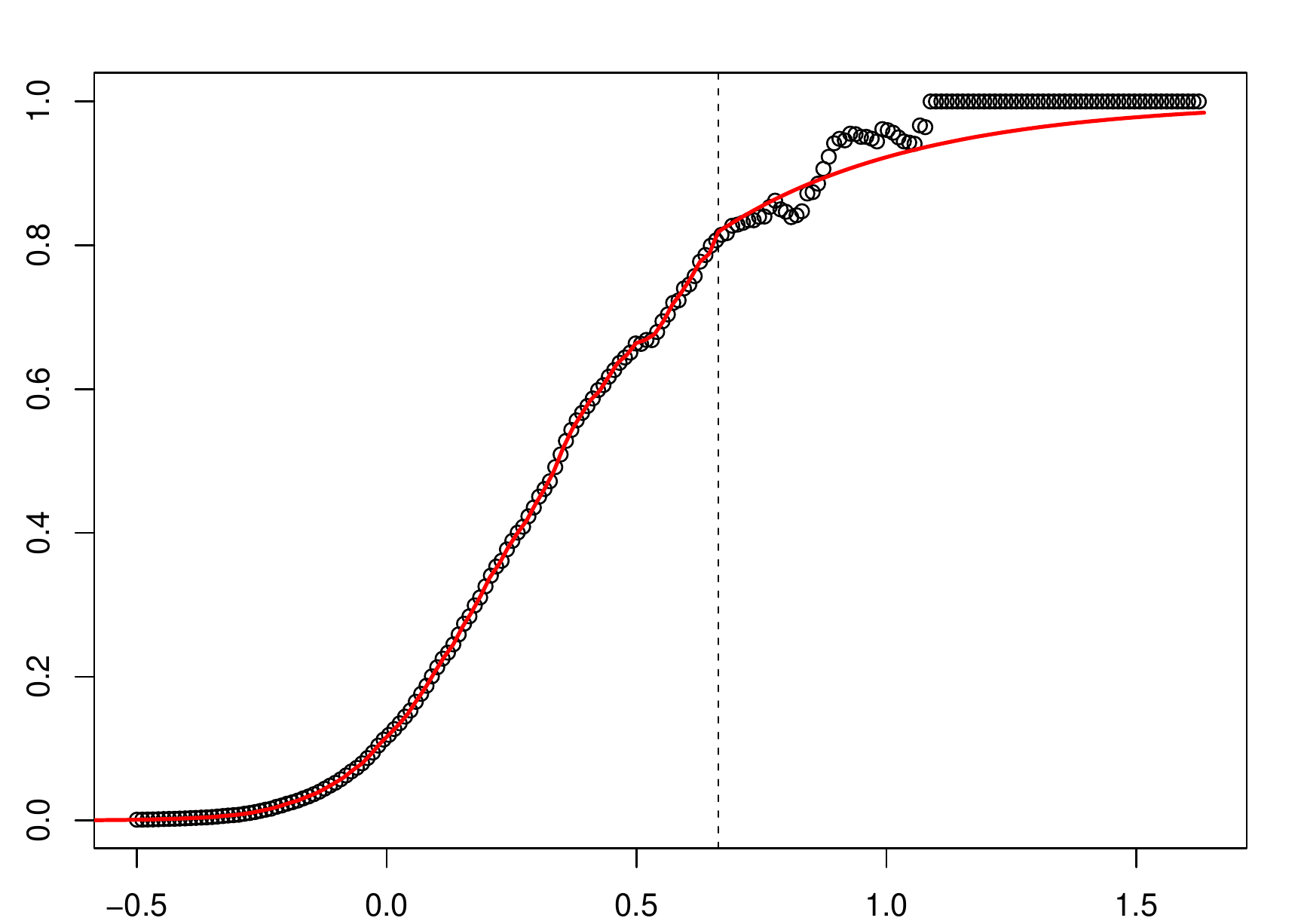}
    \caption{Estimates of the subasymptotic extremal index $\theta(y,r)$ for different skew surge levels $y$ and run length $r=2$ at Heysham using the runs estimate (black) and our model estimate (red) of expression~\eqref{eqn::exi_model}. The dashed black line is the 0.99 skew surge quantile.}
    \label{fig::HEY_exifun}
\end{figure}

So far, we have assumed that skew surges are independent. We now describe how we account for their temporal dependence across tidal cycles. As discussed in Section \ref{evmethods}, temporal dependence causes clusters of events above high thresholds; if this is ignored, sea level return levels of annual maxima will be overestimated. The best measure of dependence for the extreme values of a stationary sequence is the extremal index $\theta$. \cite{Tawn1992} and \cite{Batstone2013} use the extremal index of the highly non-stationary sea level series. We model the sub-asymptotic extremal index $\theta(y,r)$ (defined by equation~\eqref{subas_exi}) of skew surges using varying thresholds (levels) $y$ with a fixed run length $r$. Then the distribution of the monthly and annual maxima sea levels are given by
\begin{align}
    &\mathbb{P}(M_j\leq{z})=\frac{1}{K}\sum_{k=1}^{K}\prod\limits_{{i}=1}^{T^{(k)}_j}\big[F_Y^{(d,j,x)}(z- X_{j_i}^{(k)})\big]^{\hat\theta(z-X_{j_i}^{(k)},r)}\nonumber,\\
    &\mathbb{P}(M\leq z)=\frac{1}{K}\sum\limits_{k=1}^{K}\prod\limits_{j=1}^{12}\prod\limits_{{i}=1}^{T^{(k)}_j}\big[F_Y^{(d,j,x)}(z- X_{j_i}^{(k)})\big]^{\hat\theta(z-X_{j_i}^{(k)},r)}\label{eqn::sldist_exi}.
\end{align}
The empirical estimates $\tilde\theta(y,r)$ of the sub-asymptotic extremal index are shown in Figures~\ref{fig::HEY_exifun} and~\ref{fig::exifun_all} for Heysham and the remaining sites, respectively. These demonstrate substantial variation with $y$. We are interested in $\theta(y,r)$ across all ranges of skew surge, but particularly in values greater than those observed to allow extrapolation for return level estimation.

We estimate $\theta(y,r)$ in two stages. For $y\leq v$ where $v$ is a high threshold, we use the empirical runs estimate $\tilde\theta(y,r)$ as this is smooth over $y$ in this range. For computational efficiency purposes, we evaluate $\tilde\theta(y,r)$ on a regular grid of $y$ values in the range $[y_F, v]$ where $y_F$ is the minimum observed skew surge. Then we linearly interpolate these for any $y<v$ of interest. For $y>v$, the empirical estimate is increasingly variable so we adopt a parametric model for $\theta(y,r)$. We propose \begin{equation}
    \hat\theta(y,r)=\begin{cases}
        \tilde\theta(y,r) \quad\quad &\text{if }y\leq v \\
        \theta-[\theta-\tilde\theta(v,r)]\exp\Big[\scalebox{0.75}[1.0]{\( - \)}\frac{(y-v)}{\psi}\Big] \quad\quad &\text{if }y>v,
    \end{cases}\label{eqn::exi_model}
\end{equation} where $\psi>0$ and $\tilde\theta(v,r)\leq\theta\leq 1$ are parameters to be estimated. This parametric form ensures the estimate asymptotes to the extremal index $\theta$ and is continuous at $v$.



The parameters $\psi$ and $\theta$ are estimated using a weighted least squares approach with weight $w(y)=\sqrt{c(y,r)-1}$ where $c(y,r)$ is the number of clusters above $y$ separated by run length $r$. This gives a greater weight when there are more clusters, and a weight of zero when there is a single cluster. We choose $v$ to be the 0.99 quantile of all skew surges but this choice is subjective. The run length $r$ represents the approximate duration of a storm across a single site, we take $r=2,10,1$ and $10$ for Heysham, Lowestoft, Newlyn and Sheerness, respectively. These choices are supported by estimating the run length using the intervals estimator for each season, where we expect the stationary assumption to be reasonably justified~\citep{FerroSegers2003}. At Heysham, $\hat\theta=1$ and $\hat\psi=0.40$; the estimates for the other sites are given in Section~\ref{sec::suppmat_exi} of~\cite{DarcySM}. Figures~\ref{fig::HEY_exifun} and~\ref{fig::exifun_all} show this model fit for Heysham and the remaining sites, respectively, compared with the entirely empirical estimates.


\section{Results}\label{results}
\subsection{Introduction}
Using the distribution of the monthly and annual maxima sea levels derived in Section \ref{methodology}, we estimate return levels by solving \begin{equation*}
    \mathbb{P}(M_j\leq z_{j,p})=1-p \quad \text{and} \quad \mathbb{P}(M\leq z_p)=1-p,
\end{equation*} respectively for $p\in[0,1]$. We are interested in return levels up to the 10,000 year level, corresponding to annual exceedance probability $p=10^{-4}$. In the monthly case, this is the level we expect the monthly maxima $M_j$ to exceed, on average, every $1/p$ of that particular month (for example, every 10,000 Januarys). Whilst in the annual case, this is the level we expect the annual maxima to exceed every $1/p$ years, on average.

To assess the importance of each of the novel modelling steps in Section~\ref{methodology}, we derive return level estimates from each stage; accounting for non-stationarity in each component, the dependence between them and skew surge temporal dependence. Each stage is detailed below in a nested list numbered (i)-(vii), so that each model below builds on the previous method, except (iv) from (iii). We compare these to the current approach used in practice and a baseline approach, both derived under simplifying and false assumptions. We subsequently refer to each model by name, given below, using italics. The notation follows from Section~\ref{methodology}. \begin{enumerate}[label=(\roman*)]
    \item \textbf{Current}: The methodology currently used in practice, where skew surges are assumed to be iid and peak tides are stationary, with annual maxima distribution~\eqref{eqn::RJPM_ss}.
    \item \textbf{Baseline}: As in (i) but we recognise interannual variations in the tide by averaging over yearly tidal samples. The annual maxima distribution is given by \begin{equation}
        P(M\leq z)=\frac{1}{K}\sum_{k=1}^{K}\prod\limits_{i=1}^{T^{(k)}} F_Y(z-X_{i}^{(k)})\label{eqn::baselineF}
    \end{equation} where $X_i^{(k)}$, $i=1,\ldots,T^{(k)}$ represents an annual tidal sample for year $k=1,\ldots,K$, where we choose $K=100$ arbitrary but contiguous samples. 
    \item \textbf{Seasonal surge}: As in (ii) but we account for the within-year seasonality of skew surges using the skew surge distribution~\eqref{final_ss_model}. 
    \item \textbf{Seasonal tide}: As in (ii), but conversely to (iii) we account for within-year seasonality in peak tide and not in skew surge, so the skew surge model of expression~\eqref{ss_nonstat_model} is used in the monthly and annual maxima distributions~\eqref{aver_max}.
    \item \textbf{Full seasonal}: As in (ii) but we account for within-year seasonality of both components, with monthly and annual maxima distributions~\eqref{sldist_ss_stat} and~\eqref{sldist_ss_stat2}, respectively.
    \item \textbf{(Skew surge-peak tide) interaction}: As in (v) but with skew surge-peak tide dependence captured, so that the skew surge distribution~\eqref{ss_tidedep_model} is used.
    \item \textbf{(Skew surge) temporal dependence}: As in (vi) but accounting for temporal dependence in the skew surge series, with annual maxima distribution~\eqref{eqn::sldist_exi}.
\end{enumerate}

We compare our model estimates to empirical estimates. These are restricted to return periods within the range of data, but are useful for checking whether the model is capturing the distributional properties of observed sea levels. The empirical estimates act as a guide to the truth and we do not expect our model to fit these perfectly since each empirical estimate is specific to a particular annual tidal regime, whereas in our approaches (ii)-(vii) we account for tidal variations by averaging over different samples. The empirical estimates are also sensitive to missing data and so can be biased. In Sections~\ref{rl_nonstat} and \ref{rl_dep}, we make comparisons using point estimates to assess sensitivity to the model choice, in each case results in Section~\ref{methodology} identify statistically significant differences between these models. Once the model choice is finalised we present measures of uncertainty and assess model fit in Section~\ref{rl_final}. In Section~\ref{sl_seasonality} we use our model to find the probability that an extreme sea level occurs in a specific month, given it is higher than some level with a given return period of interest. See~\cite{DarcySM} for a comprehensive overview of our return level estimates.

\subsection{Return Levels: Accounting for Non-stationarity}\label{rl_nonstat}
We investigate how accounting for skew surge and peak tide seasonality influences sea level return level estimates by comparing monthly return level estimates from the \textit{baseline}, \textit{seasonal surge}, \textit{seasonal tide} and \textit{full seasonal} models. We are mainly interested in differences between the \textit{baseline} and \textit{full seasonal} estimates. We do not expect the \textit{full seasonal} model to match the empirical estimates since we have not accounted for skew surge-peak tide dependence and temporal dependence yet. The intermediate solutions (\textit{seasonal surge} and \textit{seasonal tide}) allow us to understand which components' non-stationarity is influencing the return levels most in different months.


\begin{figure}
    \centering
    \includegraphics[width=0.95\textwidth]{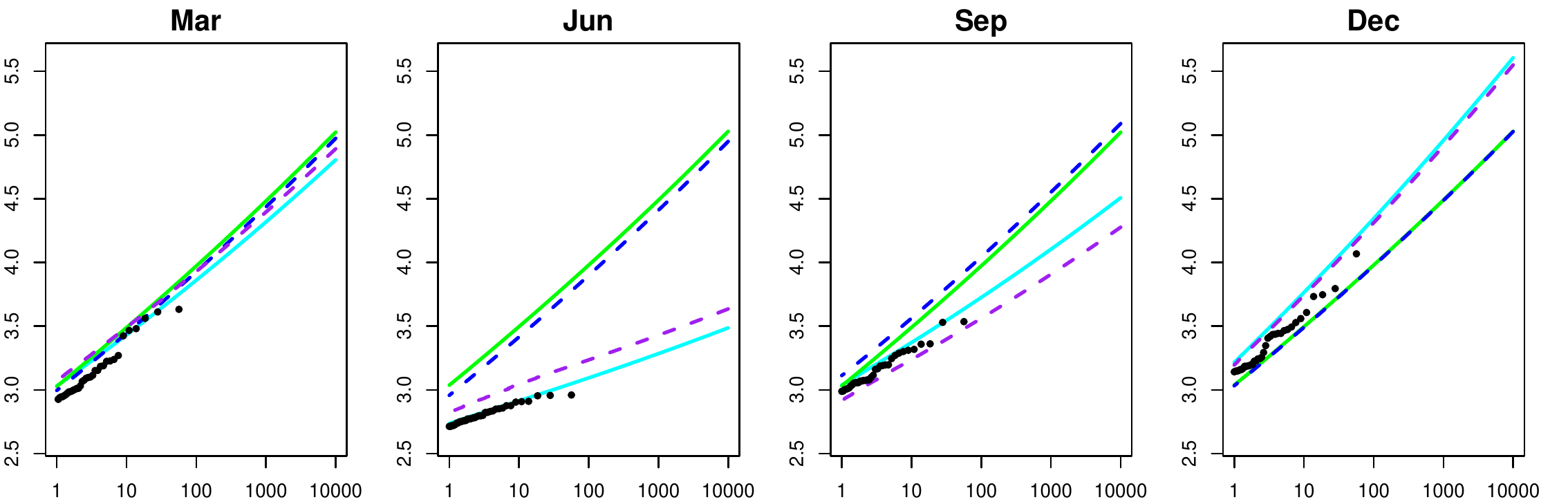}
    \caption{Monthly maxima sea level return level estimates for Lowestoft in March, June, September and December (from left to right), estimated using the \textit{baseline} (green), \textit{seasonal surge} (purple dashed), \textit{seasonal tide} (blue dashed) and \textit{full seasonal} (cyan) models. Empirical estimates are shown by black points.}
    \label{fig::monmax_ssntly_RL_LOW_1x4}
\end{figure}

\begin{figure}[h]
    \centering
    \includegraphics[width=0.4\textwidth]{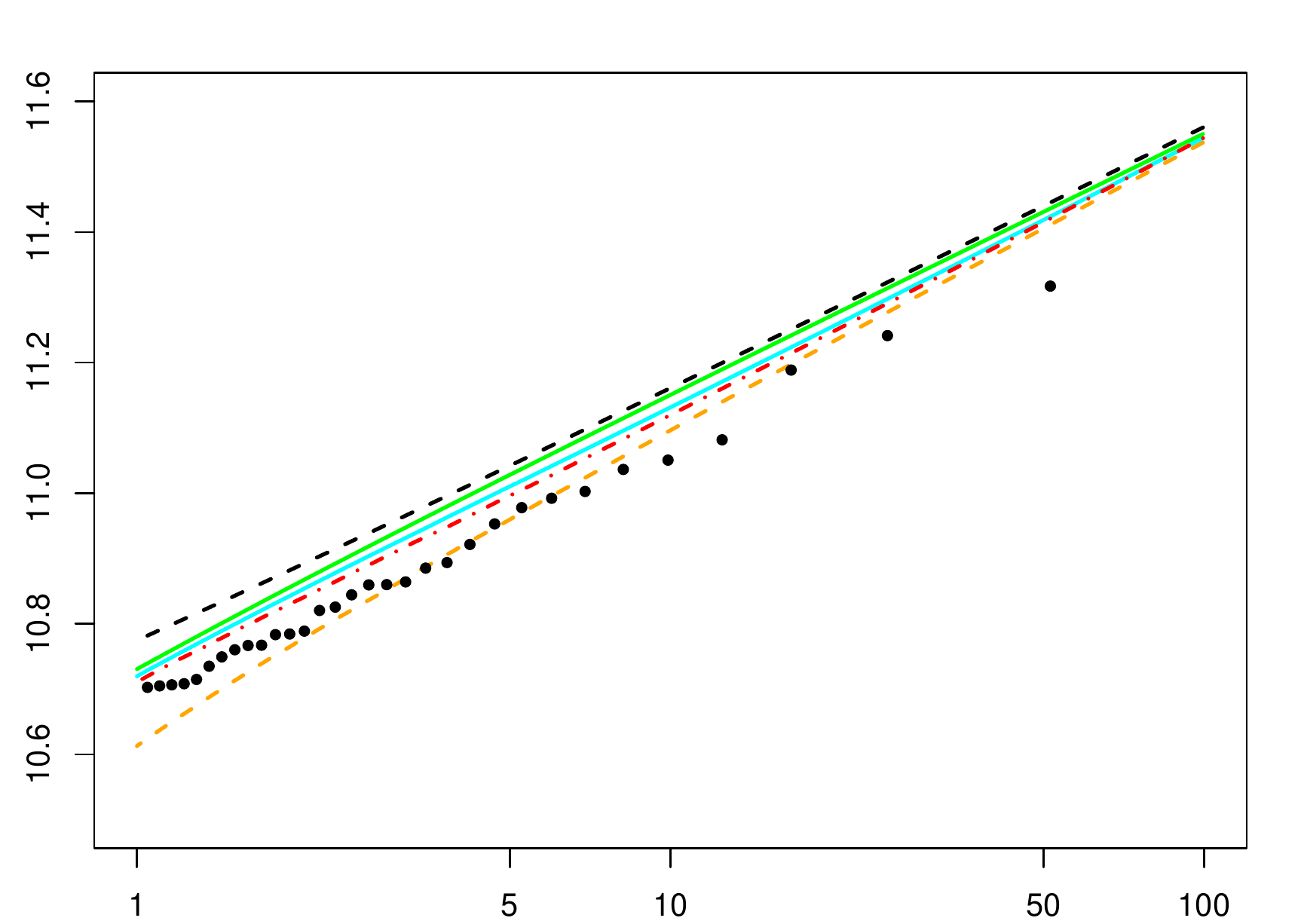}
    \includegraphics[width=0.4\textwidth]{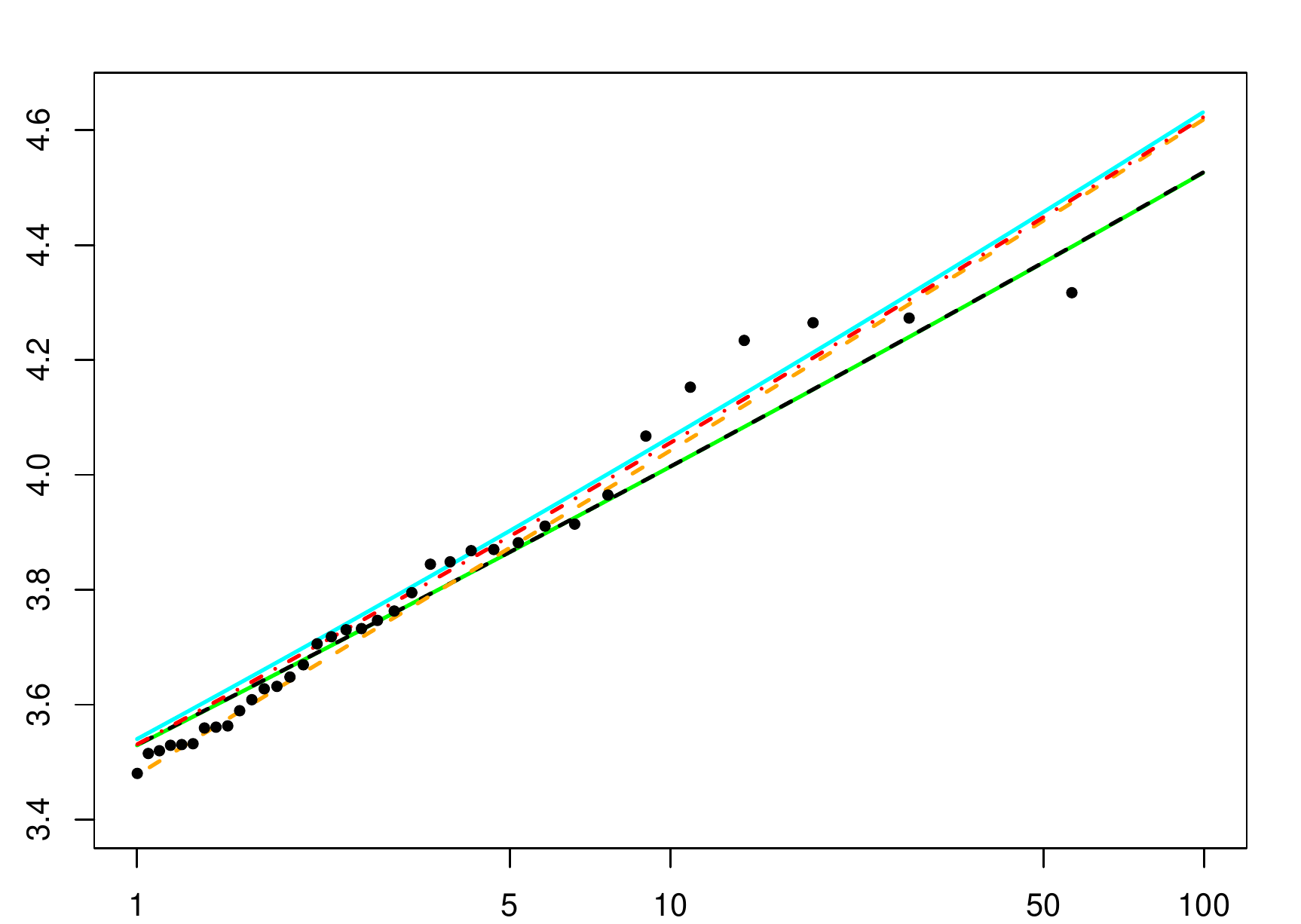}
    \includegraphics[width=0.4\textwidth]{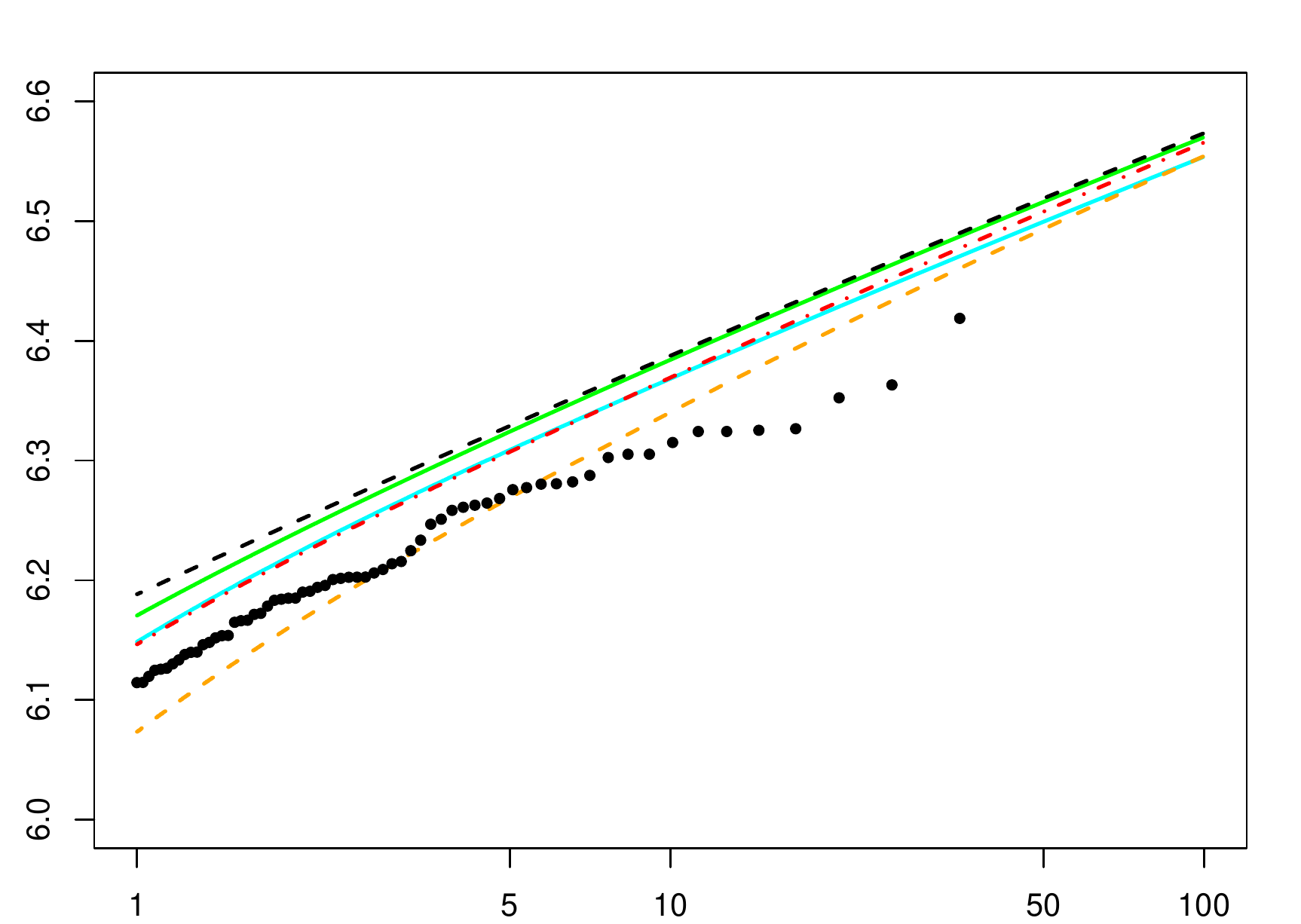}
    \includegraphics[width=0.4\textwidth]{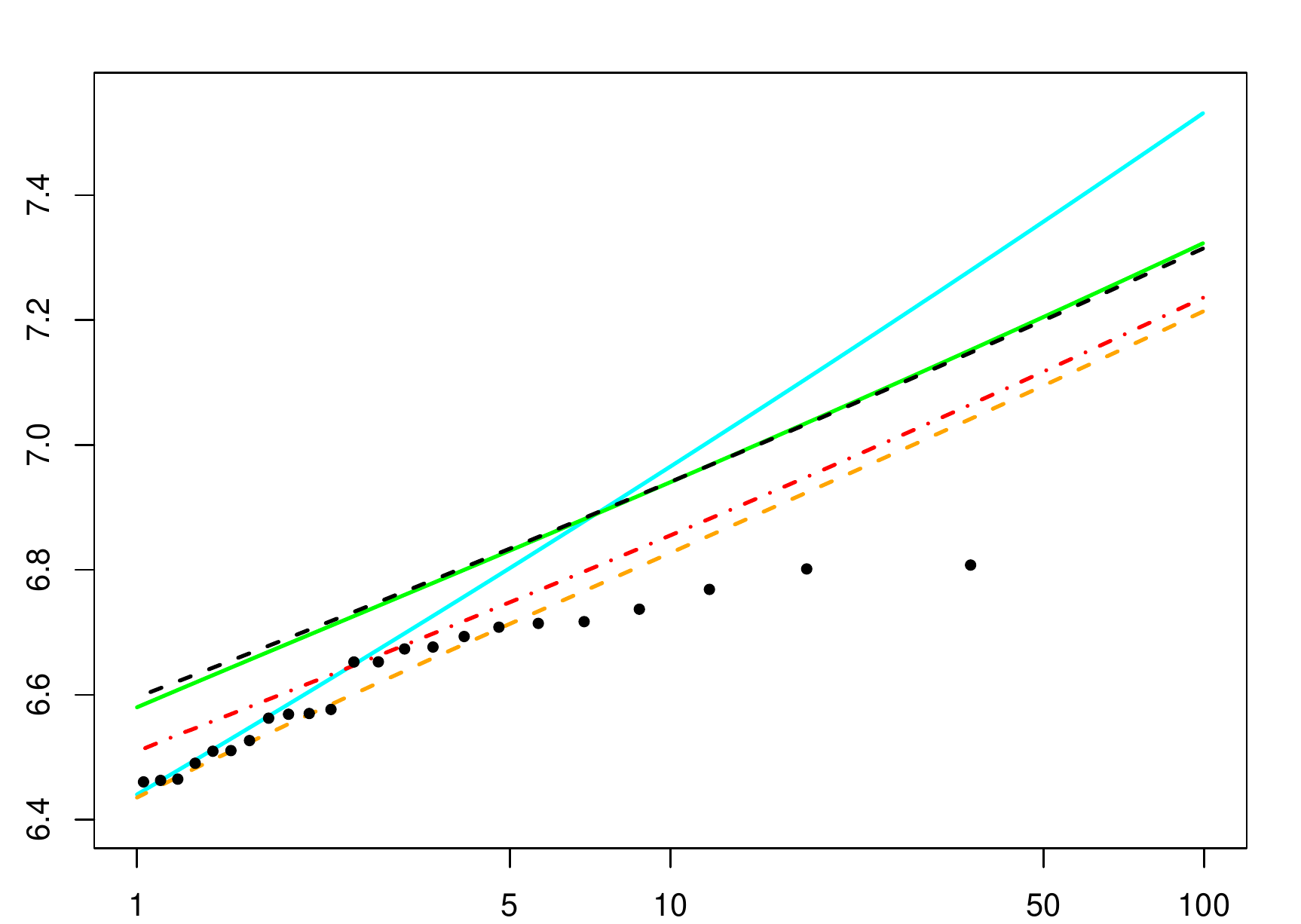}
    \caption{Annual maxima sea level return level estimates for Heysham (top left), Lowestoft (top right), Newlyn (bottom left) and Sheerness (bottom right), up to the 100 year level, estimated using the \textit{current} (black dashed), \textit{baseline} (green), \textit{full seasonal} (cyan), \textit{interaction} (red dot-dashed) and \textit{temporal dependence} (orange dashed) models. Empirical estimates are shown by the black points.}
    \label{fig::Ann_ssnlty_RL_LOWSHE}
\end{figure}

Figures~\ref{fig::monmax_ssntly_RL_LOW_1x4} and~\ref{fig::monmax_all_RL_NEW_1x4} show monthly return level estimates at Lowestoft and the other sites, respectively. At all sites and in all months, the empirical estimates lie closer to \textit{full seasonal} model than the \textit{baseline}. The most noticeable feature is the difference between the \textit{baseline} and \textit{full seasonal} model in June, this reaches 1.54m at the 10,000 year level at Lowestoft. In June skew surges are particularly low relative to the rest of the year, so ignoring their seasonality leads to significant overestimation. This overestimation increases with return period. On the other hand, in December when skew surges are relatively high compared to the other months, ignoring seasonality results in underestimates of return levels. Since the tidal range at Lowestoft is narrow compared to the range of skew surges, differences between \textit{seasonal tide} and \textit{full seasonal} are small relative to the differences between \textit{seasonal surge} and \textit{full seasonal}. For example, when the tidal range is largest at the autumn equinox in September, we observe a difference of 7cm at the 10,000 year level. Whereas at Heysham, where the tidal range is large relative to other sites, we observe significant differences between the two model estimates of 26cm at the 10,000 year level in September.

Figure~\ref{fig::Ann_ssnlty_RL_LOWSHE} shows annual return level estimates from the \textit{current}, \textit{baseline} and \textit{full sesaonal} models, compared to empirical estimates at all sites, up to the 100 year level (Figure~\ref{fig::annmax_RL_NEW} shows these up to the 10,000 year level). The \textit{current} method gives similar results to the \textit{baseline} approach for all return periods at all sites. This suggests assuming each year is identically distributed is not unreasonable. Even at Heysham, where year-to-year variations are large, the difference between the \textit{baseline} and \textit{current} method is small, with the \textit{current} method giving a 1 year return level 4cm higher than the \textit{baseline}. Since the \textit{full seasonal} model lies closer to the empirical estimates than models (i)-(iv) at all sites, this highlights the importance of accounting for both forms of seasonality. At the 10,000 year level, the \textit{baseline} gives a return level 6cm, 23cm and 58cm lower than the \textit{full seasonal} method at Heysham, Lowestoft, and Sheerness, respectively, whereas it is 1cm higher at Newlyn.



\subsection{Return Levels: Accounting for Dependence}\label{rl_dep}
Here, we build on the \textit{full seasonal} model to capture skew surge-peak tide dependence (\textit{interaction}) and skew surge temporal dependence (\textit{temporal dependence}). We compare these with the empirical and \textit{full seasonal} estimates. We present monthly maxima return levels of the \textit{interaction} model to demonstrate the changing skew surge-peak tide dependence structure throughout the year, as discussed in Section~\ref{ss_mt_indep}. Then we compare annual maxima return level estimates from both models.

\begin{figure}
    \centering
    \includegraphics[width=0.95\textwidth]{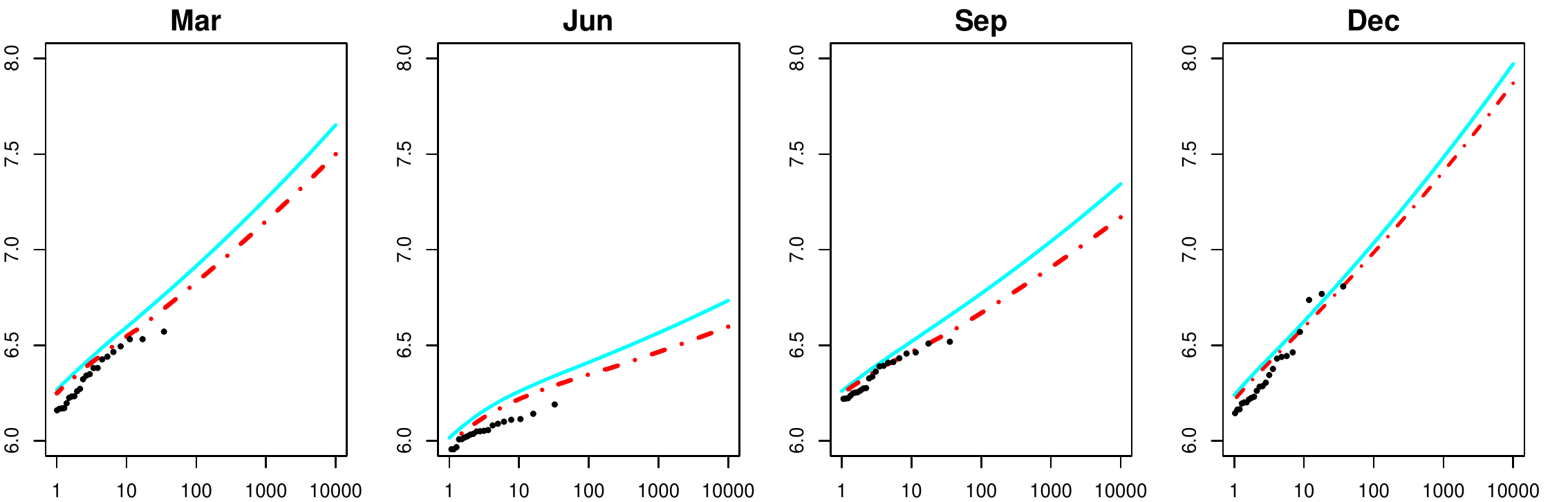}
    \caption{Monthly return level estimates for Sheerness in March, June, September and December, estimated using the \textit{full seasonal} (cyan) and \textit{interaction} (red dot-dashed) models when we account for and ignore skew surge-peak tide dependence, respectively. Empirical estimates are shown by the black points.}
    \label{fig::monmax_tidedep_RL_SHE_1x4}
\end{figure}

Figure~\ref{fig::monmax_tidedep_RL_SHE_1x4} shows monthly return level estimates in March, June, September and December for the \textit{full seasonal} and \textit{interaction} models, as well as the empirical estimates at Sheerness. We illustrate the results for Sheerness because this dependence is stronger compared to the other sites, as discussed in Section~\ref{acc_ssmt_dep}. Ignoring skew surge-peak tide dependence overestimates return levels compared to the \textit{full seasonal} model for return periods greater than 10 years. At the 10,000 year level, these are overestimated by 15, 14, 17 and 10cm in March, June, September and December, respectively, with similar values down to the 1 year level. This is slightly surprising as in Section~\ref{ss_seasonality} we found evidence that it is reasonable to assume skew surge and peak tide are independent in March, September and December. For Sheerness the results from the \textit{interaction} model tend to lie closer to the empirical estimates than those from the \textit{full seasonal} model (see Figure~\ref{fig::monmax_tidedep_RL_SHE_1x4}); this suggests that accounting for this dependence is important and better reflects the process properties. At the other sites, the return level estimates from the \textit{full seasonal} and \textit{interaction} model are almost indistinguishable, which echos that adding peak tide covariates into the skew surge model  when independence is a reasonable assumption does not alter results significantly.

As skew surge temporal dependence is not yet accounted for in the \textit{interaction} model, its return level estimates are anticipated to slightly overestimate the empiricals. Figure~\ref{fig::Ann_ssnlty_RL_LOWSHE} compares annual return level curves for the \textit{temporal dependence} and \textit{interaction} models. These are very close for high return periods (>100 years). For lower return periods (<10 years) the \textit{temporal dependence} estimates lie closer to the empiricals at all sites. This is a natural consequence of the extremal index model given by expression~\eqref{eqn::exi_model} for temporal dependence, since the estimated extremal index is closer to zero for lower skew surges, corresponding to shorter return periods. But for the highest skew surges, and for those greater than observed, $\hat\theta(y,r)\rightarrow \hat\theta\approx 1$ and $y\rightarrow y^F$, so the largest return level estimates remain unchanged. Accounting for temporal skew surge dependence has the greatest influence at Heysham, where the acf estimates are higher compared to the other sites (see Figure~\ref{fig::acf_ss}). Here, the \textit{temporal dependence} model gives a return values 10cm, 4cm and 2cm lower than the \textit{interaction} model at the 1, 5 and 10 year return periods to better match the empirical estimates.

\begin{figure}
    \centering
     \includegraphics[width=0.4\textwidth]{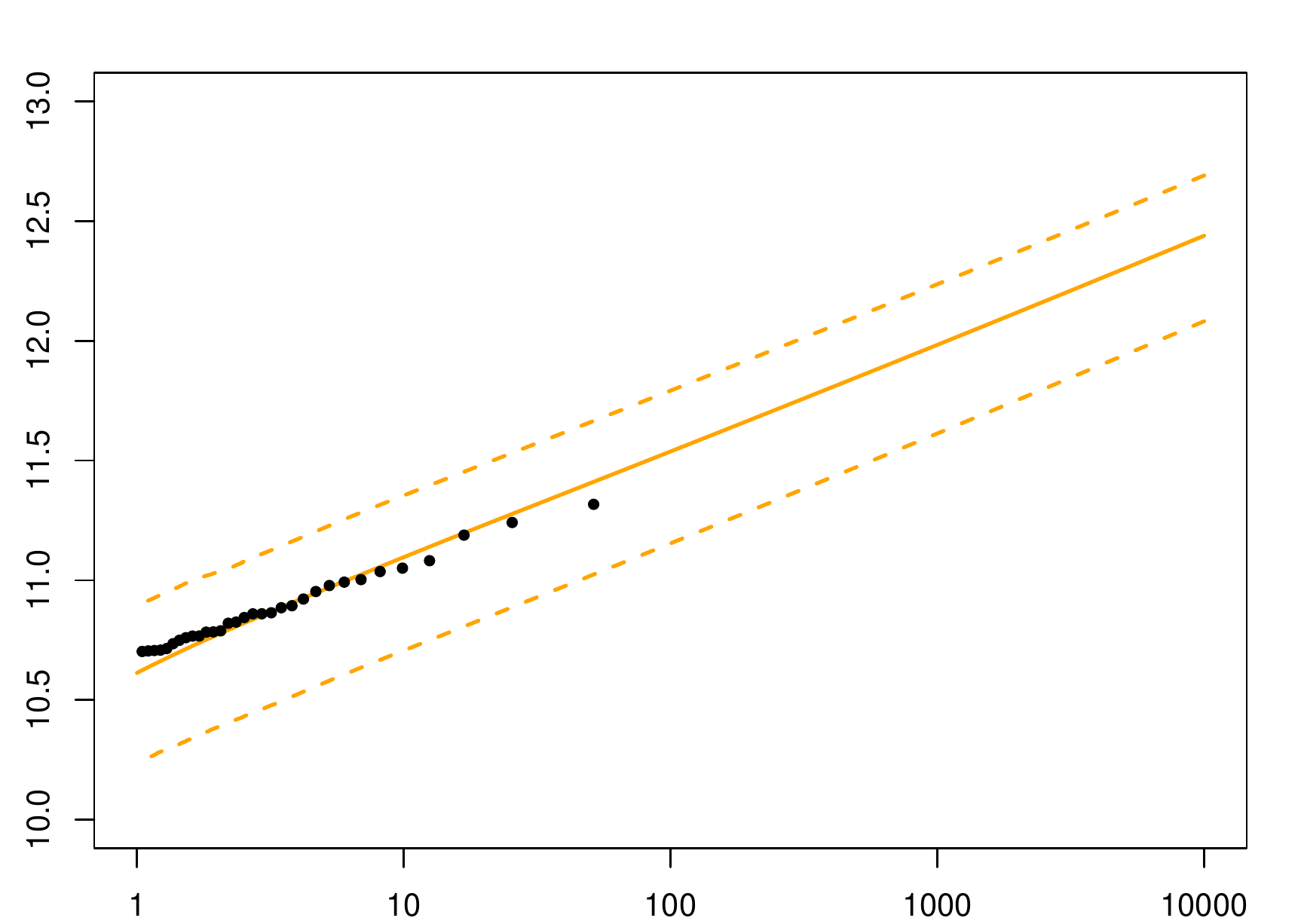}\includegraphics[width=0.4\textwidth]{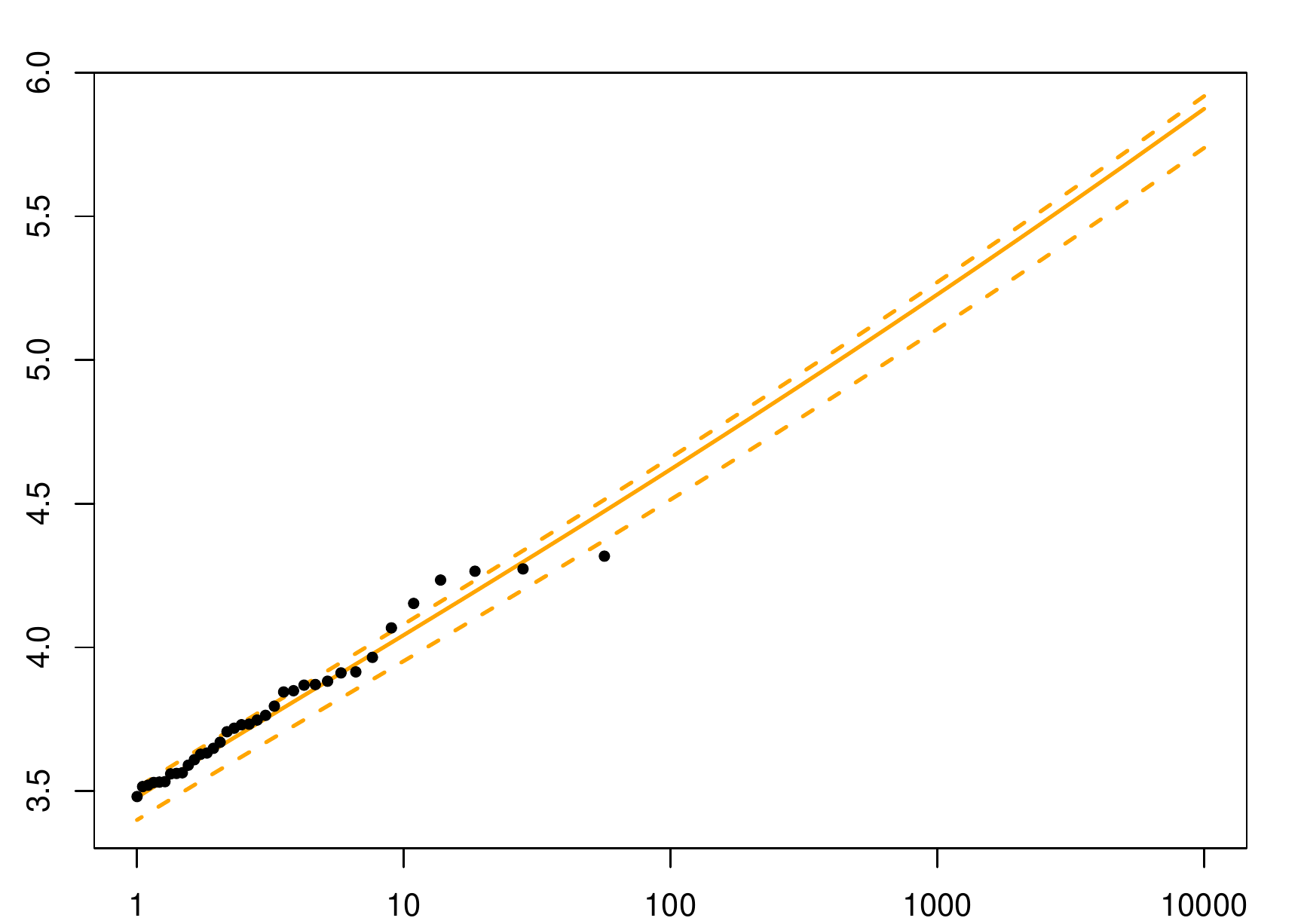}
    \caption{Return level estimates from the final model (\textit{temporal dependence}) (orange), with the maximum and minimum year-specific return level estimates (dashed orange) and empirical estimates (black) at Heysham (left) and Lowestoft (right).}
    \label{fig::final_rlest_bnd}
\end{figure}

\subsection{Assessment of Fit and Uncertainty}\label{rl_final}
Figure~\ref{fig::final_rlest_bnd} shows annual maxima return level estimates from the \textit{temporal dependence} model, judged our best model, at Heysham and Lowestoft (see Figure~\ref{fig::SMfinal_rlest_bnd} for the other sites). We also add the estimated maximum and minimum year-specific return levels, showing the effect of interannual peak tide variation, as given by \eqref{eqn::sldist_exi} without averaging over $K$ years. Figure~\ref{fig::annmax_yearly_grad} shows yearly estimates at each site. The range of yearly estimates for Lowestoft is very narrow since the tidal range is small, compared to Heysham which has the largest tidal range of the four sites. Empirical estimates may deviate away from the return level point estimates if that specific value occurs in a year with particularly low or high tides compared to the average. Figures~\ref{fig::final_rlest_bnd} and~\ref{fig::SMfinal_rlest_bnd} show that the empirical estimates lie within the bounds for each site, at most return periods. \cite{Rodriguez2022} demonstrate that the 4.4 year perigean cycle has a greater effect on modulations in return levels than the nodal cycle in the UK. Our results are less clear (see Figure~\ref{fig::annmax_yearly_overyr}), with Sheerness year-specific return levels having an $\sim{20}$ year cycle.

\begin{figure}
    \centering
    \includegraphics[width=0.85\textwidth]{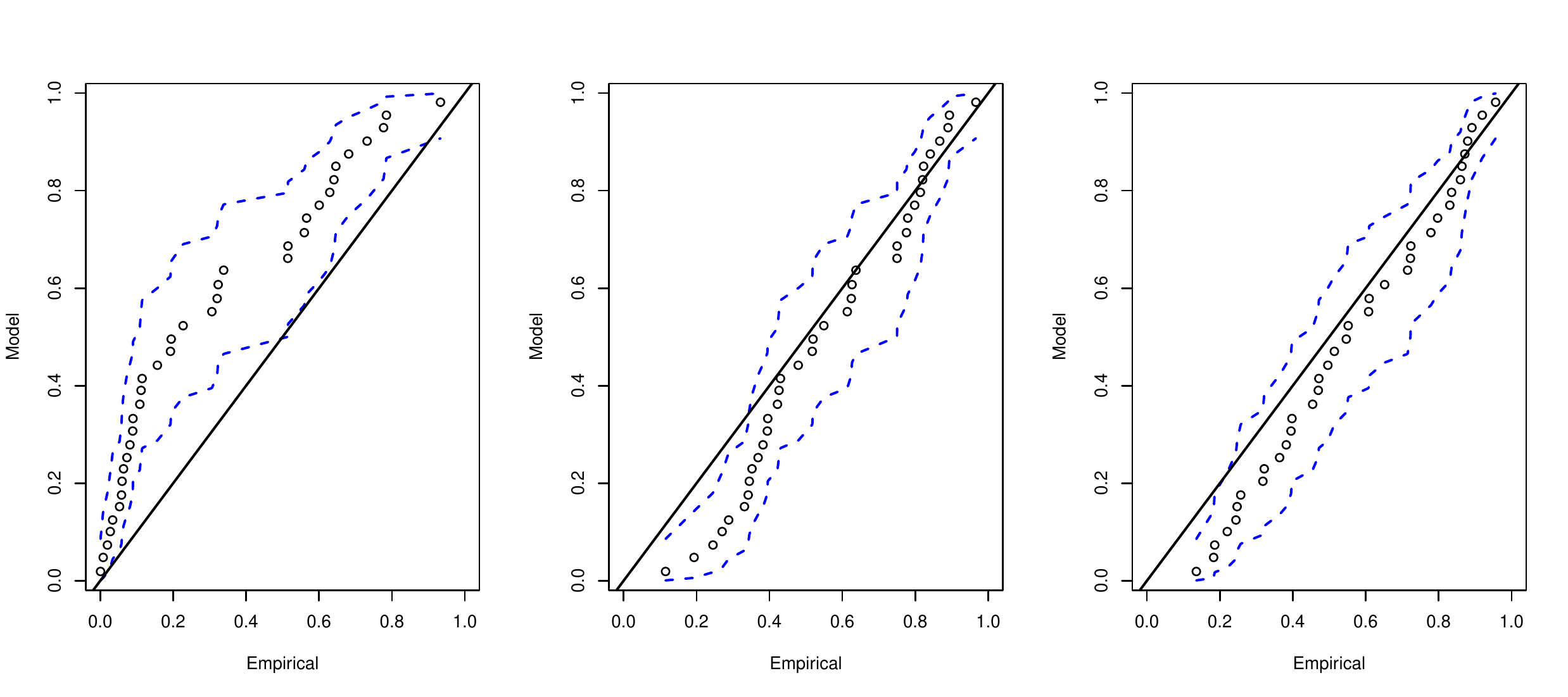}
    \caption{PP plots for the transformed annual maxima sea levels at Sheerness through the \textit{baseline} (left), final (\textit{temporal dependence}) (centre) and \textit{year specific final} (right) distribution functions. The black line shows the line of equality, $y=x$. 95\% tolerance bounds are shown by blue dashed lines.}
    \label{fig::mapannmax}
\end{figure}

We also assess goodness-of-fit by transforming the observed annual maxima to a uniform distribution using the probability integral transform with the distribution function of the annual maxima for their respective year. We do this for three cases; the \textit{baseline} approach \eqref{eqn::baselineF}, our final model \eqref{eqn::sldist_exi} and the year-specific final model. If the model is a fits well we expect the transformed annual maxima to be $\text{Uniform}(0,1)$ distributed. Figure~\ref{fig::mapannmax} shows PP plots for Sheerness, and the remaining sites are shown in Figure~\ref{fig::mapannmax_all}; a good fit is indicated by the empirical and model probabilities being equal so that the line of equality lies between the tolerance bounds. Figure~\ref{fig::mapannmax} demonstrates the improvement in fit across the three models. We formally test this using the Kolmogorov-Smirnov test at each site for the final model, with $p$ values 0.0066, 0.51, 0.0044 and 0.10 at Heysham, Lowestoft, Newlyn and Sheerness, respectively. At Heysham and Newlyn the $p$ value is significant at the 5\% level so we cannot conclude a good fit, but it is a much better than for the baseline fits.

We also assess the model fit uncertainty by estimating 95\% confidence intervals on the 1, 10, 100, 1000 and 10,000 year return levels using a stationary bootstrap procedure. We first transform the skew surges using our final model with their corresponding seasonal and tidal covariates, call this series $\{U^{Y}_i\}$ for $i=1\ldots,T$ for $T$ the total number of observations, where $U^{Y}_i=F_{Y}^{(d,j,x)}(Y_i)$. This gives the basis for another approach to assess model fit. If the model is `correct', these transformed observations will be Uniform(0,1) and we check this formally using the Kolmogorov-Smirnov test over all years, and for each year. We discuss this further in Section~\ref{sec::transform_ss} of~\citep{DarcySM}. To account for temporal dependence, we sample $\{U^{Y}_i\}$ using a stationary bootstrap with block length $L\sim\text{Geometric}(1/10)$, so the mean block length corresponds to 5 days (the maximum duration of a storm), and the total length of each sample is equal to that of the original data~\citep{Politis1994}. Call this bootstrapped series of uniform variables $\{U^{B}_i\}.$ Then we transform the bootstrap sample back to the original scale using our final model, but preserving the original covariate information of $\{Y_i\}$, i.e., $\{X_{j_i},d,j\}$, call this series $\{Y^{B}_i\}$. Then we fit our tail model to the $\{Y^B\}$ series to re-estimate all of the parameters $\alpha_\sigma, \beta_\sigma, \phi_\sigma, \gamma_\sigma^{(x)}, \xi, \beta_{\lambda}, \phi_\lambda, \alpha_\lambda^{(x)}, \beta_\lambda^{(x)}, \phi_\lambda^{(x)}, \theta$ and $\psi$, as well as the monthly empirical distributions and thresholds. These estimates are then used to derive the annual maxima distribution of sea levels and estimate return levels. 

\begin{figure}
    \centering
    \includegraphics[width=0.4\textwidth]{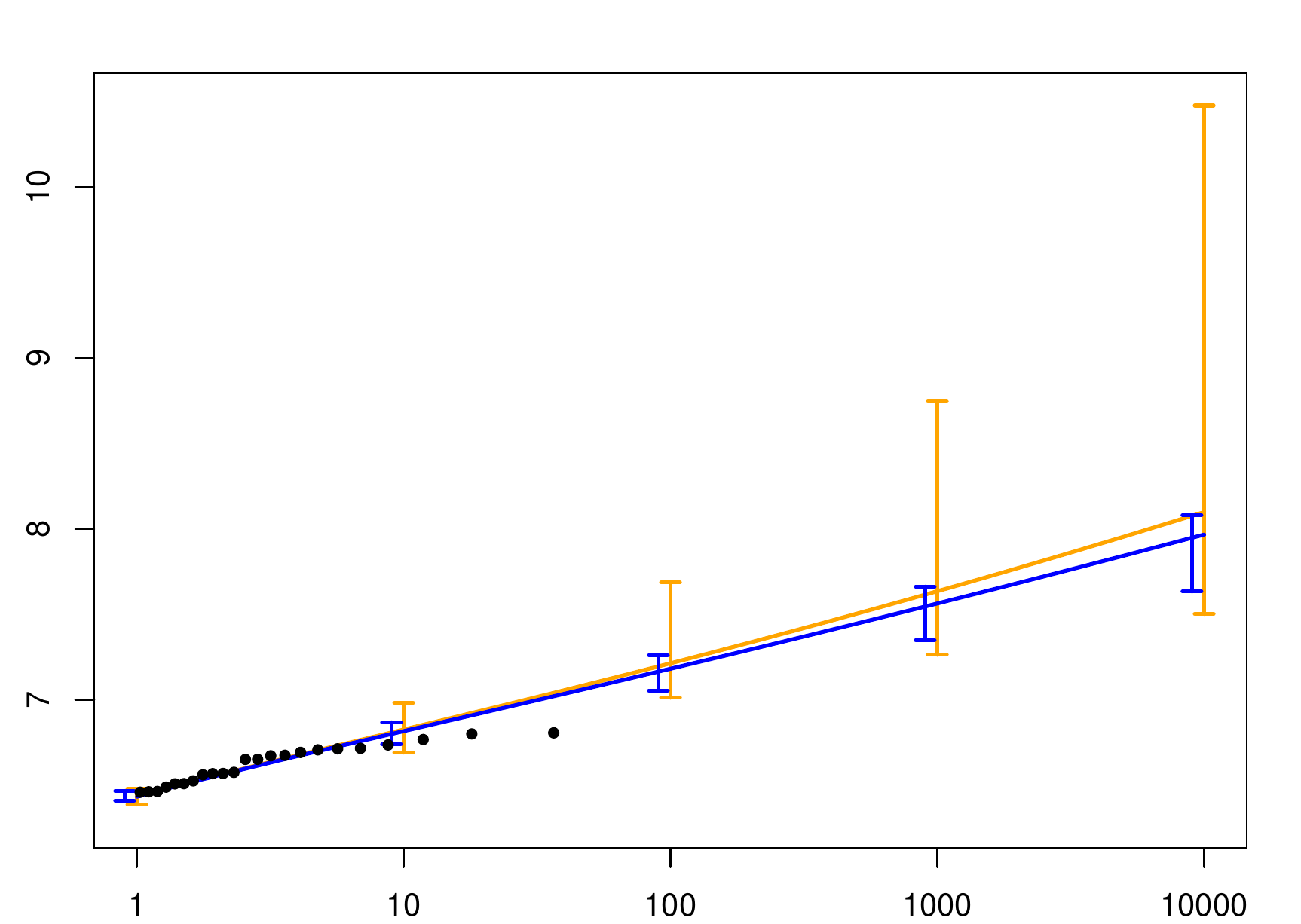}\includegraphics[width=0.4\textwidth]{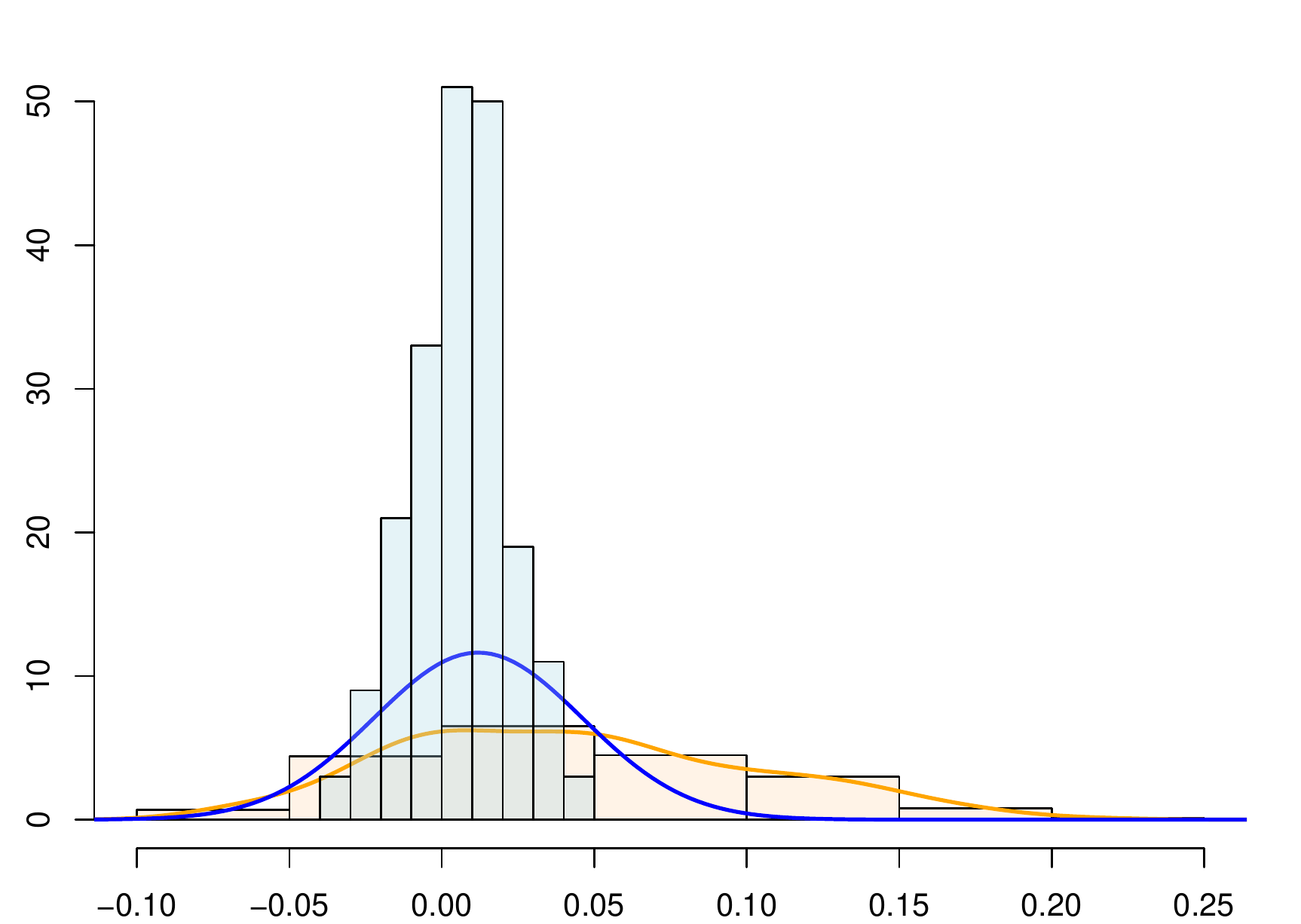}
    \caption{95\% bootstrap confidence intervals on the final (\textit{temporal dependence}) return level estimates at Sheerness (left) before (orange) and after (blue) adding a prior distribution to the shape parameter. Empirical estimates are shown by black points. Histograms of the shape parameter estimates and their densities (right) for these two models in their corresponding colours.}
    \label{fig::SHE_boots}
\end{figure}

We take 200 bootstrap samples and present 95\% confidence intervals in Figures~\ref{fig::SHE_boots} and~\ref{fig::all_boots} at Sheerness and the other sites, respectively. The uncertainty associated with each return level increases with return period, as expected, since it becomes uncertain as we extrapolate. Uncertainty at the largest return periods can be attributed to the difficulty in accurately and precisely estimating the shape parameter of the GPD. This can be reduced by specifying a prior distribution on the shape parameter.~\cite{CFB2018} use a $\text{Normal}(0.0119, 0.0343^2)$ prior, based on site specific parameter estimates. We incorporate this prior information into our model, using a penalised likelihood framework for parameter estimation. The scale and rate parameter estimates remain unchanged, but our shape estimates now lie closer to zero with narrower confidence intervals at -0.019 (-0.021, 0.059), 0.014 (-0.025, 0.052), -0.027 (-0.058, 0.004) and 0.008 (-0.039, 0.054) for Heysham, Lowestoft, Newlyn and Sheerness, respectively, with 95\% confidence intervals given in parentheses. Figure~\ref{fig::SHE_boots} demonstrates how adding this prior information reduces uncertainty, since the distribution of $\hat\xi$ over 200 bootstrap samples is much narrower. Figure~\ref{fig::SHE_boots} also shows the return level estimates and 95\% bootstrap confidence intervals for this updated model; the return level curve is only affected at large return levels, as expected, where the estimates are less bounded. There is a dramatic reduction in the uncertainty associated with high return levels; at the 10,000 year level, adding prior information has reduced the range of confidence interval by 2.5m. Clearly, adding prior information on the shape parameter is important. However, we didn't do this sooner because it is important to allow the data to speak for itself when trying to the reflect the realism of the sea level process in other aspects of our methodology.

\subsection{Sea Level Seasonality}\label{sl_seasonality}
It is helpful to understand when in the year the most extreme sea levels might occur, such as for coastal defence maintenance planning. We use our final model (\textit{temporal dependence}) to investigate the seasonality of extreme sea levels. We are interested in the probability that a randomly selected sea level annual maxima $M$ is from month $j$ given it is equal to some level $z$. We consider $z_p$, a sea level with an associated annual exceedance probability $p\in[0,1]$, so that $z_{p}$ is derived from expression~\eqref{eqn::sldist_exi}. Then the probability of interest is defined by, \begin{equation}
    \hat P_M(j;z)=\hat{\mathbb{P}}(m(M) = j|M=z),\label{eqn::annmaxprob_model}
\end{equation} 
where $m(M)$ denotes the month of occurrence of $M$ and $\hat{\mathbb{P}}(\cdot)$ is under our final model. This is a similar metric to that used for assessing peak tide seasonality (Section~\ref{ss_seasonality}), however, since we have a model for the sea levels we condition on $M=z$ because we can obtain the form of the probability density function. Increasing $z$ to rare return levels uncovers the seasonal variations in extreme sea levels. For months where the most extreme sea levels occur, $\hat P_M(j;z)$ takes its largest values, with $\hat P_M(j;z)\rightarrow 0$ for months with the least extreme sea levels. We derive the form of $\hat P_M(j;z)$ in Section~\ref{sec::annmaxprob_form} of~\cite{DarcySM}.

 \begin{figure}
     \centering
     \includegraphics[width=0.55\textwidth]{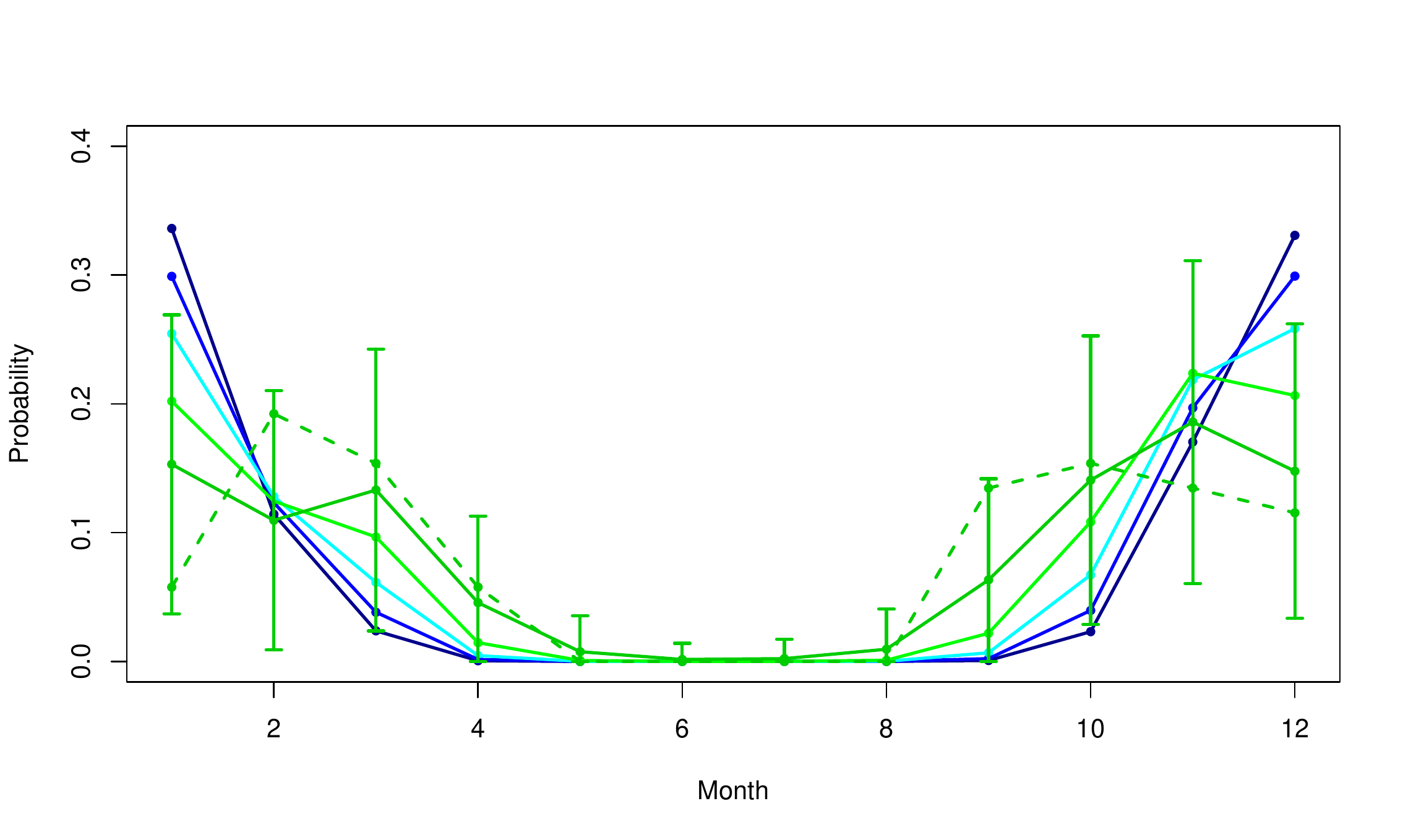}
     \caption{Estimates of $\hat P_M(j;z)$ for months $j=1-12$ at Sheerness, for $p=1$ (dark green), 0.1 (green), 0.01 (cyan), 0.001 (blue), and 0.0001 (dark blue). The dashed dark green line is the empirical estimate $\tilde P_M(j;z_1)$, 95\% confidence intervals are for $\hat P_M(j;z_1)$.}
     \label{fig::Annmaxprob_LOW}
 \end{figure}
 
Figure~\ref{fig::Annmaxprob_LOW} shows the estimates $\hat P_M(j;z)$ to demonstrate sea level seasonality of extreme levels at Sheerness, as well as the empirical estimate for sea levels equal to the 1 year return level (see Figure~\ref{fig::Annmaxprob_HEYNEW} for the remaining sites). These empirical estimates tend to lie within the 95\% confidence intervals for the corresponding model estimates, at the 1 year return level. This demonstrates that our model for extreme sea levels is capturing the seasonality well. Our model estimates allow us to extrapolate beyond the observed data. At Sheerness, the  empirical and model estimates at the 1 year return level are influenced by the tidal seasonality since they peak near the equinoxes. However, once we extrapolate to higher return levels, the sea level seasonality is almost entirely influenced by skew surge seasonality. This is the case for all sites at high return levels, since tides are bounded above by HAT but this is less than the 10 year sea level return level for all four sites. Therefore, extremely large skew surges are required to exceed the highest return levels, which are most likely to occur in winter.

\section{Discussion}\label{discussion}
By accounting for seasonality in skew surge and peak tide, the dependence between them, and temporal dependence in skew surges, we have developed a novel methodology for estimating extreme sea levels. Our results show a significant improvement on current methods, which ignored these features of the sea level processes and instead made several simplifying assumptions. Our model also allows us to study the seasonality of sea levels exceeding levels previously unobserved; this can be useful for coastal defence maintenance planning. The return levels estimated from our model present a more accurate representation of future extreme levels, and should be used for future coastal defence upgrades. Our methodology can be applied to all UK National Tide Gauge Network sites with a sufficient data record length, and is not limited to UK locations. We conclude by addressing some of the limitations with our model and suggest avenues for further exploration.

Our method assumes a steady state climate, since the existing mean sea level rise trend was removed from before analysis, and we have not included any longer-term trends in our model. Note the mean sea level trend needs to be added back onto the return level estimates presented in this paper when used in practice. Since there is growing concern with anthropogenic induced climate change and sea level rise, this should be accounted for. Whilst most of this is captured by mean sea level rise, it has been suggested there is a small, but different, trend in the extreme sea levels \citep{haigh2010, MenendezWoodworth2010}. There are two approaches to do this; the preprocessing model of \cite{EastoeTawn2009} where trends are modelled and removed before analysis, then covariates can be added to the extreme value model to explain trends in the tail of the distribution that are different from that in the body. Alternatively, covariates can be added to model parameters straight away without an attempt to remove any trends from the data \citep{davisonsmith1990}. 

Spatial pooling provides a promising framework to capture longer-term trends due to climate change; single site trends are subtle but sharing information across sites could give more significant results. Spatial pooling also enables inference at locations with limited or no data, where our current singe site model would not perform well. In Section~\ref{rl_final} we add a prior penalty for the shape parameter in our model, based on spatial information; this is a first approach to spatial pooling and drastically reduced the uncertainty associated with longer term return levels. Table~\ref{tbl:SMparam_ests} shows that the 95\% confidence intervals of $\hat\xi$ for Model~$S4$ (no penalty) overlap at all sites, therefore it may be reasonable to fix the shape parameter across sites, as an alternative approach for borrowing information. Fixing the shape parameter in homogeneous regions is a crucial step in regional frequency analysis, originally introduced by \cite{HoskingWallis1997}. This is also an assumption in the Bayesian hierarchical modelling framework presented by \cite{sharkeywinter2019} and a common assumption for spatial extreme value theory (see \cite{davison2019spatial}). We refer the reader to \cite{Batstone2013}, \cite{bernardara2011} and \cite{haigh2010} for different approaches to spatial pooling for extreme sea level estimation. Whilst spatial pooling is clearly an important aspect of extreme sea level estimation, it is also fundamental that the marginal site estimates are accurate, therefore our approach for capturing the realism of the sea level process should be adopted within a spatial framework.

We have shown that skew surges clearly exhibit within-year seasonality. We used a monthly threshold, defined as a quantile of the monthly skew surge distribution, to define extreme values and develop our non-stationary model. Using a quantile ensures there are a similar number of exceedances to model per month. This approach is similar to that of \cite{carter1981}, where we first assume stationarity within months and then build in the seasonal variation on a shorter temporal resolution through covariates in the GPD parameters. This meant that we were able to capture most of the non-stationarity, as well as skew surge-peak tide dependence, at the same stage of the modelling process. However, we could have considered a smoother threshold choice by using quantile regression \citep{northropjonathan2016} but we did not try this as our monthly threshold appeared sufficient. 


\section*{Acknowledgments} 
This paper is based on work completed while Eleanor D'Arcy was part of the EPSRC funded STOR-i centre for doctoral training (EP/S022252/1). Thanks to Jenny Sansom of the Environment Agency and Joanne Williams of National Oceanography Centre for providing the data.

\bibliography{ref.bib}

\begin{thebibliography}{}

\bibitem[Coles et~al., 1999]{Coles1999}
Coles, S.~G., Heffernan, J., and Tawn, J.~A. (1999).
\newblock Dependence measures for extreme value analyses.
\newblock {\em Extremes}, 2(4):339--365.

\bibitem[D'Arcy and Tawn, 2021]{darcytawn21}
D'Arcy, E. and Tawn, J.~A. (2021).
\newblock Discussion of ``{T}owards using state-of-the-art climate models to
  help constrain estimates of unprecedented {UK} storm surges'' {(by Howard, T.
  and Williams, S. D. P.)}.
\newblock {\em Natural Hazards and Earth System Sciences Discussions},
  21(12):3693--3712.

\bibitem[Howard and Williams, 2021]{howardWilliams2021}
Howard, T. and Williams, S. D.~P. (2021).
\newblock Towards using state-of-the-art climate models to help constrain
  estimates of unprecedented {UK} storm surges.
\newblock {\em Natural Hazards and Earth System Sciences}, 21(12):3693--3712.

\bibitem[Politis and Romano, 1994]{Politis1994}
Politis, D.~N. and Romano, J.~P. (1994).
\newblock The stationary bootstrap.
\newblock {\em Journal of the American Statistical Association},
  89(428):1303--1313.

\end{thebibliography}


\begin{thebibliography}{}

\bibitem[Asadi et~al., 2018]{Asadi2018}
Asadi, P., Engelke, S., and Davison, A.~C. (2018).
\newblock Optimal regionalization of extreme value distributions for flood
  estimation.
\newblock {\em Journal of Hydrology}, 556:182--193.

\bibitem[Baranes et~al., 2020]{baranes2020}
Baranes, H., Woodruff, J., Talke, S., Kopp, R., Ray, R., and DeConto, R.
  (2020).
\newblock {Tidally driven interannual variation in extreme sea level
  frequencies in the Gulf of Maine}.
\newblock {\em Journal of Geophysical Research: Oceans}, 125(10):e2020JC016291.

\bibitem[Bashtannyk and Hyndman, 2001]{bashtannyk2001}
Bashtannyk, D.~M. and Hyndman, R.~J. (2001).
\newblock Bandwidth selection for kernel conditional density estimation.
\newblock {\em Computational Statistics \& Data Analysis}, 36(3):279--298.

\bibitem[Batstone et~al., 2013]{Batstone2013}
Batstone, C., Lawless, M., Tawn, J.~A., Horsburgh, K., Blackman, D., McMillan,
  A., Worth, D., Laeger, S., and Hunt, T. (2013).
\newblock {A UK best-practice approach for extreme sea-level analysis along
  complex topographic coastlines}.
\newblock {\em Ocean Engineering}, 71:28--39.

\bibitem[Bernardara et~al., 2011]{bernardara2011}
Bernardara, P., Andreewsky, M., and Benoit, M. (2011).
\newblock Application of regional frequency analysis to the estimation of
  extreme storm surges.
\newblock {\em Journal of Geophysical Research: Oceans}, 116(C2).

\bibitem[Carter and Challenor, 1981]{carter1981}
Carter, D. and Challenor, P. (1981).
\newblock Estimating return values of environmental parameters.
\newblock {\em Quarterly Journal of the Royal Meteorological Society},
  107(451):259--266.

\bibitem[Chavez-Demoulin and Davison, 2005]{ChavesDavison2005}
Chavez-Demoulin, V. and Davison, A.~C. (2005).
\newblock Generalized additive modelling of sample extremes.
\newblock {\em Journal of the Royal Statistical Society: Series C},
  54(1):207--222.

\bibitem[Coles, 2001]{Coles2001}
Coles, S.~G. (2001).
\newblock {\em An Introduction to Statistical Modeling of Extreme Values}.
\newblock Springer, London.

\bibitem[Coles et~al., 1999]{Coles1999}
Coles, S.~G., Heffernan, J., and Tawn, J.~A. (1999).
\newblock Dependence measures for extreme value analyses.
\newblock {\em Extremes}, 2(4):339--365.

\bibitem[Coles and Tawn, 1990]{ColesTawn1990}
Coles, S.~G. and Tawn, J.~A. (1990).
\newblock Statistics of coastal flood prevention.
\newblock {\em Philosophical Transactions of the Royal Society of London.
  Series A: Physical and Engineering Sciences}, 332(1627):457--476.

\bibitem[Coles et~al., 1994]{ColesTawnSmith1994}
Coles, S.~G., Tawn, J.~A., and Smith, R.~L. (1994).
\newblock A seasonal {M}arkov model for extremely low temperatures.
\newblock {\em Environmetrics}, 5(3):221--239.

\bibitem[D'Arcy et~al., 2022]{DarcySM}
D'Arcy, E., Tawn, J.~A., Joly-Laugel, A., and Sifnioti, D.~E. (2022).
\newblock Supplement to ``{A}ccounting for seasonality in extreme sea level
  estimation''.

\bibitem[Davison et~al., 2019]{davison2019spatial}
Davison, A.~C., Huser, R., and Thibaud, E. (2019).
\newblock Spatial extremes.
\newblock In {\em Handbook of Environmental and Ecological Statistics}, pages
  711--744. Chapman and Hall/CRC.

\bibitem[Davison and Smith, 1990]{davisonsmith1990}
Davison, A.~C. and Smith, R.~L. (1990).
\newblock Models for exceedances over high thresholds (with discussion).
\newblock {\em Journal of the Royal Statistical Society: Series B},
  52(3):393--425.

\bibitem[Dixon and Tawn, 1994]{dixontawn1994}
Dixon, M. and Tawn, J. (1994).
\newblock Extreme sea levels: modelling interaction between tide and surge.
\newblock {\em Statistics for the Environment 2: Water Related Issues}, pages
  221--232.

\bibitem[Dixon and Tawn, 1999]{DixonTawn1999}
Dixon, M.~J. and Tawn, J.~A. (1999).
\newblock The effect of non-stationarity on extreme sea-level estimation.
\newblock {\em Journal of the Royal Statistical Society: Series C},
  48(2):135--151.

\bibitem[Dixon et~al., 1998]{dixontawnvassie1998}
Dixon, M.~J., Tawn, J.~A., and Vassie, J.~M. (1998).
\newblock Spatial modelling of extreme sea-levels.
\newblock {\em Environmetrics}, 9(3):283--301.

\bibitem[Eastoe and Tawn, 2009]{EastoeTawn2009}
Eastoe, E.~F. and Tawn, J.~A. (2009).
\newblock Modelling non‐stationary extremes with application to surface level
  ozone.
\newblock {\em Journal of the Royal Statistical Society: Series C},
  58(1):25--45.

\bibitem[Egbert and Ray, 2017]{egbert2017}
Egbert, G.~D. and Ray, R.~D. (2017).
\newblock Tidal prediction.
\newblock {\em Journal of Marine Research}, 75(3):189--237.

\bibitem[{Environment Agency}, 2018]{CFB2018}
{Environment Agency} (2018).
\newblock {Coastal Flood Boundary Conditions for the UK: update 2018. Technical
  summary report.}
\newblock
  \url{https://environment.data.gov.uk/dataset/6e856bda-0ca9-404f-93d7-566a2378a7a8}.
\newblock Accessed 01/10/21.

\bibitem[Fawcett and Walshaw, 2007]{FawcettWalshaw2007}
Fawcett, L. and Walshaw, D. (2007).
\newblock Improved estimation for temporally clustered extremes.
\newblock {\em Environmetrics}, 18(2):173--188.

\bibitem[Ferro and Segers, 2003]{FerroSegers2003}
Ferro, C. A.~T. and Segers, J. (2003).
\newblock Inference for clusters of extreme values.
\newblock {\em Journal of the Royal Statistical Society: Series B},
  65(2):545--556.

\bibitem[Graff, 1978]{graff1978concerning}
Graff, J. (1978).
\newblock Concerning the recurrence of abnormal sea levels.
\newblock {\em Coastal Engineering}, 2:177--187.

\bibitem[Haigh et~al., 2010]{haigh2010}
Haigh, I.~D., Nicholls, R., and Wells, N. (2010).
\newblock A comparison of the main methods for estimating probabilities of
  extreme still water levels.
\newblock {\em Coastal Engineering}, 57(9):838--849.

\bibitem[Hosking and Wallis, 1997]{HoskingWallis1997}
Hosking, J. R.~M. and Wallis, J.~R. (1997).
\newblock {\em Regional Frequency Analysis: An Approach Based on L-moments}.
\newblock Cambridge University Press, Cambridge; New York.

\bibitem[Howard and Williams, 2021]{howardWilliams2021}
Howard, T. and Williams, S. D.~P. (2021).
\newblock Towards using state-of-the-art climate models to help constrain
  estimates of unprecedented {UK} storm surges.
\newblock {\em Natural Hazards and Earth System Sciences}, 21(12):3693--3712.

\bibitem[Jonathan et~al., 2014]{Jonathan2014}
Jonathan, P., Randell, D., Wu, Y., and Ewans, K. (2014).
\newblock Return level estimation from non-stationary spatial data exhibiting
  multidimensional covariate effects.
\newblock {\em Ocean Engineering}, 88:520--532.

\bibitem[Leadbetter et~al., 1983]{Leadbetter1983}
Leadbetter, M., Lindgren, G., and Rootz\'{e}n, H. (1983).
\newblock {\em {Extremes and Related Properties of Random Sequences and
  Processes}}.
\newblock Springer-Verlag, New York.

\bibitem[Ledford and Tawn, 2003]{LedfordTawn2003}
Ledford, A.~W. and Tawn, J.~A. (2003).
\newblock Diagnostics for dependence within time series extremes.
\newblock {\em Journal of the Royal Statistical Society: Series B},
  65(2):521--543.

\bibitem[Men{\'e}ndez and Woodworth, 2010]{MenendezWoodworth2010}
Men{\'e}ndez, M. and Woodworth, P.~L. (2010).
\newblock Changes in extreme high water levels based on a quasi-global
  tide-gauge data set.
\newblock {\em Journal of Geophysical Research: Oceans}, 115(C10).

\bibitem[{NOC}, 2021]{nationaltidegaugenetwork}
{NOC} (2021).
\newblock {National Tidal and Sea Level Facility}.
\newblock \url{https://www.ntslf.org/}.

\bibitem[Northrop et~al., 2016]{northropjonathan2016}
Northrop, P.~J., Jonathan, P., and Randell, D. (2016).
\newblock Threshold modeling of nonstationary extremes.
\newblock In {\em Extreme Value Modeling and Risk Analysis}, pages 107--128.
  Chapman and Hall/CRC.

\bibitem[Politis and Romano, 1994]{Politis1994}
Politis, D.~N. and Romano, J.~P. (1994).
\newblock The stationary bootstrap.
\newblock {\em Journal of the American Statistical Association},
  89(428):1303--1313.

\bibitem[Prandle and Wolf, 1978]{PrandleWolf1978}
Prandle, D. and Wolf, J. (1978).
\newblock Surge-tide interaction in the southern {N}orth {S}ea.
\newblock {\em Elsevier Oceanography Series}, 23:161--185.

\bibitem[Pugh and Vassie, 1978]{PughVassie1978}
Pugh, D. and Vassie, J. (1978).
\newblock Extreme sea levels from tide and surge probability.
\newblock {\em Coastal Engineering}, 16:911--930.

\bibitem[Pugh and Woodworth, 2014]{pughwoodworth2014}
Pugh, D. and Woodworth, P. (2014).
\newblock {\em Sea-Level Science: Understanding Tides, Surges, Tsunamis and
  Mean Sea-Level Changes}.
\newblock Cambridge University Press.

\bibitem[Rodríguez~Enríquez et~al., 2021]{Rodriguez2022}
Rodríguez~Enríquez, A., Wahl, T., Baranes, H., Talke, S.~A., Orton, P.~M.,
  Booth, J.~F., and D~Haigh, I. (2021).
\newblock Predictable changes in extreme sea levels and coastal flood risk due
  to nodal and perigean astronomical tidal cycles.
\newblock {\em Earth and Space Science Open Archive}.

\bibitem[Rohrbeck and Tawn, 2021]{Rohrbeck2020}
Rohrbeck, C. and Tawn, J.~A. (2021).
\newblock Bayesian spatial clustering of extremal behavior for hydrological
  variables.
\newblock {\em Journal of Computational and Graphical Statistics},
  30(1):91--105.

\bibitem[Rootz{\'e}n and Katz, 2013]{rootzen2013}
Rootz{\'e}n, H. and Katz, R.~W. (2013).
\newblock Design life level: quantifying risk in a changing climate.
\newblock {\em Water Resources Research}, 49(9):5964--5972.

\bibitem[Sharkey and Winter, 2019]{sharkeywinter2019}
Sharkey, P. and Winter, H.~C. (2019).
\newblock A {B}ayesian spatial hierarchical model for extreme precipitation in
  {G}reat {B}ritain.
\newblock {\em Environmetrics}, 30(1):e2529.

\bibitem[Smith and Weissman, 1994]{SmithWeissman1994}
Smith, R.~L. and Weissman, I. (1994).
\newblock Estimating the extremal index.
\newblock {\em Journal of the Royal Statistical Society: Series B},
  56(3):515--528.

\bibitem[Tawn and Vassie, 1989]{TawnVassie1989}
Tawn, J. and Vassie, J. (1989).
\newblock Extreme sea levels: the joint probabilities method revisited and
  revised.
\newblock {\em Proceedings of the Institution of Civil Engineers},
  87(3):429--442.

\bibitem[Tawn, 1988]{tawn1988sealevels}
Tawn, J.~A. (1988).
\newblock An extreme-value theory model for dependent observations.
\newblock {\em Journal of Hydrology}, 101(1-4):227--250.

\bibitem[Tawn, 1992]{Tawn1992}
Tawn, J.~A. (1992).
\newblock Estimating probabilities of extreme sea-levels.
\newblock {\em Journal of the Royal Statistical Society: Series C},
  41(1):77--93.

\bibitem[Wadey et~al., 2015]{wadey2015}
Wadey, M.~P., Haigh, I., Nicholls, R.~J., Brown, J.~M., Horsburgh, K., Carroll,
  B., Gallop, S.~L., Mason, T., Bradshaw, E., et~al. (2015).
\newblock {A comparison of the 31 January--1 February 1953 and 5--6 December
  2013 coastal flood events around the UK}.
\newblock {\em Frontiers in Marine Science}, 2:84.

\bibitem[Williams et~al., 2016]{Williams2016}
Williams, J., Horsburgh, K.~J., Williams, J.~A., and Proctor, R.~N. (2016).
\newblock Tide and skew surge independence: new insights for flood risk.
\newblock {\em Geophysical Research Letters}, 43(12):6410--6417.

\end{thebibliography}
\bibliographystyle{apalike}

\newpage
\setcounter{page}{1}
\setcounter{figure}{0}
\setcounter{table}{0}
\setcounter{equation}{1}
\setcounter{section}{0} 


\renewcommand{\thefigure}{S\arabic{figure}}
\renewcommand{\thetable}{S\arabic{table}}
\renewcommand{\theequation}{S.\arabic{equation}}

\title{Supplementary Material for ``Accounting for Seasonality in Extreme Sea Level Estimation''}


\maketitle

\section{Introduction}
This document outlines the supplementary material for ``Accounting for Seasonality in Extreme Sea Level Estimation'', subsequently referred to as the main paper. Firstly in Section~\ref{ss_mt_indep_supp}, we present the exploratory analysis to assess skew surge-peak tide dependence, discussed in Section~\ref{ss_mt_indep} of the main paper; we demonstrate a time-varying relationship at Sheerness. We fit the tide dependent skew surge model presented in Section~\ref{acc_ssmt_dep} of the main paper to 483-year climate model data in Section~\ref{suppmat_climmodel}, to illustrate the physical justification of modelling skew surge-peak tide dependence. Then in Section~\ref{sec::suppmat_exi} we investigate temporal dependence in the skew surge series, this was discussed in Section~\ref{ss_seasonality} of the main paper and then accounted for in our methodology in Section~\ref{ss_dep}. In Section~\ref{sec::annmaxprob_form} we derive an analytical expression for the probability that a randomly selected sea level annual maxima is from a particular month, given it is equal to a return level (see equation~\eqref{eqn::annmaxprob_model} of the main paper); this is evaluated in Section~\ref{sl_seasonality} of the main paper to understand the seasonality of extreme sea levels. In Section~\ref{sec::transform_ss} we detail the process of transforming skew surges to uniform margins using the final model presented in the main paper, and we use this as a means of assessing skew surge model fit; this was used in the bootstrap procedure for uncertainty quantification on return level estimations in Section~\ref{rl_final} of the main paper. Lastly, we present supplementary figures in Section~\ref{suppfig}.

\section{Skew Surge-Peak Tide Dependence}\label{ss_mt_indep_supp}
In this section, we present our exploratory analysis to demonstrate that it is reasonable to assume skew surge and peak tide are independent at Heysham, Lowestoft and Newlyn, but not at Sheerness. We perform various statistical tests at all sites. Firstly, we formally test if there is a relationship between extreme skew surges and their associated ranked peak tide, where extreme skew surges are defined as exceedances of different thresholds. Then we investigate if all peak tides come from the same distribution as those associated with extreme skew surges. We also do this on a monthly scale at Sheerness to understand how the dependence structure changes within a year. Lastly, we use a simple quantile regression technique to test if the quantile of skew surges associated with different ordered tidal bands varies.

\begin{figure}[h]
    \centering
    \includegraphics[width=0.45\textwidth]{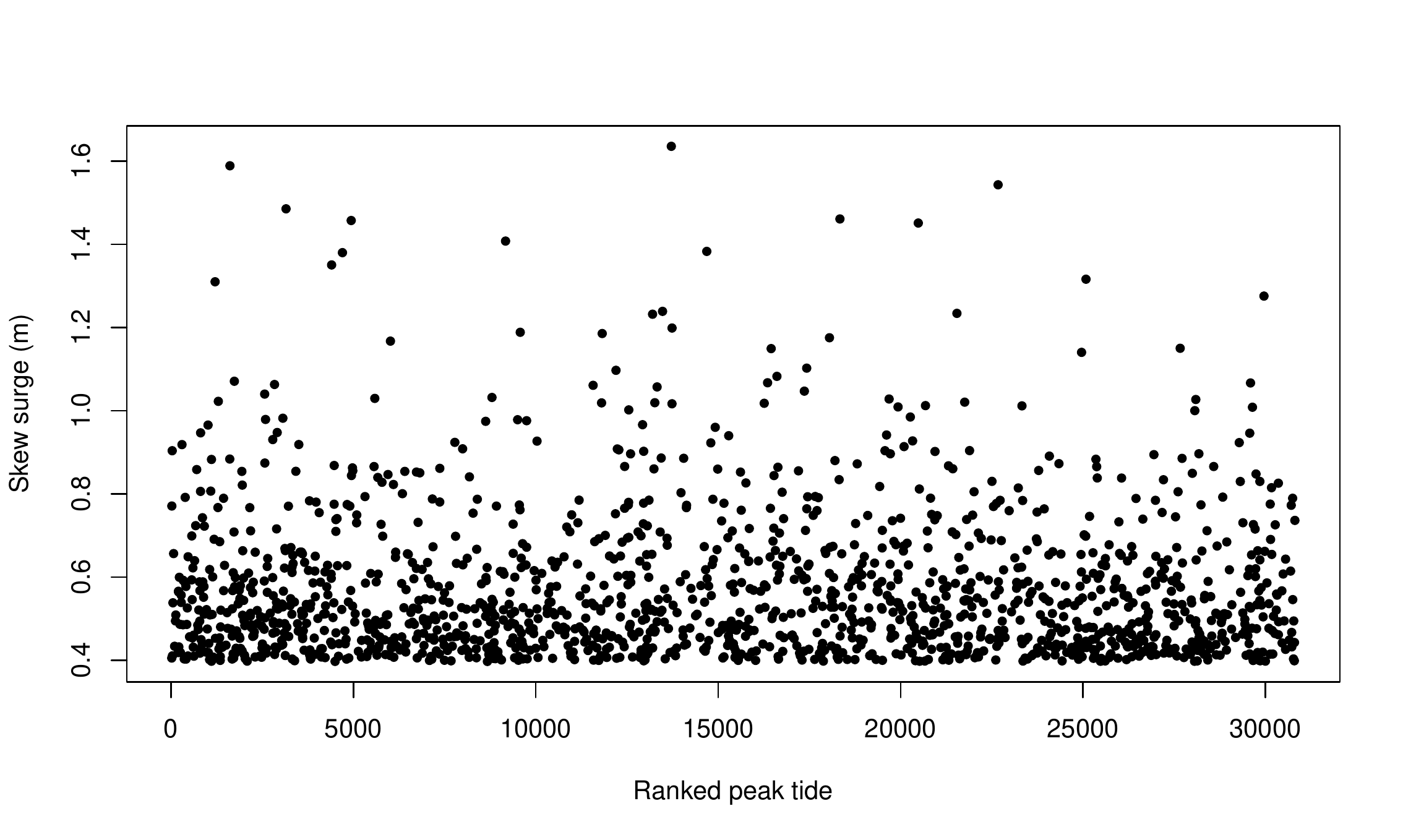}\includegraphics[width=0.45\textwidth]{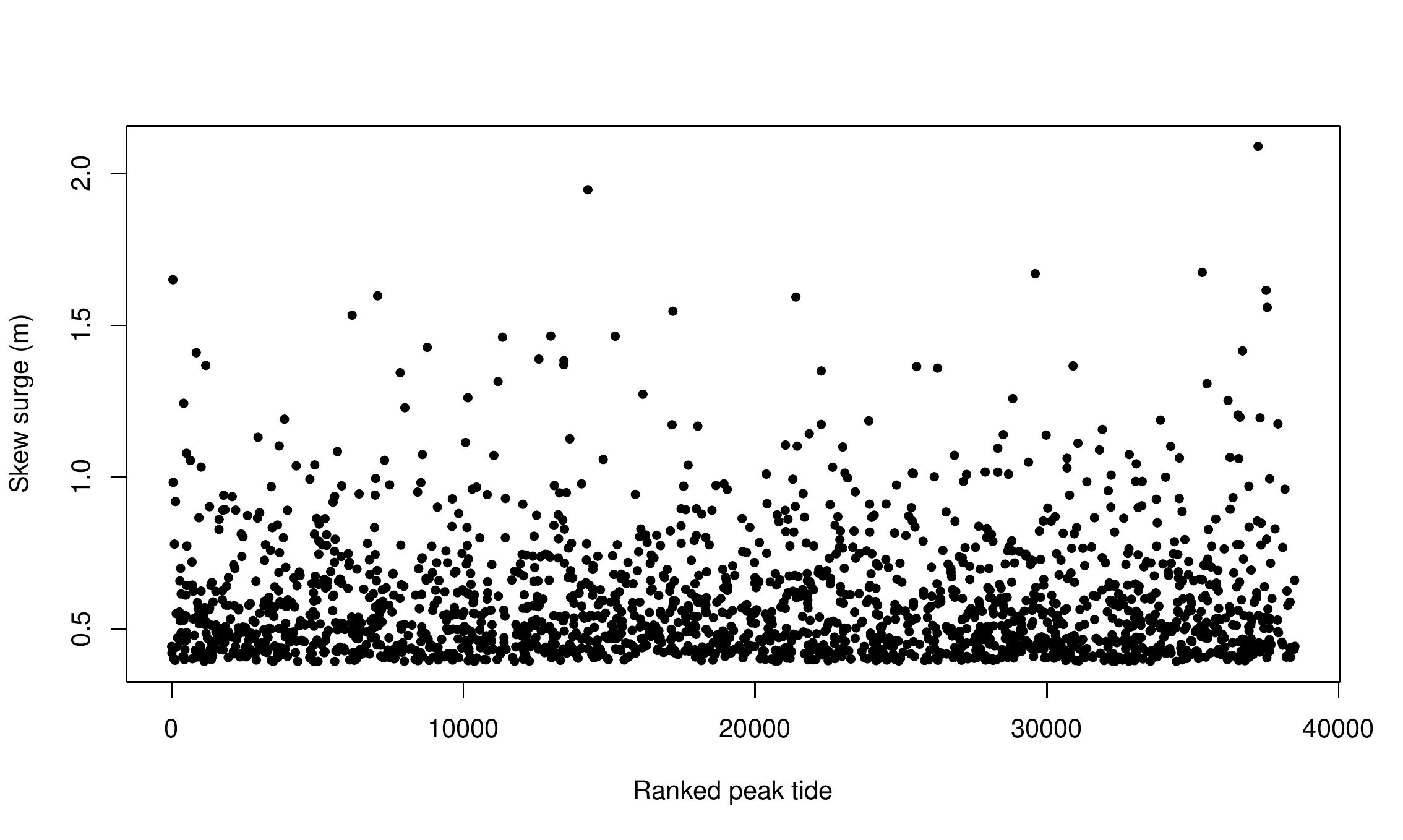}\\
    \includegraphics[width=0.45\textwidth]{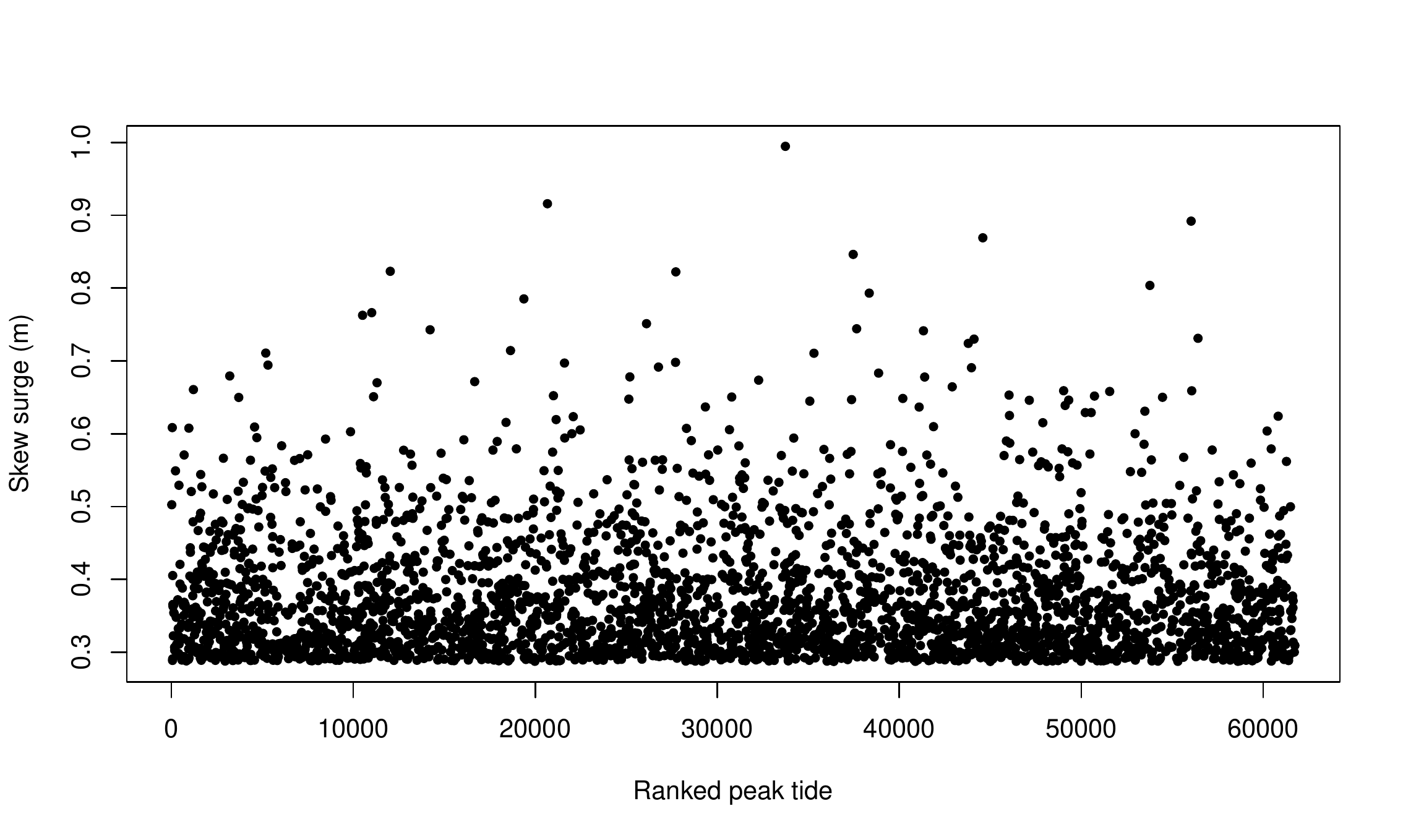}\includegraphics[width=0.45\textwidth]{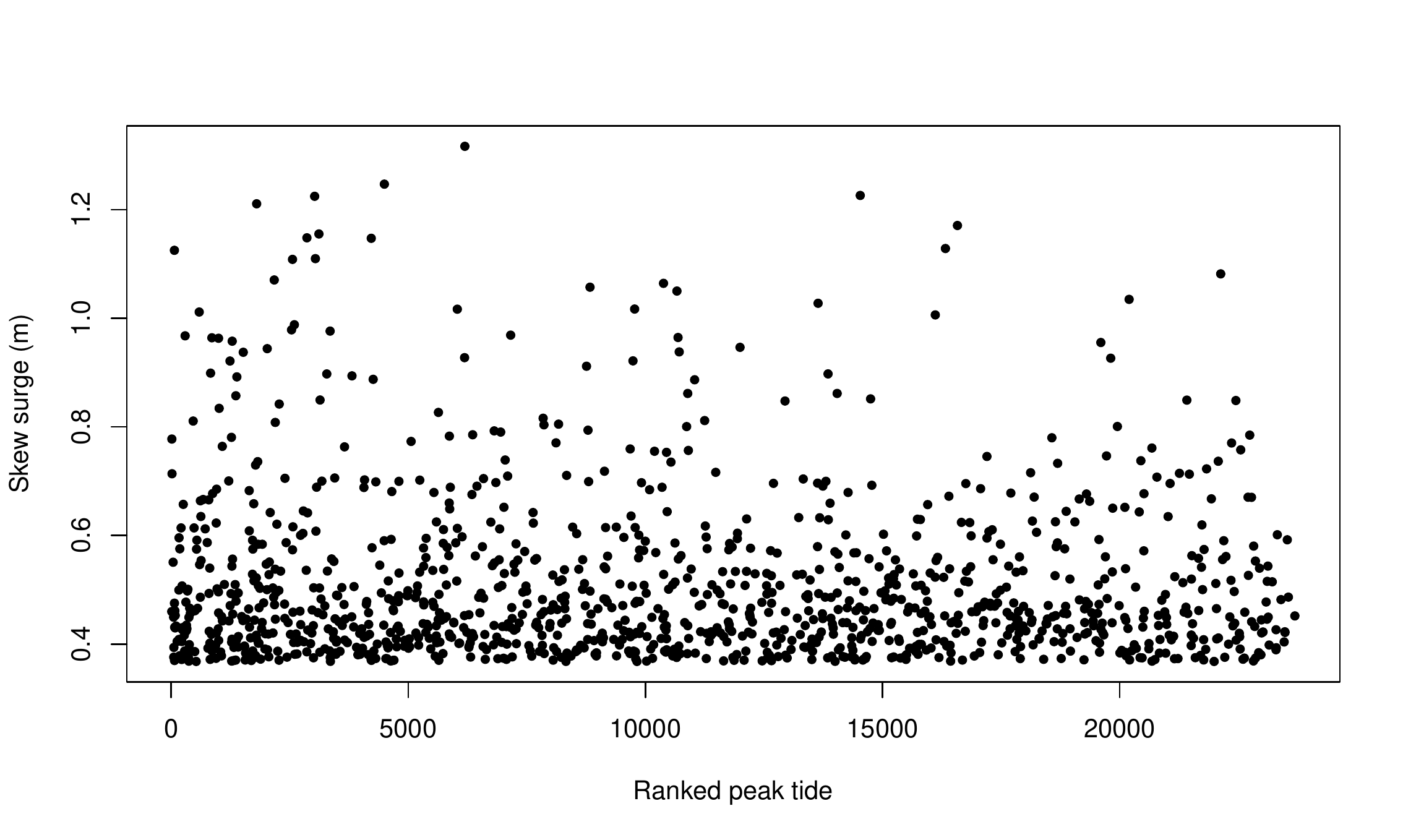}
    \caption{Scatter plot of extreme skew surge observations (exceedances of 0.95 quantile) against associated ranked peak tides at Heysham (top right), Lowestoft (top left), Newlyn (bottom left) and Sheerness (bottom right).}
    \label{ss_mt_scatter}
\end{figure}

We test if ranked peak tides associated with extreme skew surge (defined as exceedances of the 0.95 quantile) are uniformly distributed, using a Kolmogorov-Smirnov test (see Figure~\ref{ss_mt_scatter}). If the two components are independent, these will be distributed $\text{Uniform}(0,T)$ where $T$ is the total number of tidal cycles. The standard version of the Kolmogorov-Smirnov test falsely assumes that peak tides are temporally independent; we use the bootstrap procedure of~\citeSM{Politis1994} to account for this. Average $p$ values over 100 iterations are reported in Table \ref{KStest_threshold}, with expected block sizes inferred from the site specific autocorrelation function (acf) plots (see Figure \ref{fig::acf_ss}). At Sheerness, we find strong evidence to reject the null hypothesis that ranked peak tides associated with extreme skew surges are uniformly distributed, with $p$ value $2.89\times{10^{-3}}$. At the other three sites, we find sufficient evidence to reject this claim at the $5\%$ level. We also explore the sensitivity of this test to the choice of threshold used to define extreme skew surges, the $p$ values are reported in Table~\ref{KStest_threshold}. Lower thresholds typically correspond to lower $p$ values, suggesting we are more likely to reject the independence hypothesis when there are more exceedances.

\begin{table}
    \centering
    \resizebox{\textwidth}{!}{
    \begin{tabular}{c||c|c|c|c|c}
         & Block size & 0.9 & 0.95 & 0.975 & 0.99   \\ \hline
         HEY & 19 & 0.0052 (0.012) & 0.044 (0.083) & 0.1813 (0.11) & 0.32 (0.12) \\
         LOW & 5 & 0.63 (0.17) & 0.72 (0.22) & 0.78 (0.27) & 0.33 (0.23) \\
         NEW & 20 & 0.070 (0.075) & 0.091 (0.087) & 0.15 (0.17) & 0.12 (0.11) \\
         SHE & 6 & $1.5\times10^{-8}$ ($4.1\times10^{-8})$ & $1.1\times{10}^{-4}$ ($2.9\times10^{-4})$ & $9.9\times10^{-4}$ (0.0011) & 0.0013 (7$\times10{-4}$) \\
    \end{tabular}
    }
    \caption{Kolmogorov-Smirnov test $p$ values for uniformity of ranked peak tides associated with extreme skew surges, defined exceedances of different quantiles of the data (0.9, 0.95, 0.975, 0.999). Average $p$ values, after repeated bootstrapping, are shown in parentheses.}
    \label{KStest_threshold}
\end{table}

\begin{figure}[h]
    \centering
    \includegraphics[width=0.4\textwidth]{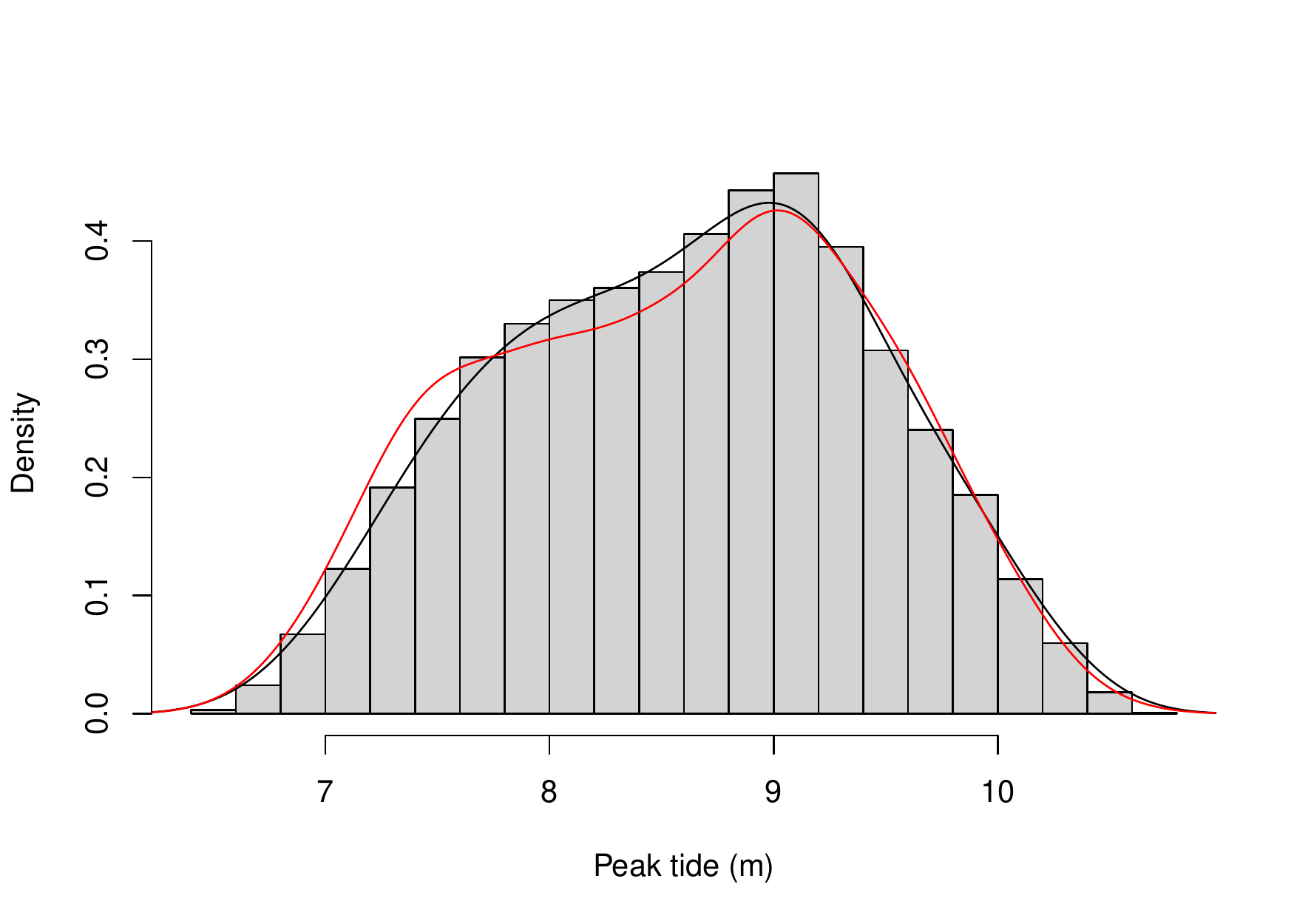}\includegraphics[width=0.4\textwidth]{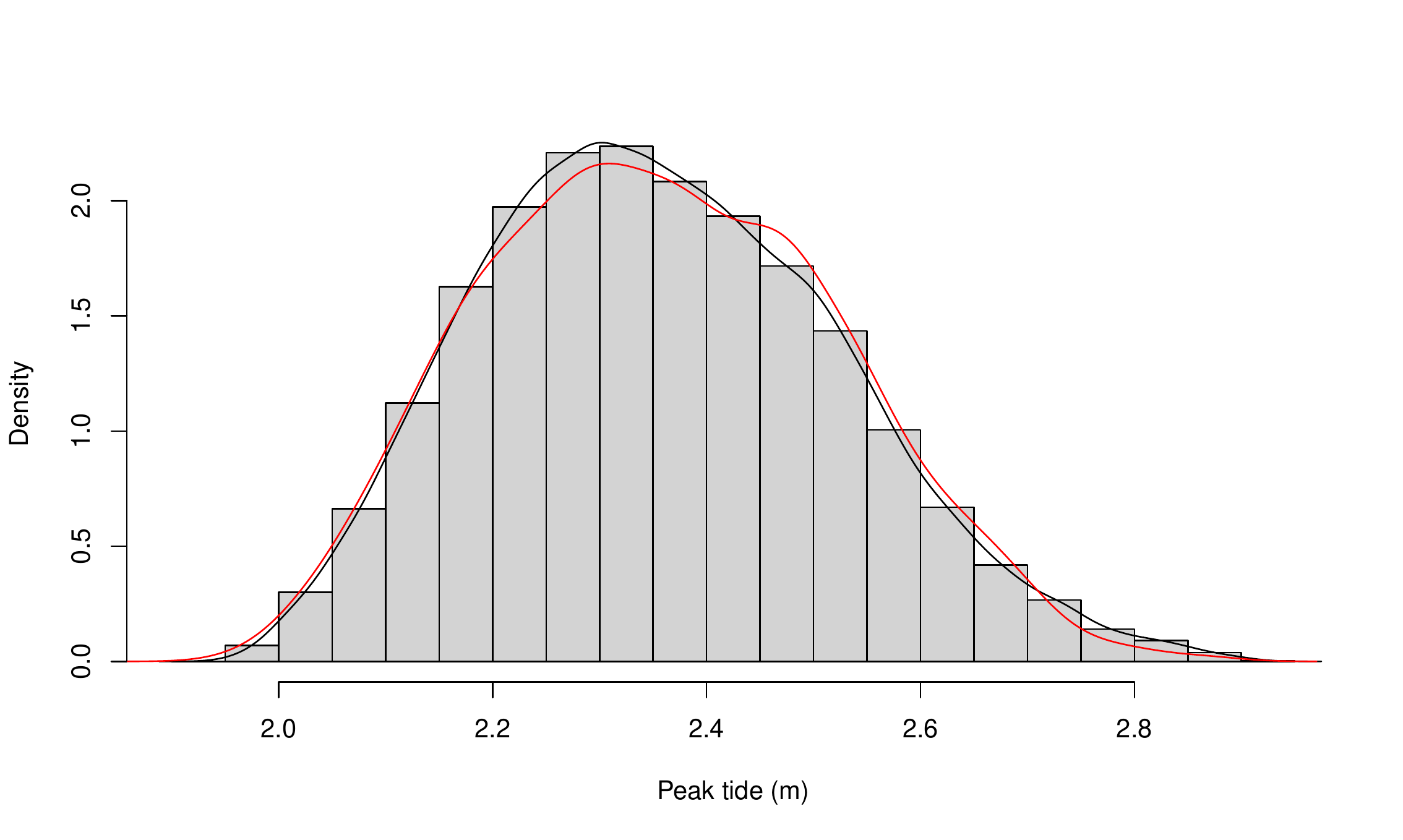}\\
    \includegraphics[width=0.4\textwidth]{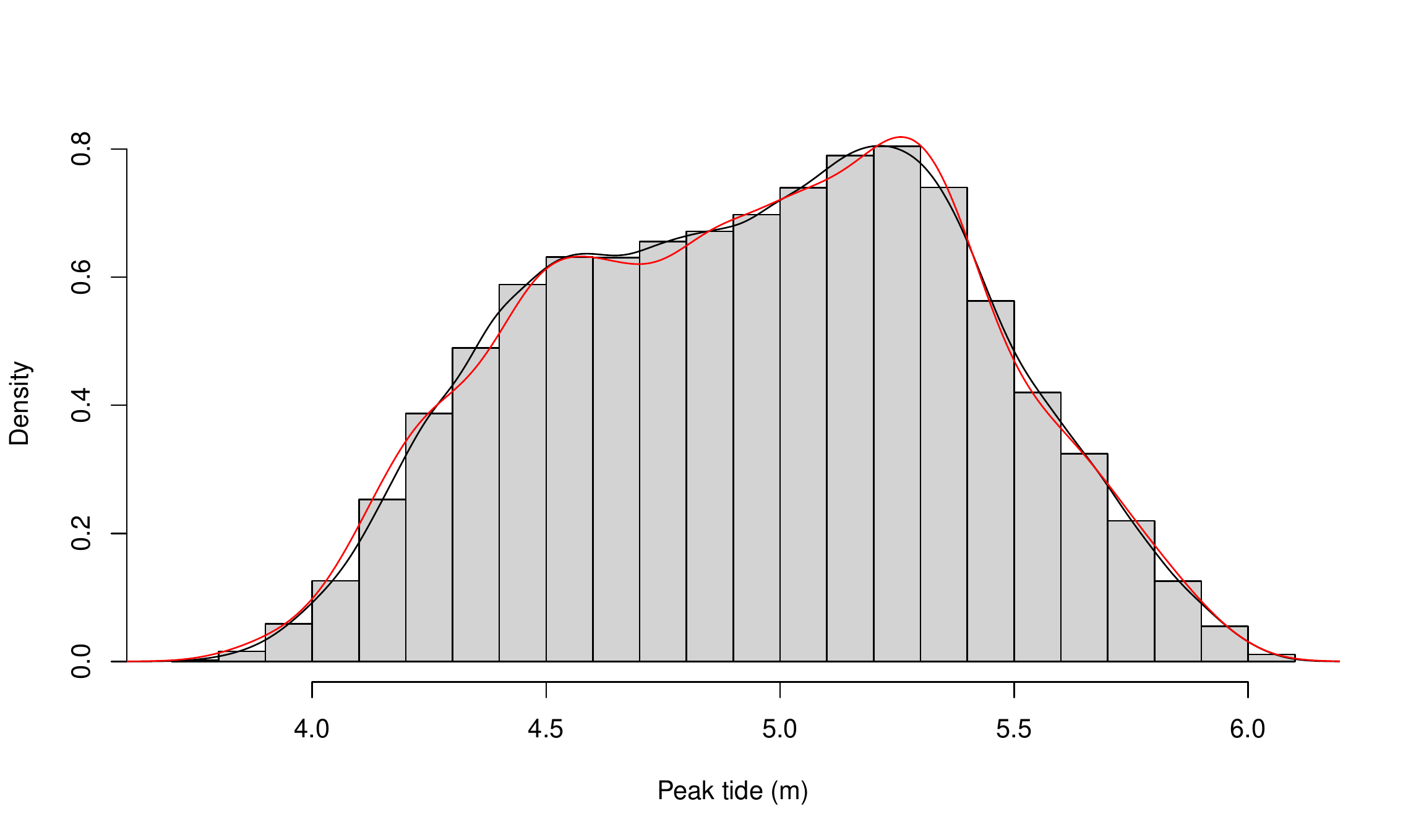}\includegraphics[width=0.4\textwidth]{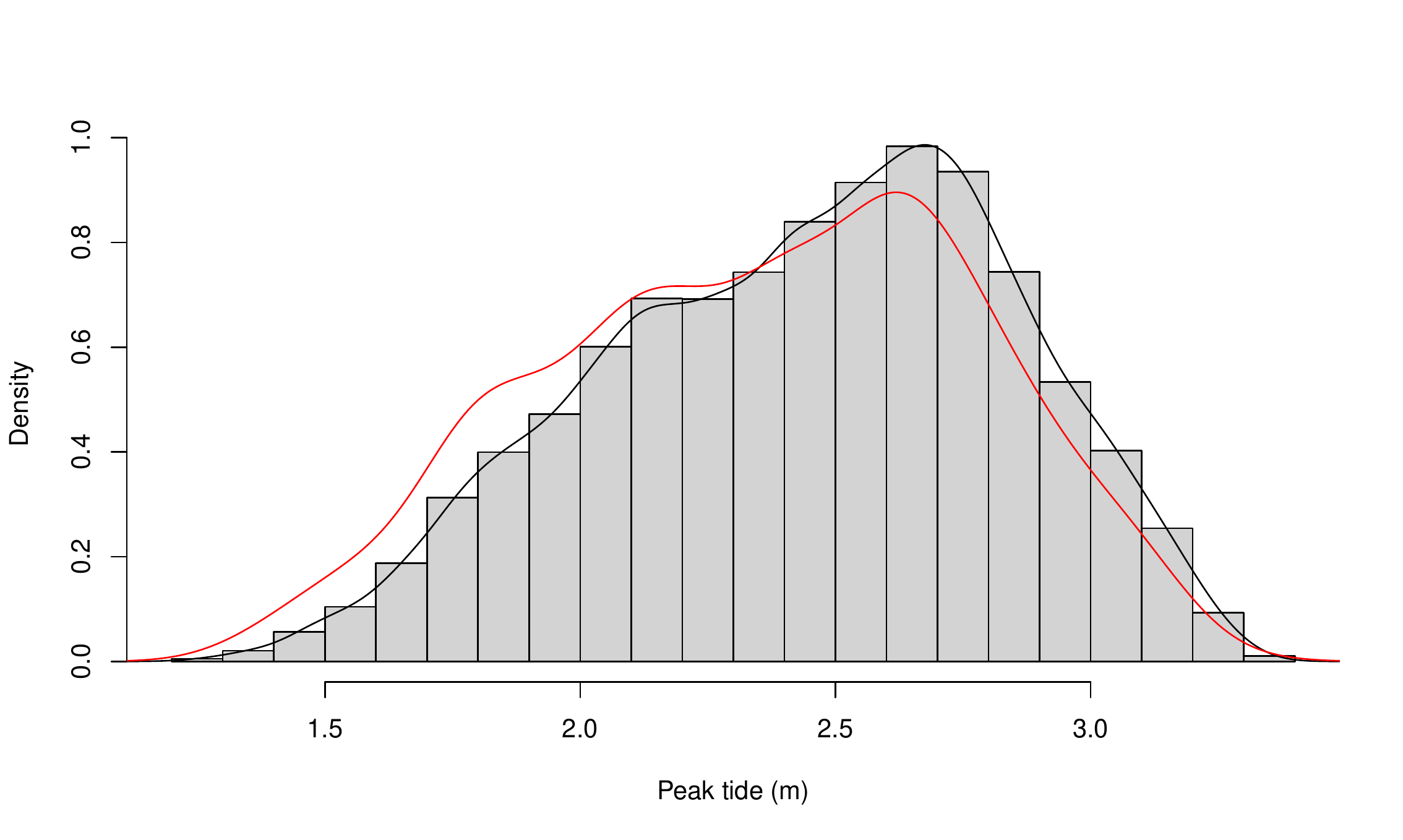}
    \caption{Histogram of all peak tides at Heysham (top left), Lowestoft (top right), Newlyn (bottom left) and Sheerness (bottom right). Probability density function of all peak tides (black) and peak tides associated with extreme skew surges (red) are interpolated onto each distribution.}
    \label{pdf_mt}
\end{figure}

We also test if the distribution of peak tides associated with extreme skew surges is the same as the distribution of all peak tides, using the Anderson-Darling test. Figure~\ref{pdf_mt} shows these distributions at all sites with their associated probability density functions, estimated using a Gaussian kernel density estimator. If peak tide and skew surge are independent, these two distributions should be identical up to sampling variation. We find insufficient evidence to reject the null hypothesis that these are from the same distribution at Heysham, Lowestoft and Newlyn at the 0.01 significance level ($p$ values are 0.014, 0.083 and 0.215, respectively). However, this is not the case at Sheerness, with $p$ value 0.00025. 

\begin{figure}[h]
    \centering
    \includegraphics[width=0.99\textwidth]{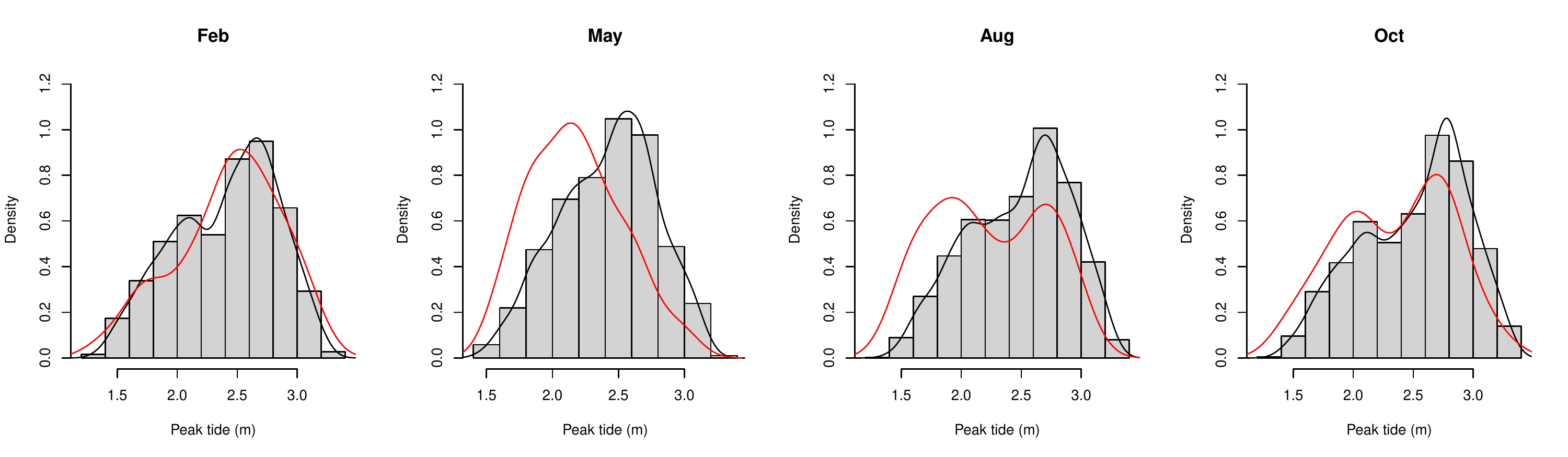}
    \caption{Monthly distributions of peak tides at Sheerness in February, May, August and October. The probability density function of all peak tides (black) and peak tides associated with extreme skew surge (red) are interpolated onto each distribution.}
    \label{monthly_pdf_SHE}
\end{figure}

We investigate this relationship at Sheerness further by studying the dependence on a monthly basis to understand how the skew surge-peak tide dependence changes throughout the year. Again, we compare the distributions of all peak tides and extreme skew surge-related peak tides. Figure~\ref{monthly_pdf_SHE} shows this for February, May, August and October. In May, the mode of the distribution of peak tides associated with extreme skew surges has shifted to a lower value than the distribution of all peak tides. Results from the Anderson-Darling test suggest there is significant evidence, at the 1\% level, to reject the null hypothesis that peak tides and the peak tides associated with extreme skew surge come from different distributions in 7 months (including May, August and October from Figure~\ref{monthly_pdf_SHE}). Therefore, we conclude the skew surge-peak tide independence assumption is not valid for these months, at least. When modelling the dependence of skew surge and peak tide in Section~\ref{acc_ssmt_dep} of the main paper, we recognise that this relationship is changing across the year. 

\begin{figure}
    \centering
    \includegraphics[width=0.4\textwidth]{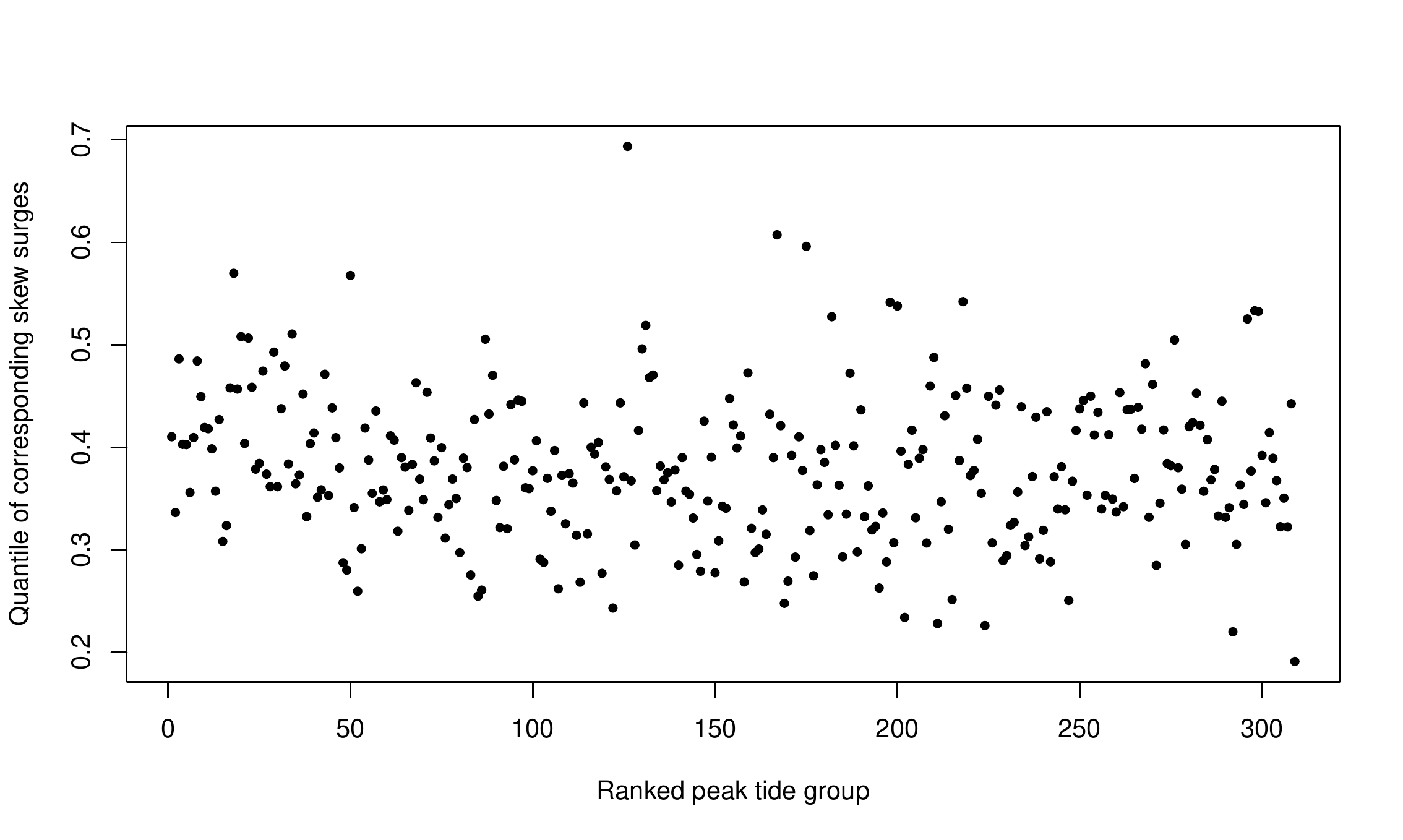}\includegraphics[width=0.4\textwidth]{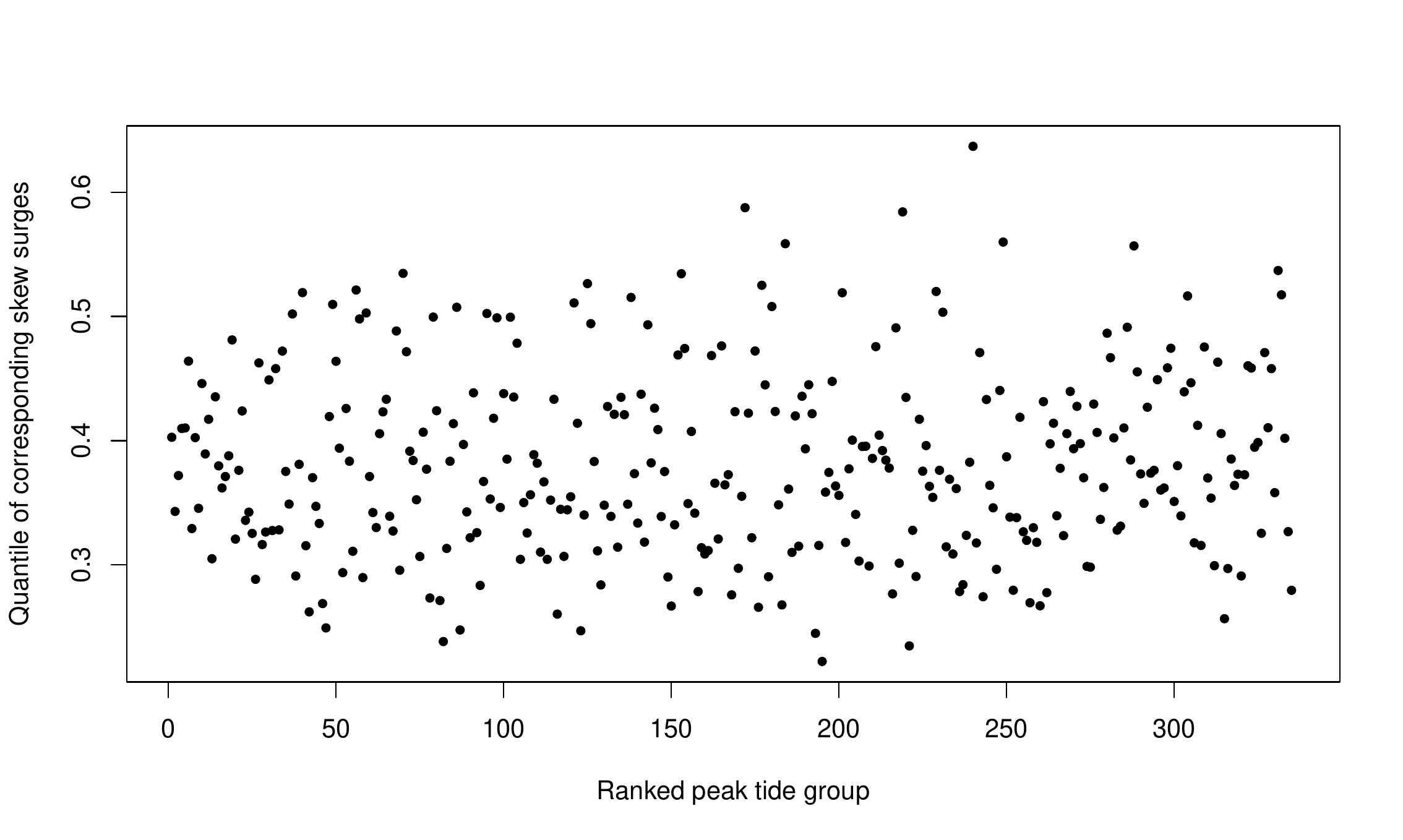}\\
    \includegraphics[width=0.4\textwidth]{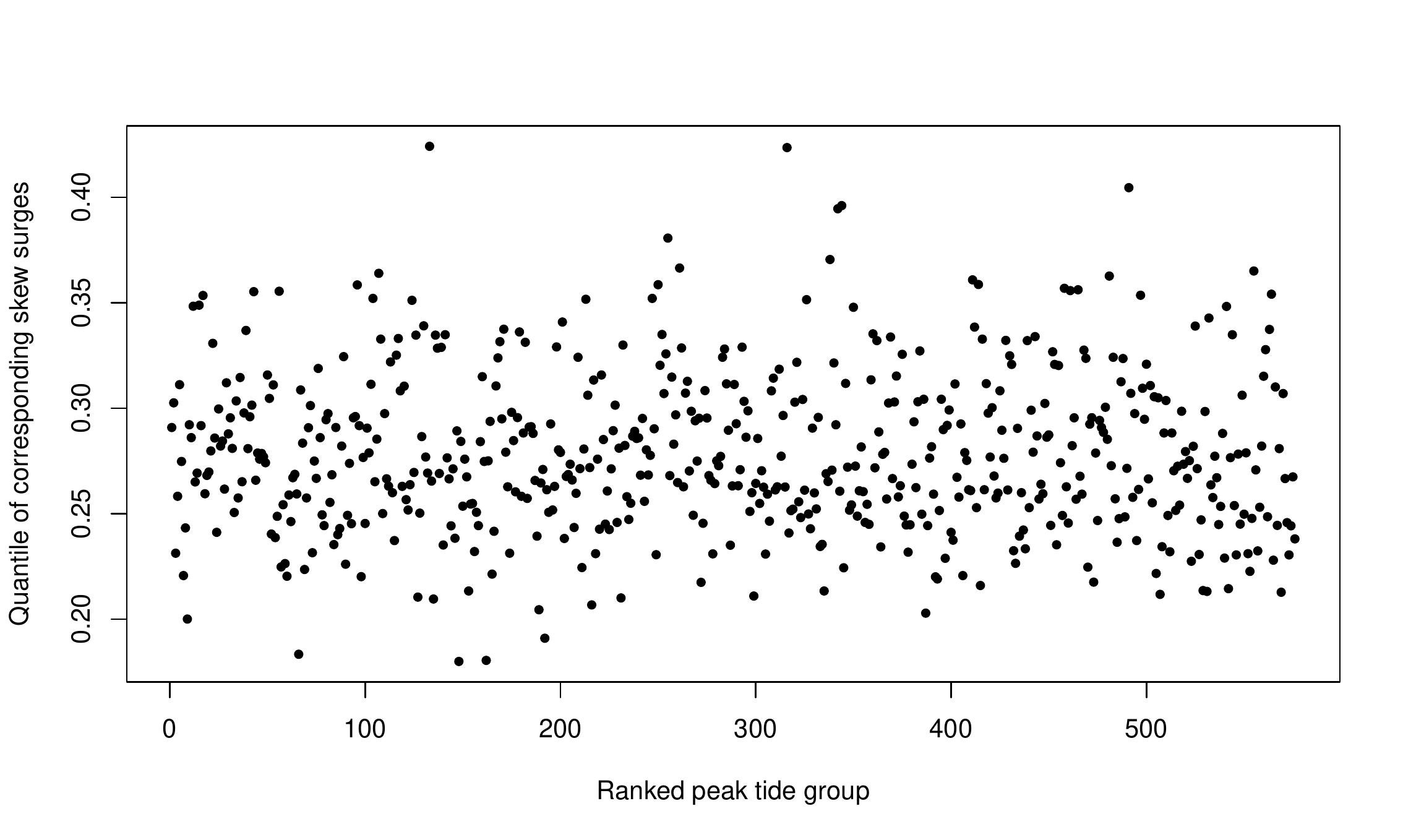}\includegraphics[width=0.4\textwidth]{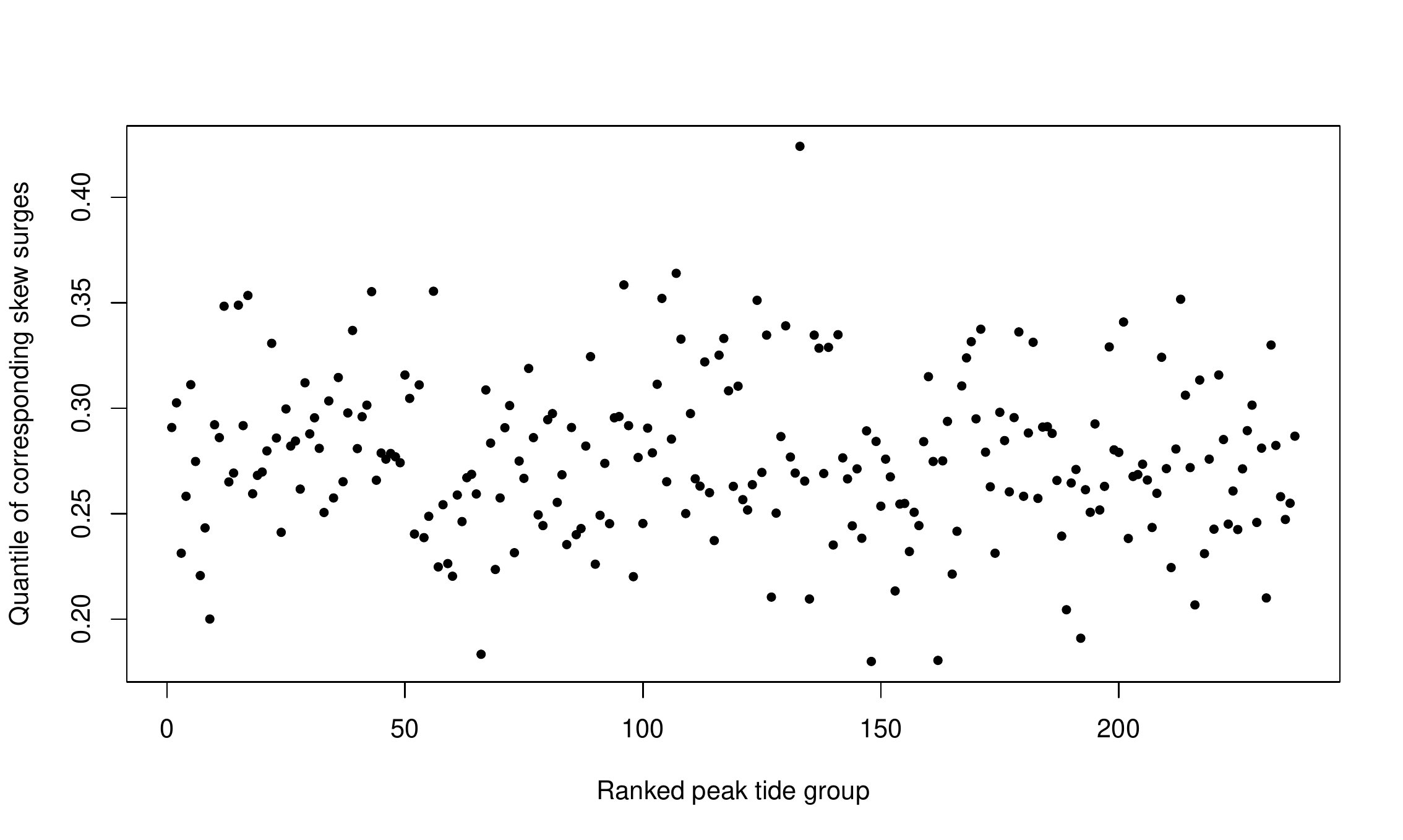}
    \caption{0.95 quantile estimates of skew surges associated with ranked peak tide groups of size 100, at Heysham (top left), Lowestoft (top right), Newlyn (bottom left) and Sheerness (bottom right).}
    \label{quantile_reg}
\end{figure}

To further investigate skew surge-peak tide dependence, we partition the ordered peak tide series into blocks of 100 (so that block 1 corresponds to the 100 smallest peak tide observations). For each block we estimate the 0.95 quantile of the associated skew surges. The estimated quantiles are shown in Figure \ref{quantile_reg}. There is no immediate relationship between the skew surge quantile and block number. However, when we fit a linear model, there is a significant trend at the $<10^{-4}\%$ level at Sheerness and the 0.1\% level at Newlyn, but no significant trend at Lowestoft and Heysham. Since our other tests have not found dependence at Newlyn, this relationship is likely to be a physically small relationship but statistically significant due to the length of the data series. However, at Sheerness, it is likely this is due to skew surge-peak tide dependence based on our other findings. 

\section{Climate Model Data}\label{suppmat_climmodel}
In Section~\ref{ss_mt_indep_supp} here (and Section~\ref{ss_mt_indep} of the main paper) we identify weak dependence between skew surge and peak tide at Sheerness; we account for this in our skew surge model. Our results show that incorporating peak tide as a covariate on the rate and scale parameter of the GPD for extreme skew surges improved the model fit. In this section, we make the same comparisons on a 483-year data set from a hydrodynamical model driven by a regional climate model (HadGEM3-GC3-MM). \citeSM{howardWilliams2021} present this model to generate a dataset of 483-year present-day surges at sites on the UK National Tide Gauge Network. They use their simulations to review the skew surge-peak tide independence assumption at Sheerness, and demonstrate that extreme skew surges are more likely to occur on larger peak tides. \citeSM{darcytawn21} evaluate this assumption using a simplified version of the method presented in the main paper. For the scale parameter, we compare the models given in equations~\eqref{sinescale} and~\eqref{eqn::ss_scale_tide1}, and we refer to these as Models $S2$ (without a tide covariate) and $S4$ (with a tide covariate), respectively, as in the main paper. Models for the rate parameter are given by equations~\eqref{rate_day} and \eqref{eqn::ss_rate_tide}, we refer to these as Model $R0$ (without a tide covariate) and $R1$ (with a tide covariate), respectively.

For the climate model skew surge data, we begin by comparing models for the scale parameter only. Model $S4$ reduces both AIC and BIC by 28.75 and 21.03, respectively, relative to Model~$S2$ suggesting that the extra parameter that captures variation with peak tide tide is necessary. The likelihood ratio test strongly agrees with this, giving a significant $p$ value of the order $10^{-8}$. We estimate the tidal coefficient to be $\hat\gamma_\sigma = -0.013$ with 95\% confidence interval (-0.018, -0.0081). This suggests that more extreme skew surges occur on lower peak tides, as found in our exploratory analysis. Since the confidence interval doesn't contain 0 the tidal coefficient is significant. This parameter estimate is reassuringly close to the corresponding estimate for the observed data of $\hat\gamma_\sigma=-0.012$ $(-0.026, 0.0011)$ estimated in Section~\ref{acc_ssmt_dep} of the main paper. Whilst this suggests threshold excesses occur on lower peak tides, the confidence interval contains 0 so this result for the observed data is not statistically significant. However, this result is supported by the climate model data at Sheerness, which are based on physical reasoning with no data measurement issues and no issues with changes in the tide gauge and estuary over time.

Next, we compare models for the rate parameter on the climate model data. We find that Model $R1$ reduces AIC by 88.4 and BIC by 56.3 when compared with Model $R0$, this suggests that adding a peak tidal covariate to the rate parameter is important. Therefore, we make the same conclusions on 483-years of climate model data, as we do on 37 years of observed data. This shows the rate model parameterisation is supported empirically and physically. When we cannot reasonably assume skew surge and peak tide are independent, accounting for their dependence is important as it can lead to practical differences in the sea level return level estimates.

\section{Temporal Skew Surge Dependence}\label{sec::suppmat_exi}
\begin{figure}
    \centering
    \includegraphics[width=0.83\textwidth,height=5cm]{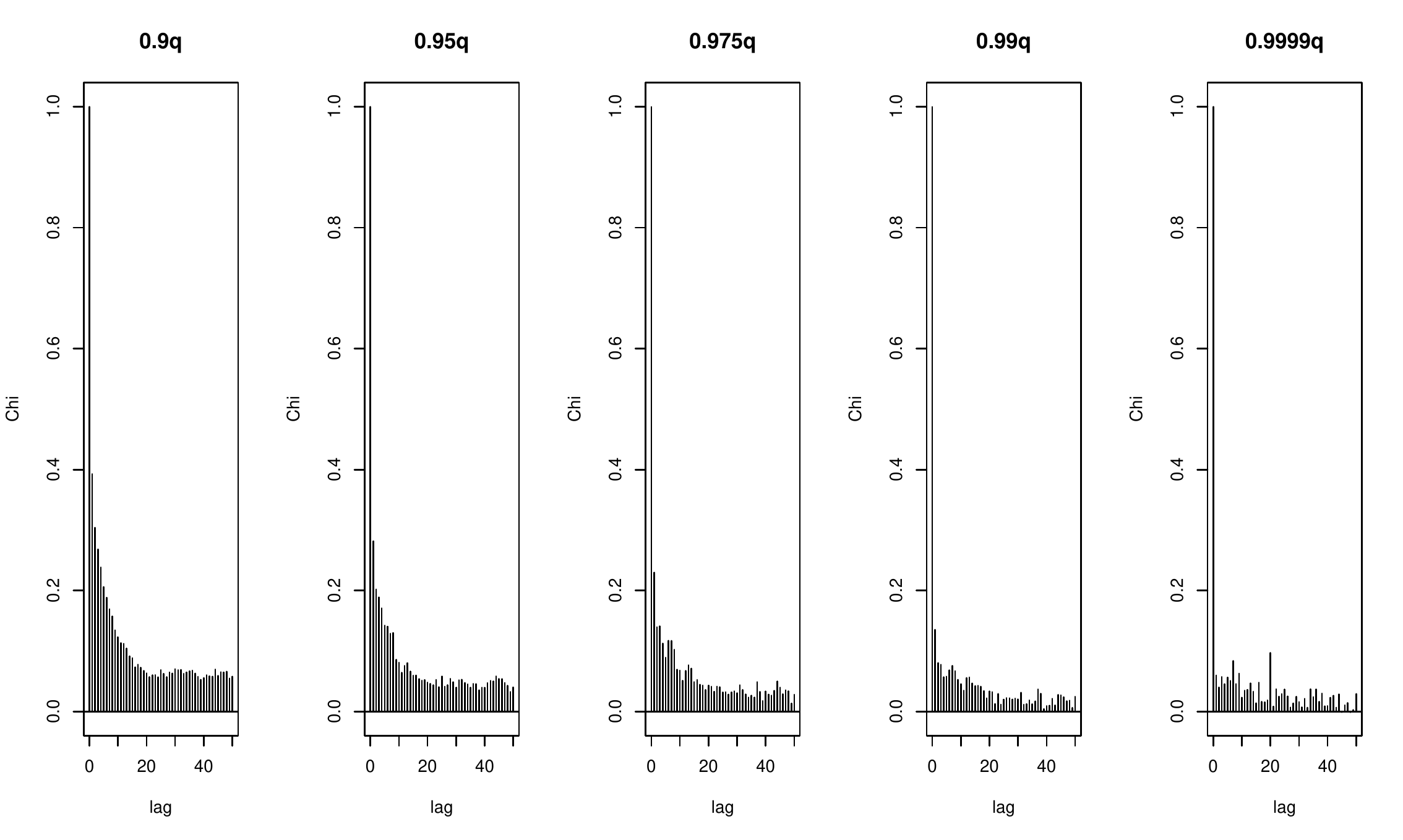} \\
        \includegraphics[width=0.83\textwidth,height=5cm]{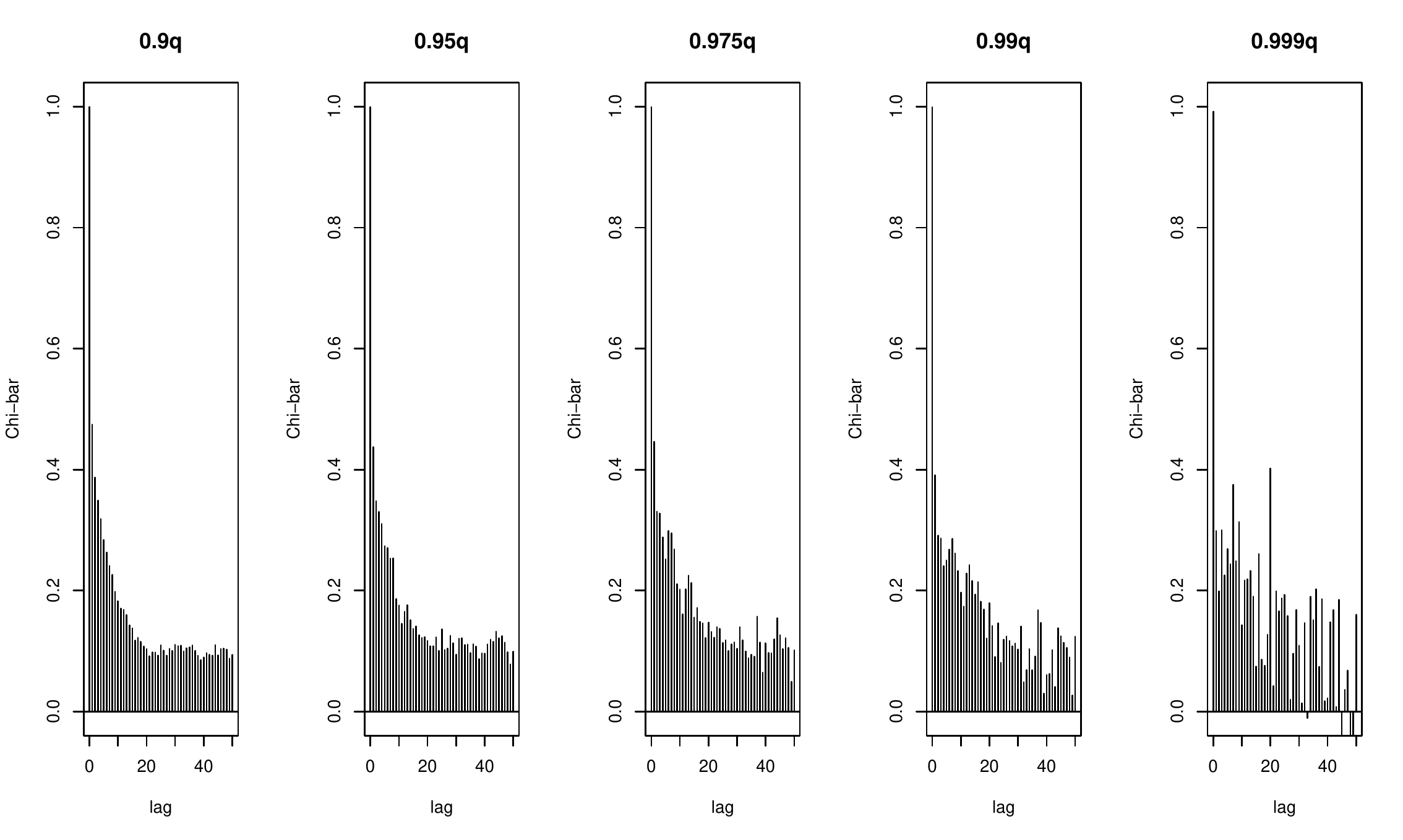}
    \caption{Estimates of $\chi$ (top row) and $\bar\chi$ (bottom row) for Heysham for exceedances of the 0.9, 0.95, 0.975, 0.99 and 0.999 quantiles (from left to right column) at various lags.}
    \label{fig::chiHEY}
\end{figure}

\begin{figure}[h]
    \centering
    \includegraphics[width=0.83\textwidth,height=5cm]{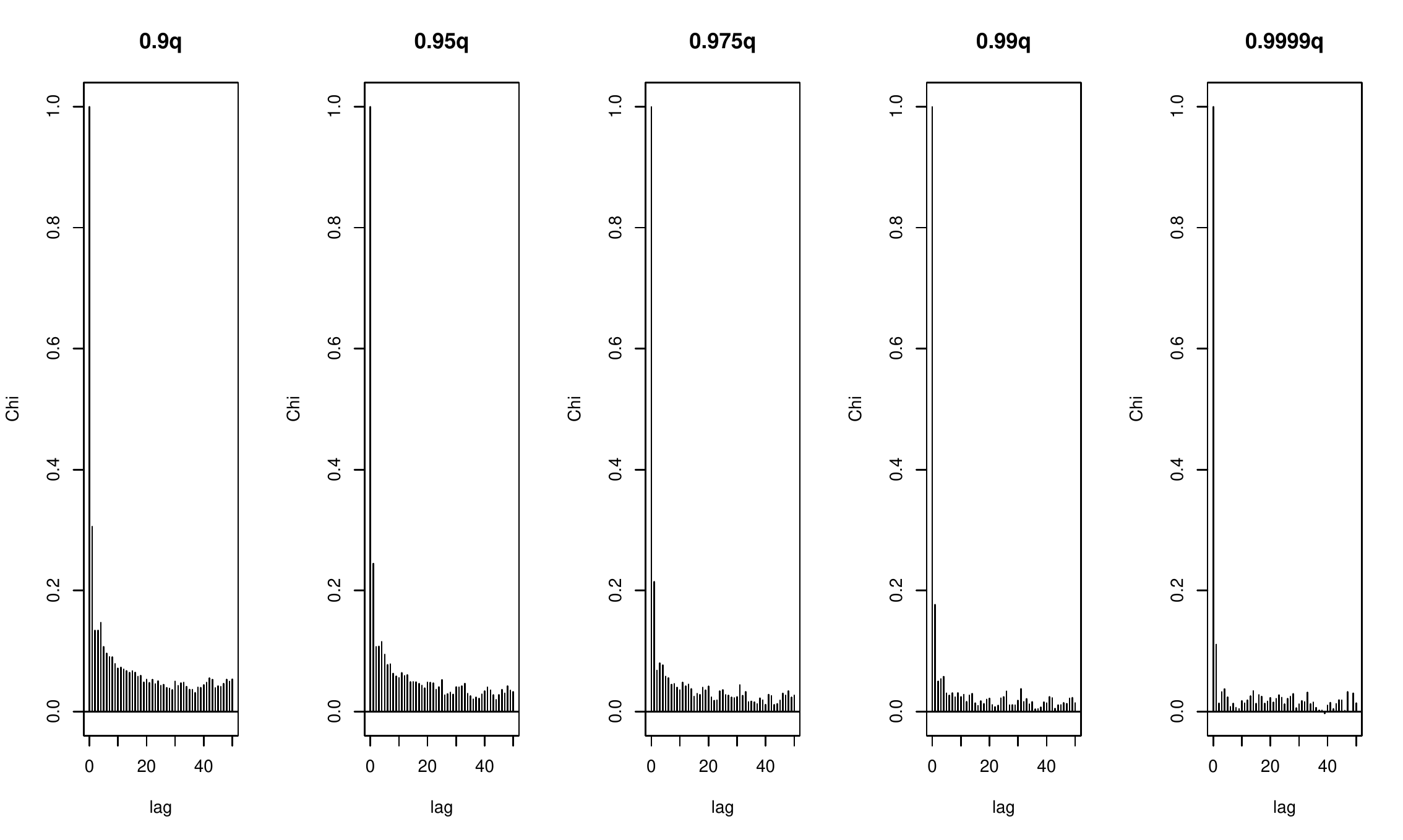} \\
    \includegraphics[width=0.83\textwidth,height=5cm]{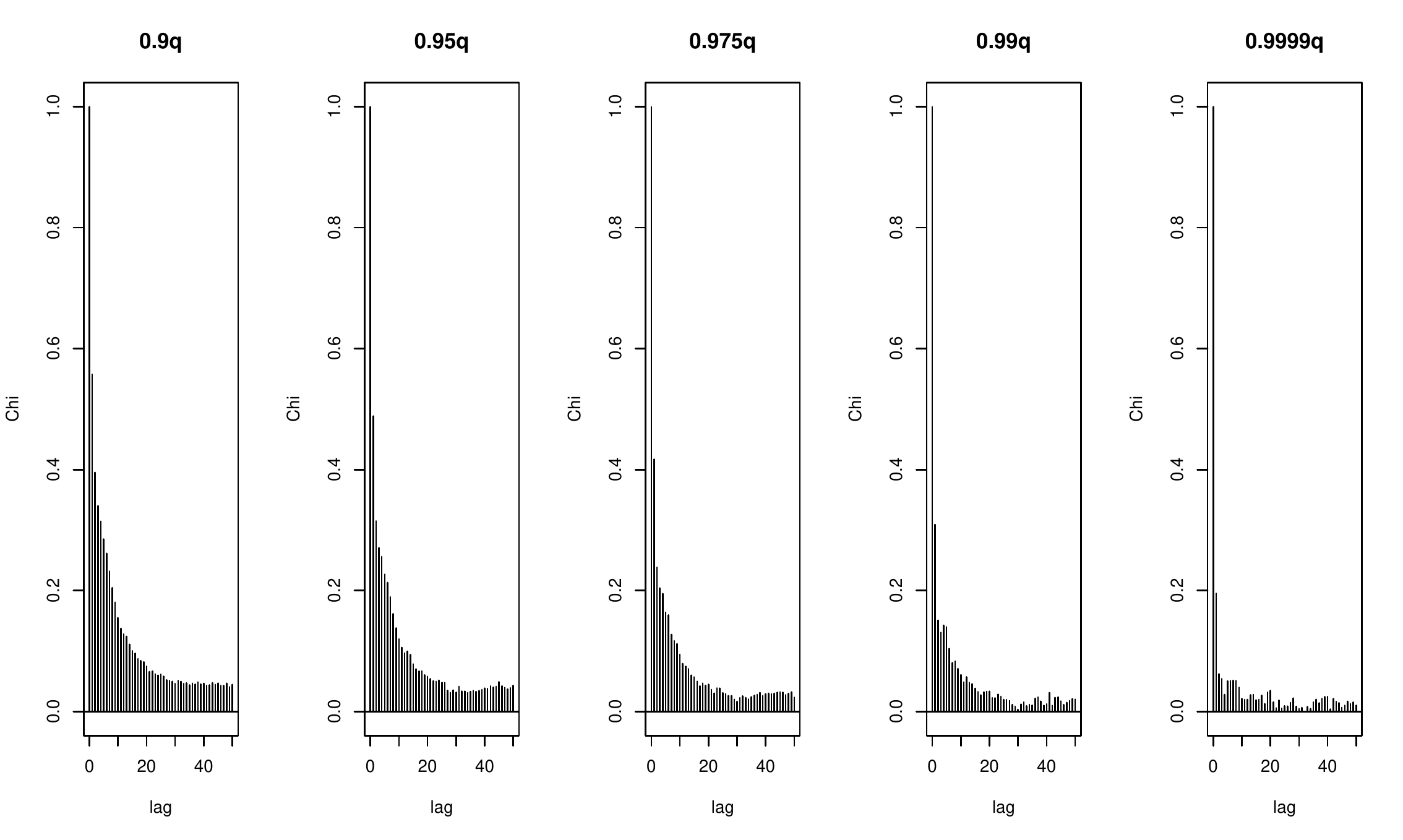} \\
    \includegraphics[width=0.83\textwidth,height=5cm]{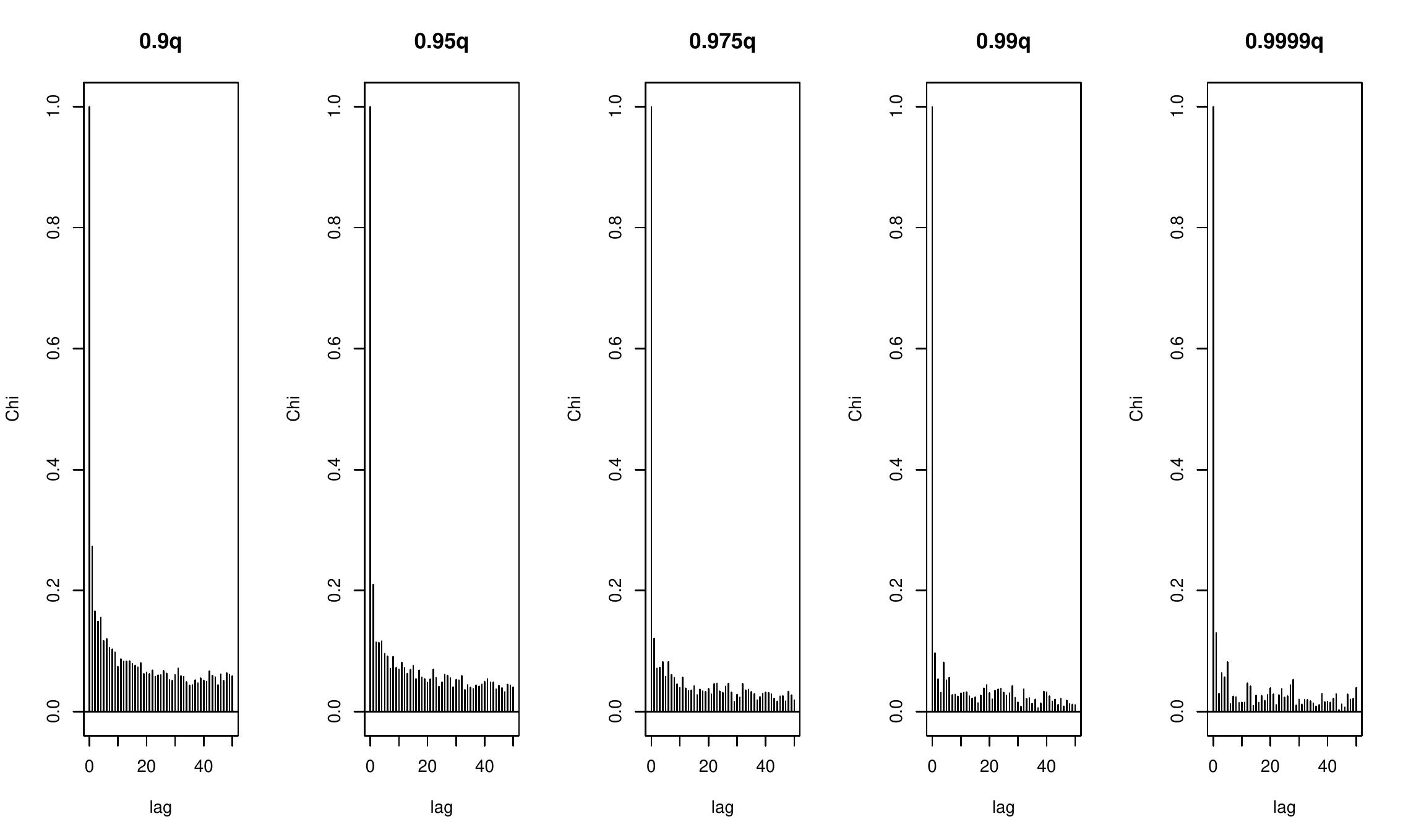}
    \caption{Estimates of $\chi$ for Lowestoft, Newlyn and Sheerness (from top to bottom row) for exceedances of the 0.9, 0.95, 0.975, 0.99 and 0.999 quantiles (from left to right column) at various lags.}
    \label{fig::chi}
\end{figure}

\begin{figure}[h]
    \centering
    \includegraphics[width=0.83\textwidth,height=5cm]{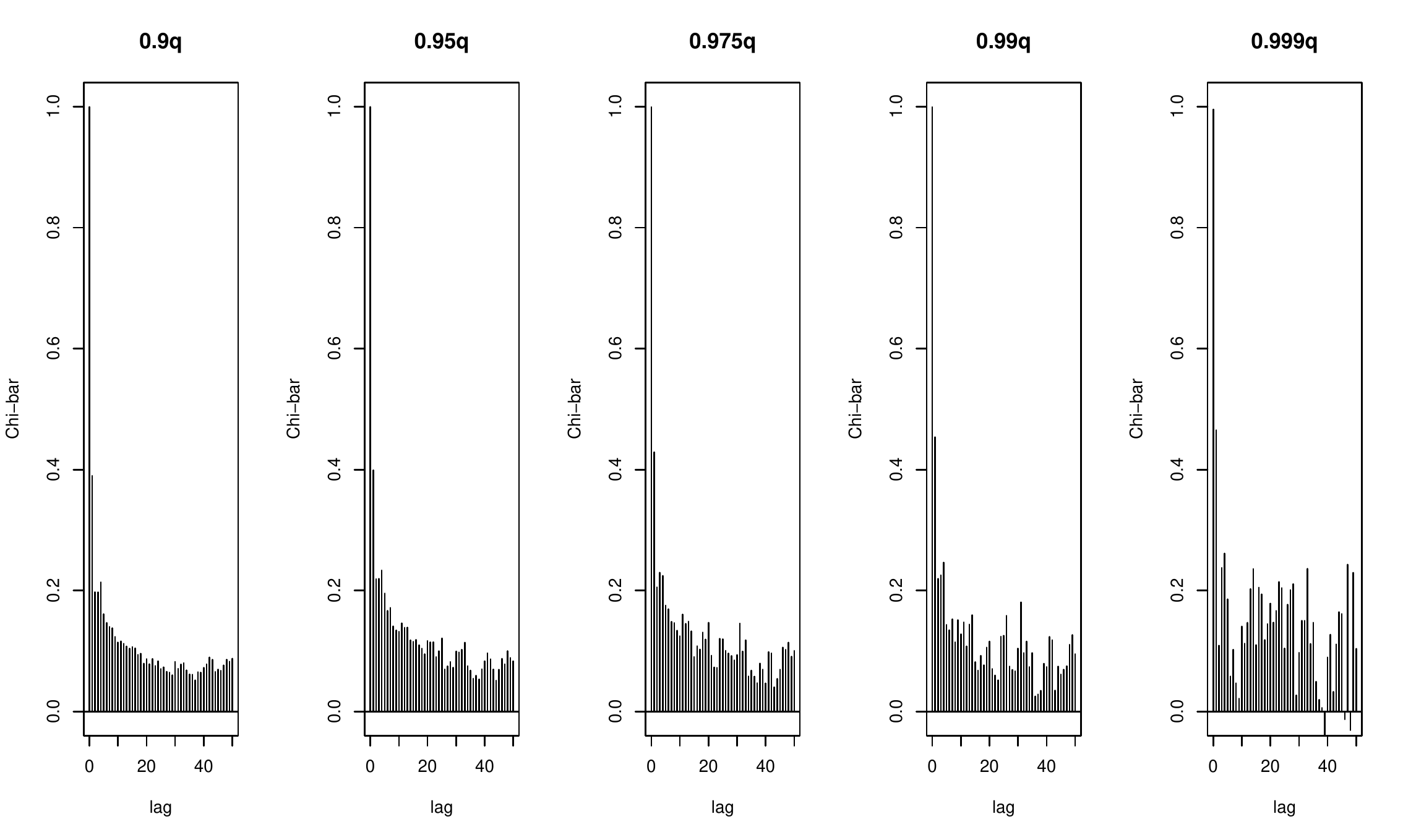} \\
    \includegraphics[width=0.83\textwidth,height=5cm]{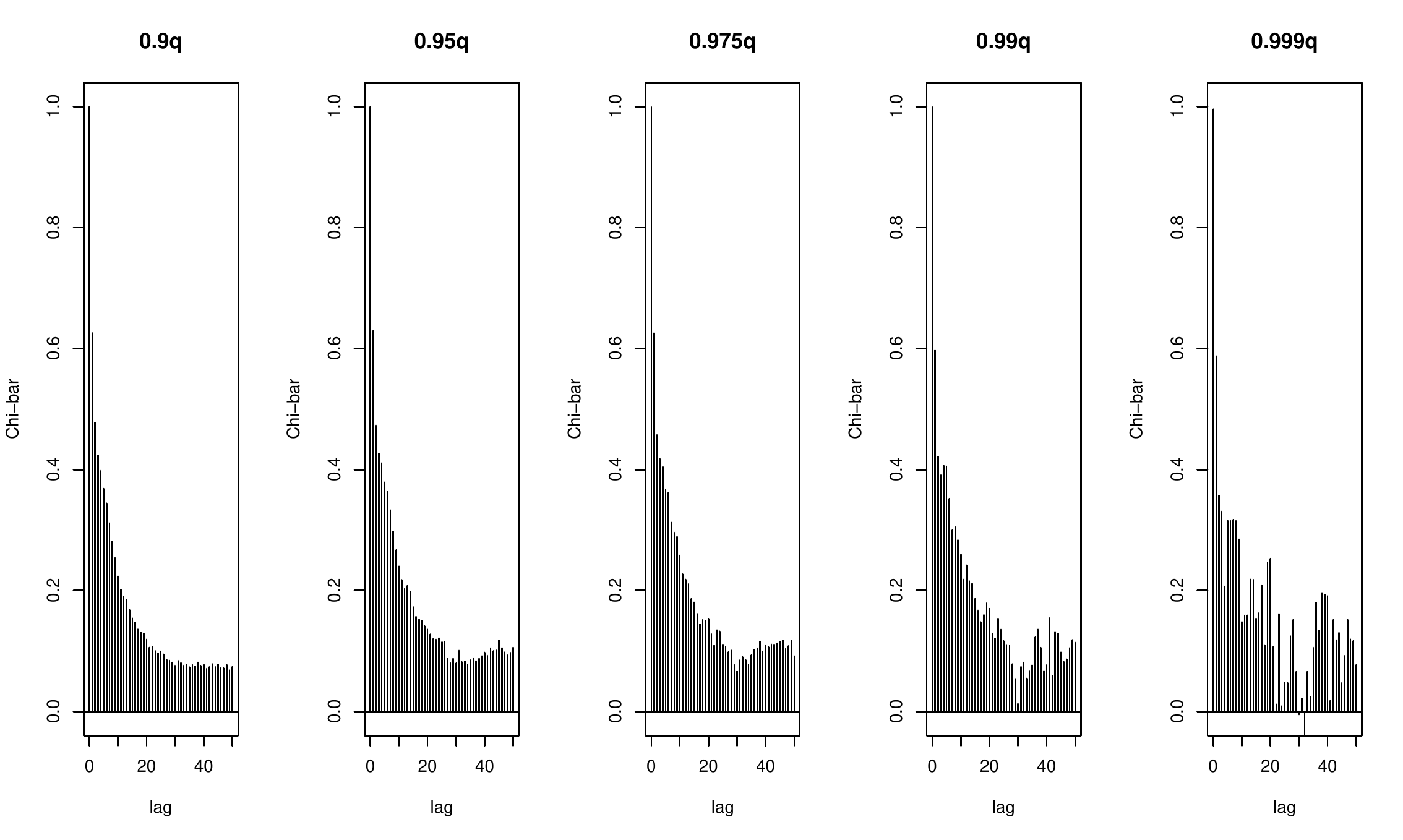} \\
    \includegraphics[width=0.83\textwidth,height=5cm]{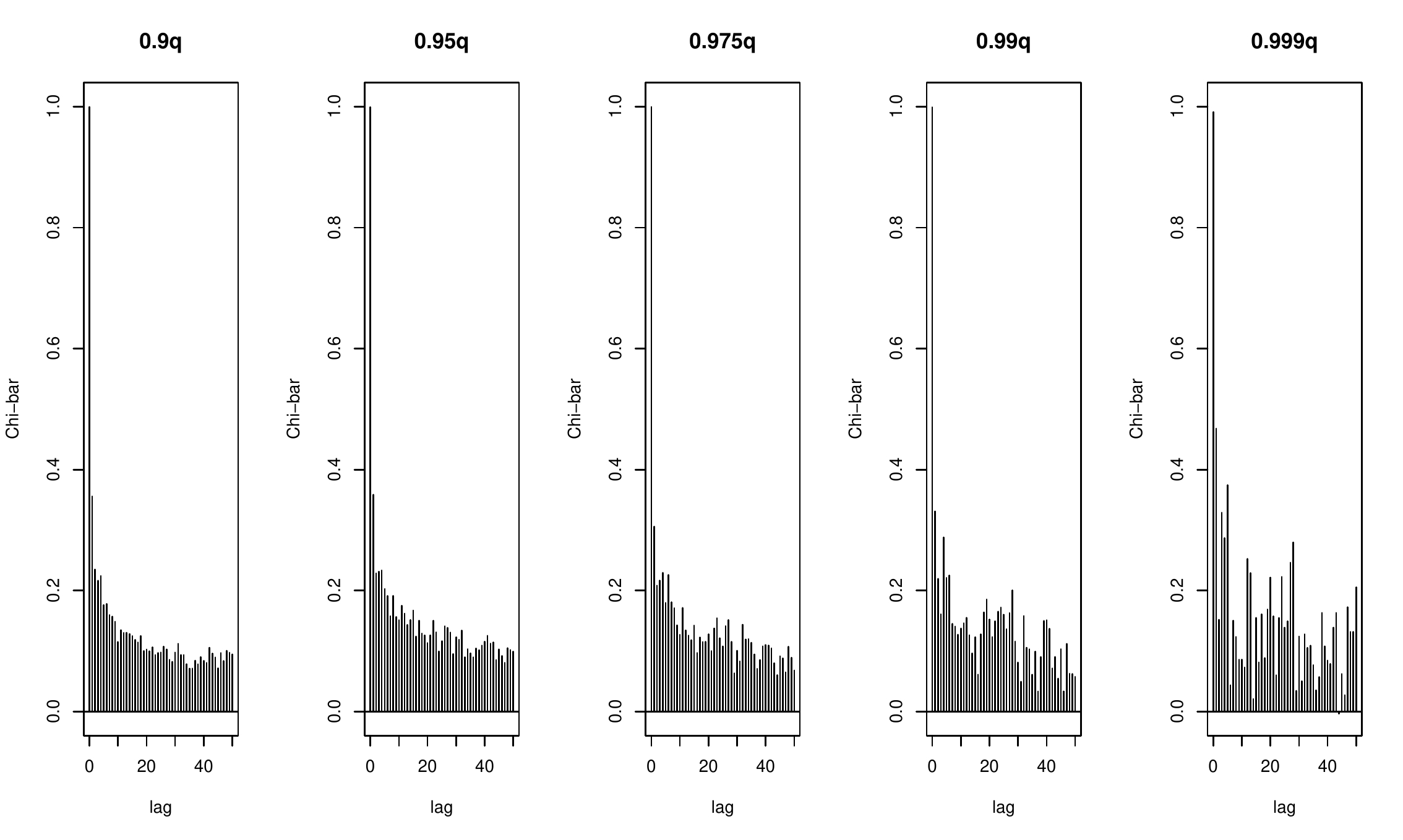}
    \caption{Estimates of $\bar\chi$ for Lowestoft, Newlyn and Sheerness (from top to bottom row) for exceedances of the 0.9, 0.95, 0.975, 0.99 and 0.999 quantiles (from left to right column) at various lags.}
    \label{fig::chibar}
\end{figure}

Here we provide further details on skew surge temporal dependence discussed in Section~\ref{ss_seasonality} of the main paper. We study the two key measures of extremal dependence, these are a measure of asymptotic dependence $\chi$ and of asymptotic independence $\bar\chi$ for each site. For random variables $Y_i$ and $Y_{i+\tau}$ separated by lag $\tau$ from a stationary sequence, with $y^{F}$ the upper endpoint of their distribution $F$, \citeSM{Coles1999} define \begin{equation}
    \chi_\tau=\lim_{y\rightarrow y^F} \mathbb{P}(Y_{i+\tau}>y|Y_{i}>y)
\end{equation} as a measure of asymptotic dependence where $\chi_\tau\in[0,1]$. If $\chi_\tau\in(0,1]$, we say that $Y_i$ and $Y_{i+\tau}$ are asymptotically dependent; this means there is non-zero probability of $Y_{i+\tau}$ being large when $Y_{i+\tau}$ is large at all extreme levels. Whereas $\chi_\tau=1$ and $\chi_\tau=0$ corresponds to perfect dependence and asymptotic independence, respectively. Therefore, $\chi_\tau$ fails to signify the level of asymptotic independence, so~\citeSM{Coles1999} also define the measure $\bar\chi_\tau$ as \begin{equation}
    \bar\chi_\tau=\lim_{y\rightarrow y^F}\frac{2\log\mathbb{P}(Y_i>y)}{\log\mathbb{P}(Y_i>y, Y_{i+\tau}>y)}-1,
\end{equation} where $\bar\chi_\tau\in(-1,1]$. Asymptotic dependence and asymptotic independence correspond to $\bar\chi=1$ and $\bar\chi<1$, respectively, whilst $0<\bar\chi_\tau<1$ and $-1<\bar\chi_\tau<0$ correspond to positive and negative association, respectively, and when $\bar\chi_\tau=0$ corresponds to near independence. We evaluate both measures with $y$ at different quantiles of the distribution in Figure~\ref{fig::chiHEY} for Heysham, and Figures~\ref{fig::chi} and~\ref{fig::chibar} for the remaining sites.

\begin{figure}
    \centering
    \includegraphics[width=0.4\textwidth]{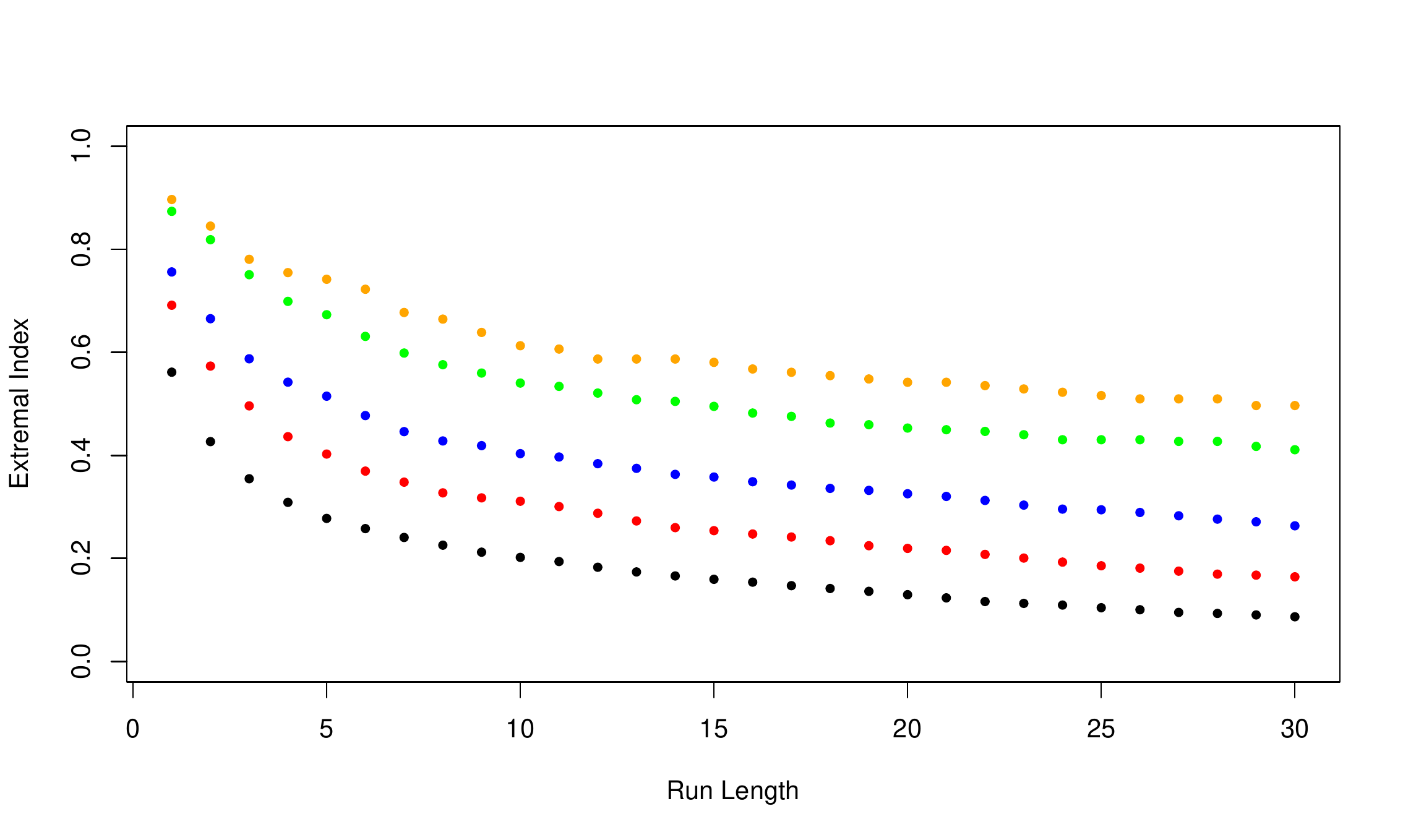}    \includegraphics[width=0.4\textwidth]{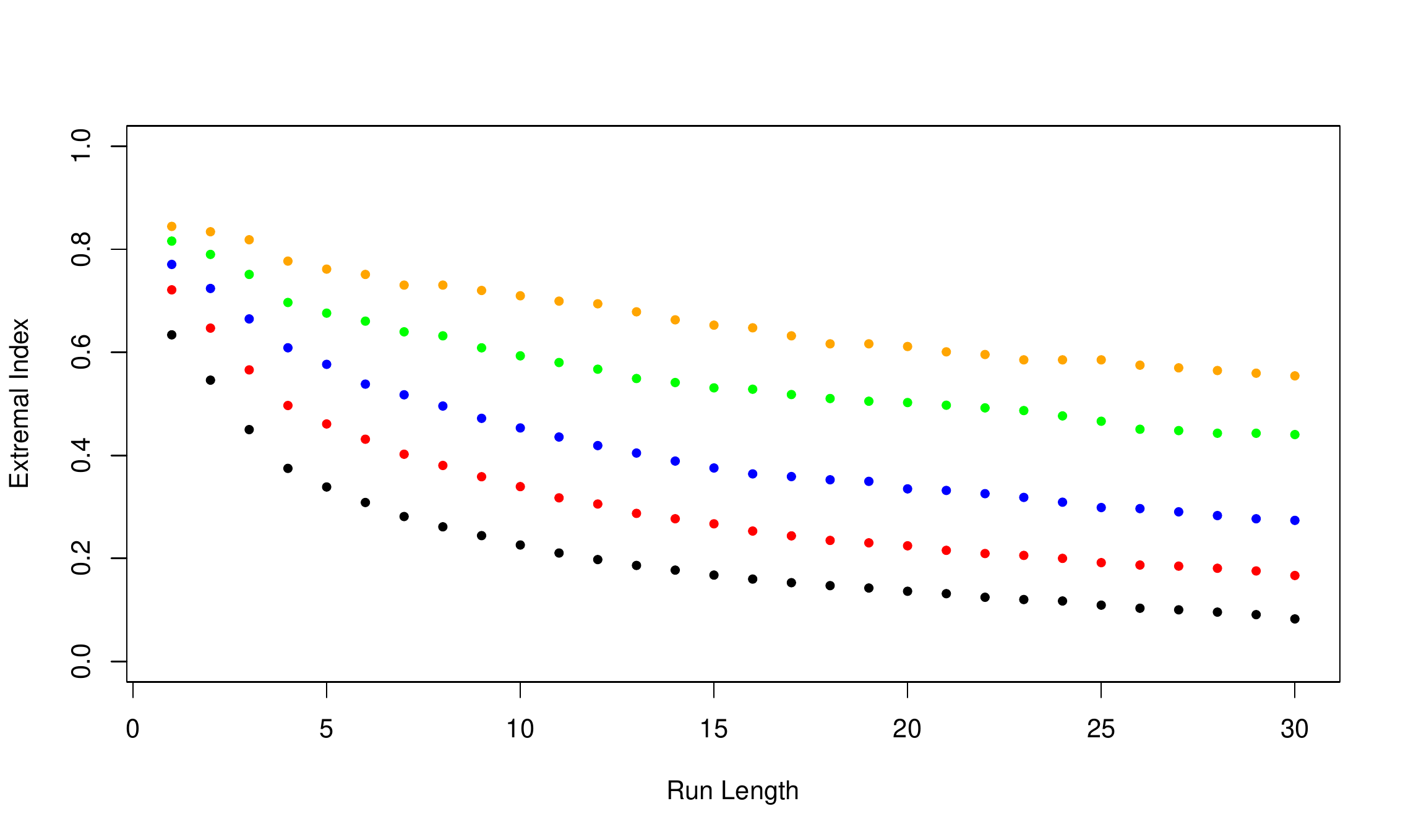}\\
    \includegraphics[width=0.4\textwidth]{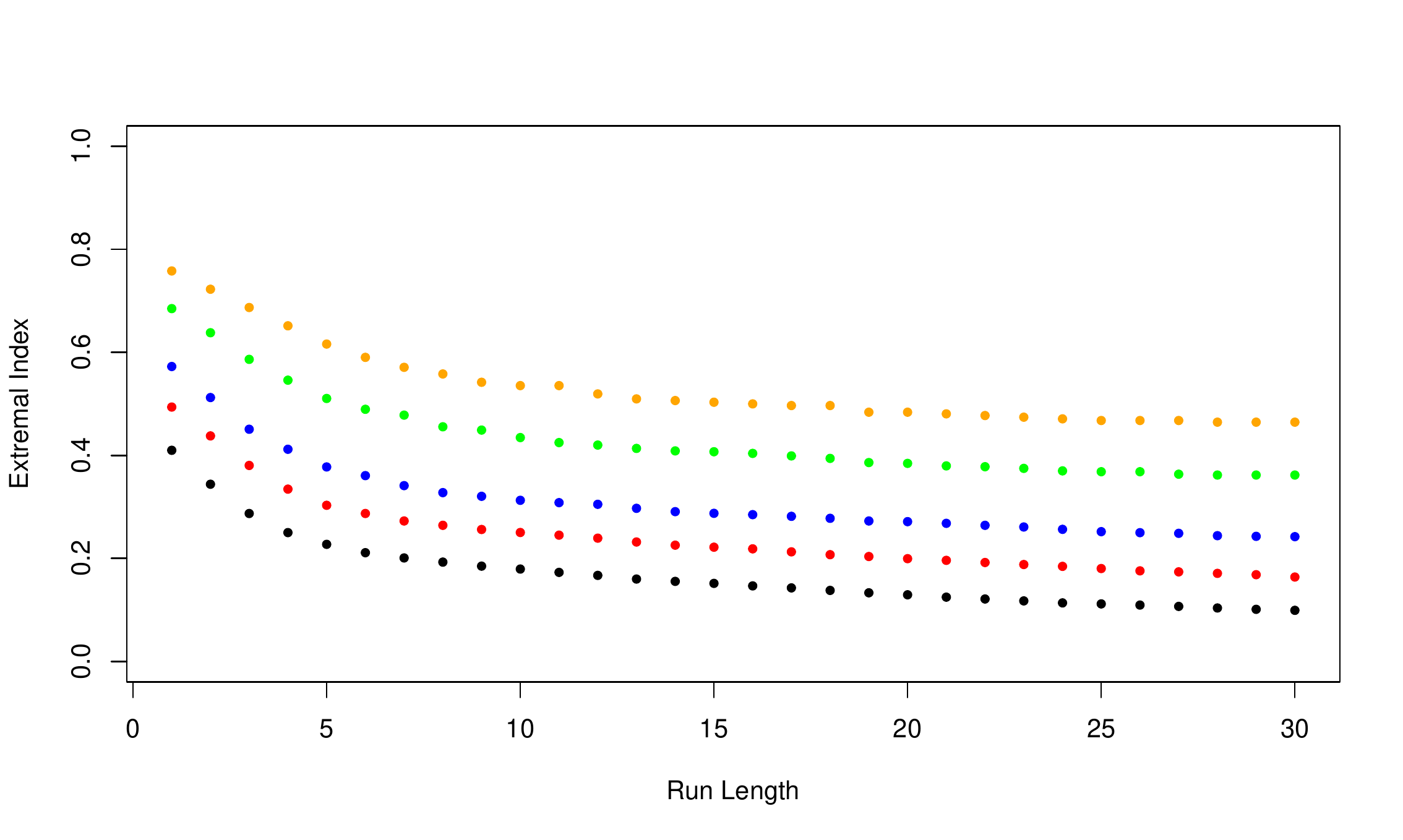}
    \includegraphics[width=0.4\textwidth]{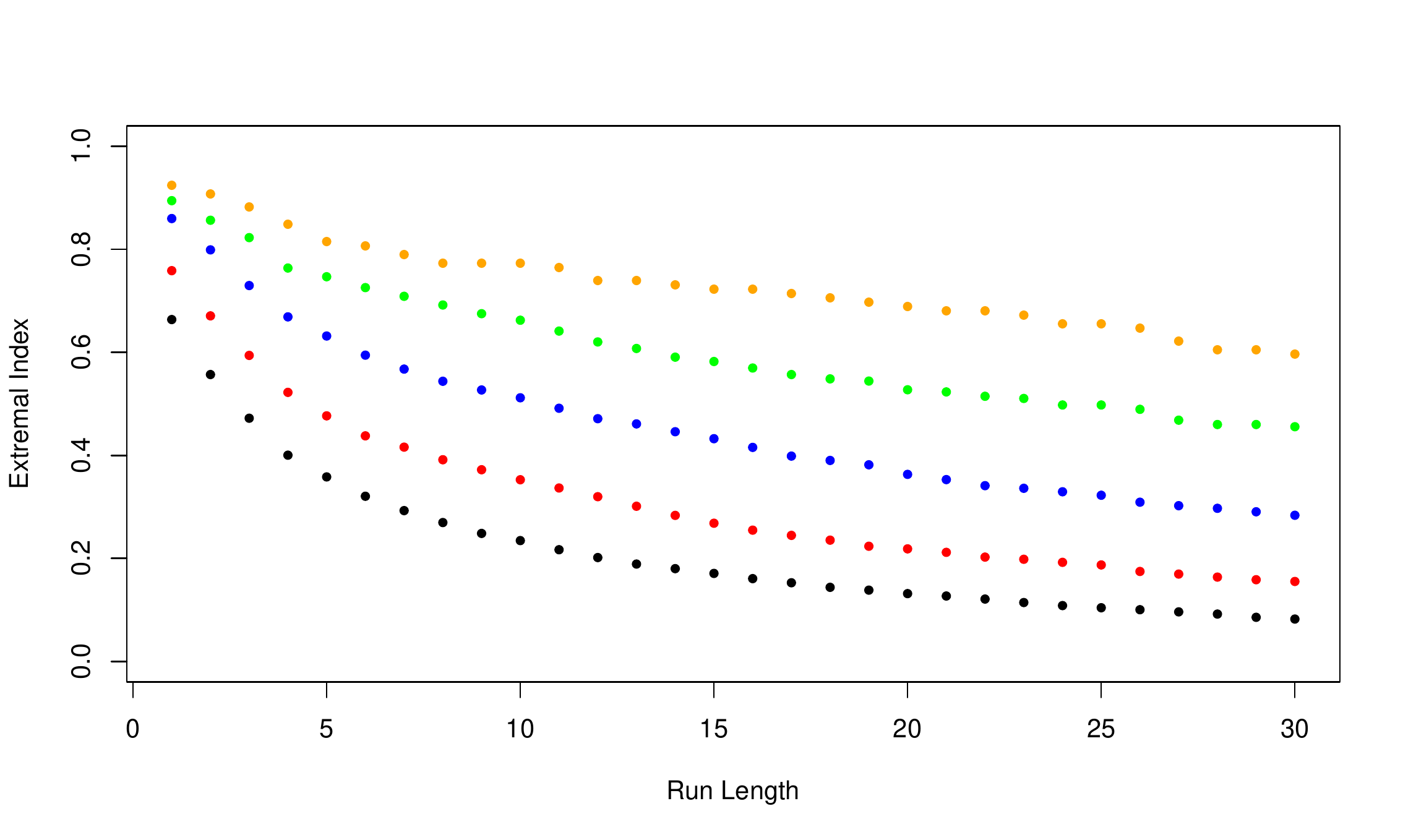}
    \caption{Empirical (runs) estimates of the extremal index $\theta$ for skew surge, at various quantiles and run lengths at Heysham (top left), Lowestoft (top right), Newlyn (bottom left) and Sheerness (bottom right). The estimates use thresholds which are taken to be quantiles 0.9 (black), 0.95 (red), 0.975 (blue), 0.99 (green) and 0.999 (orange).}
    \label{fig::exi_vartrl}
\end{figure}

\begin{figure}[h]
    \centering
    \includegraphics[width=0.45\textwidth]{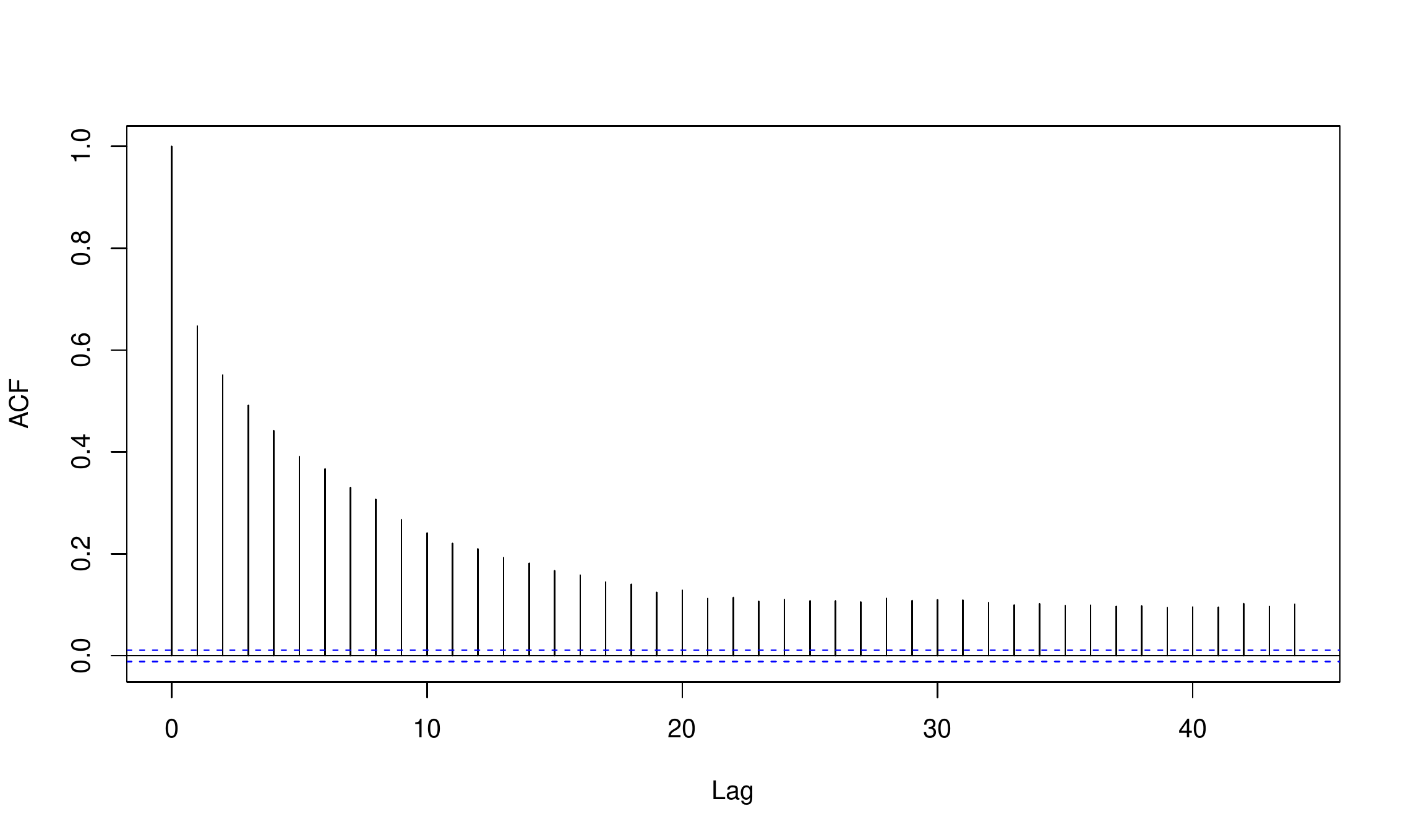}\includegraphics[width=0.45\textwidth]{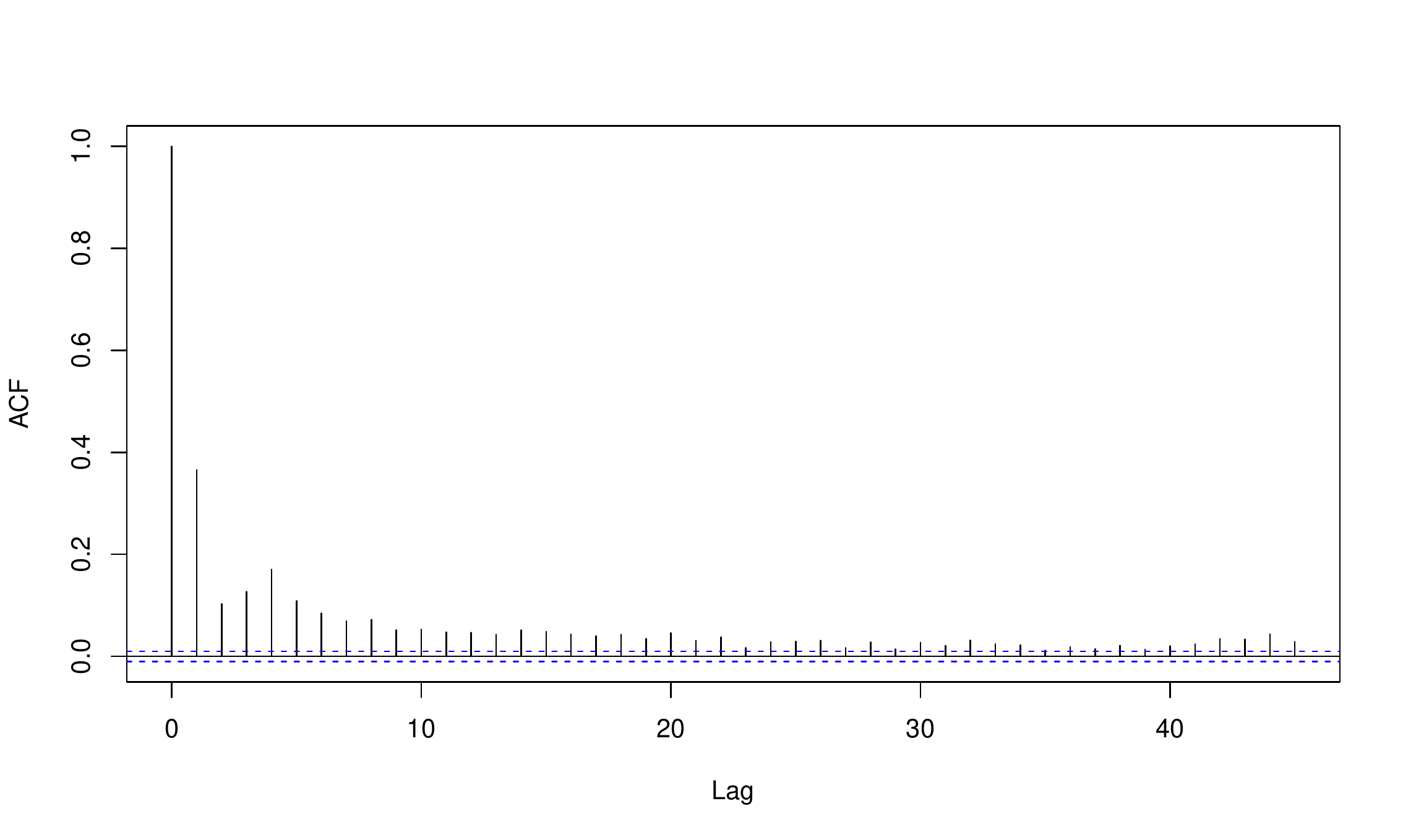} \\
    \includegraphics[width=0.45\textwidth]{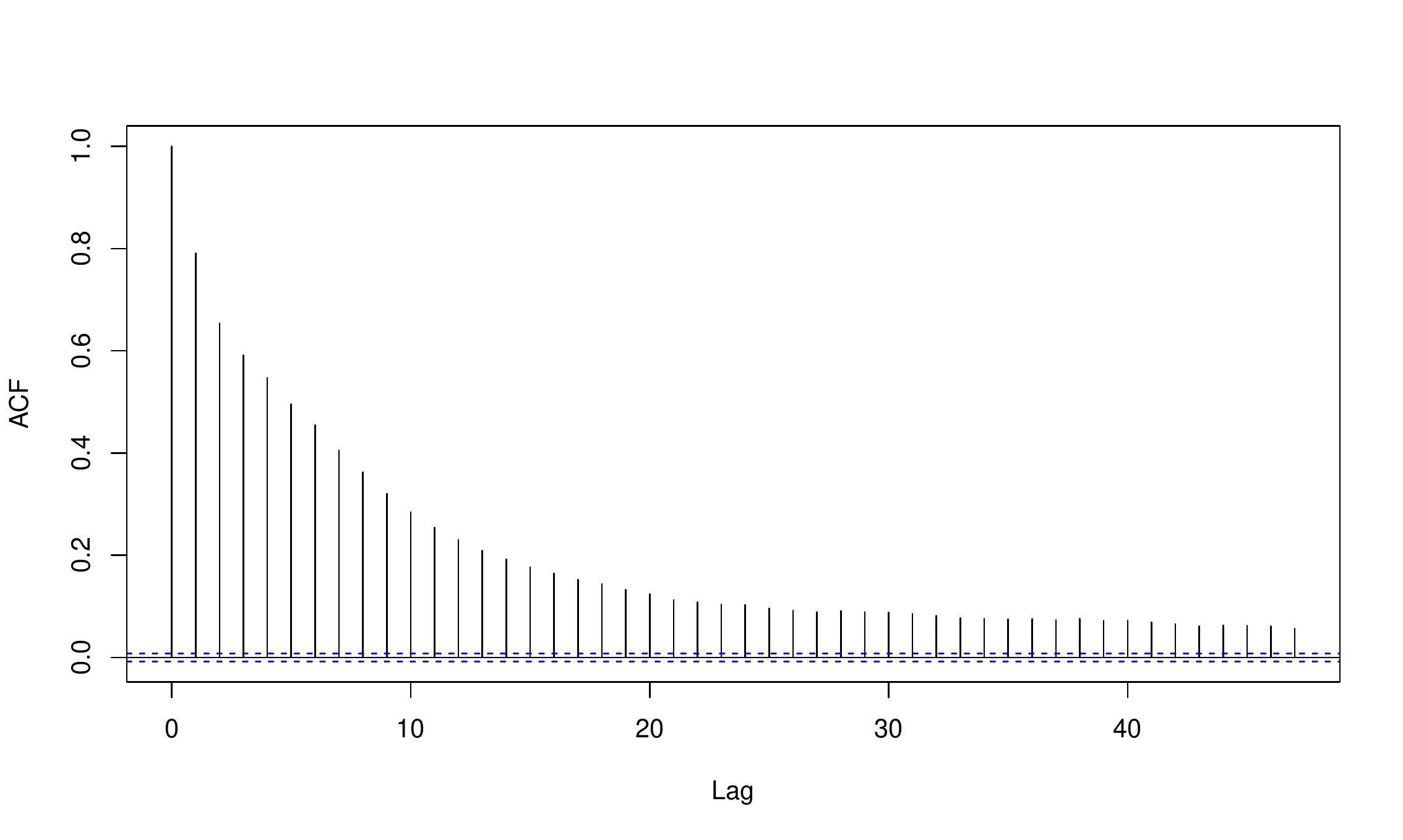}\includegraphics[width=0.45\textwidth]{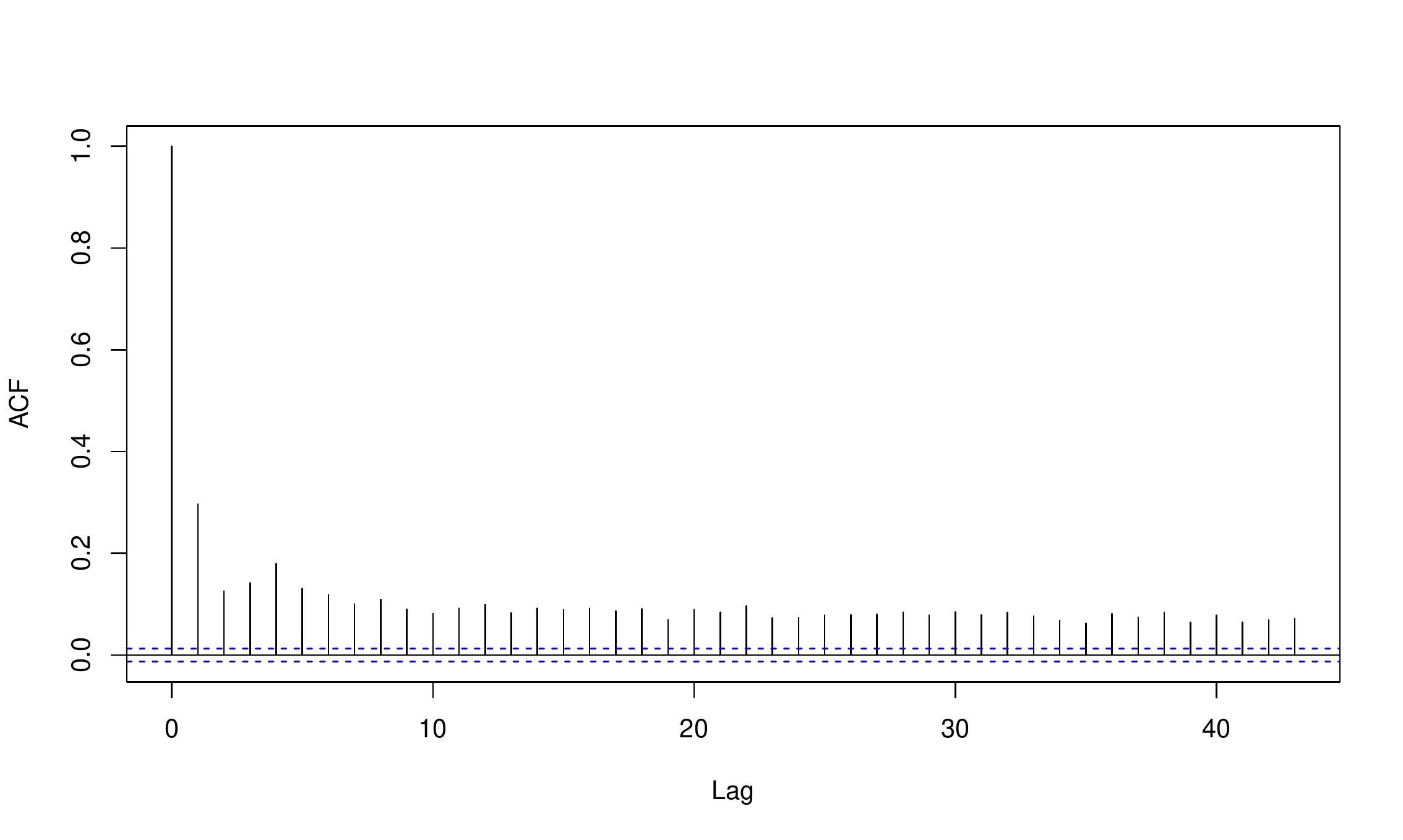}
    \caption{Autocorrelation function (acf) plots for Heysham (top left), Lowestoft (top right), Newlyn (bottom left) and Sheerness (bottom right).}
    \label{fig::acf_ss}
\end{figure}

Firstly, we use estimates of $\chi$ and $\bar\chi$ to choose a high threshold $y$ to empirically estimate the extremal index $\theta$ using the runs method. We are interested in choosing a threshold where these measures tend to zero without significant noise; the 0.95 quantile is sufficient for each site. We explore the sensitivity of our estimate to this threshold choice in Figure~\ref{fig::exi_vartrl}; as we increase the threshold, the estimate of $\theta$ increases. This is what we expect since exceedances of lower thresholds are likely to exhibit more dependence than exceedances of higher thresholds. We also explore the sensitivity to run length $r$ in Figure~\ref{fig::exi_vartrl}. Estimates of the extremal index are generally higher at Newlyn and Sheerness, which agrees with the acf values in Figure~\ref{fig::acf_ss}. The choice of run length and threshold level are important because they have a significant influence on the $\theta$ estimate, so we choose these carefully.

Since our empirical estimates are sensitive to the threshold level, we develop a parametric model for the subasymptotic extremal index $\hat\theta(y,r)$, that is dependent on the skew surge level $y$; this is given in equation~\eqref{eqn::exi_model} of the main paper. Figure~\ref{fig::HEY_exifun} of the main paper shows a consistent model fit with the empiricals at Heysham, with parameter estimates $\hat\theta=1$ and $\hat\psi=0.33$ for $r=2$. Figure~\ref{fig::exifun_all} shows the fit at the remaining sites and these also match closely with the empiricals. The parameter estimates are $\hat\theta=1,0.95,0.89$ and $\hat\psi=0.42,0.17,0.21$ for $r=10,1,10$ at Lowestoft, Newlyn and Sheerness, respectively. As expected, $\hat\theta$ lies close to 1 in all cases, this represents the limiting extremal index and the case of independence. The estimate of $\psi$ tells us about the speed of convergence to $\hat\theta$, so that Newlyn converges fastest whilst Lowestoft is the slowest.

\section{Derivation of Expression for Probability~(\ref{eqn::annmaxprob_model})}\label{sec::annmaxprob_form}
We are interested in deriving an analytical expression for the probability $\hat P_{M}(j;z)$ that a randomly selected sea level (annual maxima) $M$ is from month $j$ given it equal to some level $z$, given by, 
\begin{equation}
    \hat P_M(j;z)=\hat{\mathbb{P}}(m(M) = j|M=z),
\end{equation} 
where $m(M)$ denotes the month of occurrence of the variable $M$ and $\hat{\mathbb{P}}(\cdot)$ is under our final model of Section~\ref{ss_dep} in the main paper. This is probability~\eqref{eqn::annmaxprob_model} in the main paper and we use this in Section~\ref{sl_seasonality} to evaluate extreme sea level seasonality. We are specifically interested when $z=z_p$, a level with an associated annual exceedance probability $p\in[0,1]$, so that $z$ is a return level derived from expression~\eqref{eqn::sldist_exi}.

As the distribution of $M$ varies with year $k$ due to the tidal variations, we begin by conditioning on a fixed year $k$, so that we only look at sea levels within a specific year and consider the probability $\hat P_{M^{(k)}}(j;z)$. We rewrite this in terms of the distribution and density of the month $j$ maxima sea level, $F_{M_j^{(k)}}$ and $f_{M_j^{(k)}}$, respectively, and the density of the annual maxima sea level $f_{M^{(k)}}$, each conditional on some year $k=1,\ldots,K$, where $K$ is the total number of years of observation, \begin{align*}
    \hat P_{M^{(k)}}(j;z) &= \frac{f_{M_j^{(k)}}(z)\prod\limits_{\substack{J=1,\ldots,12 \\ J\neq j}}\mathbb{P}(M_{J}^{(k)}<z)}{f_{M^{(k)}}(z)} \\ \vspace{-0.2cm}\\
    &= \frac{1}{f_{M^{(k)}}(z)}\Bigg[\Bigg(\frac{f_{M_{j}^{(k)}}(z)}{F_{M_j^{(k)}}(z)}\Bigg)F_{M^{(k)}}(z)\Bigg].
\end{align*}
Therefore, it follows to find an expression for each of these terms to simplify the above expression. To find the form of the density of the monthly maxima sea levels in a given year $f_{M_j^{(k)}}$, we must differentiate the corresponding distribution function, \begin{equation*}
    F_{M_{j}^{(k)}}(z)=\mathbb{P}(M_j^{(k)}\leq{z})=\prod\limits_{{i}=1}^{T^{(k)}_j}\big[F_Y^{(d,j,x)}(z- X_{j_i}^{(k)})\big]^{\hat\theta(z-X_{j_i}^{(k)},r)},\label{eqn::sldist_exi_mon}
\end{equation*} where $F_{Y}^{(d,j,x)}$ is the skew surge distribution function given by equation~\eqref{ss_tidedep_model} and $\hat\theta(\cdot,r)$ is our extremal index model given by equation~\eqref{eqn::exi_model} for fixed run length $r$. Differentiating this gives \begin{equation}
    f_{M_j^{(k)}}(z) 
    \approx F_{M_{j}^{(k)}}(z)\sum\limits_{i=1}^{T_{j}^{(k)}}\frac{{f_Y}^{(d,j,x)}(z-X_{j_i}^{(k)})\hat\theta(z-X_{j_i}^{(k)},r)}{F_{Y}^{(d,j)}(z-X_{j_i}^{(k)})},\label{monmax_sldensity}
\end{equation} when we ignore smaller order terms arising from $\hat\theta'(\cdot,r)$. The density of skew surges $f_{Y}^{(d,j,x)}$ is given by \begin{align}
    f_{Y}^{(d,j,x)}(y)=\begin{cases}  \hat f_j(y) \quad\quad &\text{if } y\leq u_j \\ \frac{\lambda_{d,x}}{\sigma_{d,x}}\Big[1+\xi\Big(\frac{y-u_j}{\sigma_{d,x}}\Big)\Big]^{-\frac{1}{\xi}-1}_{+} \quad &\text{if }y> u_j \label{ss_den_abu}
        \end{cases}
\end{align} where $\lambda_{d,x}$ and $\sigma_{d,x}$ are given by expressions~\eqref{eqn::ss_rate_tide} and \eqref{eqn::ss_scale_tide1}, respectively. Below the monthly threshold $u_j$, we estimate the derivative of the monthly empirical distribution using a Gaussian kernel density estimator and denote this $\hat f_j$. We find the density of the annual maximum sea levels, for a fixed year $k$, by differentiating in similar way and ignoring smaller order terms, so that, \begin{equation}
    f_{M^{(k)}}(z)\approx F_{M^{(k)}}(z)\sum\limits_{J=1}^{12}\sum\limits_{i=1}^{T_{J}^{(k)}}\frac{f_Y^{(d,J,x)}(z-X_{J_i}^{(k)})\hat\theta(z-X_{J_i}^{(k)},r)}{F_{Y}^{(d,J,x)}({z-X_{J_i}^{(k)}})}\label{annmax_sldensity}.
\end{equation}
Using equations~\eqref{monmax_sldensity} and \eqref{annmax_sldensity}, we can simplify $\hat P_{M^{(k)}}(j;z)$ to \begin{equation*}
    \hat P_{M^{(k)}}(j;z) =\frac{A^{(i,j,k)}(z)}{\sum\limits_{J=1}^{12}A^{(i,J,k)}(z)} \quad \text{ where } \quad A^{(i,j,k)}(z)=\sum\limits_{i=1}^{T_{j}^{(k)}}\frac{f_Y^{(d,j,x)}(z-X_{j_i}^{(k)})\hat\theta(z-X_{j_i}^{(k)},r)}{F_{Y}^{(d,j,x)}({z-X_{j_i}^{(k)}})}.
\end{equation*} Clearly $\sum_{j=1}^{12}\hat P_{M^{(k)}}(j;z)=1$. Then the probability over all $K$ years is given by $\hat P_{M}(j;z)=\frac{1}{K}\sum_{k=1}^{K}\hat P_{M^{(k)}}(j;z)$.

\section{Transforming Skew Surges to Uniform Margins}\label{sec::transform_ss}
In Section~\ref{rl_final} of the main paper we assess the fit of our final model for sea level annual maxima~\eqref{eqn::sldist_exi} by looking at year-specific distributions and bootstrap confidence intervals on the return level estimates. Here, we test the goodness-of-fit for the final skew surge model (expression~\eqref{ss_tidedep_model}), using the probability integral transform; if our model fits well, transforming the observations through the fitted distribution function will give a sample of identically distributed Uniform$(0,1)$ values. This was also a step in our stationary bootstrap procedure, but here we check if these transformed values $\{U^Y_i\}$ are Uniform using a Kolmogorov-Smirnov test. 

\begin{figure}
    \centering
    \includegraphics[width=0.45\textwidth]{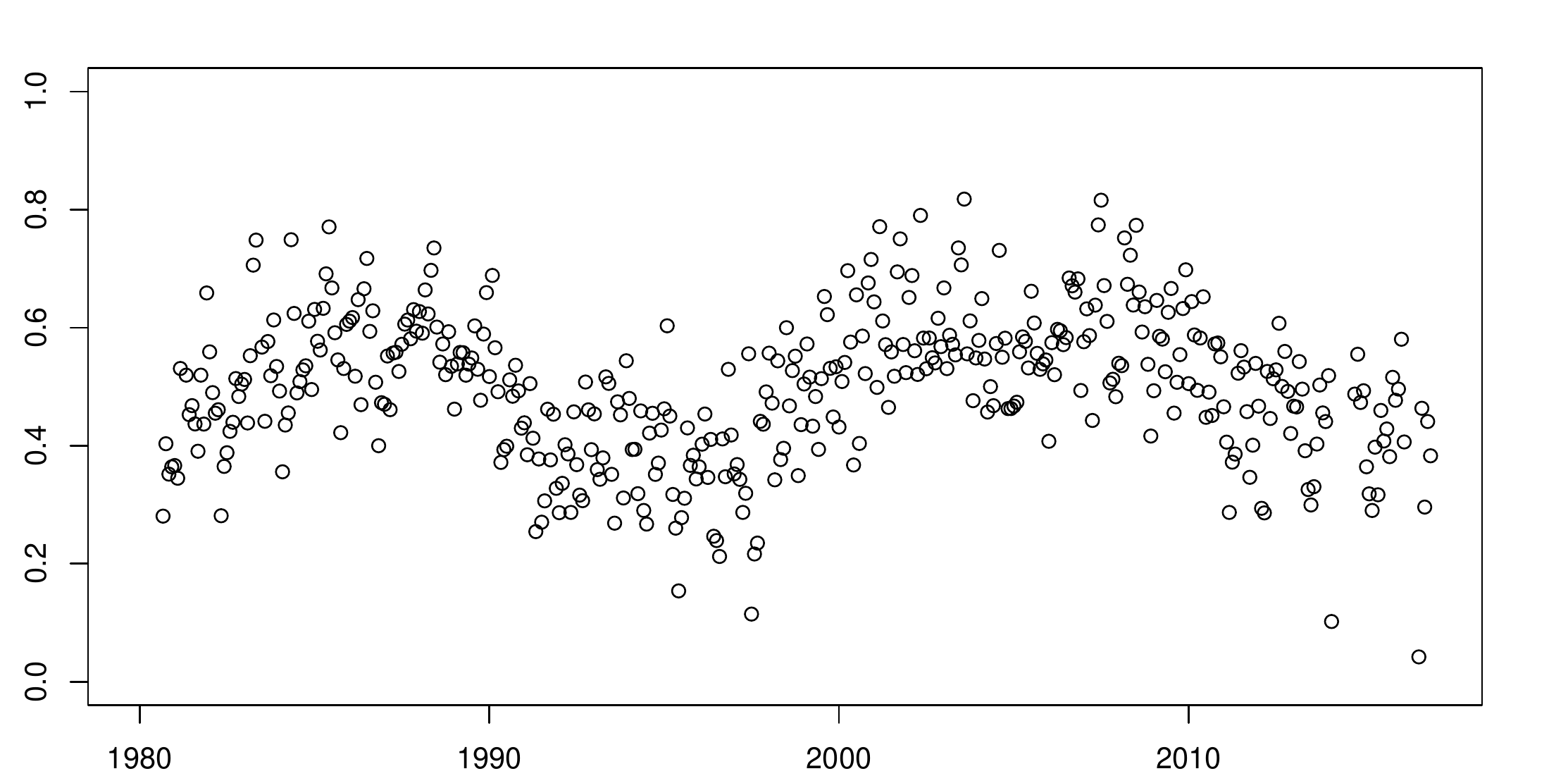}
    \includegraphics[width=0.45\textwidth]{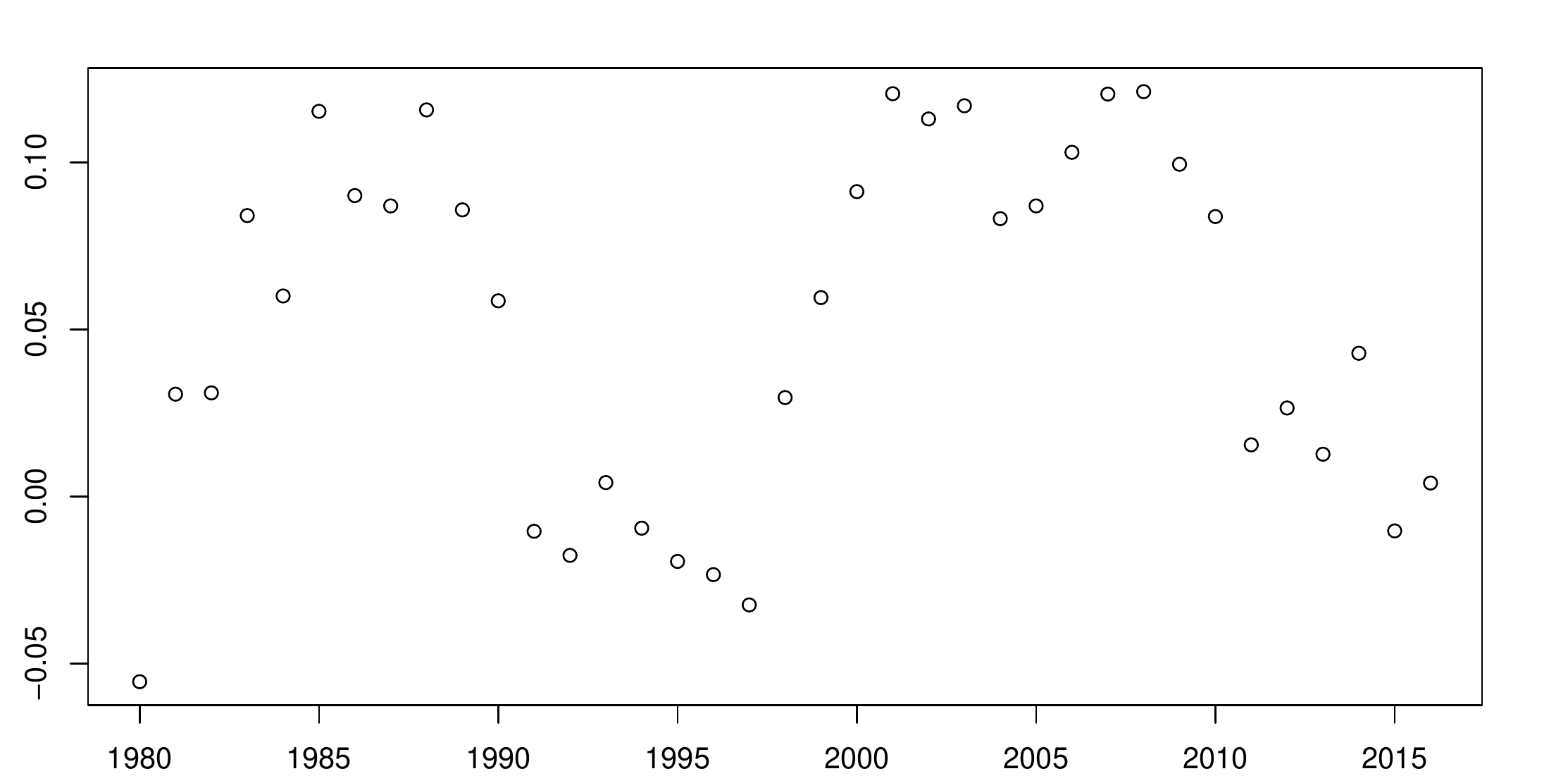}
    \caption{Monthly means of transformed skew surge observations through the final skew surge distribution function~\eqref{ss_tidedep_model} (left) and annual mean skew surges (right), both at Sheerness.}
    \label{fig::UniformSHE}
\end{figure}

\begin{figure}[h]
    \centering
    \includegraphics[width=0.4\textwidth]{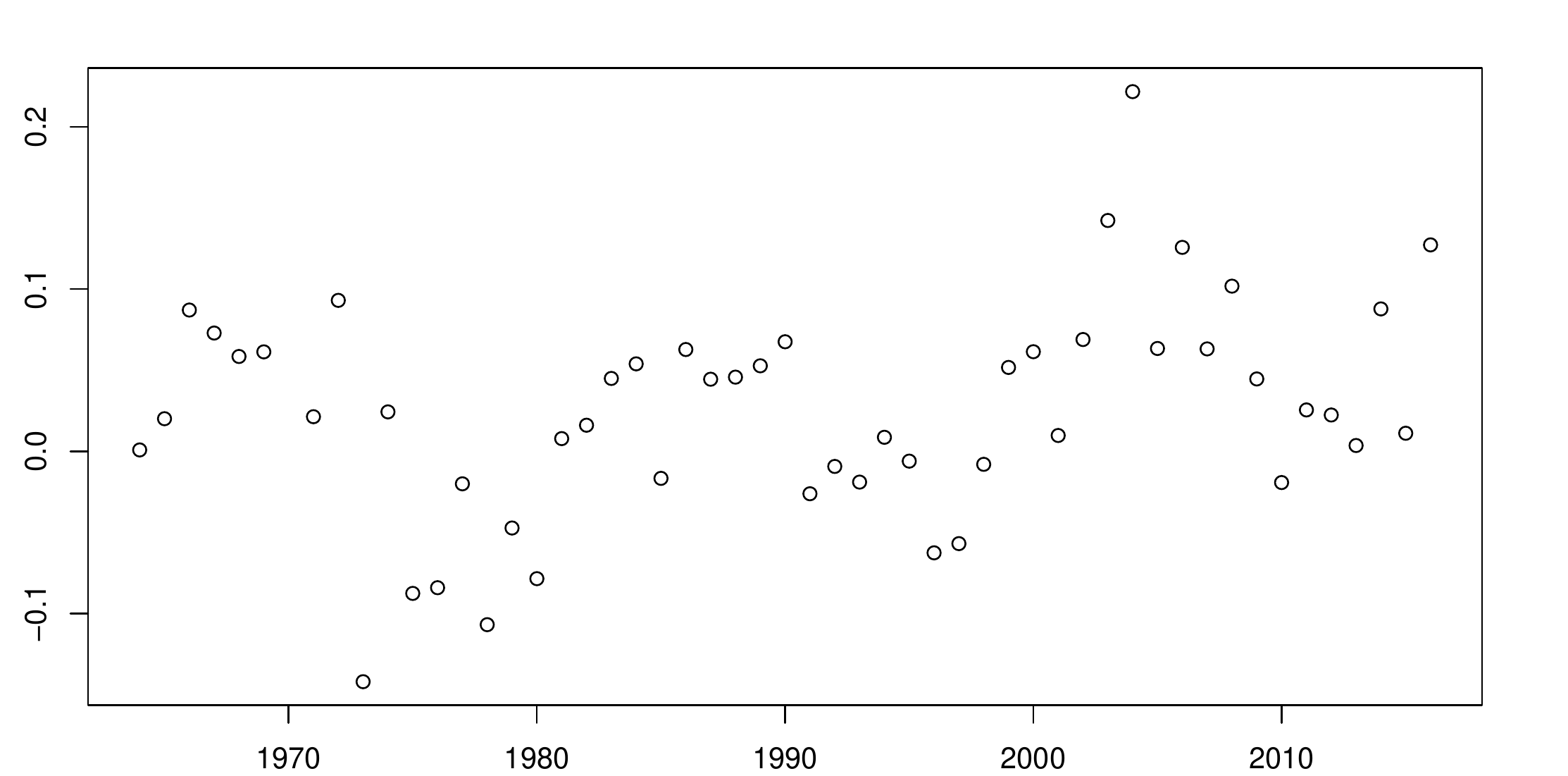}
    \includegraphics[width=0.4\textwidth]{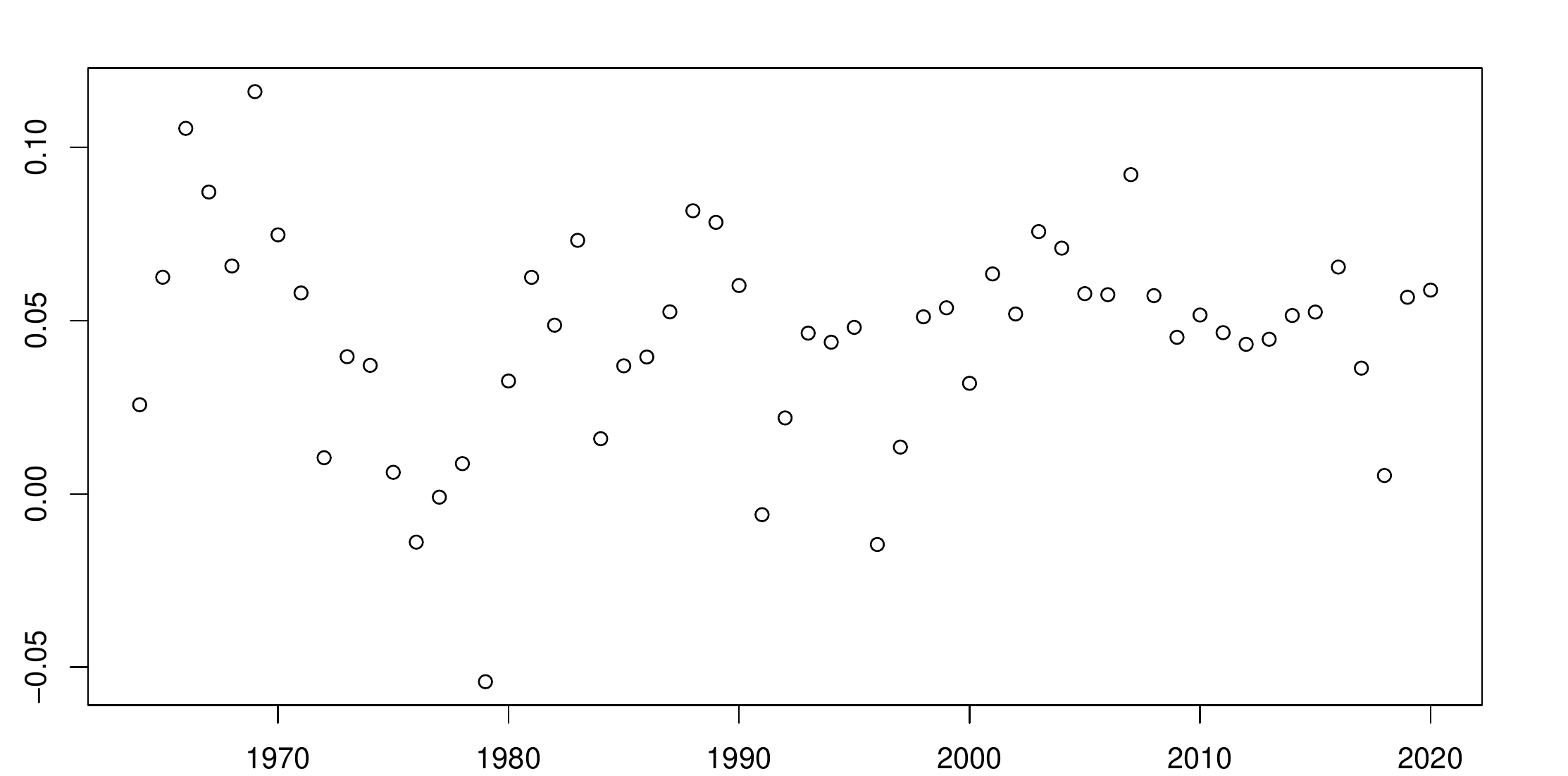} \\
    \includegraphics[width=0.4\textwidth]{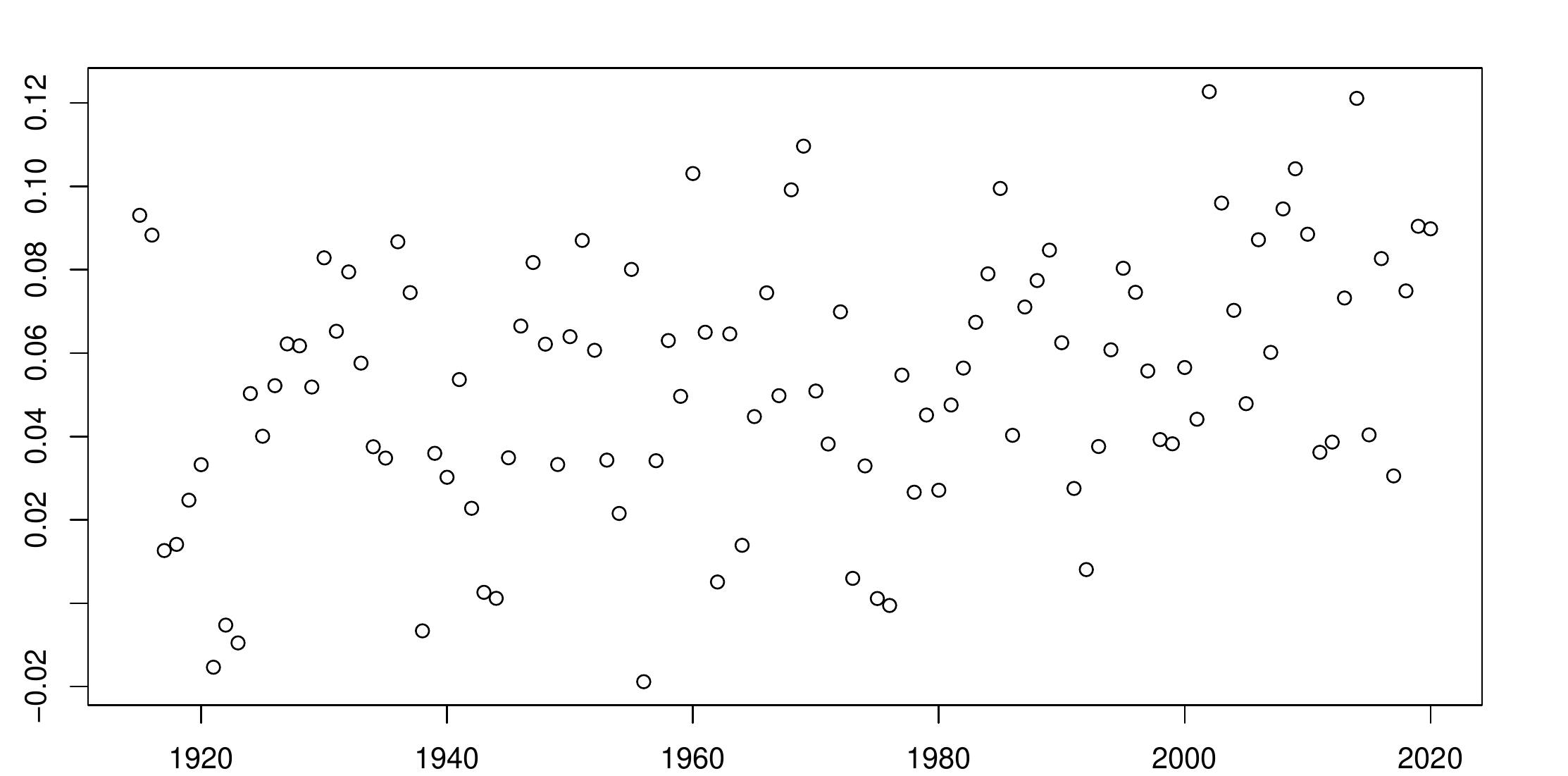}
    \caption{Annual mean skew surges at Heysham (top left), Lowestoft (top right) and Newlyn (bottom).}
    \label{fig::annmeans}
\end{figure}

\begin{figure}[h]
    \centering
    \includegraphics[width=0.49\textwidth]{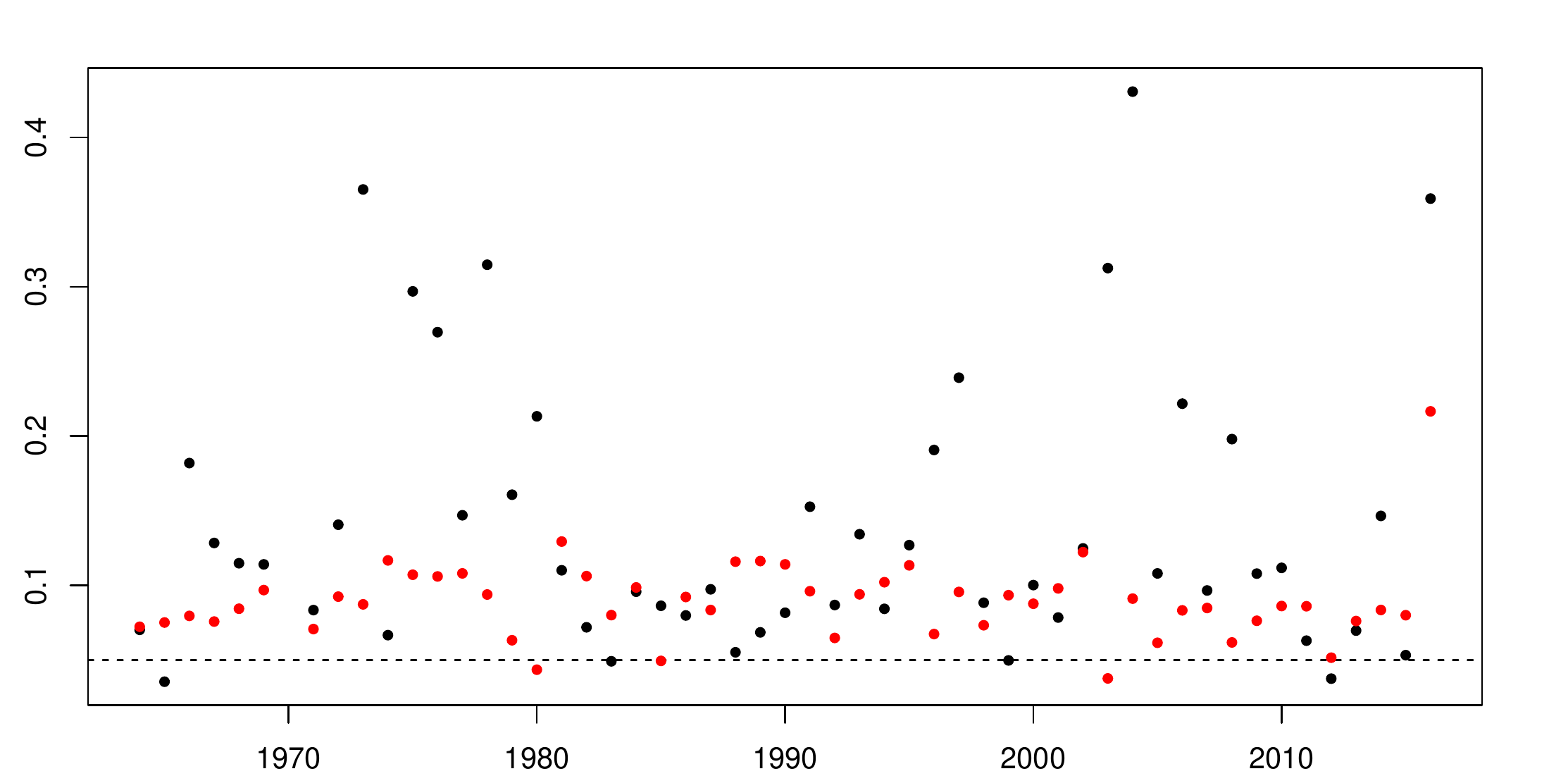}
    \includegraphics[width=0.49\textwidth]{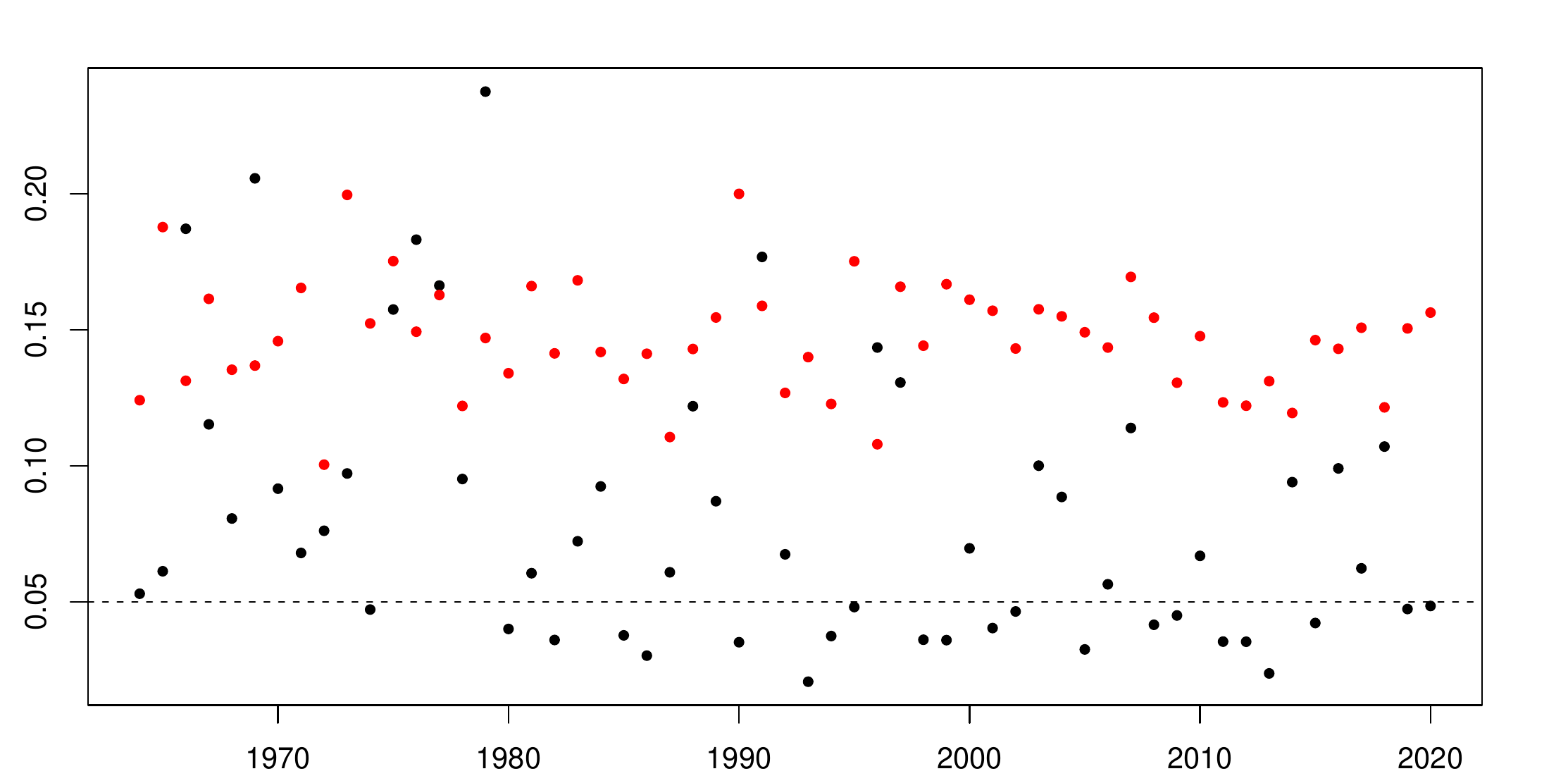}
    \includegraphics[width=0.49\textwidth]{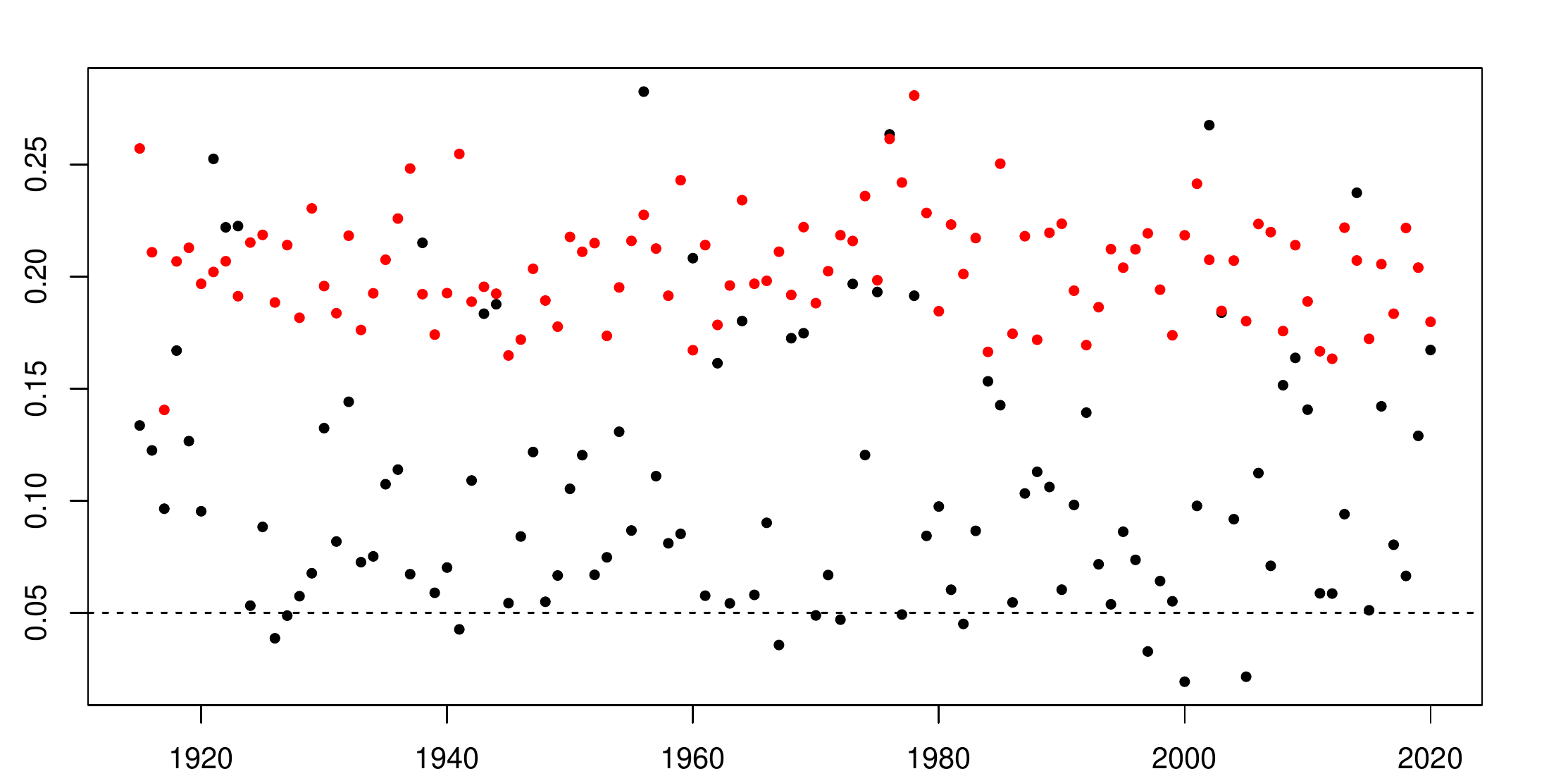}
    \includegraphics[width=0.49\textwidth]{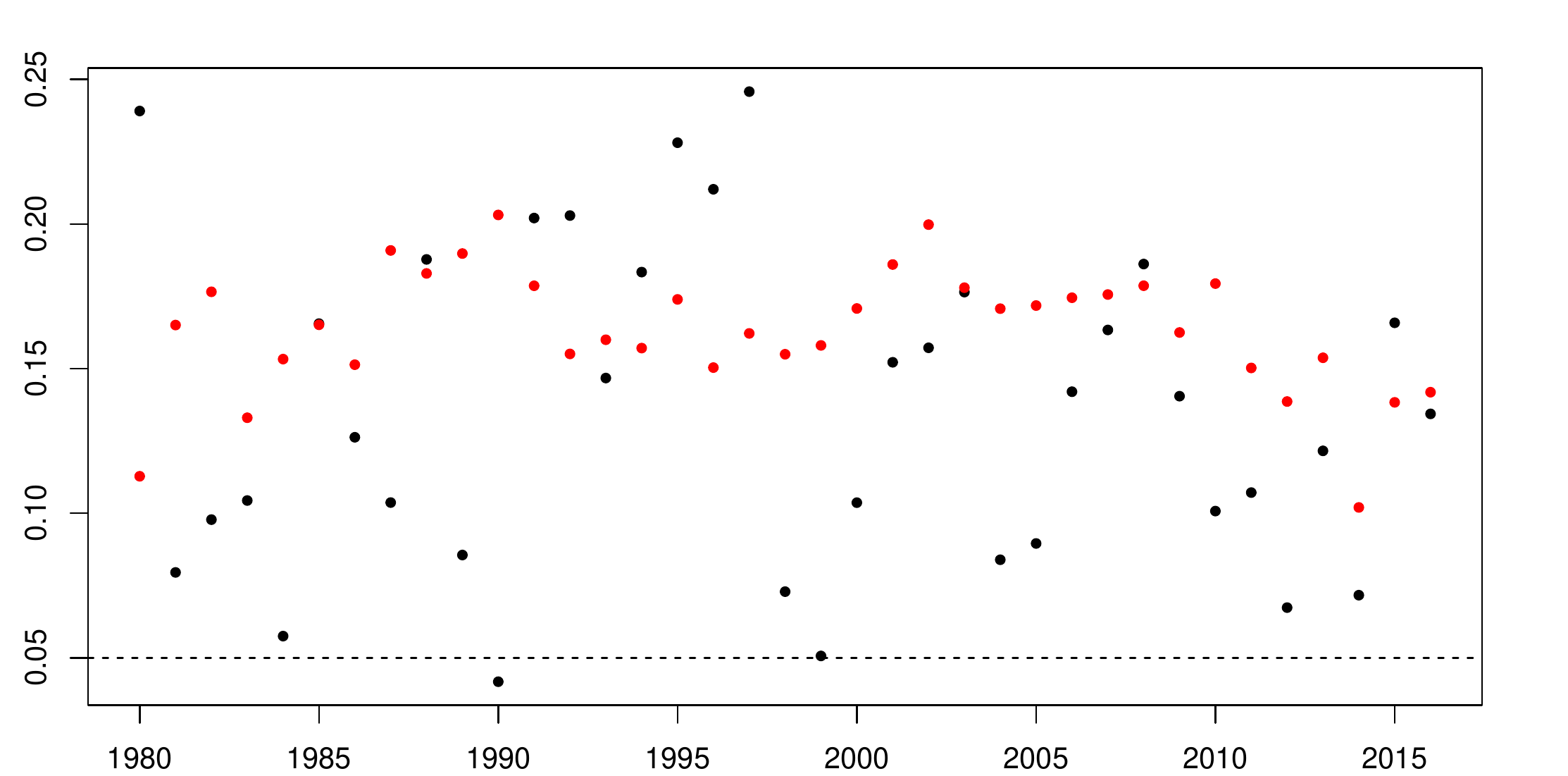}
    \caption{$p$ values from the Kolmogorov-Smirnov test for uniformity of yearly samples of the transformed skew surge observations through the final skew surge distribution function (expression~\ref{ss_tidedep_model}) at Heysham (top right), Lowestoft (top left), Newlyn (bottom left) and Sheerness (bottom right) before (black) and after (red) we remove the annual mean trends of the skew surge series. The 5\% significance level is shown by the black dashed line.}
    \label{fig::Uniform_kstests}
\end{figure}

Figure~\ref{fig::UniformSHE} shows the transformed skew surges at Sheerness. We can immediately see these are not Uniformly distributed. This is supported by results of yearly Kolmogorov-Smirnov tests for uniformity, where the $p$ values are almost all $>0.05$. Instead there appears to be a cyclic sinusoidal trend, following the trend of the 18.6 year nodal cycle. We suspect this is a data issue, perhaps the tidal series was not correctly removed from the sea level observations when the skew surges are obtained. Figure~\ref{fig::UniformSHE} also shows the annual mean skew surges at Sheerness, we can see there is a similar cyclic trend here and that the means are not centred at zero (see Figure~\ref{fig::annmeans} for the remaining sites). To correct for this we re-centre the data at zero, by removing the corresponding annual mean from the observations. This is an ad hoc approach and the data should be investigated further. We did not correct for this trend at an earlier stage in the modelling process because it does not have a significant effect on the extreme values. However, once we correct for this trend, we find that the transformed data are uniformly distributed in 32 years (out of 37) at Sheerness, indicating a good model fit for skew surges. We find similar results at the remaining sites, where a cyclic trend is first observed in $\{U^Y_i\}$ but once the annual means are removed, the transformed data can be reasonably assumed as Uniform(0,1). Figure~\ref{fig::Uniform_kstests} shows the $p$ values for each year at each site. 

\newpage
\section{Supplementary Figures}\label{suppfig}


\begin{figure}[h]
    \centering
    \includegraphics[width=0.49\textwidth]{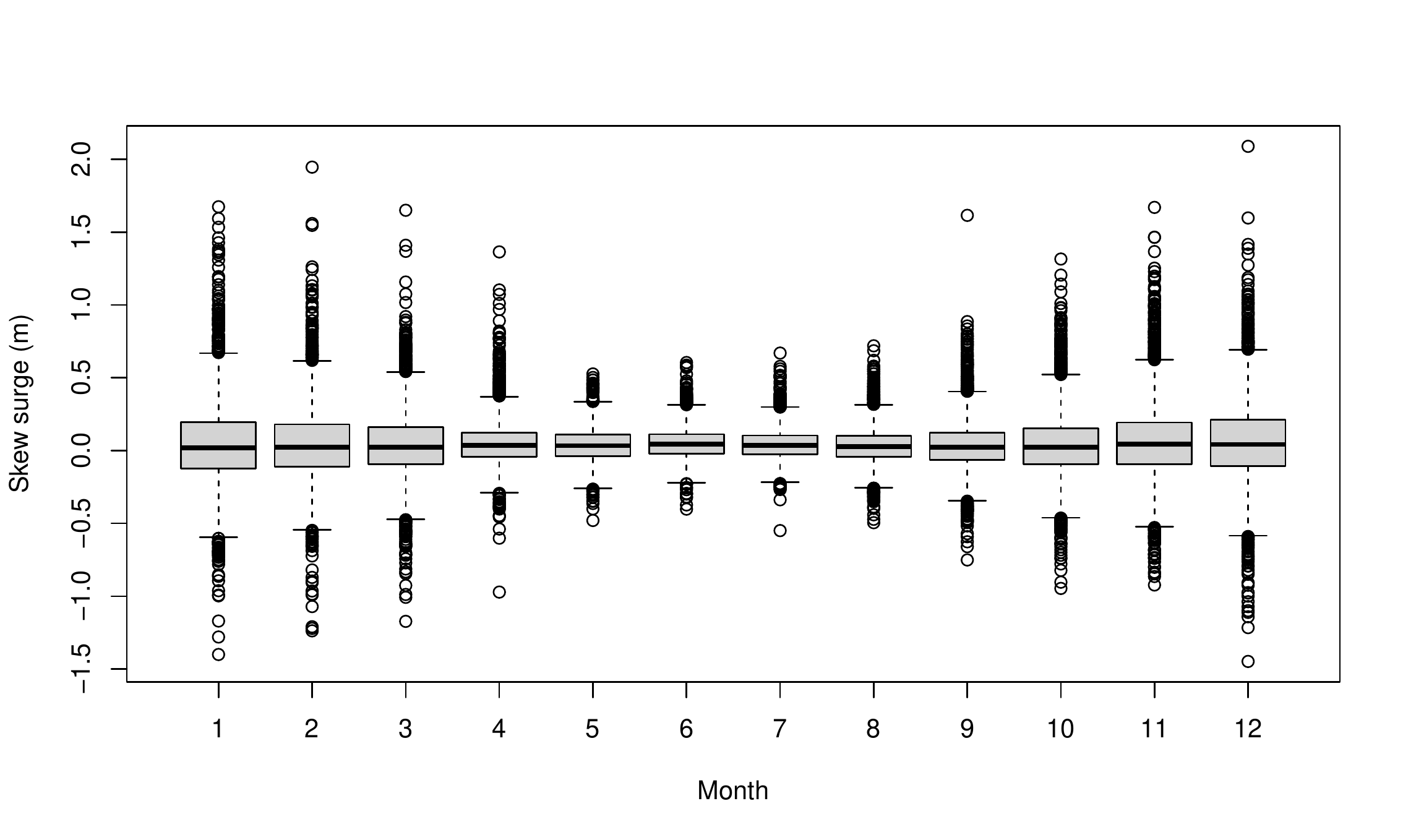}    \includegraphics[width=0.49\textwidth]{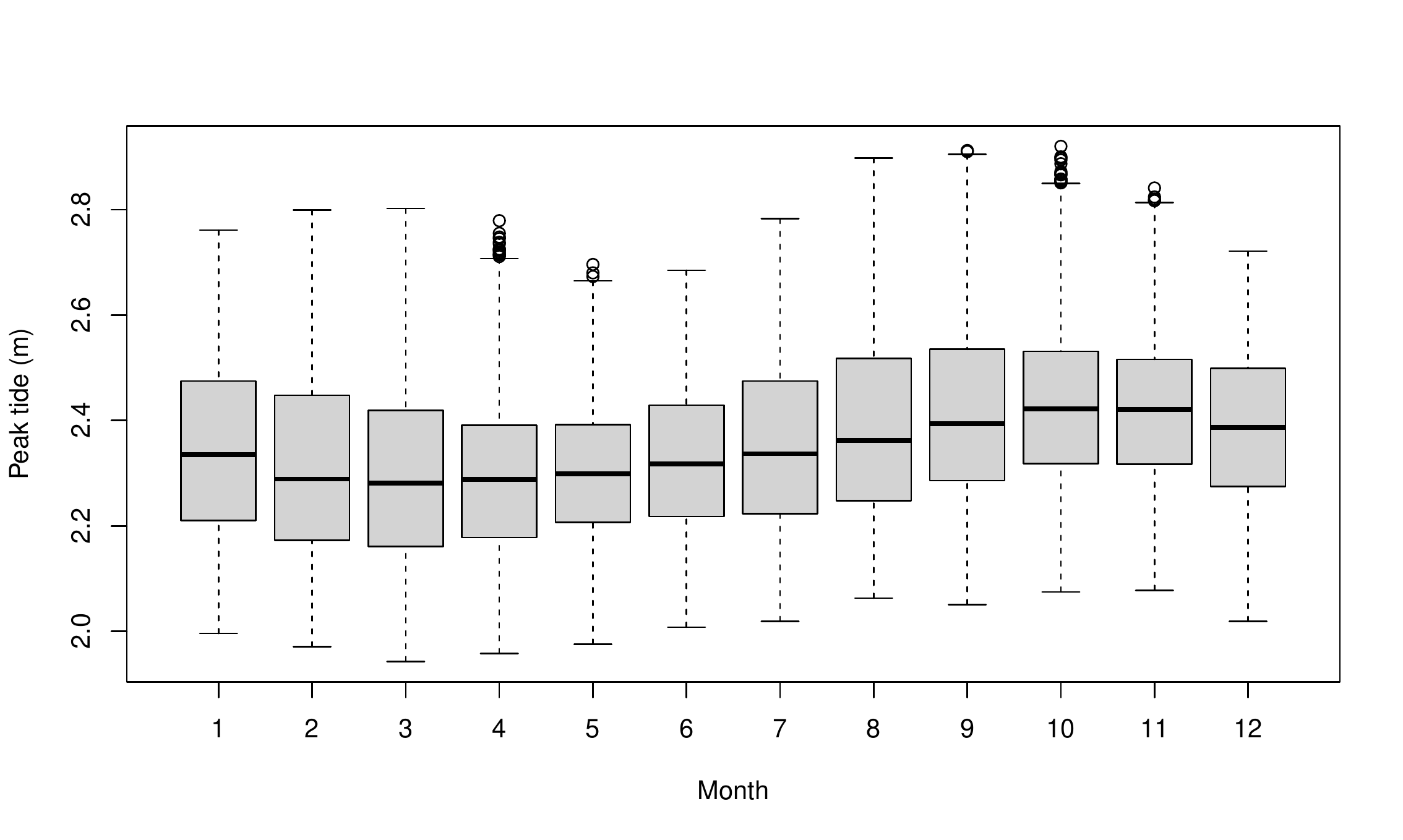}    \includegraphics[width=0.49\textwidth]{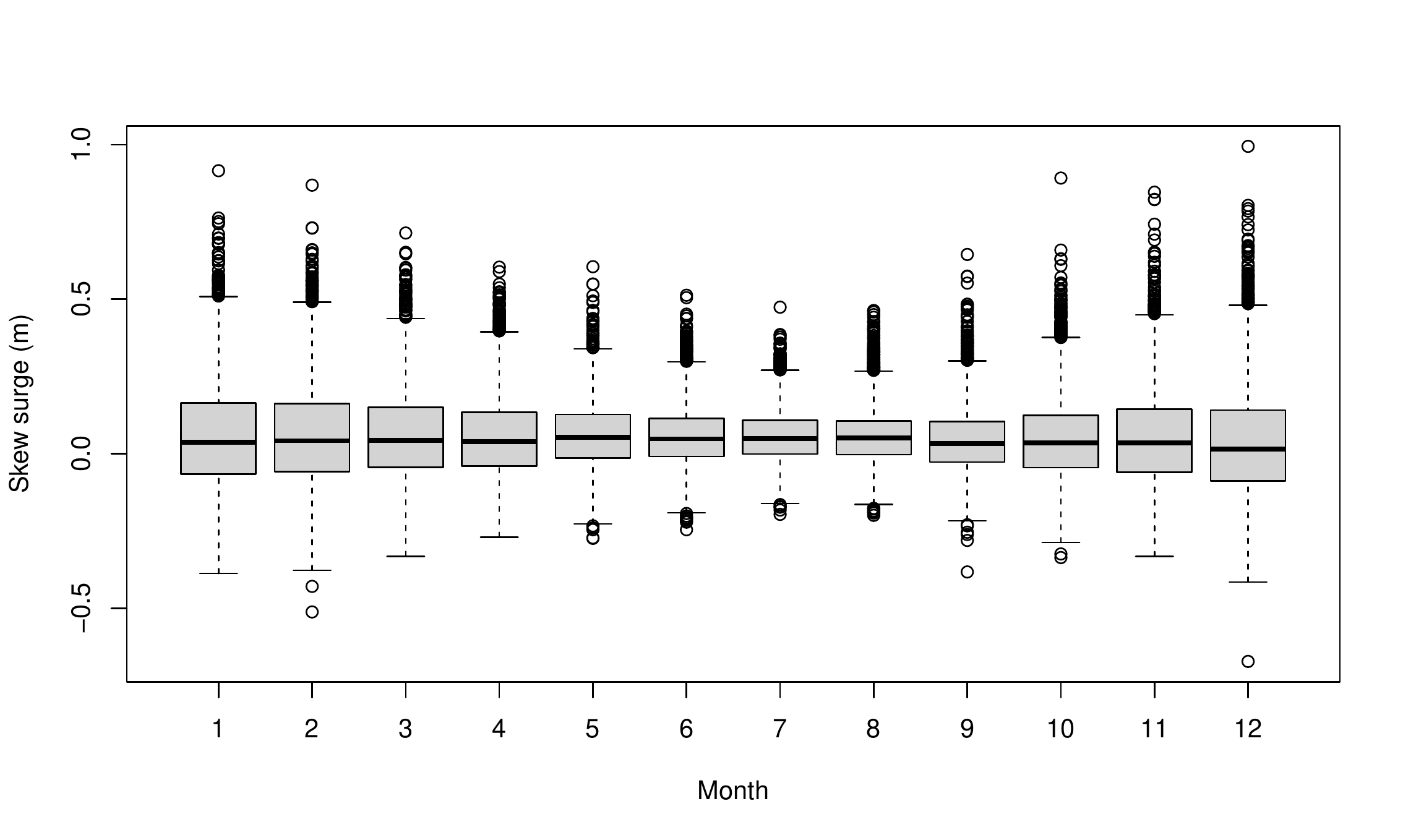}    \includegraphics[width=0.49\textwidth]{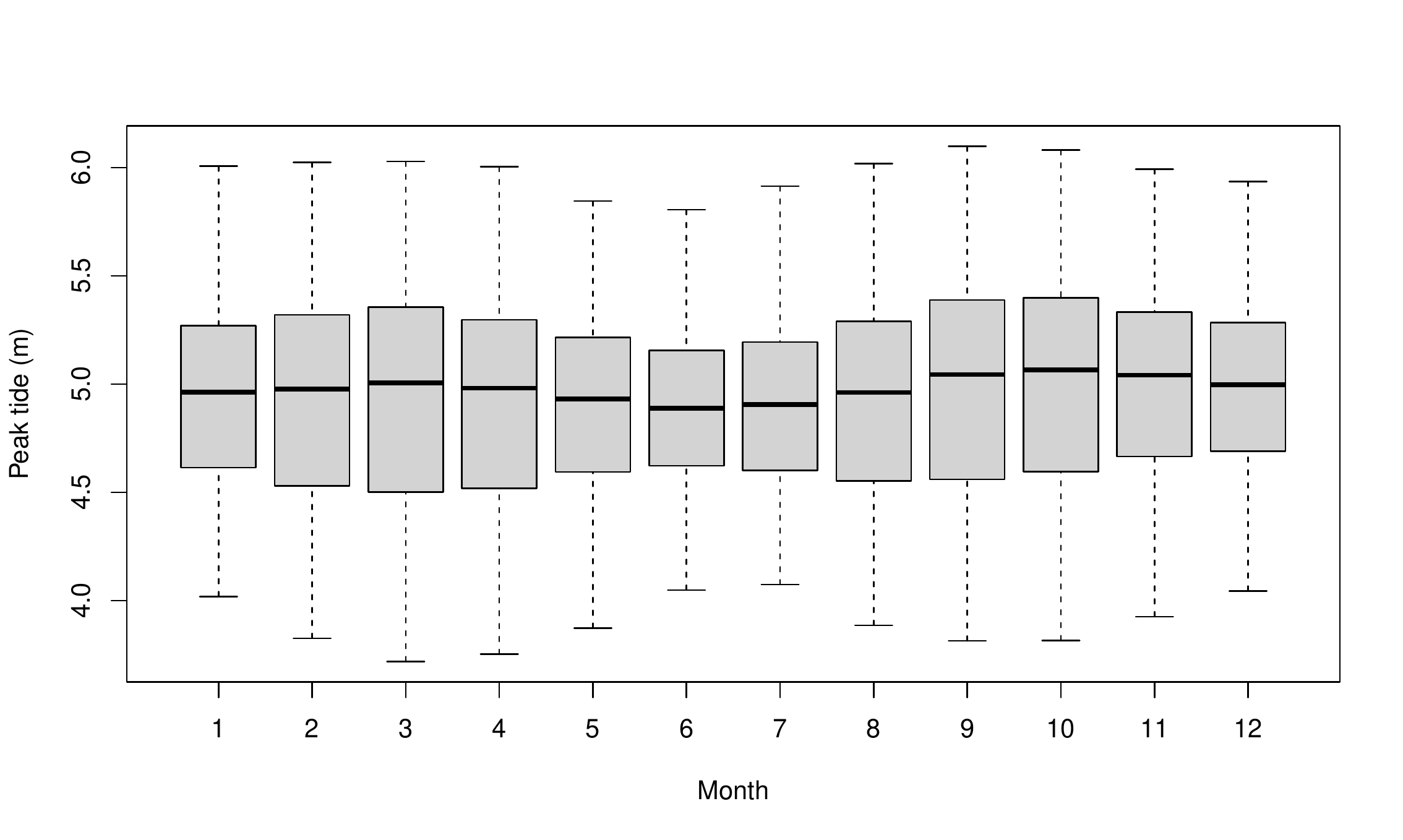}
    \includegraphics[width=0.49\textwidth]{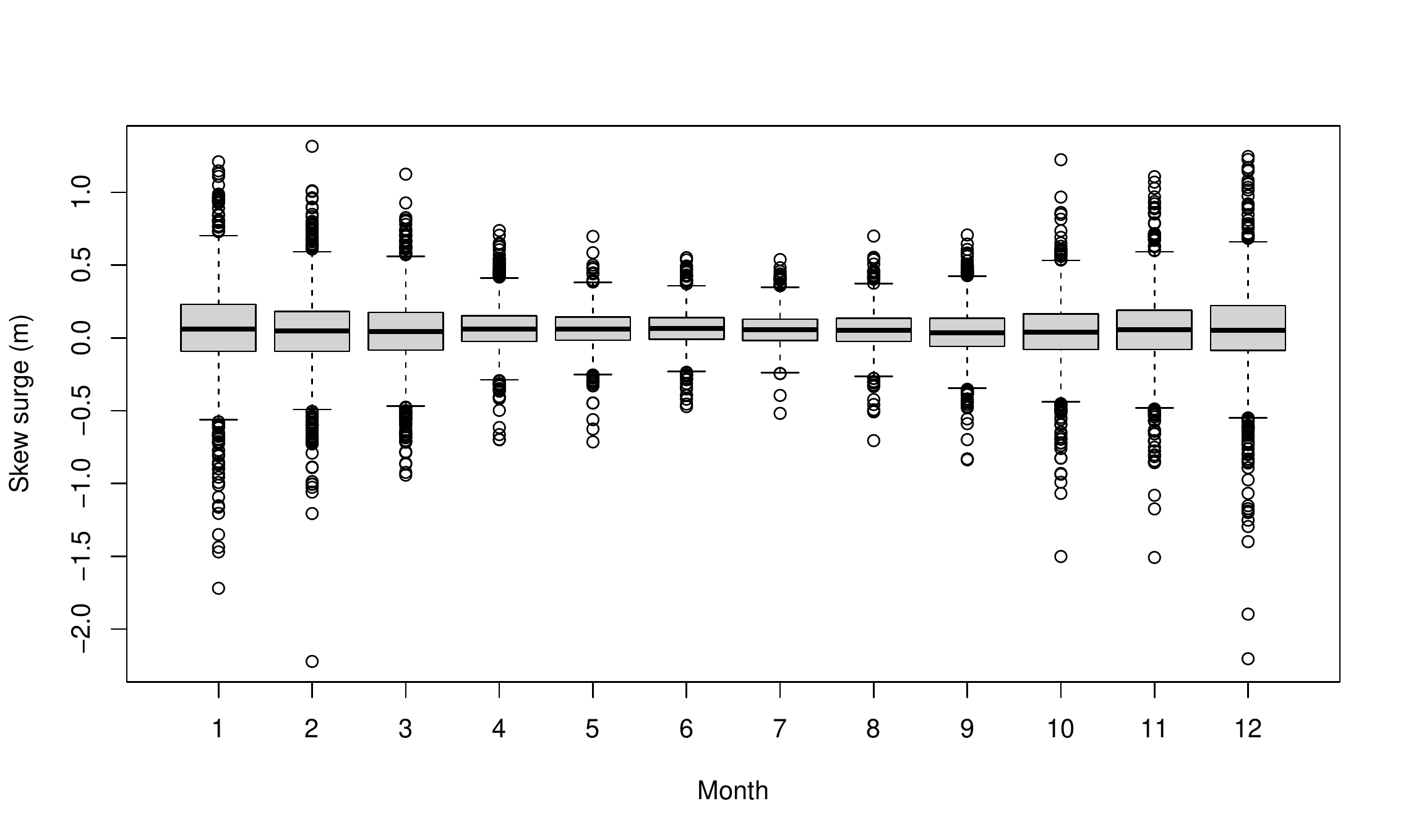}    \includegraphics[width=0.49\textwidth]{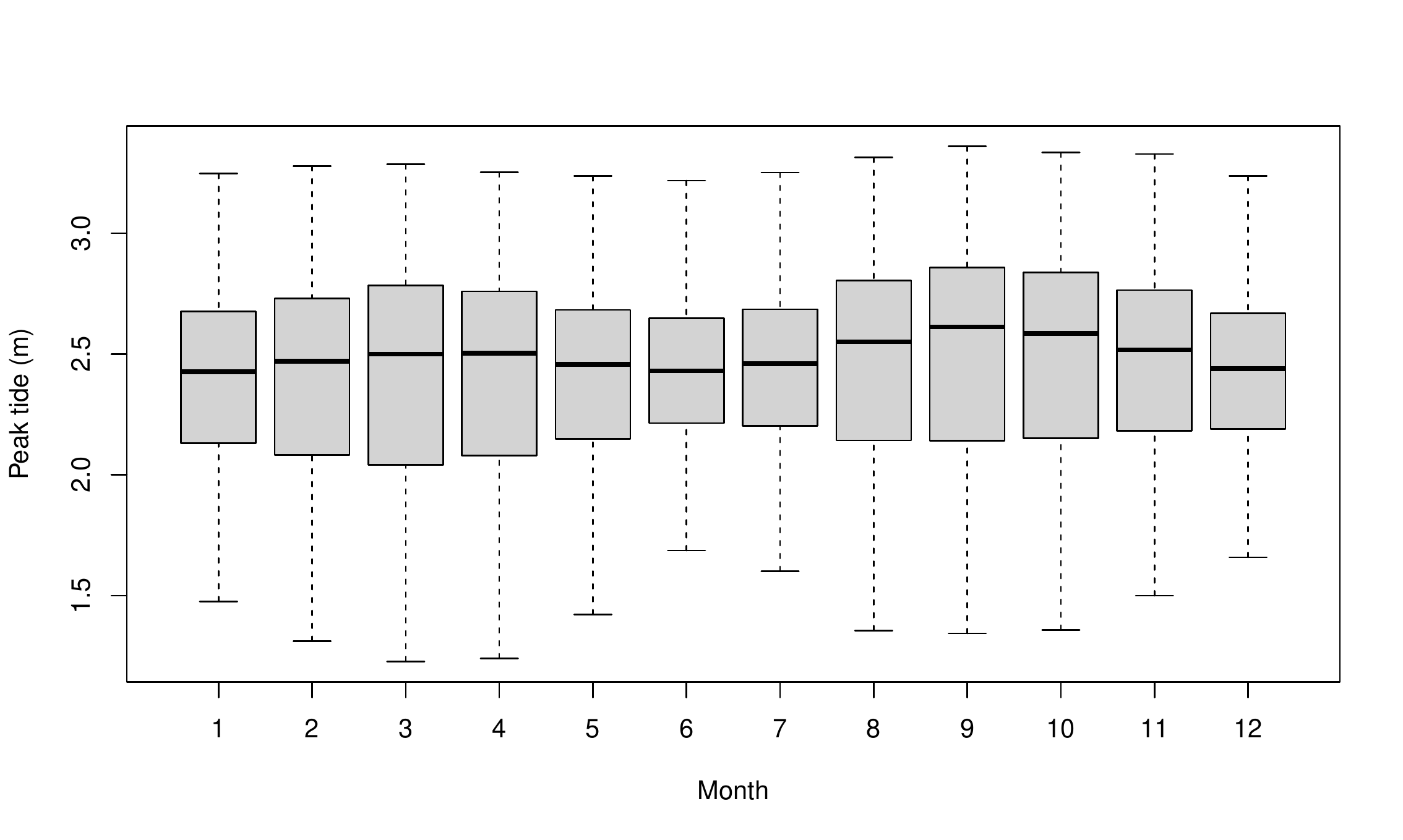}
    \caption{Monthly box plots of skew surge (left column) and peak tide (right column) at Lowestoft (top row), Newlyn (middle row) and Sheerness (bottom row).}
    \label{ss_mt_bps_all}
\end{figure}



\begin{figure}[h]
    \centering
    \includegraphics[width=0.49\textwidth]{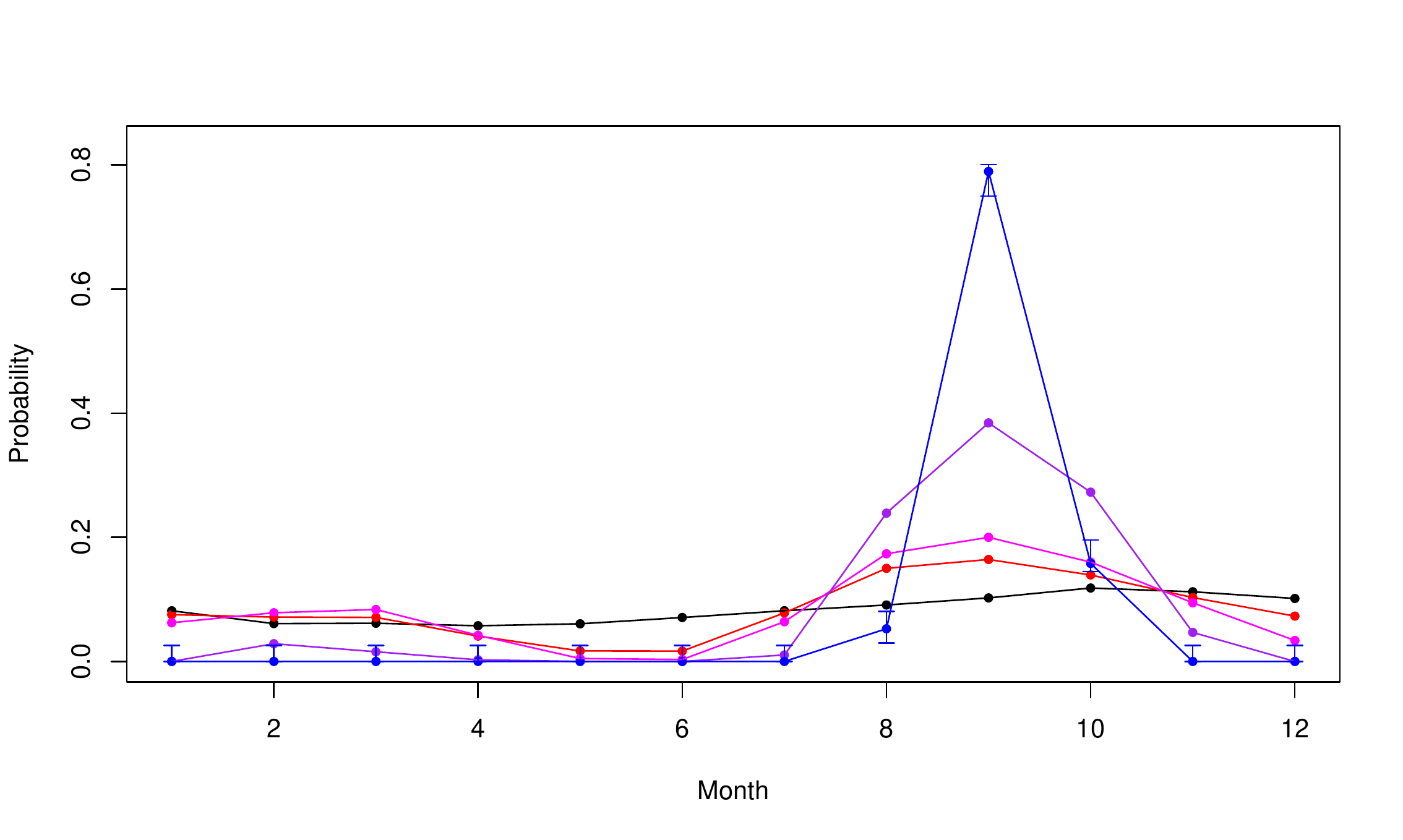}
    \includegraphics[width=0.49\textwidth]{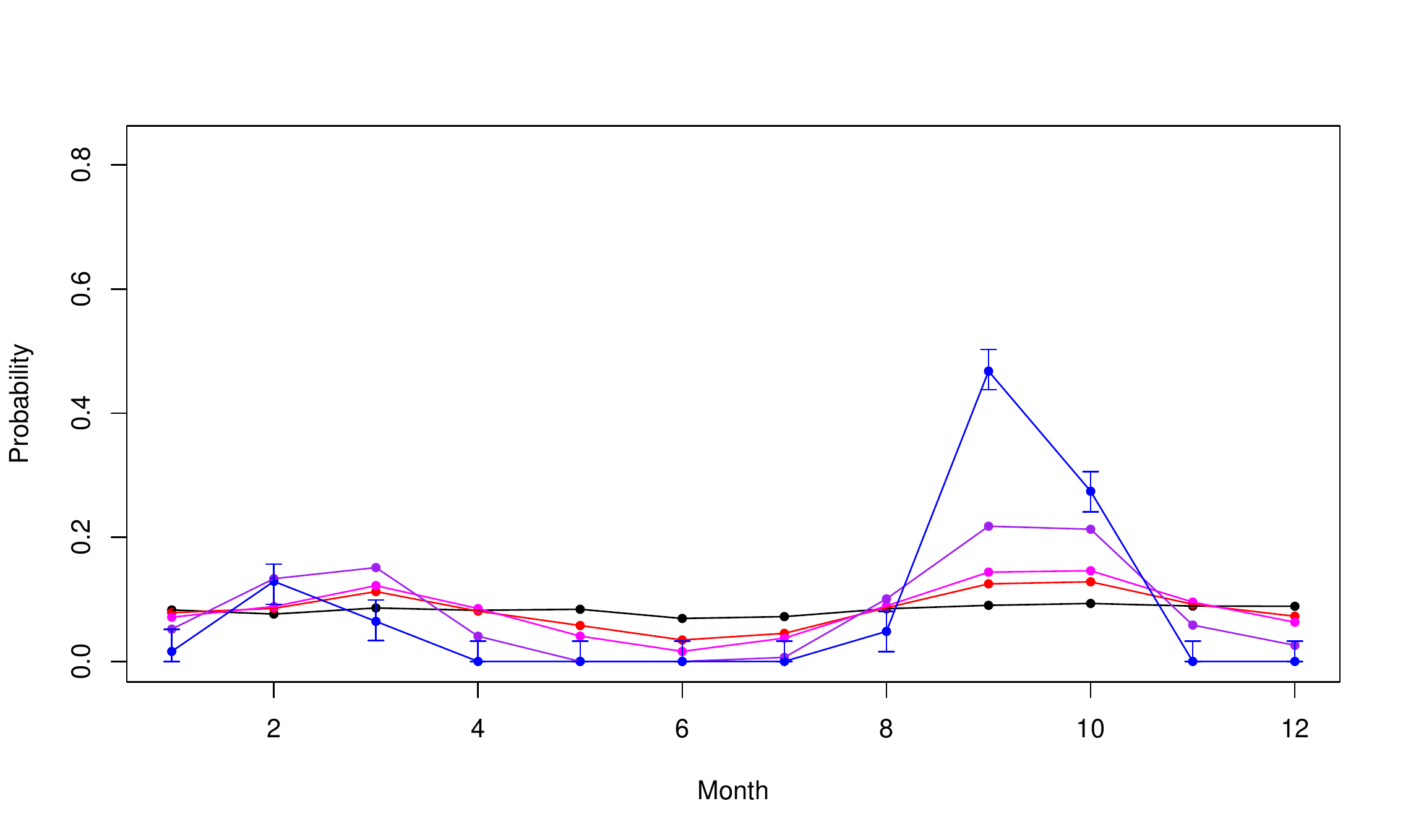}
    \includegraphics[width=0.49\textwidth]{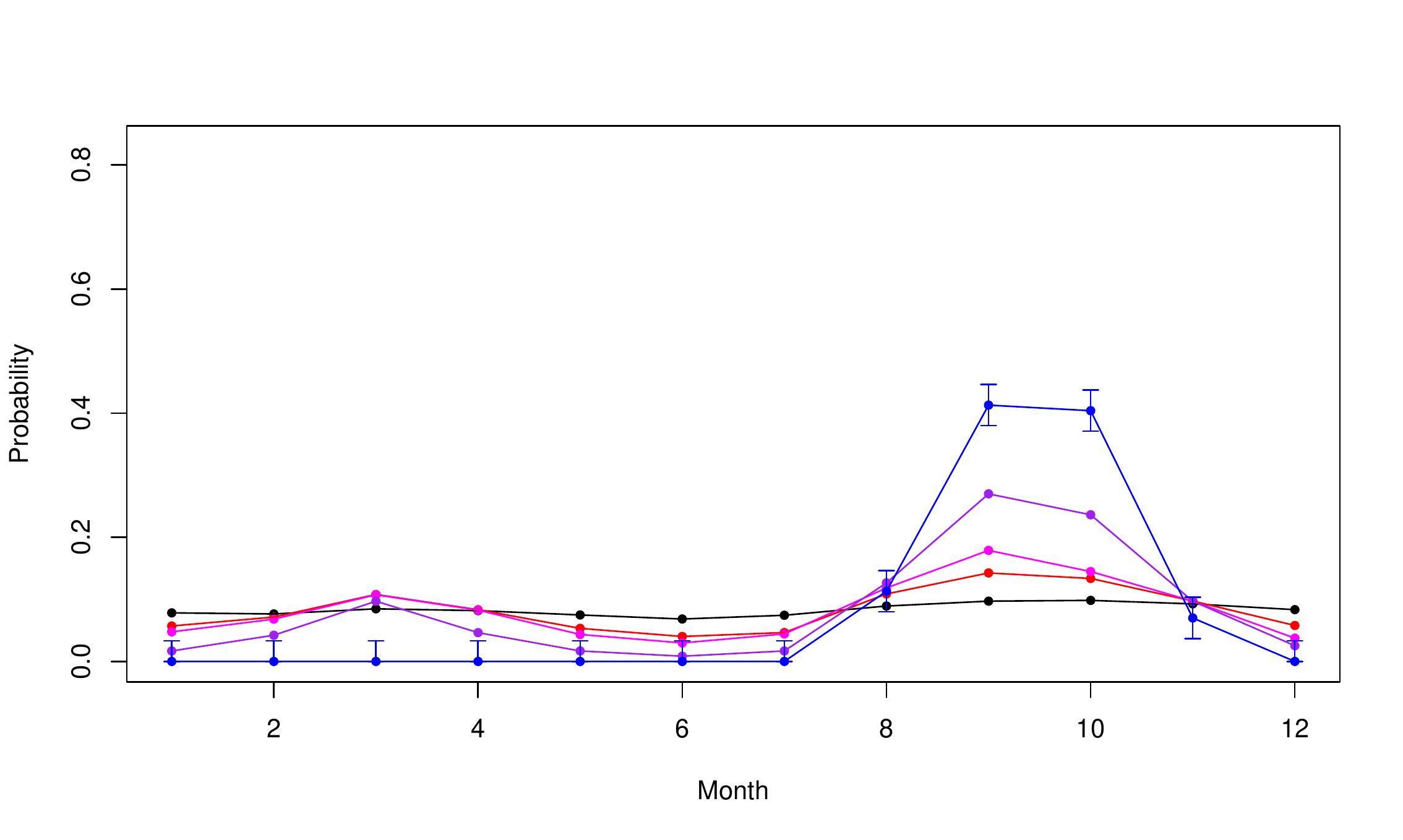}
    \caption{Estimates of $\tilde P_{X}(j;x_{q})$ ( expression~\eqref{prob} of the main paper) for months $j=1-12$ and $q=0.5$ (black), $0.9$ (red), 0.95 (magenta), 0.99 (purple), 0.999 (blue) at Lowestoft (top left), Newlyn (top right) and Sheerness (bottom), with 95\% confidence intervals when $q=0.99$.}
    \label{mt_seasonality_plot}
\end{figure}


\begin{figure}[h]
    \centering
    \includegraphics[width=0.5\textwidth]{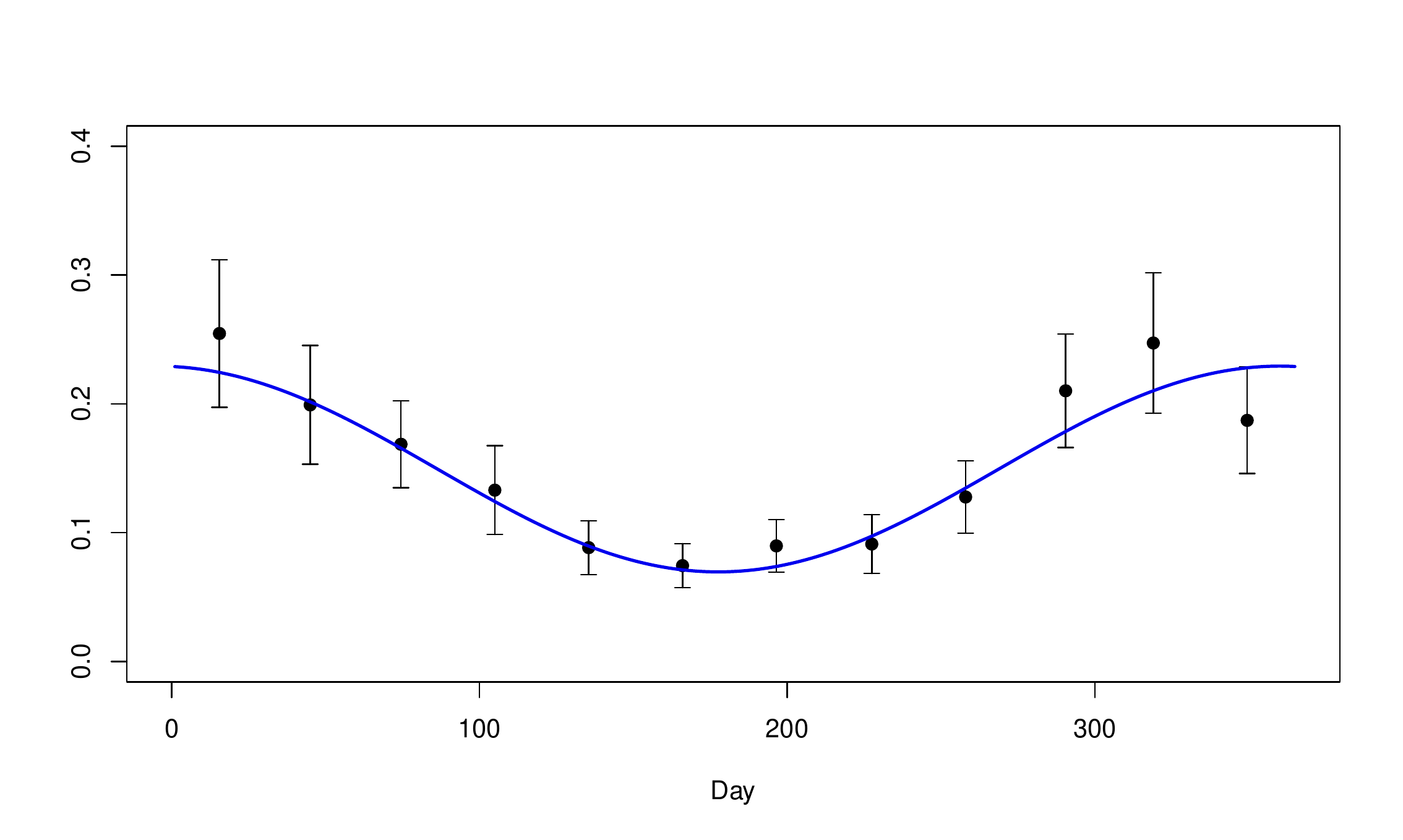}\includegraphics[width=0.5\textwidth]{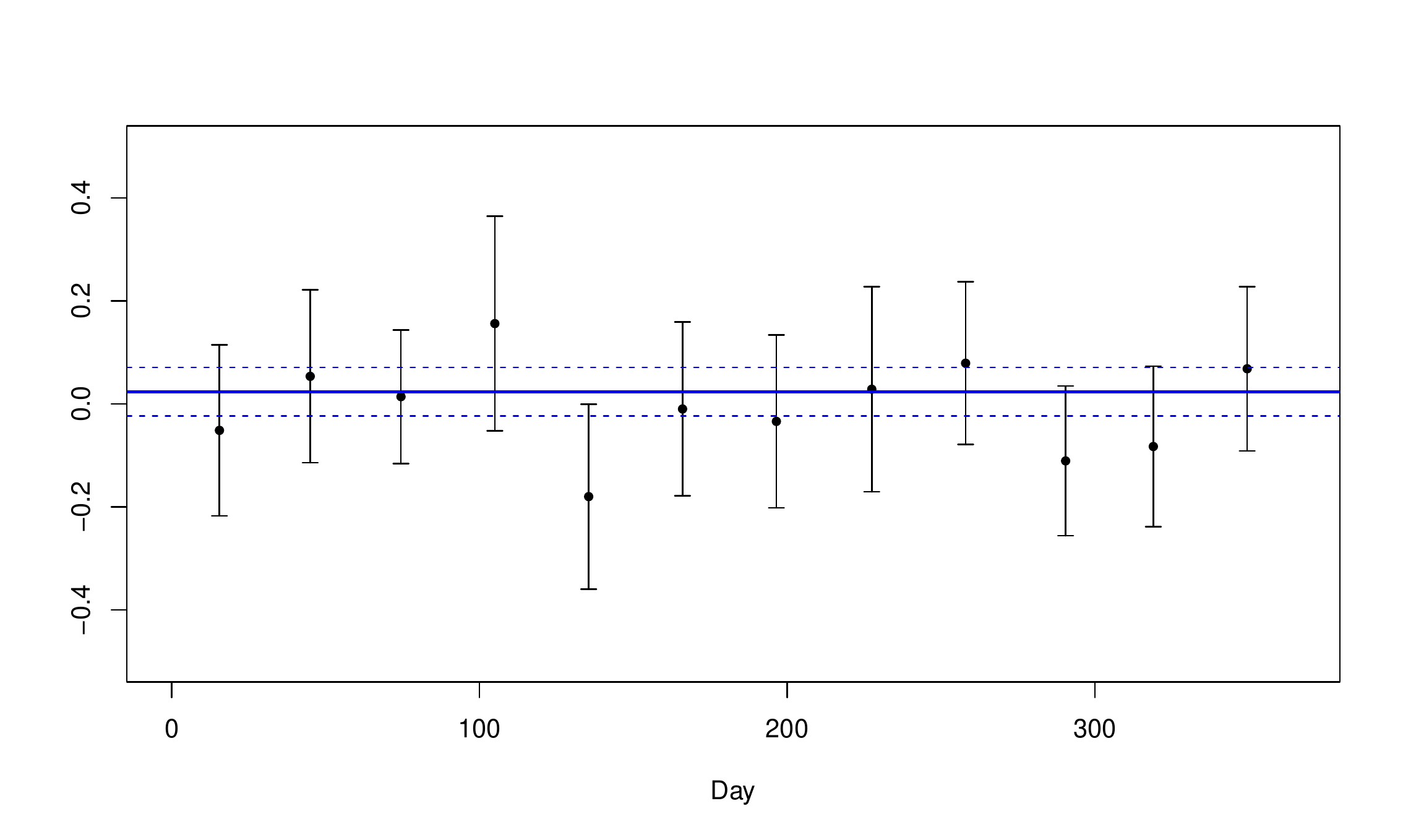}
    \includegraphics[width=0.5\textwidth]{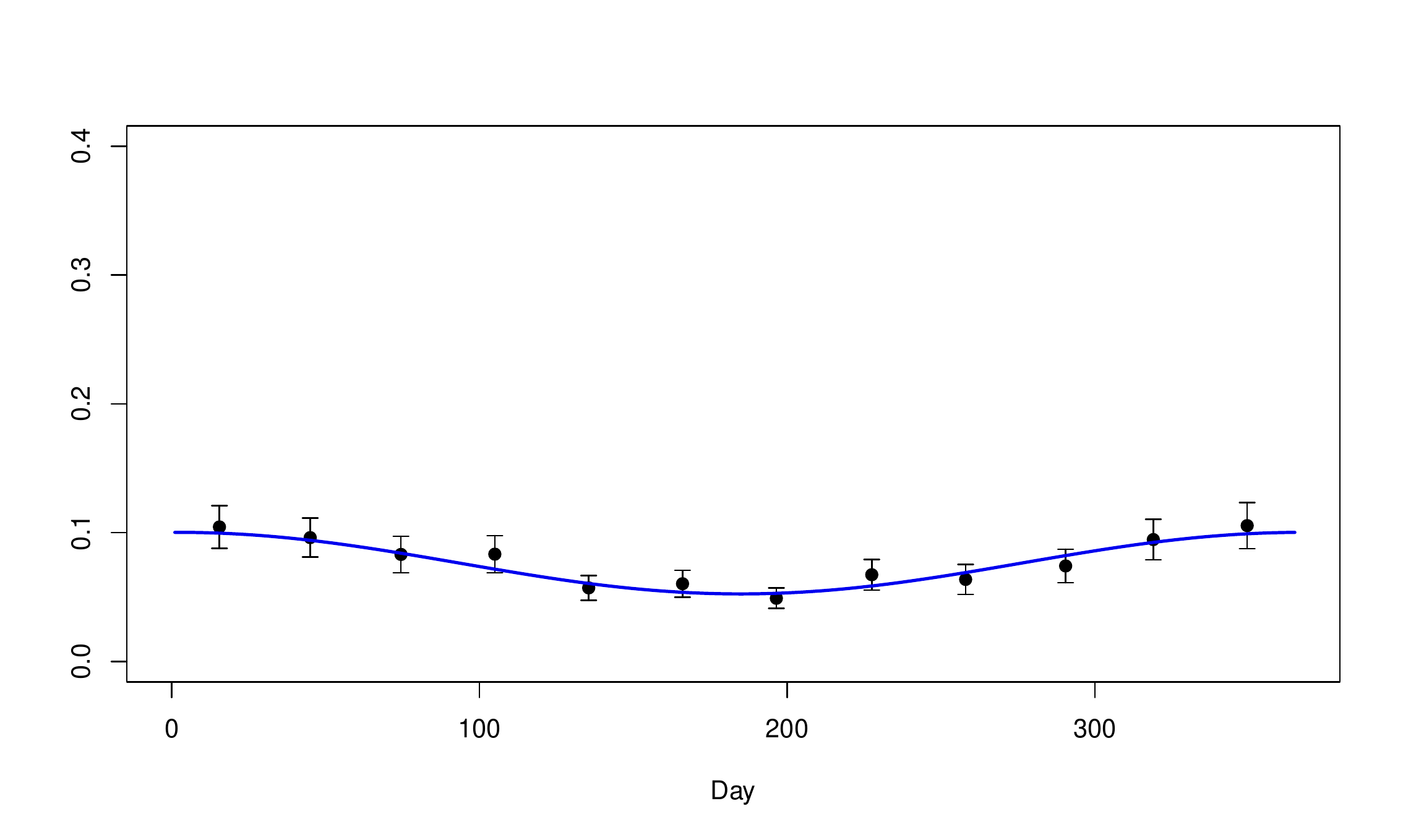}\includegraphics[width=0.5\textwidth]{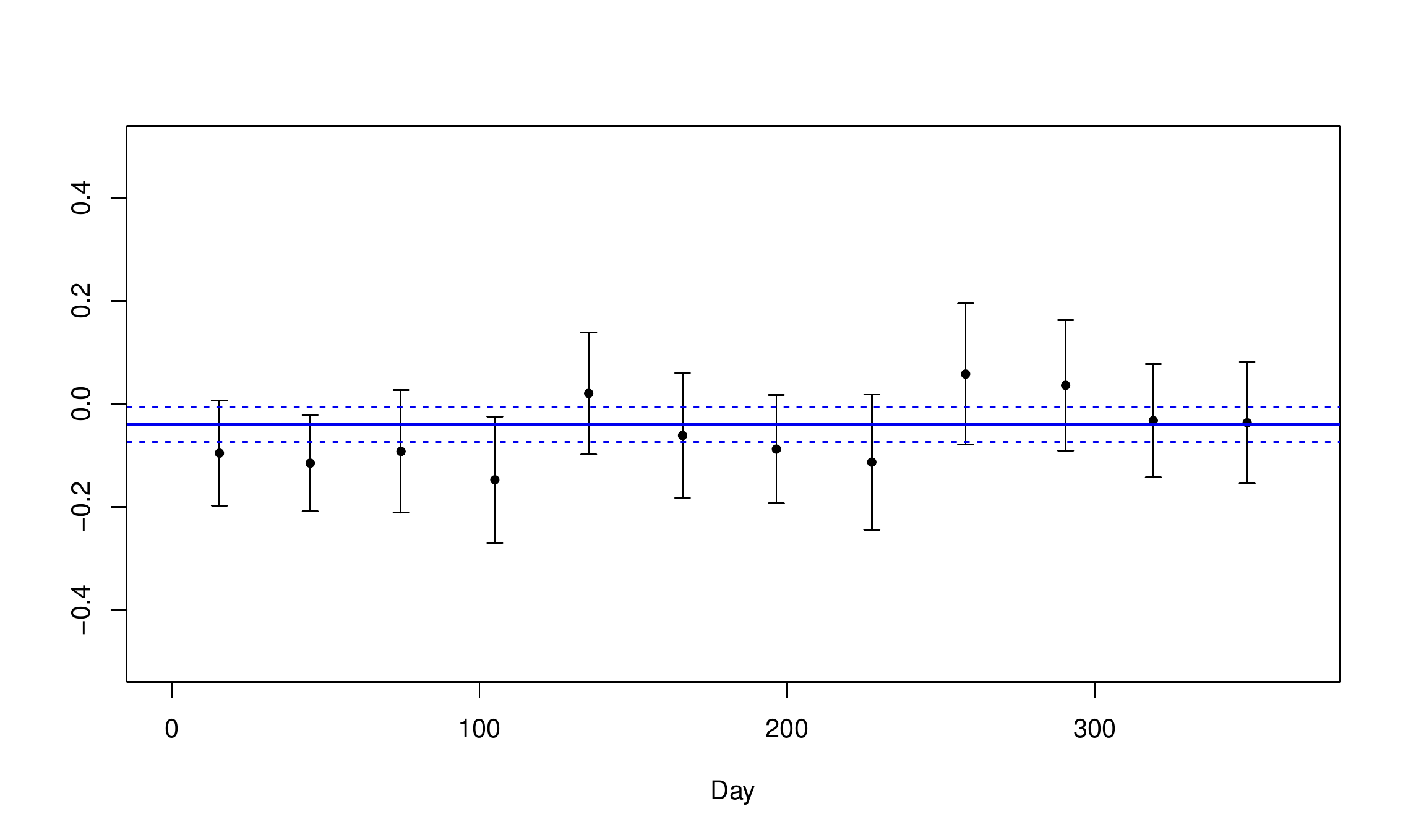}
    \includegraphics[width=0.5\textwidth]{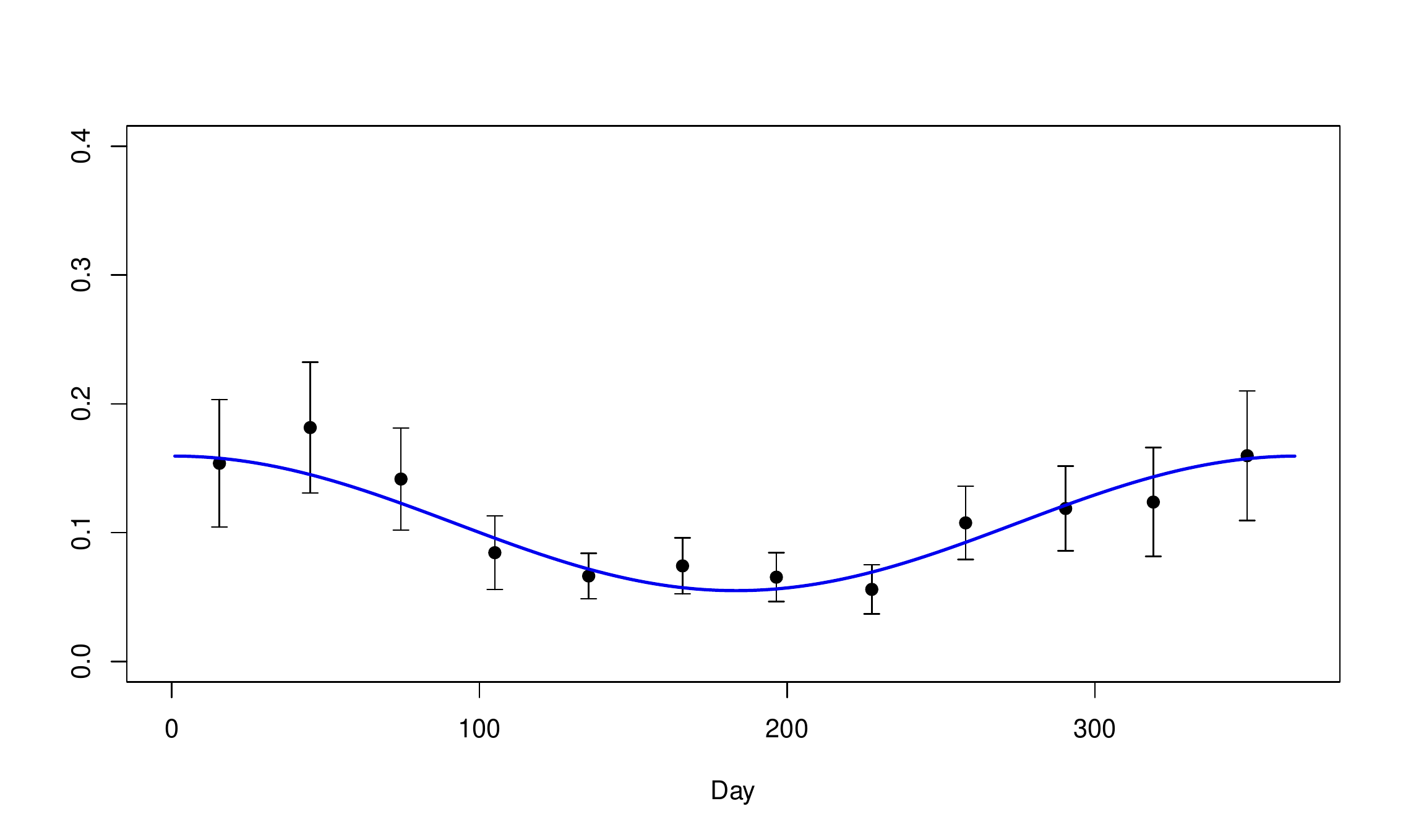}\includegraphics[width=0.5\textwidth]{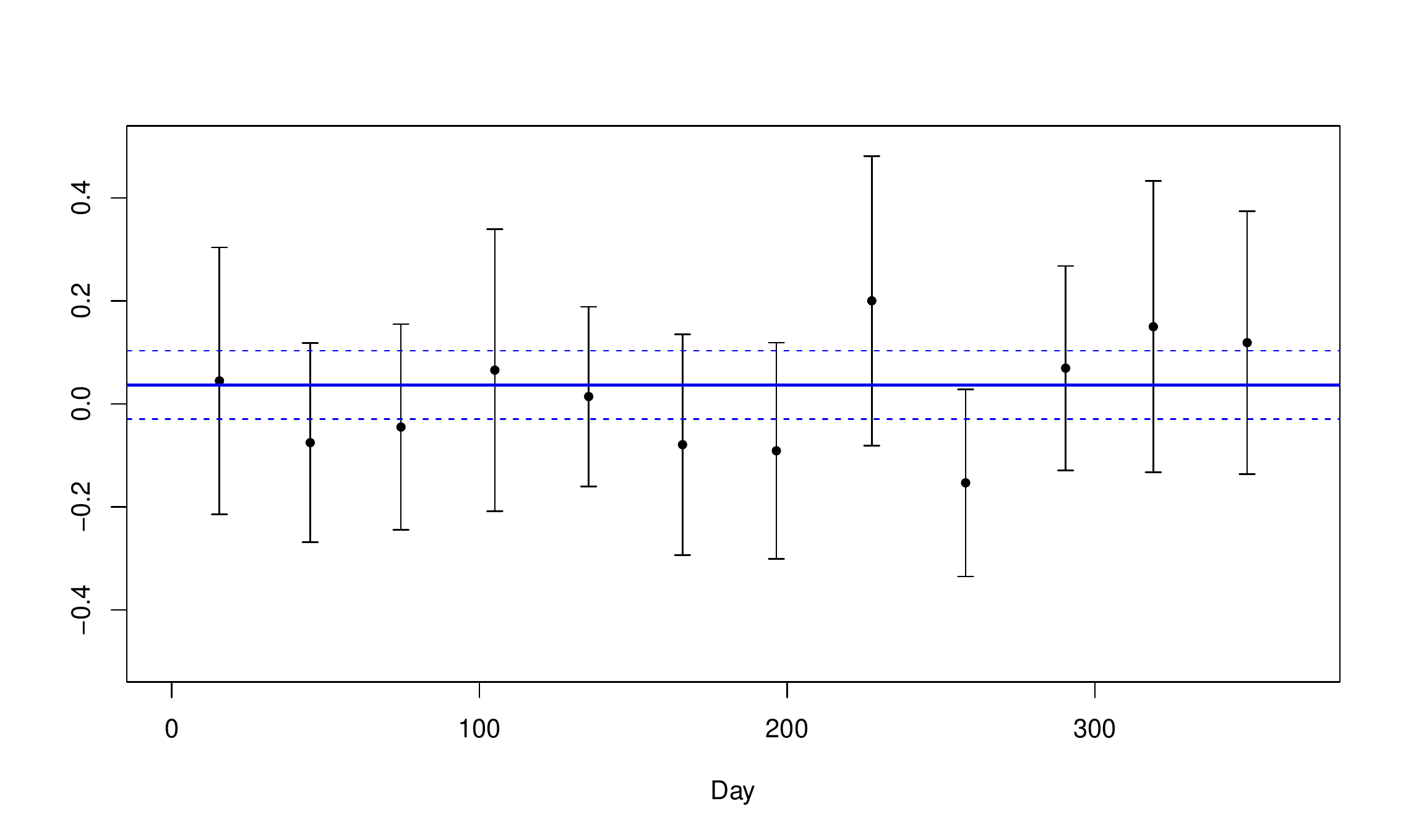}
    \caption{Scale (left column) and shape (right column) parameter estimates for Model $S2$ (blue) and Model~$S0$ (black) at Lowestoft (top row), Newlyn (middle row) and Sheerness (bottom row). 95\% confidence intervals are added to Model~$S0$ parameter estimates (shown by black error bars) and to the shape parameter estimate of Model $S2$ (blue dashed lines).}
    \label{ss_param_ests_fig_all}
\end{figure}


\begin{figure}[h]
    \centering
    \includegraphics[width=0.4\textwidth]{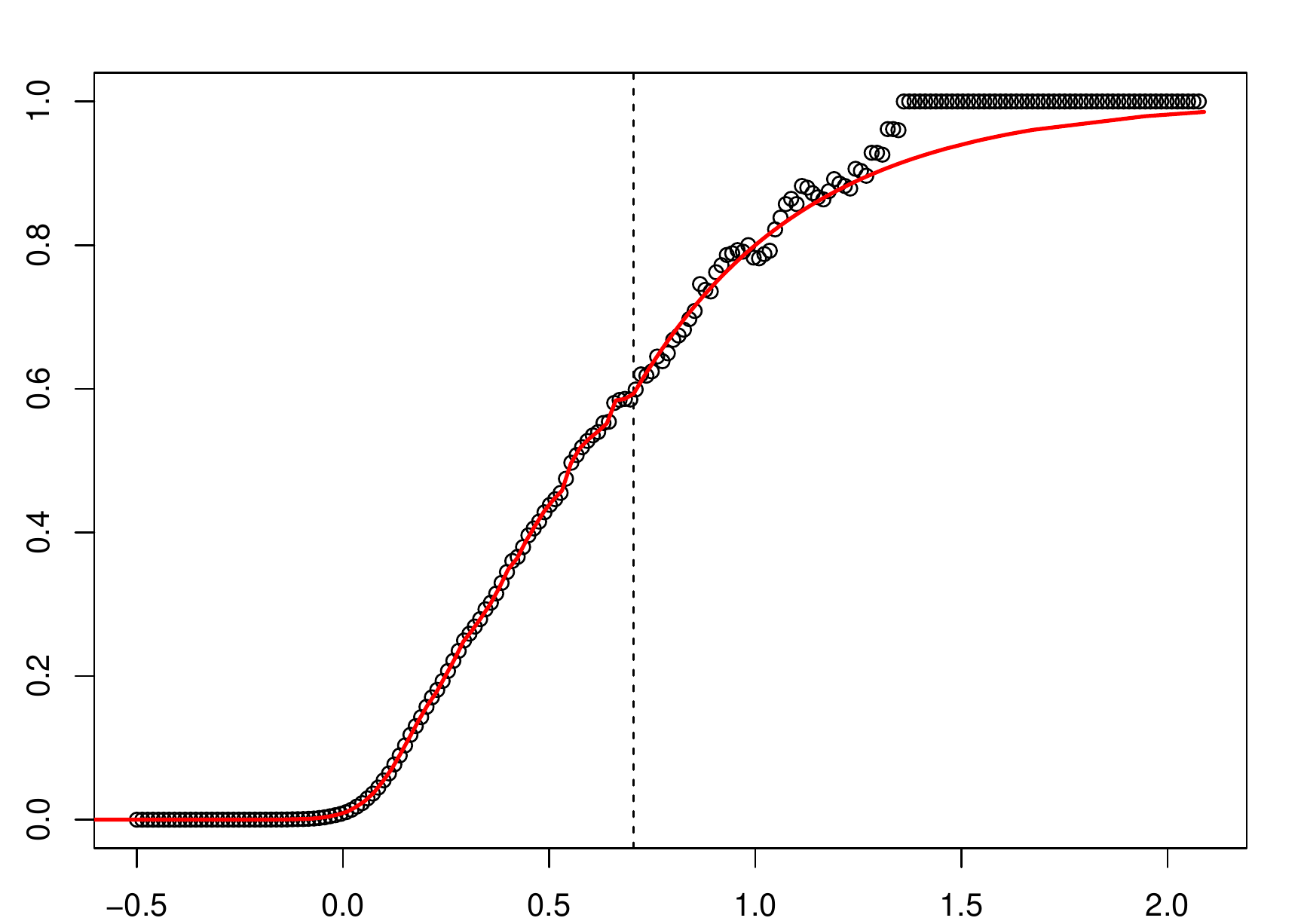}
    \includegraphics[width=0.4\textwidth]{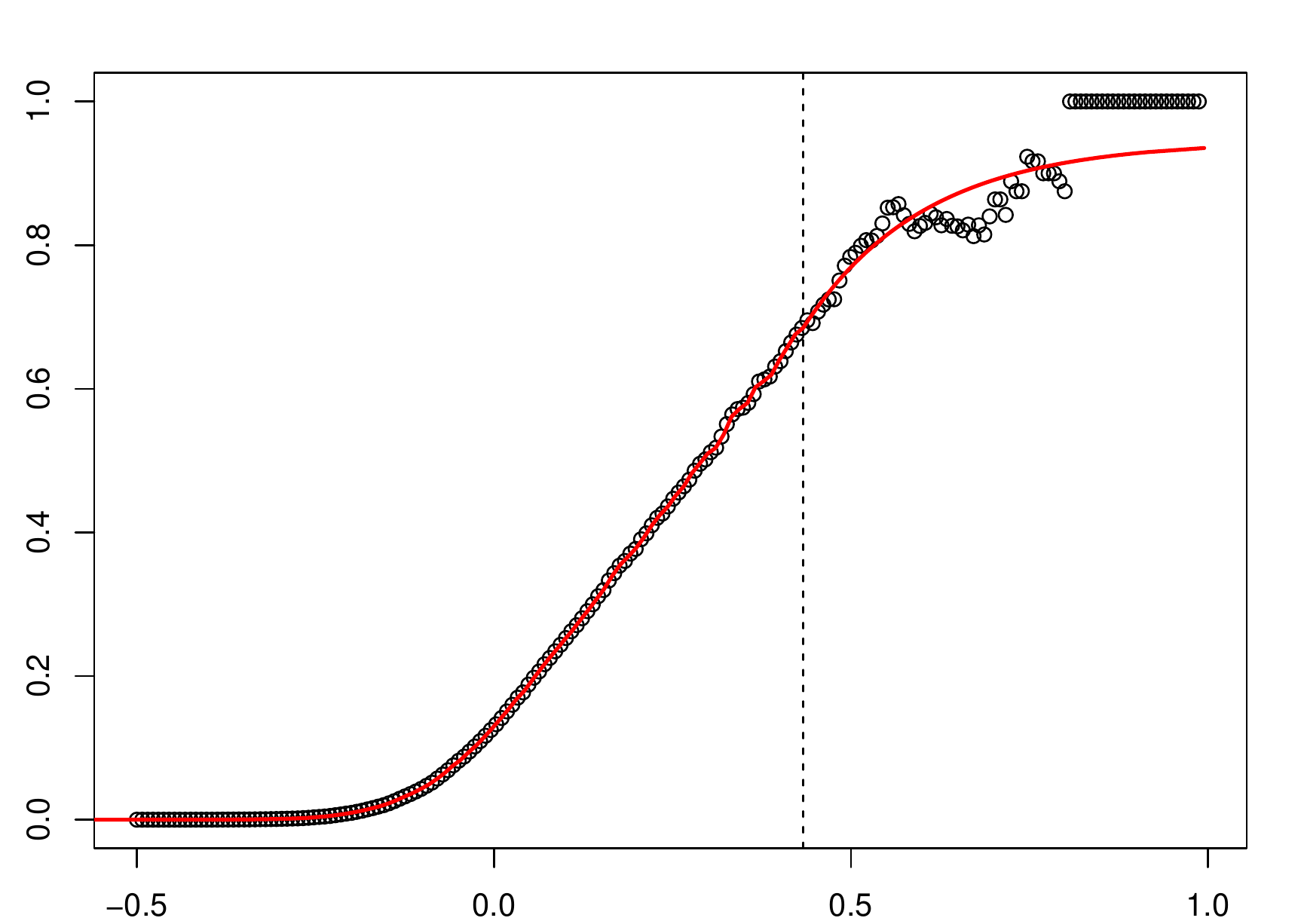}
    \includegraphics[width=0.4\textwidth]{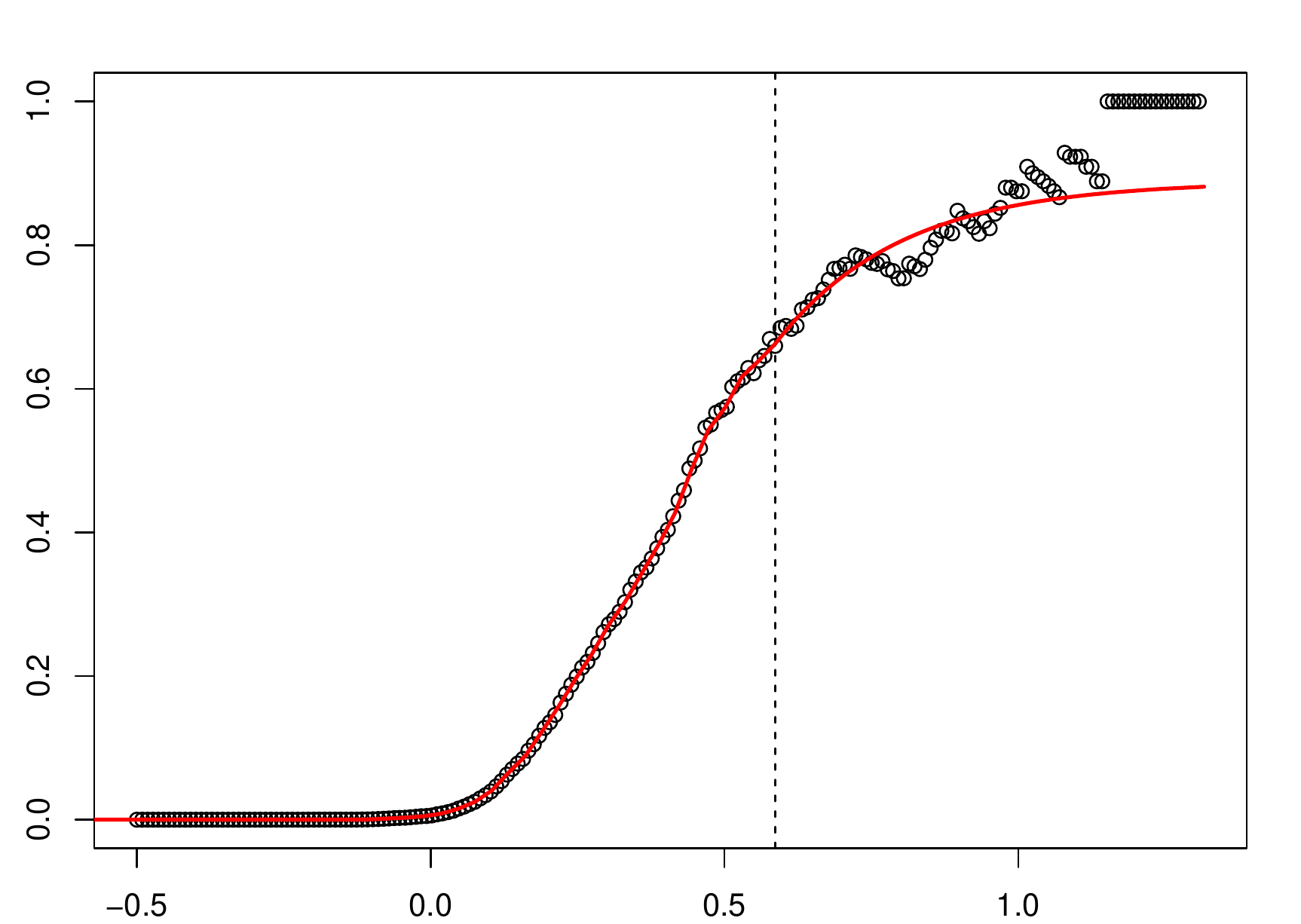}
    \caption{Estimates of the subasymptotic extremal index $\theta(y,r)$ for different skew surge levels using the runs estimate (black) and our model estimate (red) (expression~\eqref{eqn::exi_model} of the main paper) at Lowestoft (top left), Newlyn (top right) and Sheerness (bottom). Run lengths are chosen as 10, 2 and 10, respectively. The threshold $v$ is chosen as the 0.99 skew surge quantile for all sites (black dashed line).}
    \label{fig::exifun_all}
\end{figure}


\begin{figure}[h]
    \centering
    \includegraphics[width=0.49\textwidth]{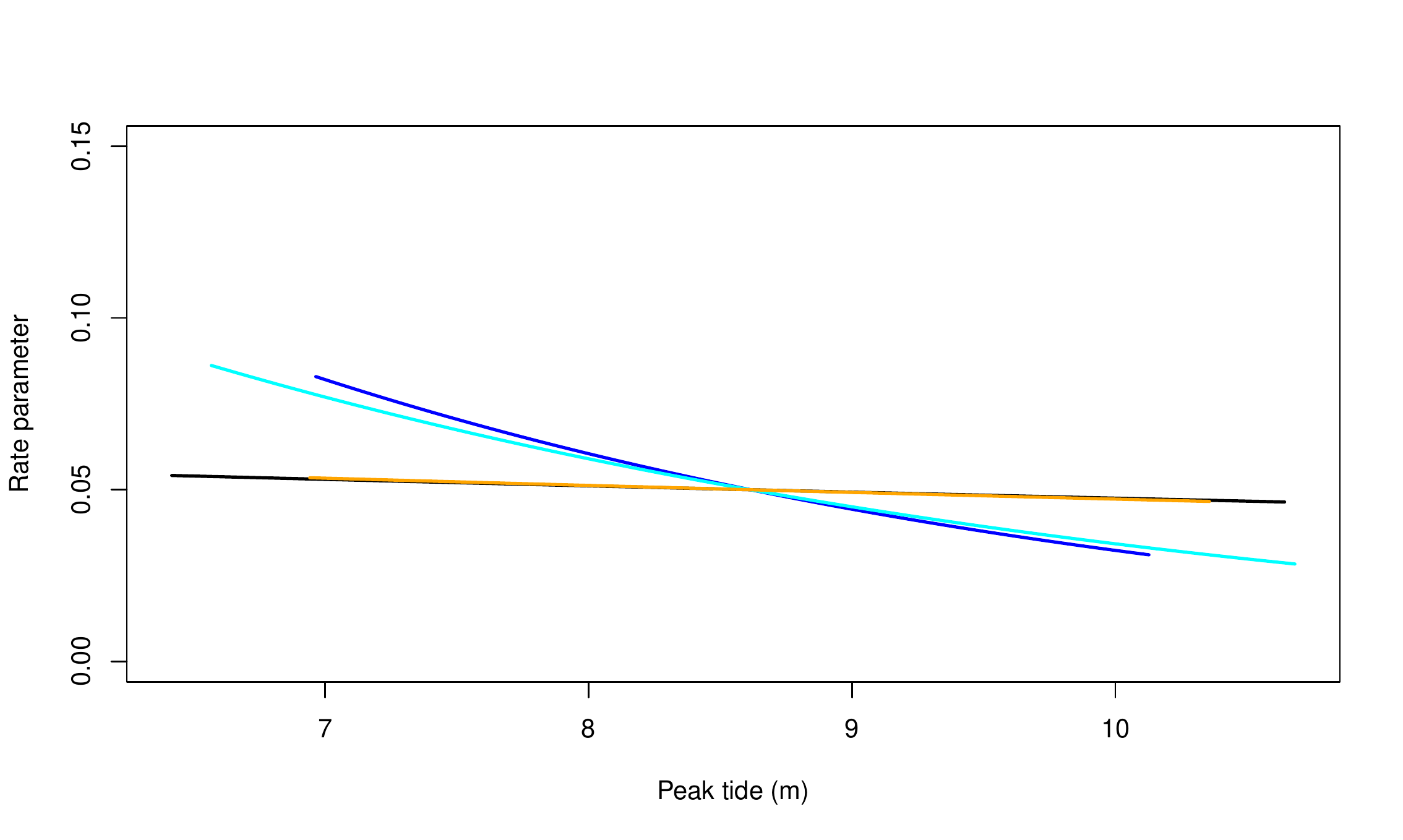}
    \includegraphics[width=0.49\textwidth]{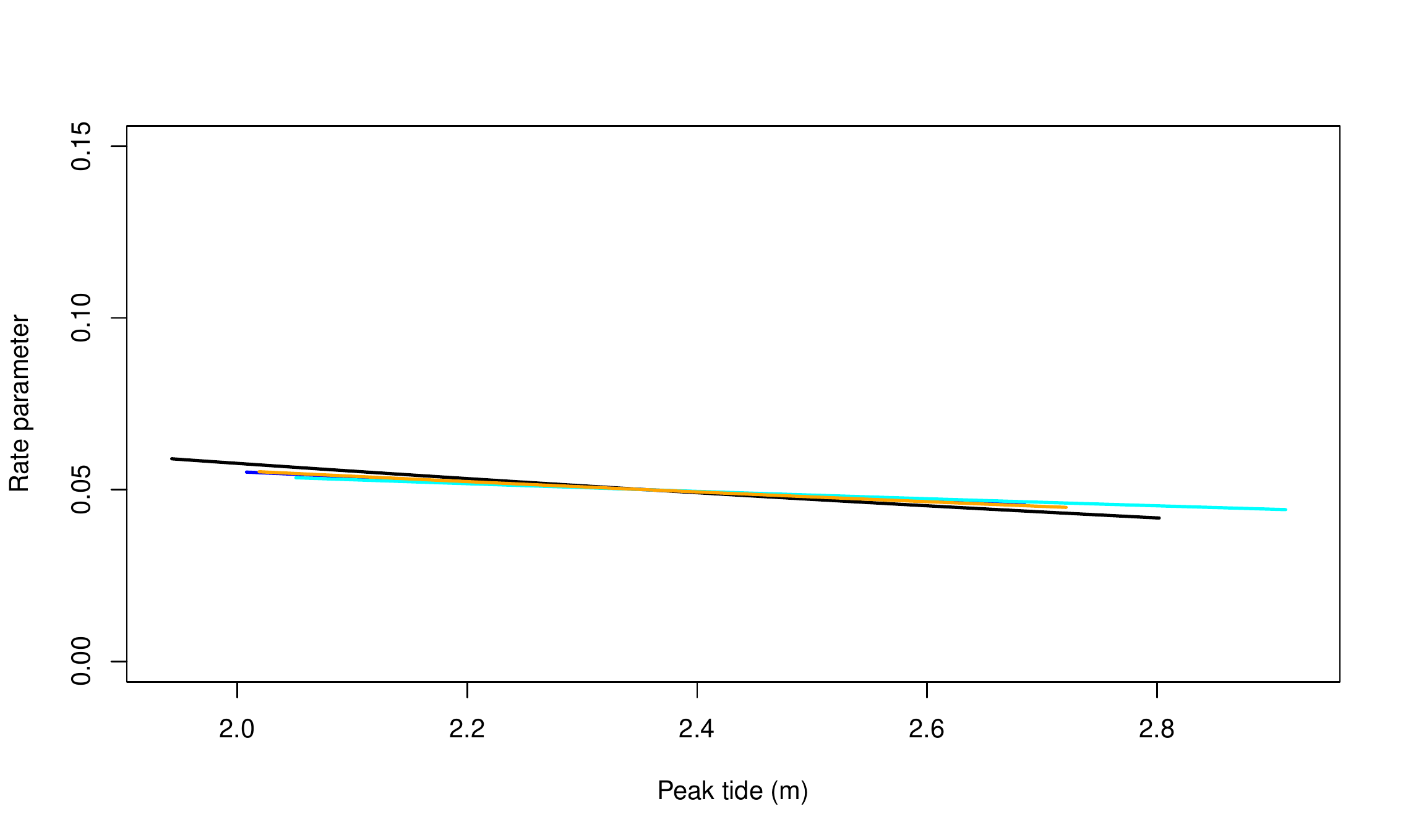}
    \includegraphics[width=0.49\textwidth]{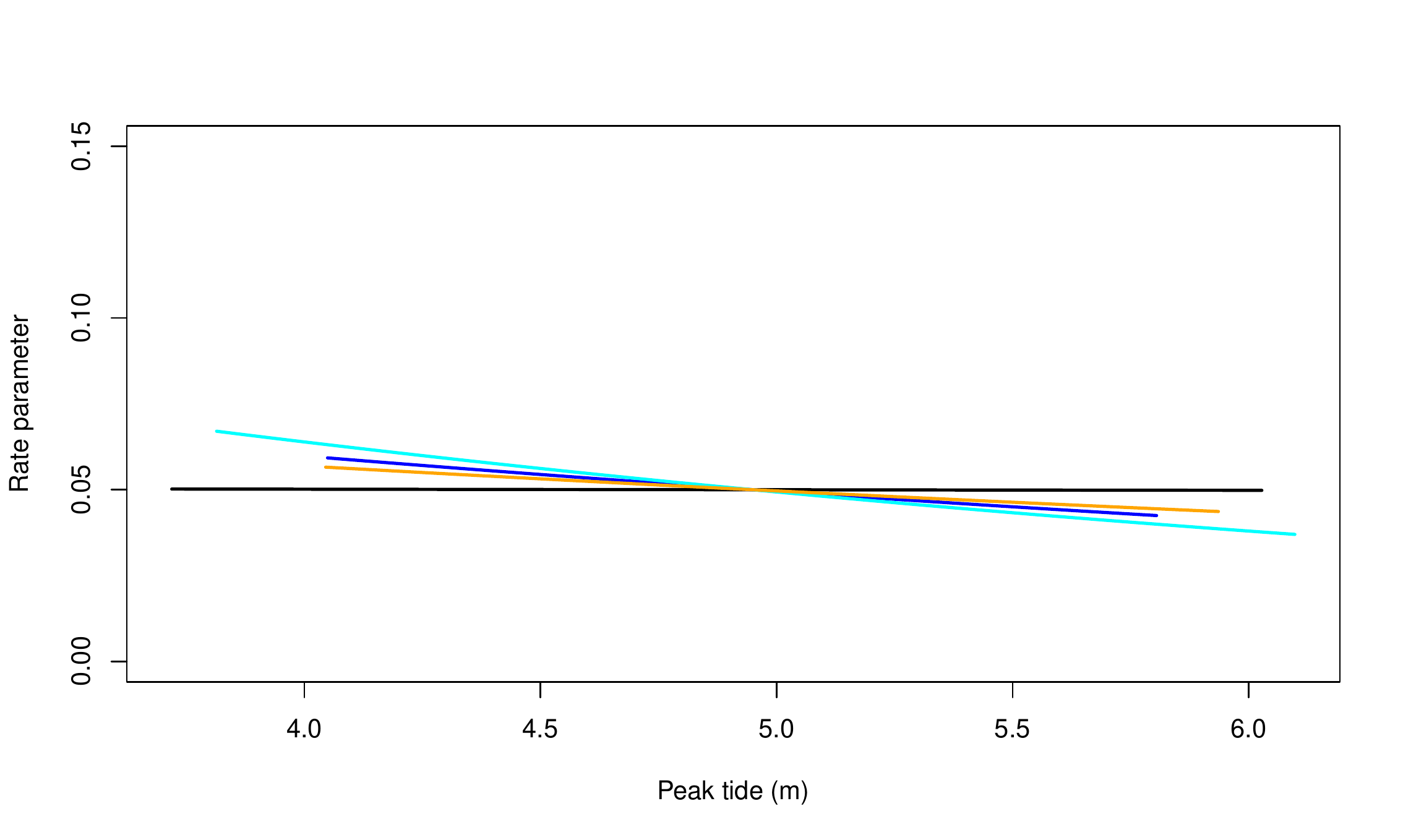}        \includegraphics[width=0.49\textwidth]{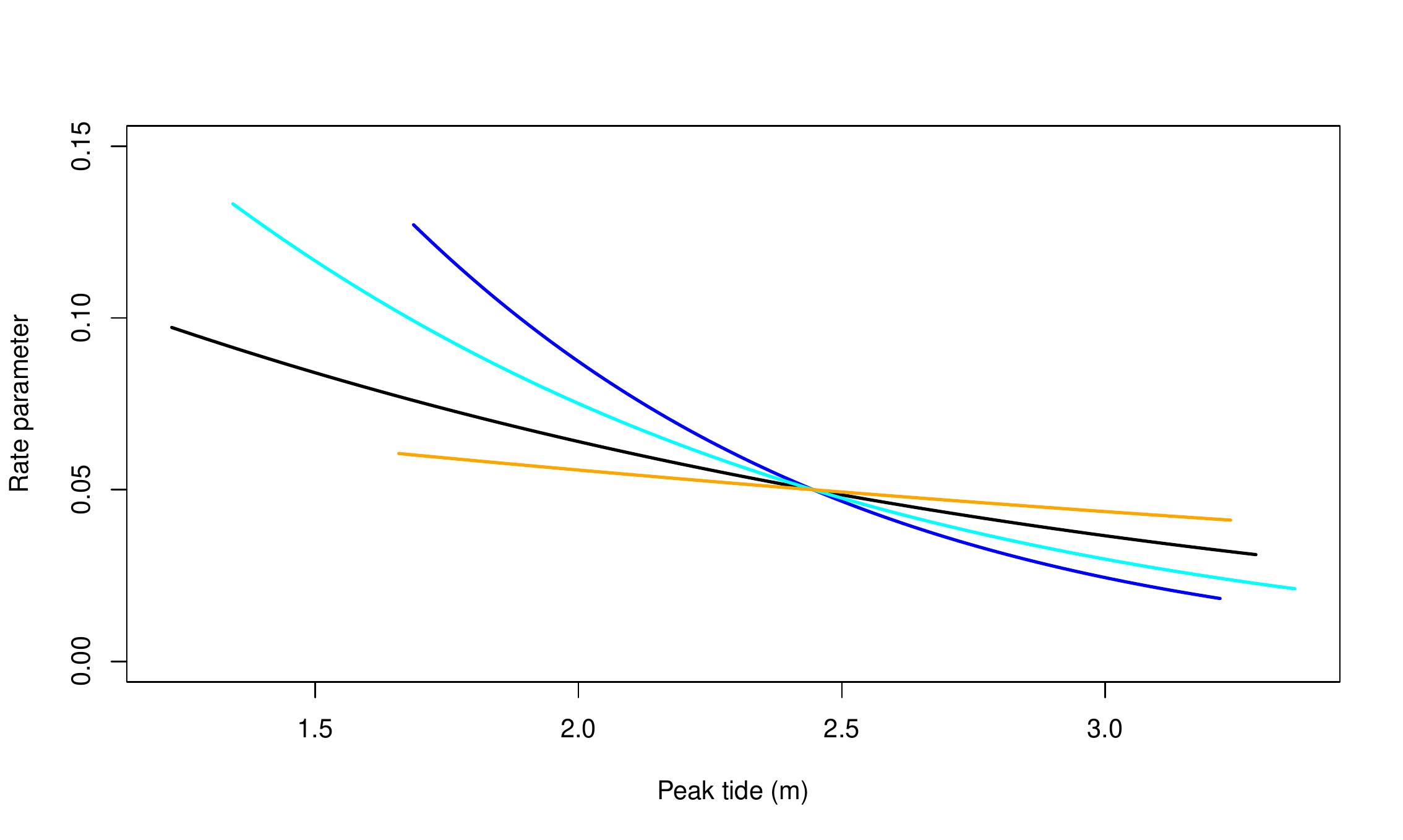}
    \caption{Estimated exceedance probability $\lambda_{d,x}$ (expression~\eqref{eqn::ss_rate_tide}) per month with respect to peak tide $x$ and day in month $d_j$, averaged over day at Heysham (top left), Lowestoft (top right), Newlyn (bottom left) and Sheerness (bottom right). Trends for March (blue), June (black), September (cyan) and December (orange) are shown here.}
    \label{fig::exprobtide_av_all}
\end{figure}

\begin{figure}[h]
    \centering
    \includegraphics[width=0.75\textwidth]{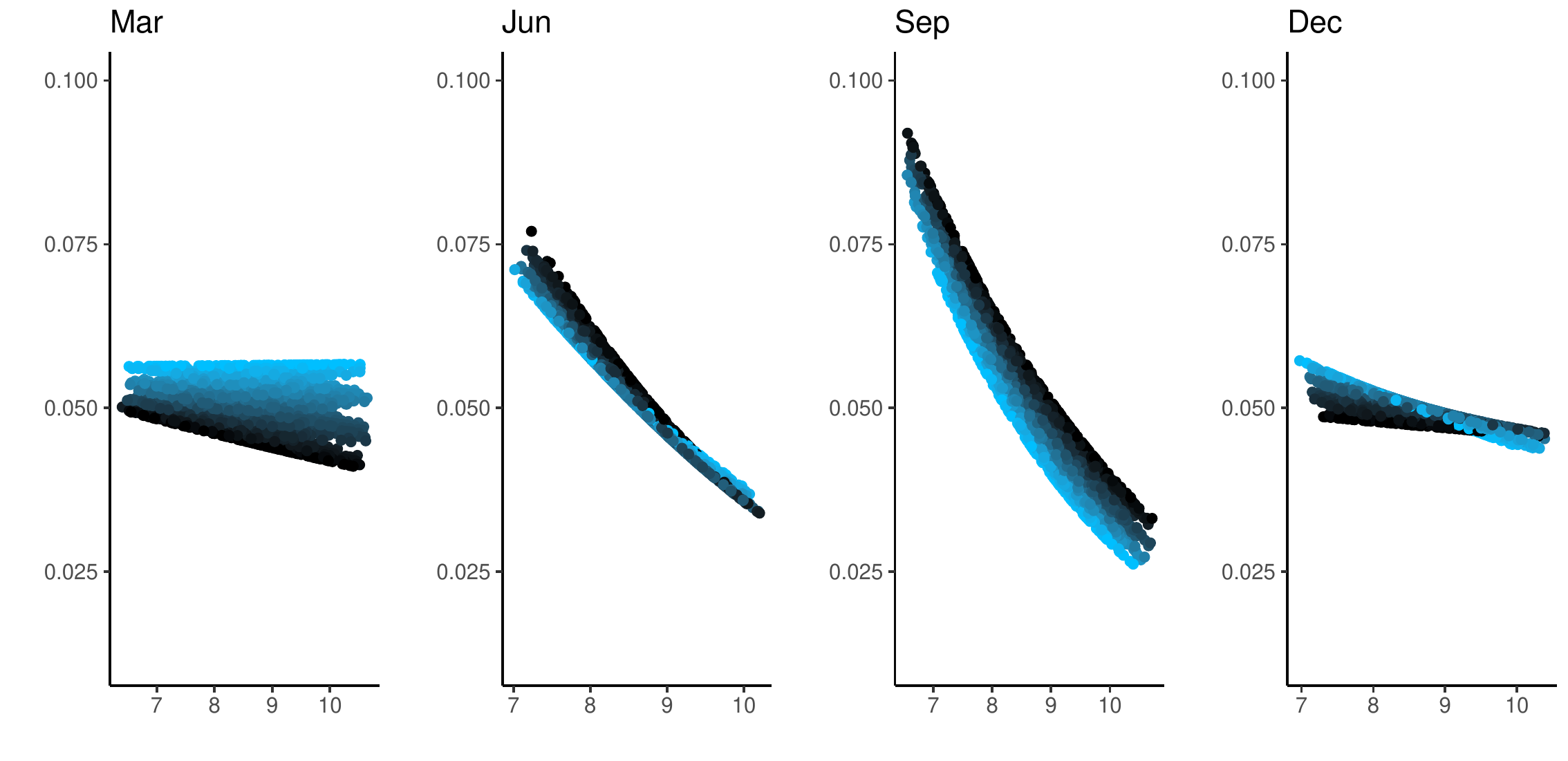}
    \includegraphics[width=0.75\textwidth]{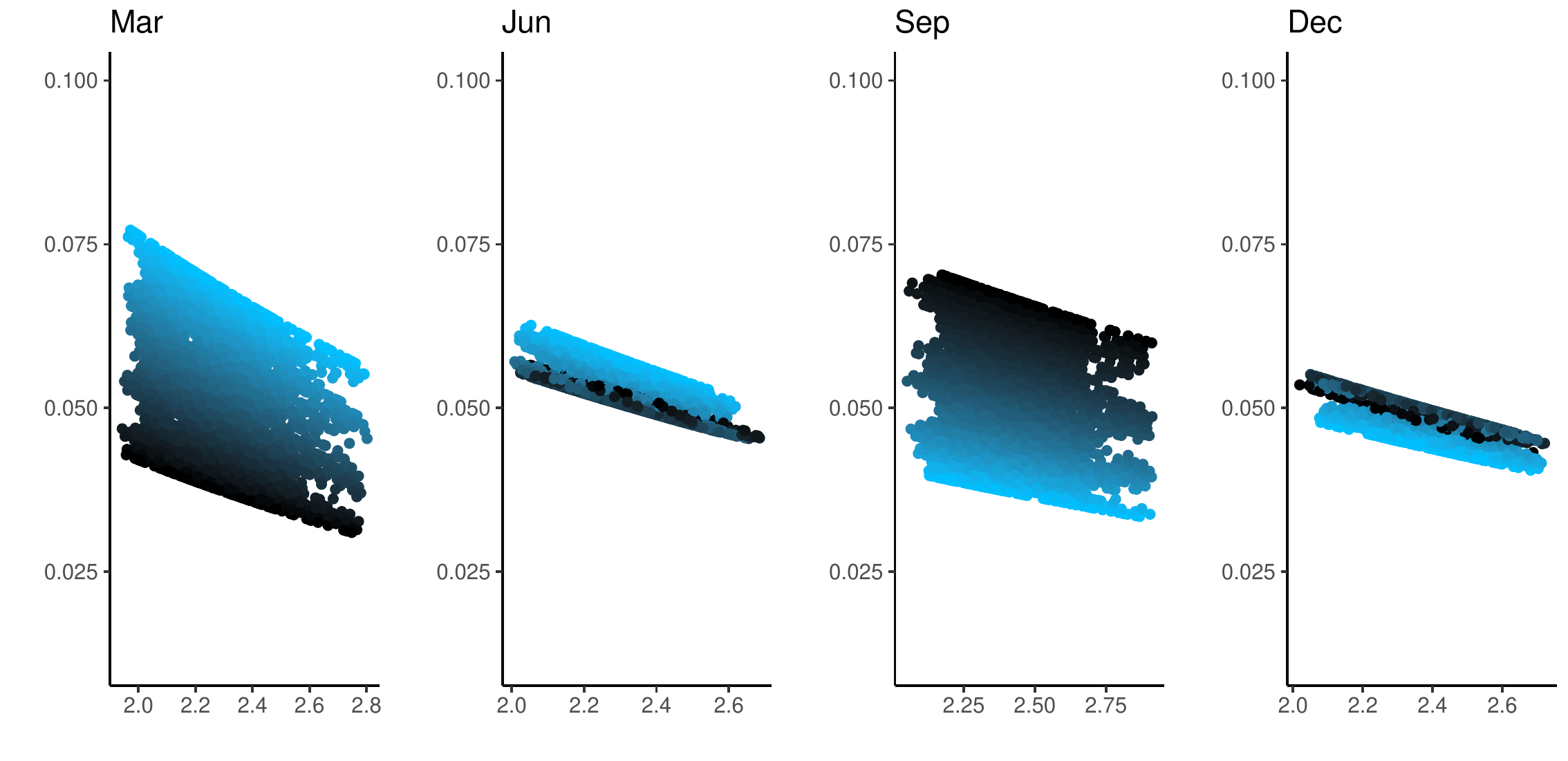}
    \includegraphics[width=0.75\textwidth]{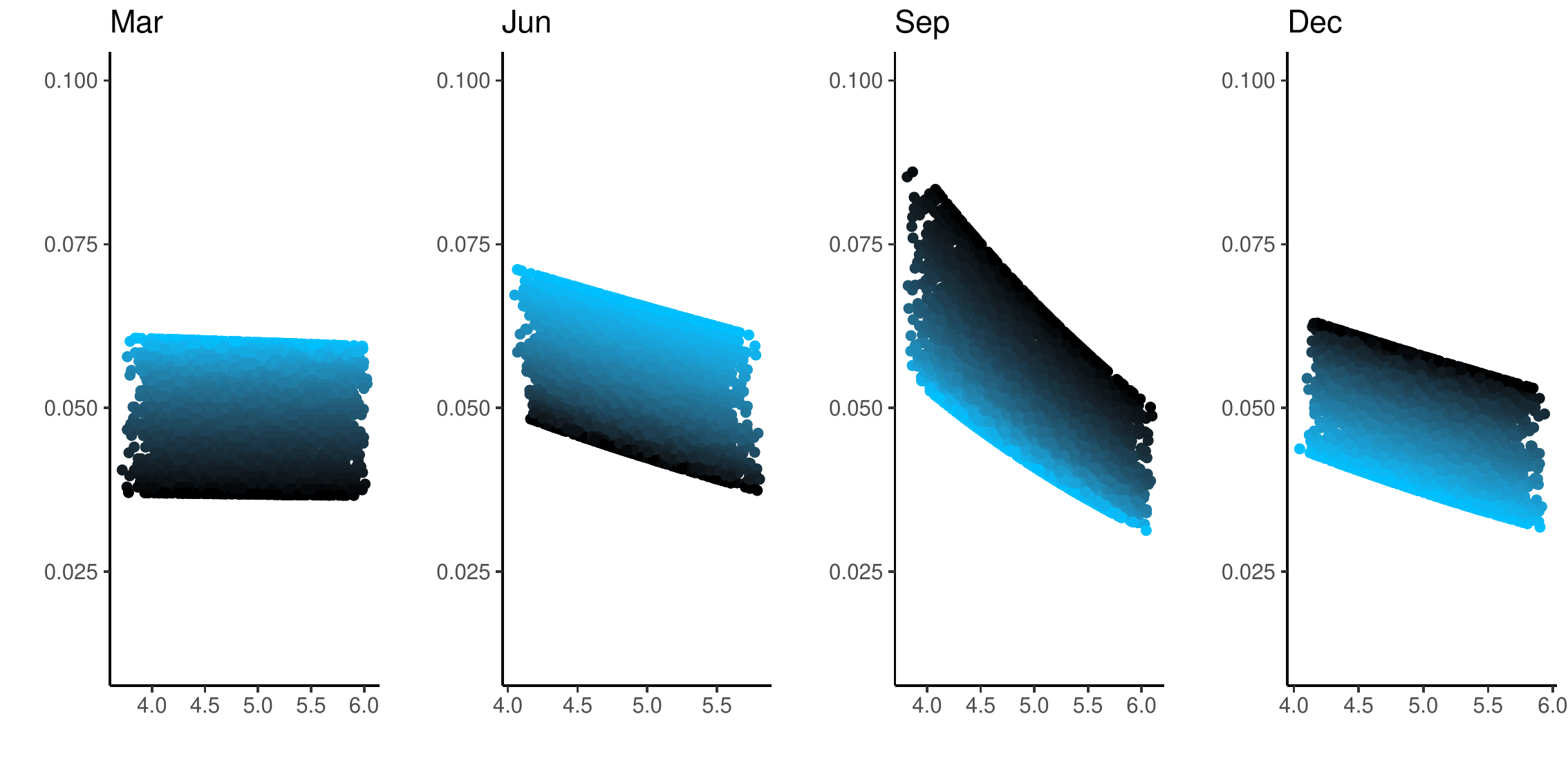}
    \caption{Estimated exceedance probability $\lambda_{d,x}$ (expression~\eqref{eqn::ss_rate_tide}) at Heysham (top row), Lowestoft (middle row) and Newlyn (bottom row), in March, June, September and December (from left to right by column) with respect to $x$ being peak tide (m) and $d_j$ being day in month at Sheerness. Darker (lighter) points represent days later (earlier) in the month.}
    \label{fig::exprobtide_MJSD_all}
\end{figure}

\begin{table}[h]
    \centering
    \resizebox{\textwidth}{!}{
    \begin{tabular}{c|c|c|c|c}
         & Heysham & Lowestoft & Newlyn & Sheerness\\ \hline
         \multicolumn{5}{l}{Model $S2$} \\ \hline
         $\alpha_\sigma$ & 0.14 (0.13, 0.15) & 0.15 (0.14, 0.16) & 0.076 (0.073, 0.080) & 0.11 (0.10, 0.12)\\
         $\beta_\sigma$ & 0.060 (0.050, 0.070) & 0.080 (0.070, 0.090) & 0.024 (0.020, 0.028) & 0.052 (0.043, 0.061) \\
         $\phi_\sigma$ & 271.51 (262.77, 280.23) & 266.32 (260.01, 272.63) & 273.58 (265.06, 282.10) & 272.11 (262.66, 281.56) \\
         $\xi$ & 0.002 (-0.042, 0.051) & 0.024 (-0.023, 0.071) & -0.040 (-0.074, 0.006) & 0.037 (-0.029, 0.10)\\ \hline
         \multicolumn{5}{l}{Model $S4$} \\ \hline
         $\alpha_\sigma$ & 0.13 (0.063, 0.19) & 0.16 (0.078, 0.24) & 0.053 (0.026, 0.79) & 0.14 (0.074, 0.14) \\
         $\beta_\sigma$ & 0.060 (0.049, 0.071) & 0.080 (0.070, 0.090) & 0.024 (0.020, 0.027) & 0.053 (0.043, 0.061) \\
         $\phi_\sigma$ & 272.20 (263.31, 281.09) & 266.10 (259.65, 272.56) & 278.97 (270.32, 287.61) & 271.37 (262.91, 281.32) \\
         $\gamma_\sigma^{(x)}$ & 0.002 (-0.005, 0.009) & -0.0051 (-0.040, 0.030) & 0.0048 (-0.00048, 0.010) & -0.012 (-0.026, 0.0011)\\
         $\xi$ & 0.0049 (-0.044, 0.054) & 0.024 (-0.023, 0.071) & -0.037 (-0.071, -0.003) & 0.033 (-0.033, 0.099)\\ \hline
         \multicolumn{5}{l}{Model $S4$ with prior on shape} \\ \hline
         $\xi$ & 0.019 (-0.021, 0.059) & 0.014 (-0.024, 0.052) & -0.027 (-0.058, 0.004) & 0.008 (-0.039, 0.054) \\ \hline
         \multicolumn{5}{l}{Model $R1$} \\ \hline
         $\beta_\lambda$ & 0.0087 (0.0004, 0.017) & 0.022 (0.015, 0.030) & 0.024 (0.018, 0.030) & 0.022 (0.014,0.032) \\
         $\phi_\lambda$ & 155.66 (100.74, 210.59) & 175.16 (155.86, 194.46) & 209.50 (195.48, 223.52) & 184.31 (160.94,207.69) \\
         $\alpha_\lambda^{(x)}$ & -0.13 (-0.18, -0.079) & -0.055 (-0.101, 0.009) & -0.063 (-0.099, 0.108) & -0.32 (-0.37,-0.26) \\
         $\beta_\lambda^{(x)}$ & 0.14 (0.068, 0.21) & -0.016 (-0.084, 0.051) & 0.061 (0.014, 0.108) & 0.23 (0.14, 0.31) \\
         $\phi_\lambda^{(x)}$ & 311.86 (281.78, 341.94) & 359.95 (265.77, 454.15) & 352.38 (299.32, 405.44) & 278.54 (260.44, 293.63)\\
    \end{tabular}
    }
    \caption{Parameter estimates for the scale parameter Models $S2$ and $S4$, and the rate parameter Model $R1$ with 95\% confidence intervals, at each site.}
    \label{tbl:SMparam_ests}
\end{table}





\begin{figure}[h]
    \centering
    \includegraphics[width=0.8\textwidth]{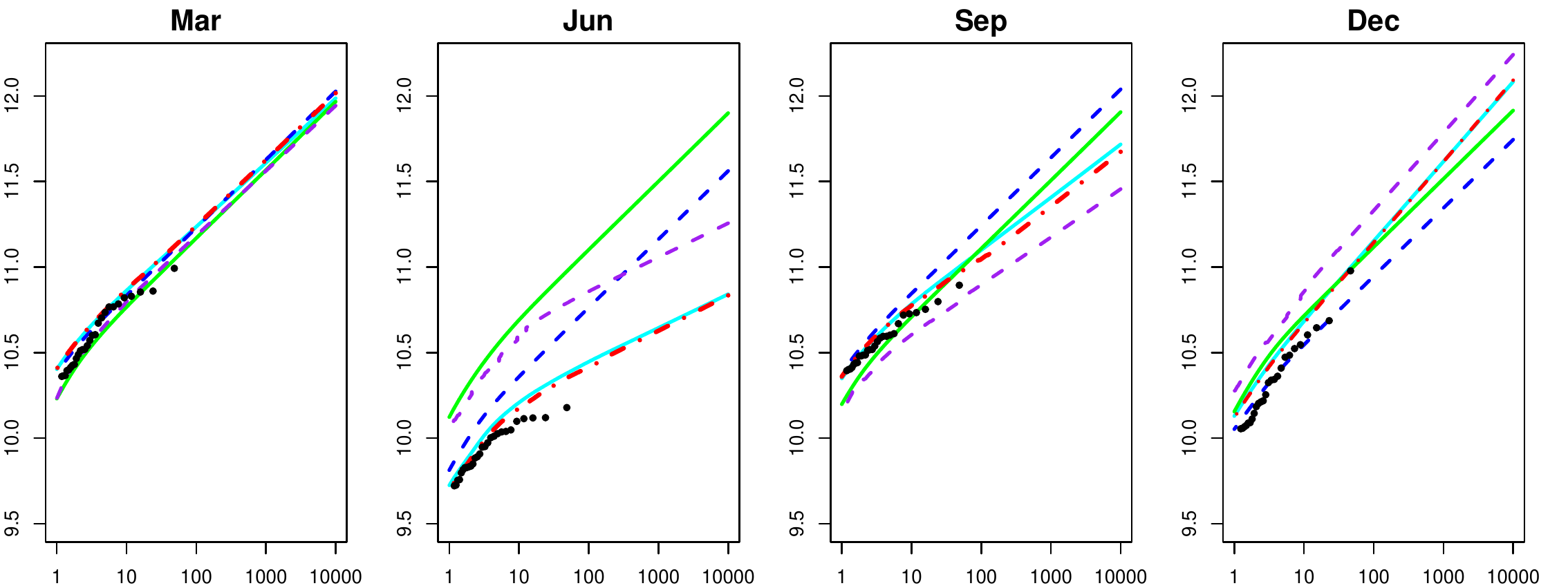} \\
    \includegraphics[width=0.8\textwidth]{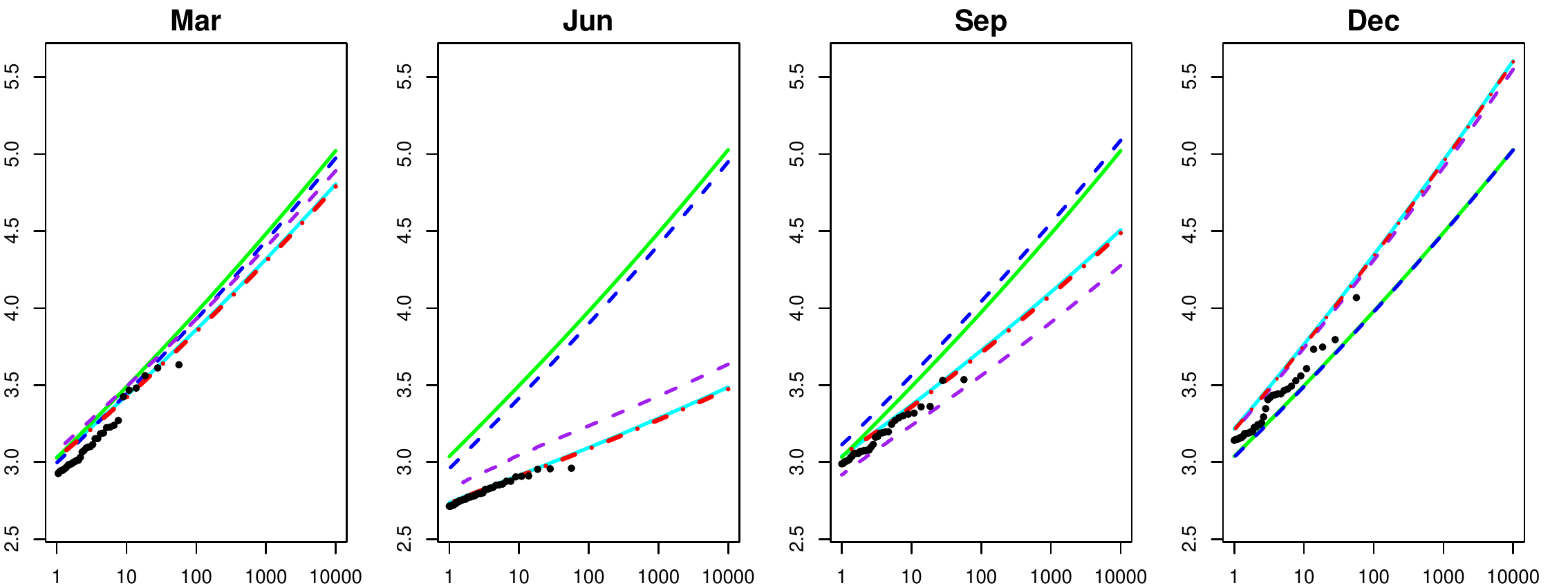} \\
    \includegraphics[width=0.8\textwidth]{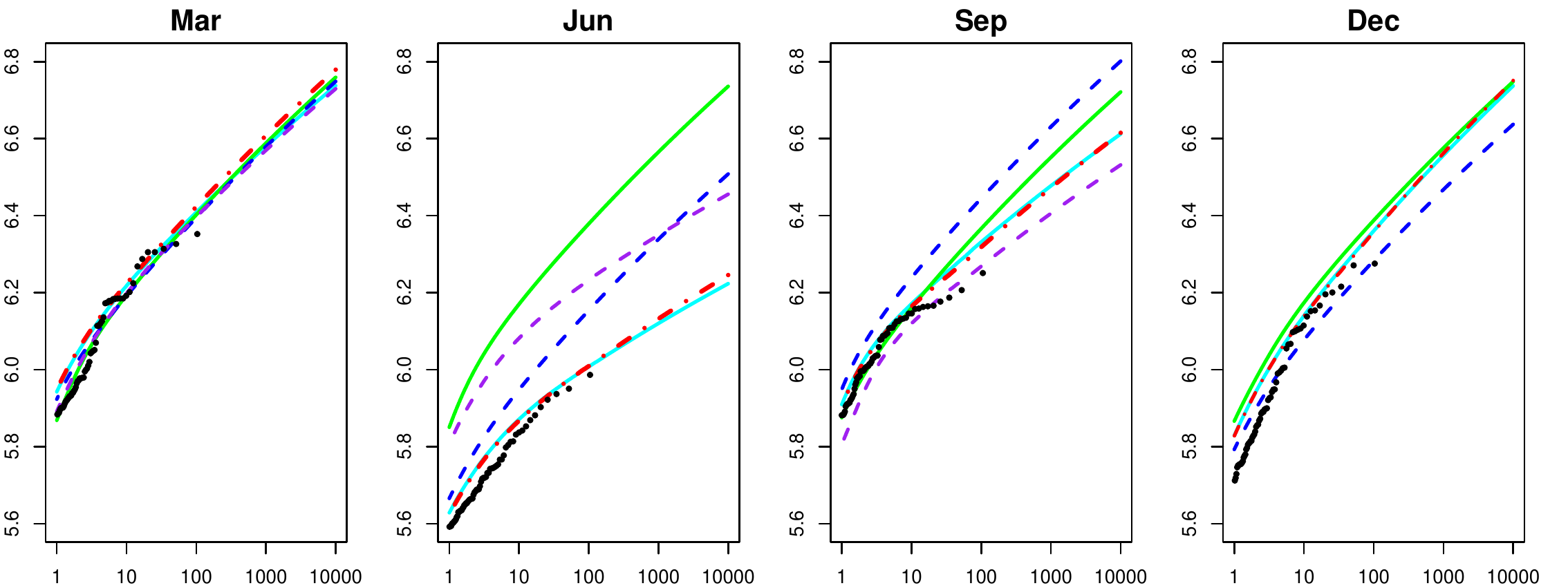} \\
    \includegraphics[width=0.8\textwidth]{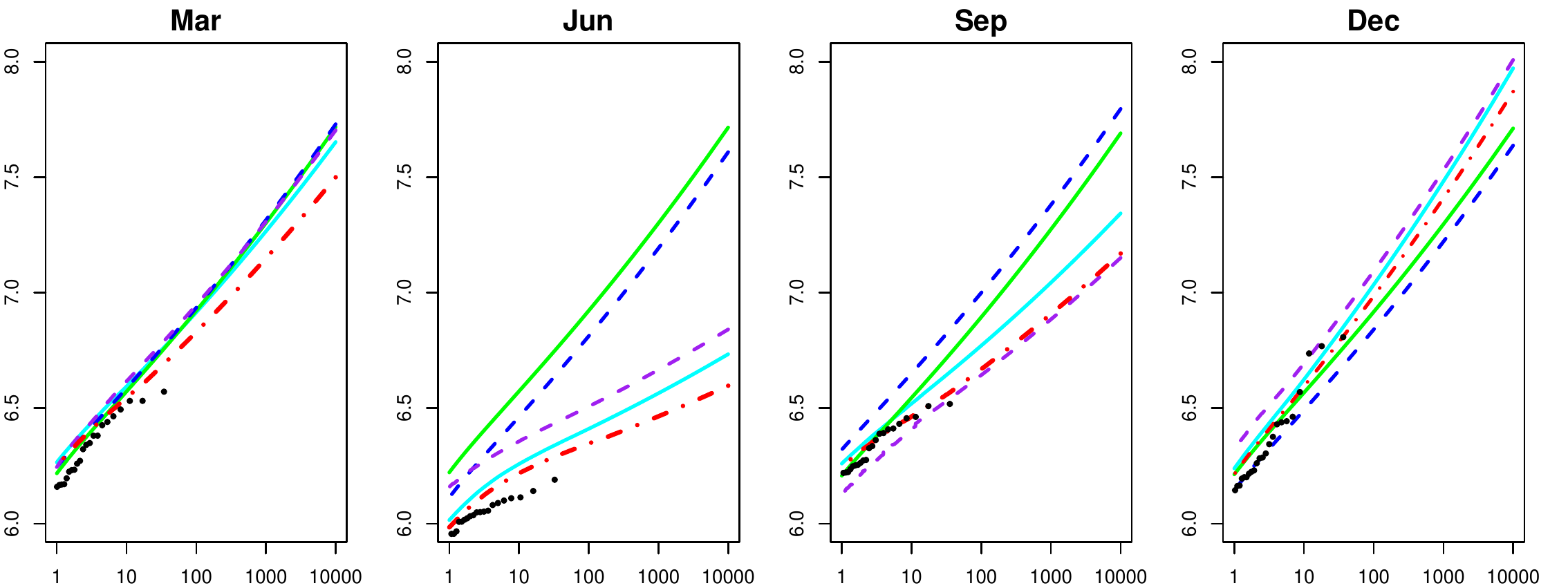} 
    \caption{Monthly maxima sea level return level estimates for Heysham, Lowestoft, Newlyn and Sheerness (from top to bottom row) in March, June, September and December (from left to right by column) estimated using the \textit{baseline} (green), \textit{seasonal surge} (purple dashed), \textit{seasonal tide} (blue dashed), \textit{full seasonal} (cyan) and \textit{interaction} (red dot-dashed) models. Empirical estimates are shown by black points.}
    \label{fig::monmax_all_RL_NEW_1x4}
\end{figure}

\begin{figure}[h]
    \centering 
    \includegraphics[width=0.4\textwidth]{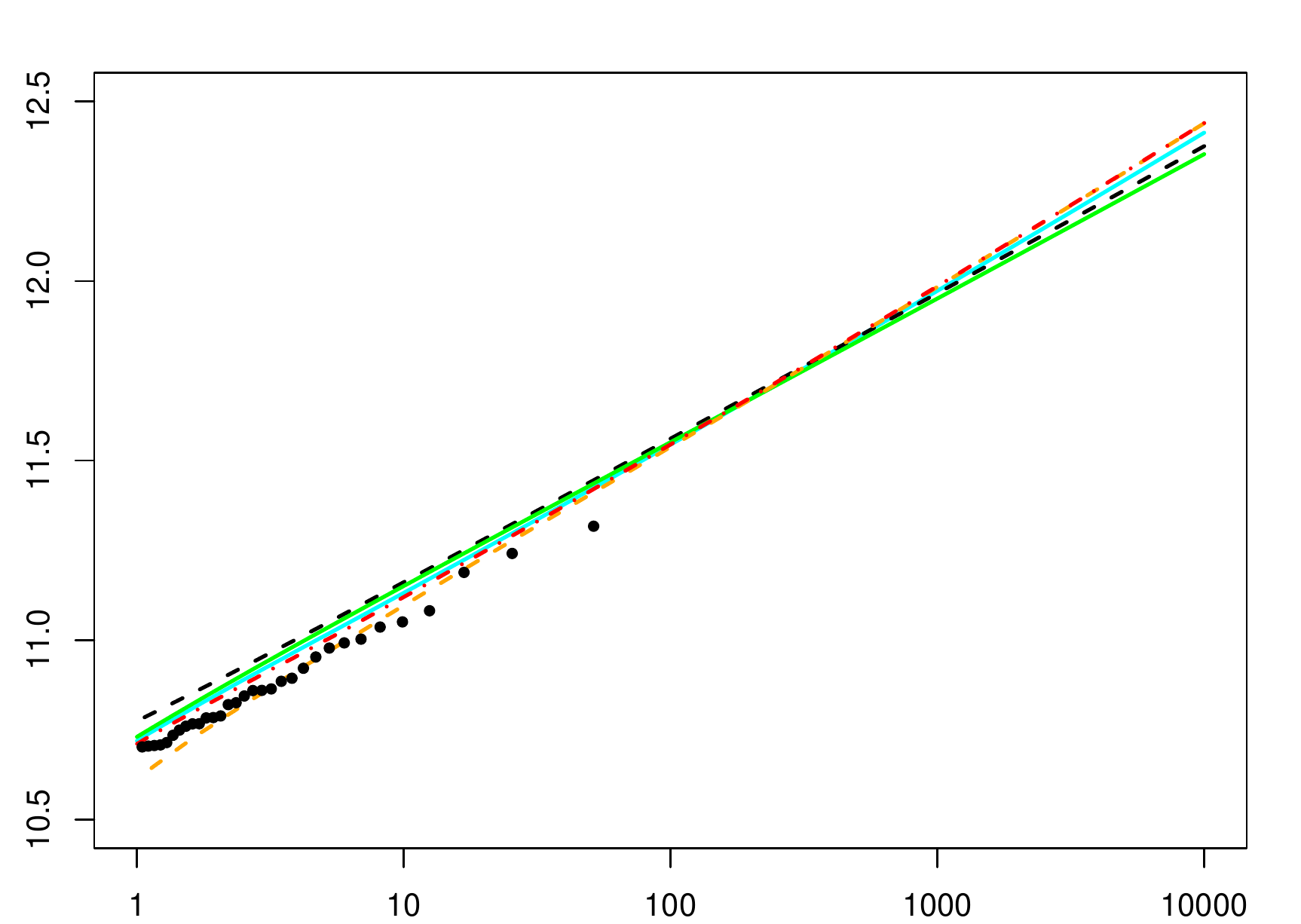}
    \includegraphics[width=0.4\textwidth]{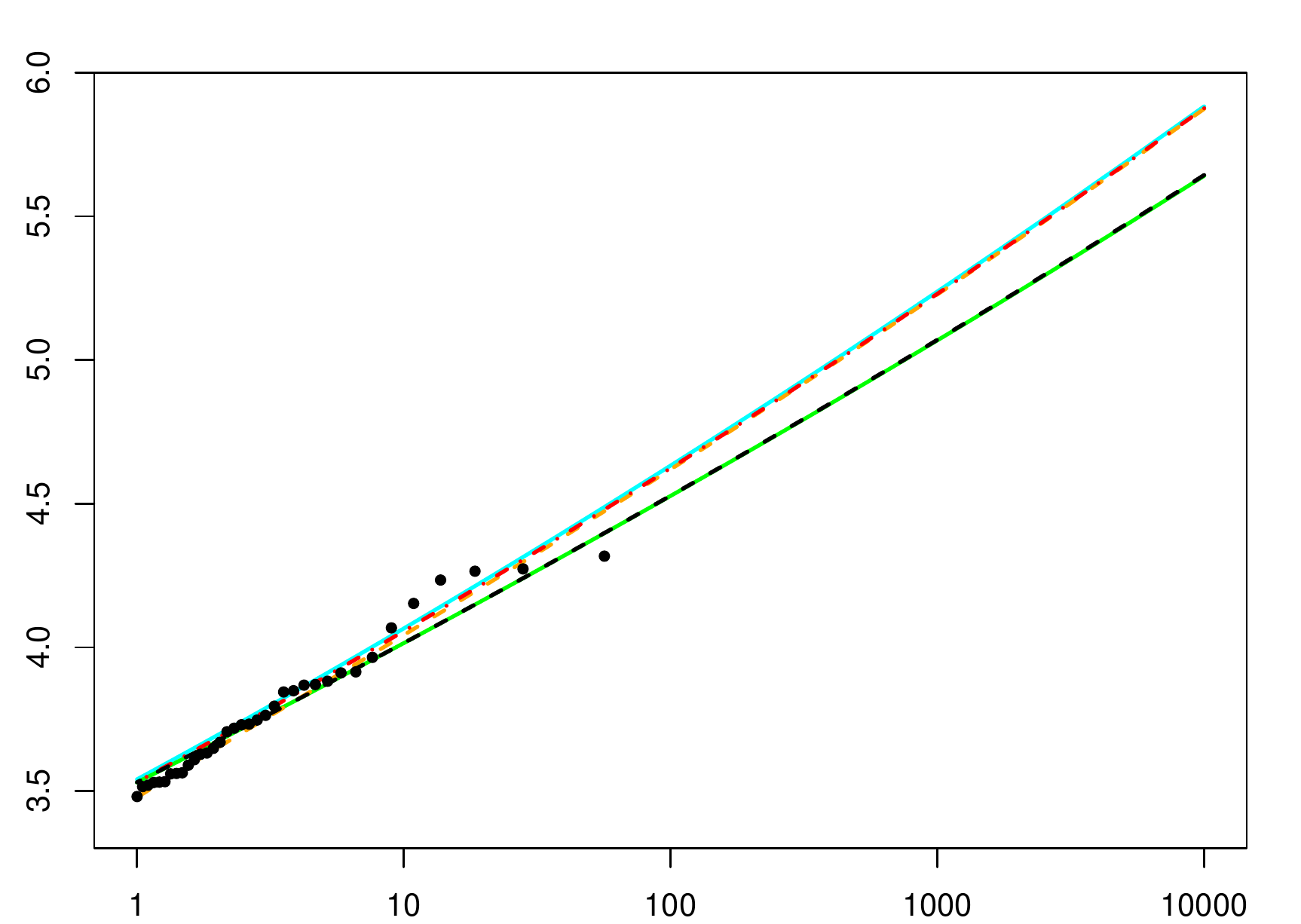} \\
    \includegraphics[width=0.4\textwidth]{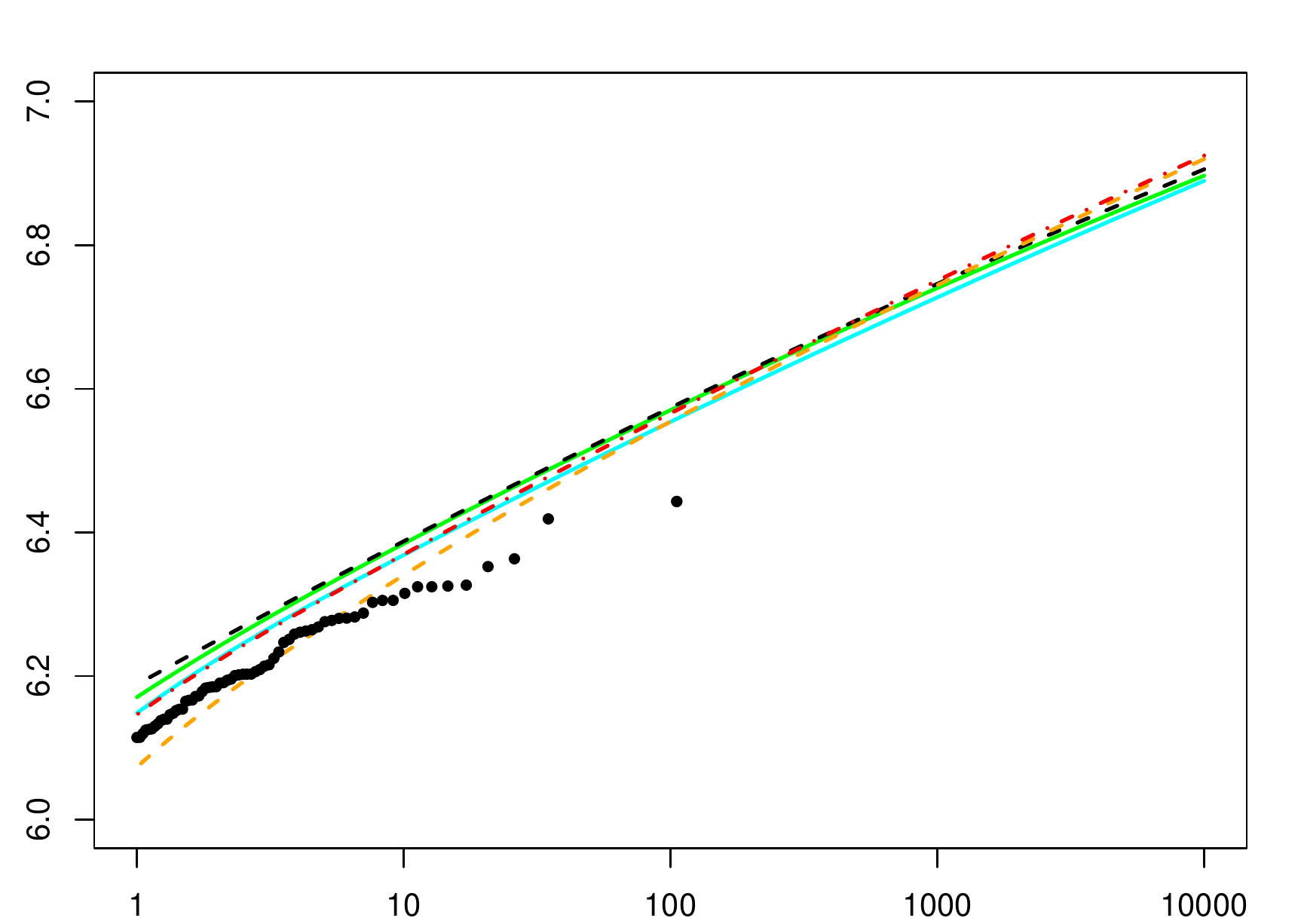}
    \includegraphics[width=0.4\textwidth]{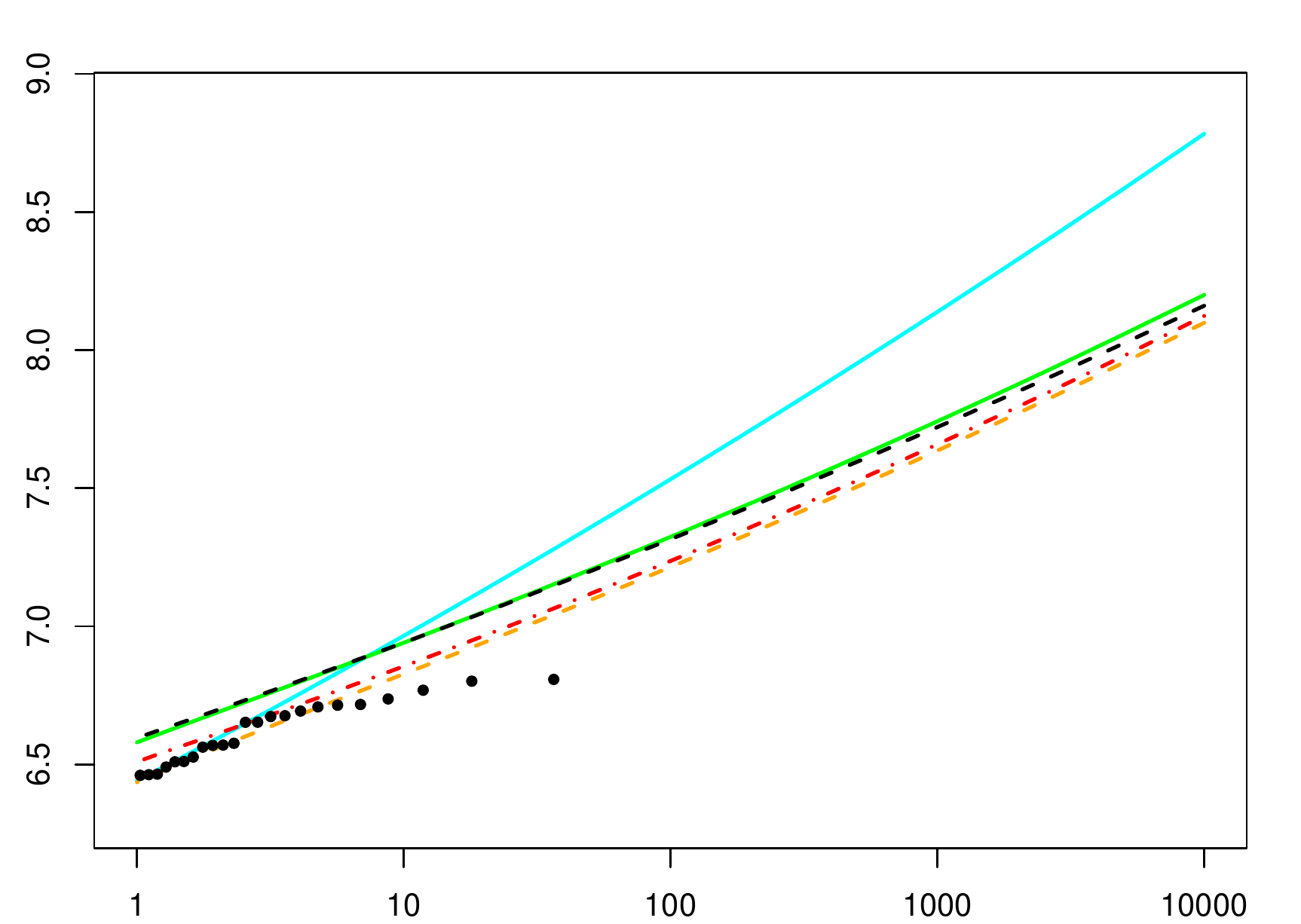} 
    \caption{Annual maxima sea level return level estimates for Heysham (top left), Lowestoft (top right), Sheerness (bottom left) and Newlyn (bottom right), estimated using the \textit{current} (black dashed), \textit{baseline} (green), \textit{full seasonal} (cyan), \textit{interaction} (red dot-dashed) and \textit{temporal dependence} (orange dashed) methods. Empirical estimates are shown by black points.}
    \label{fig::annmax_RL_NEW}
\end{figure}

\begin{figure}[h]
    \centering
     \includegraphics[width=0.4\textwidth]{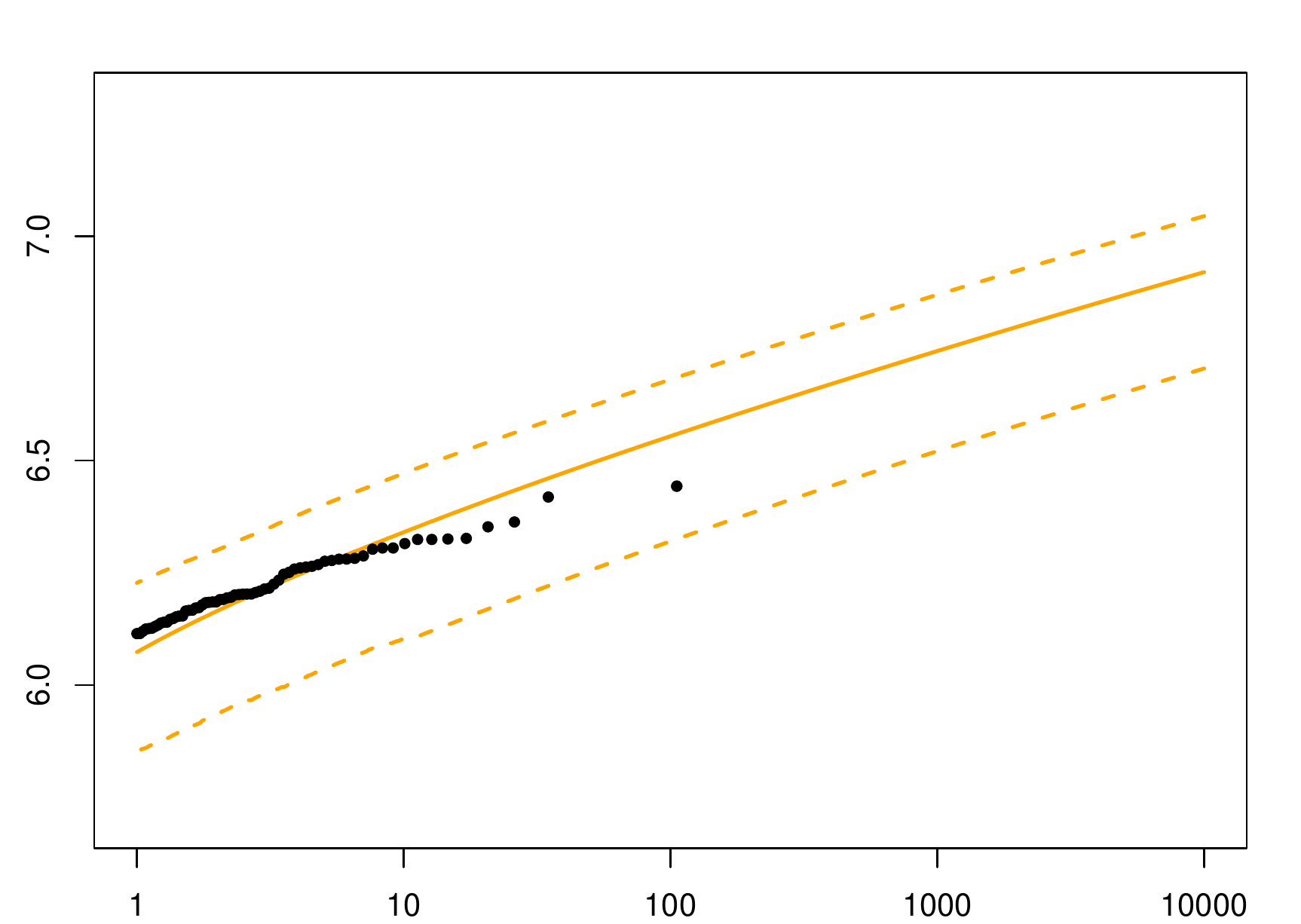}\includegraphics[width=0.4\textwidth]{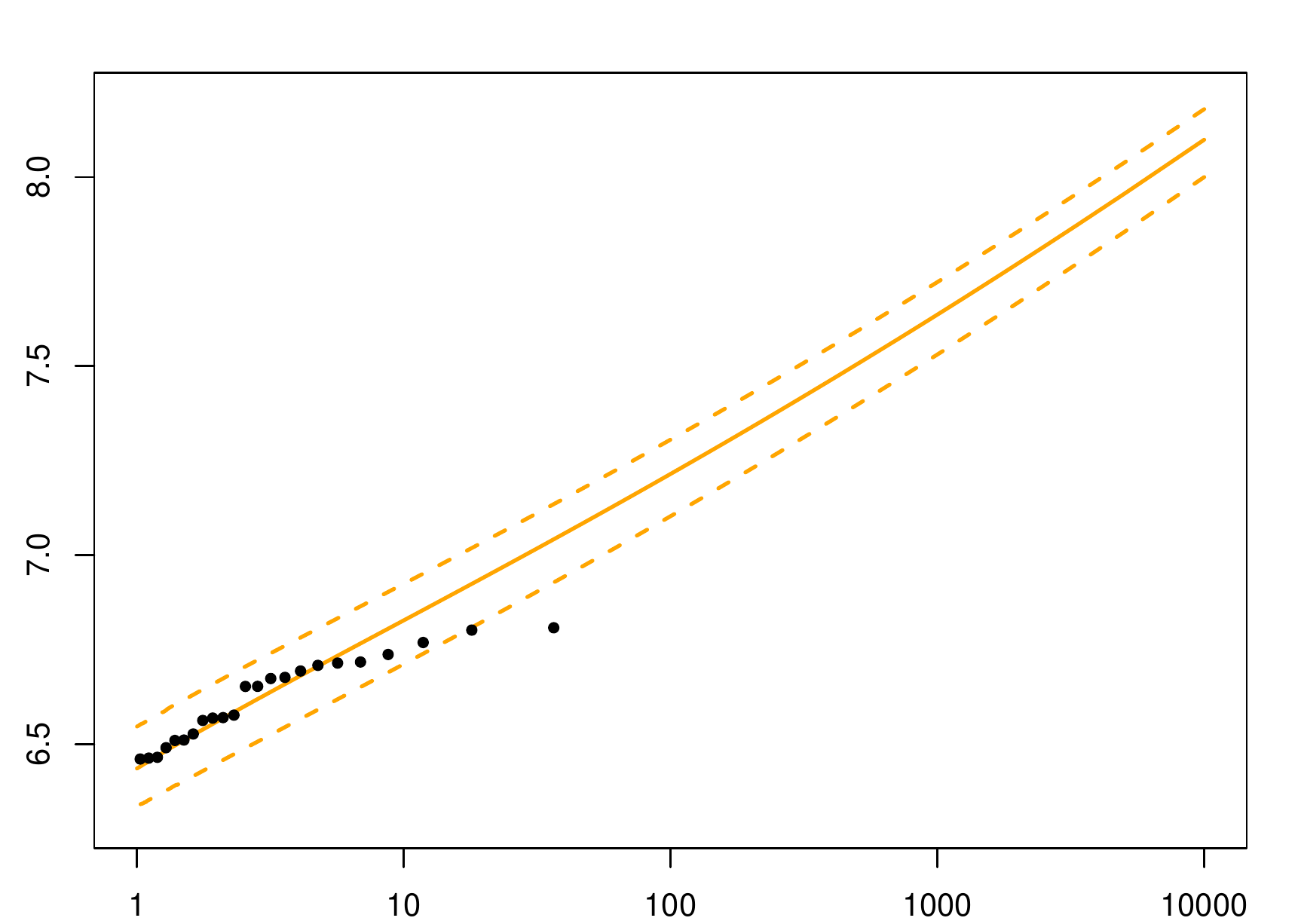}
    \caption{Return level estimates from the final model (\textit{temporal dependence}) (orange), with the maximum and minimum year-specific return level estimates (dashed orange) and empirical estimates (black) at Newlyn (left) and Sheerness (right).}
    \label{fig::SMfinal_rlest_bnd}
\end{figure}

\begin{figure}[h]
    \centering 
    \includegraphics[width=0.4\textwidth]{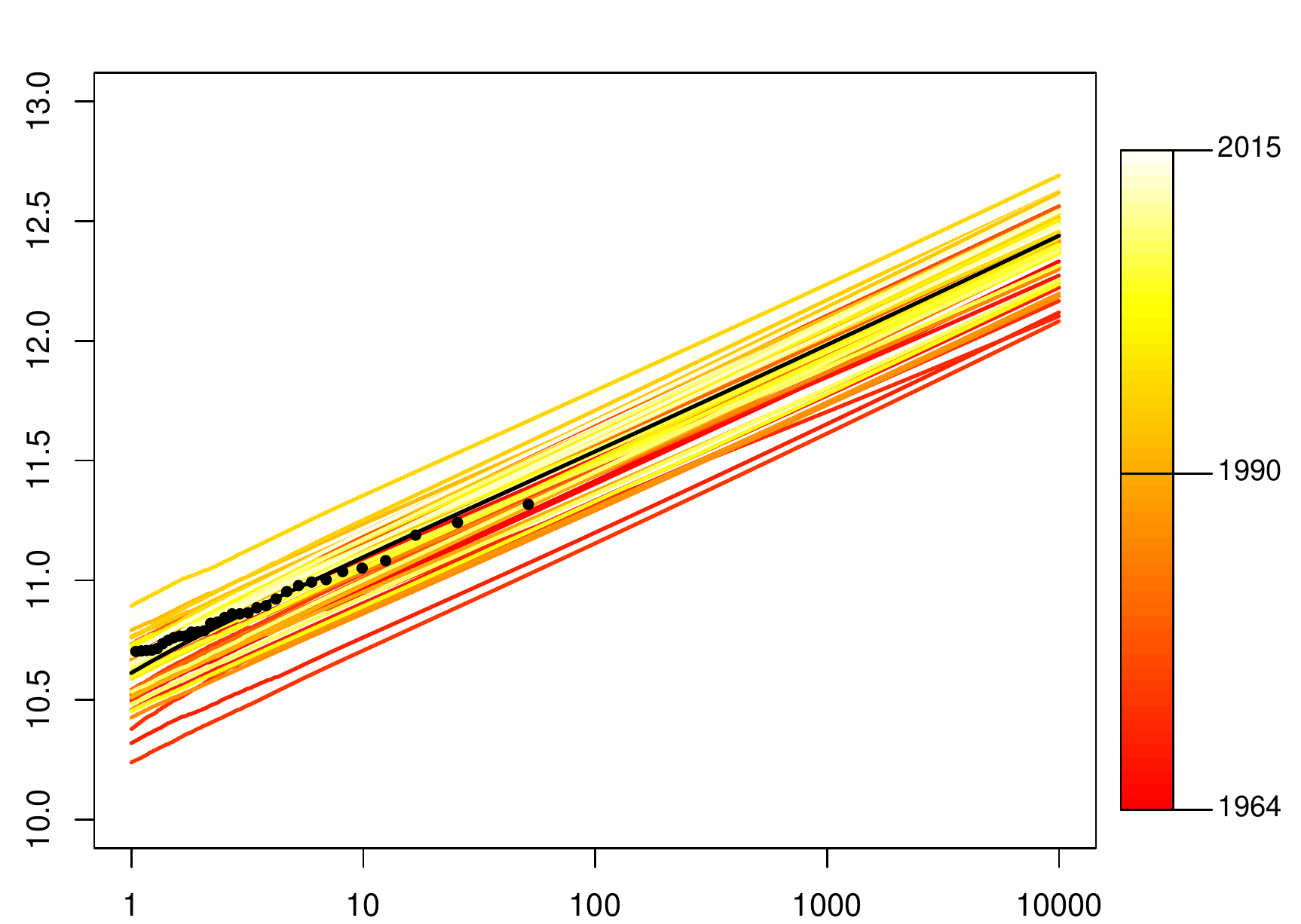}
    \includegraphics[width=0.4\textwidth]{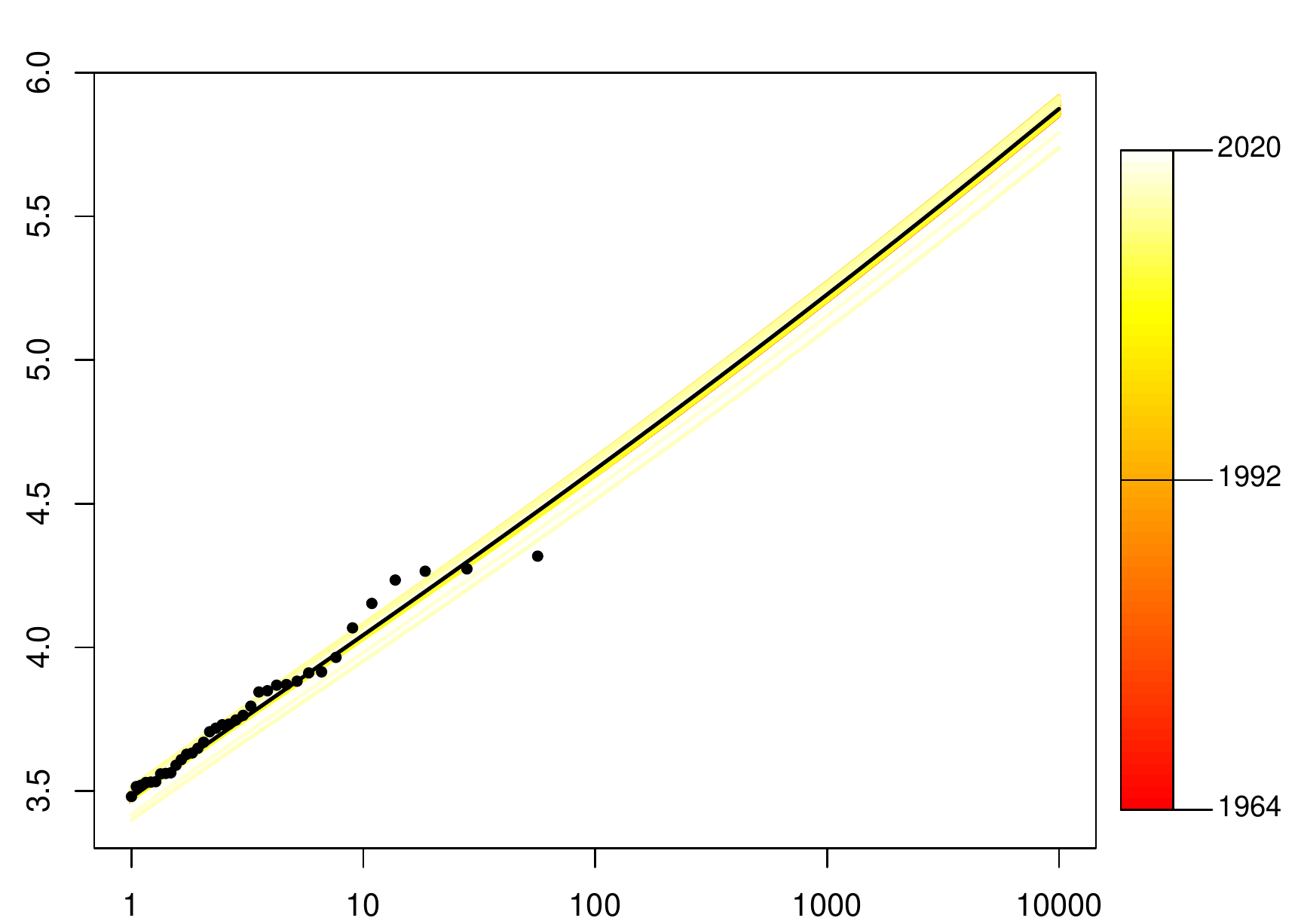} \\
    \includegraphics[width=0.4\textwidth]{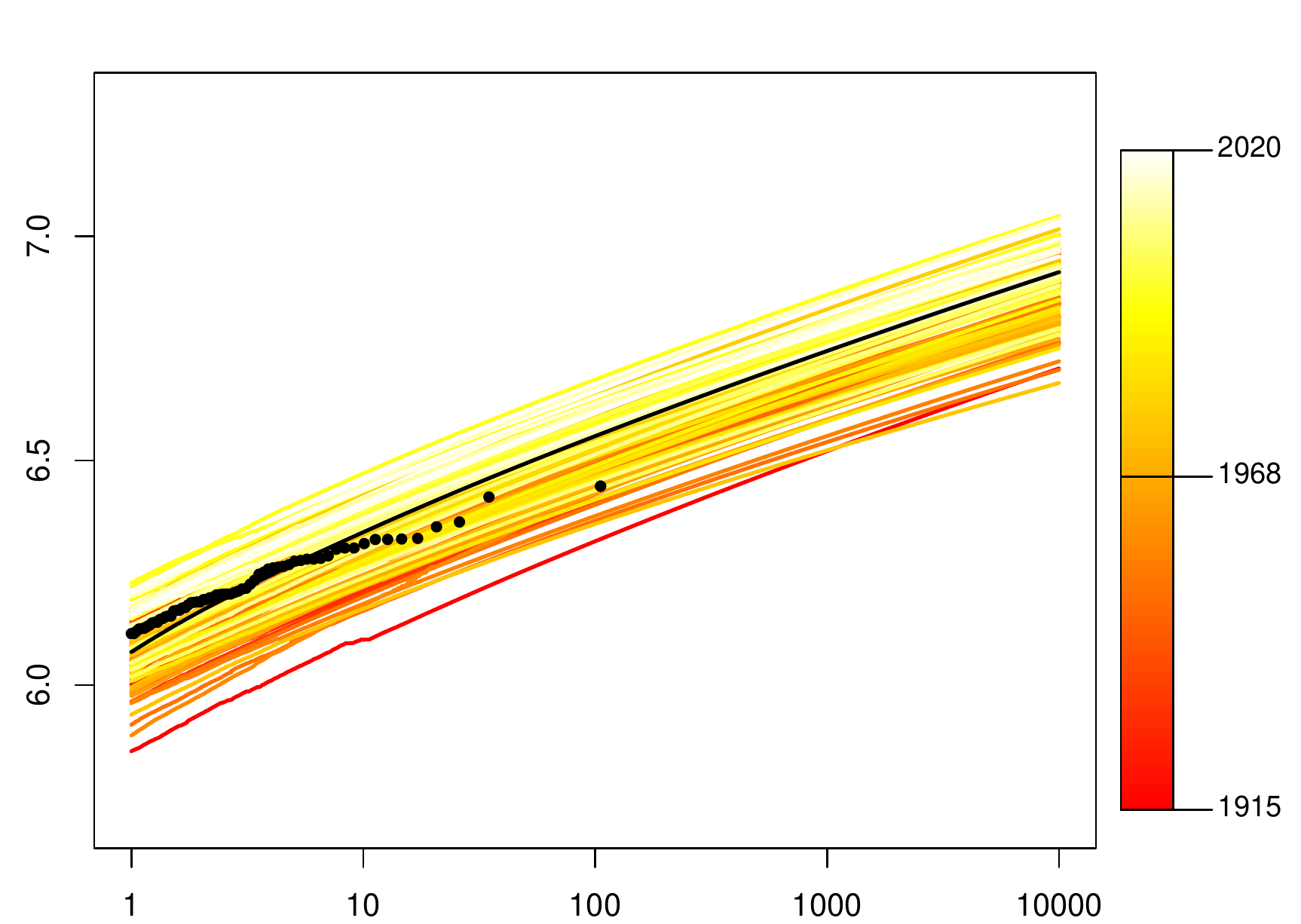} 
    \includegraphics[width=0.4\textwidth]{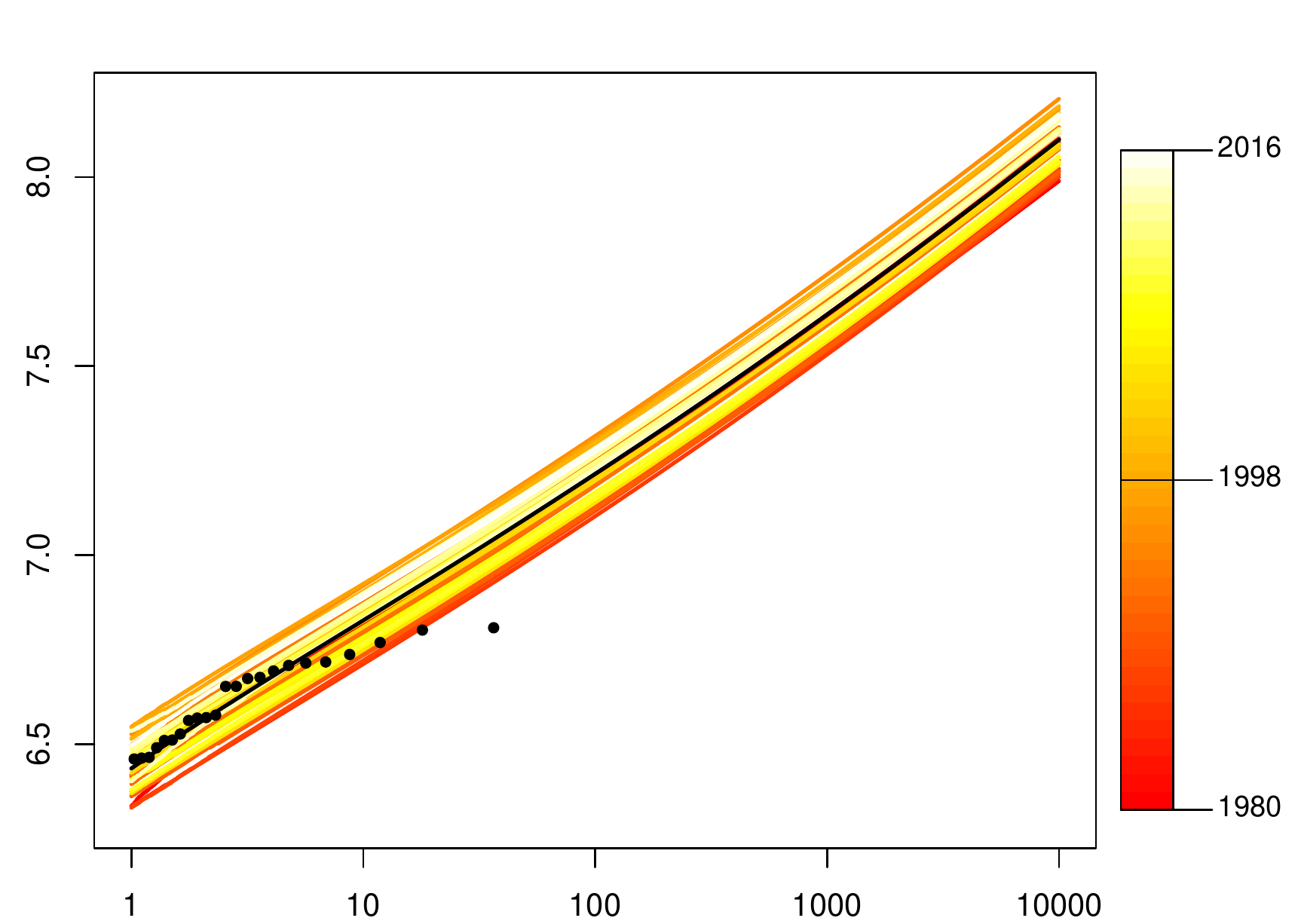}
    \caption{Annual maxima sea level return level estimates for Heysham (top left), Lowestoft (top right), Newlyn (bottom left) and Sheerness (bottom right) estimated using the final model (\textit{temporal dependence}) shown by the solid black line, with year-specific return levels shown by the red-yellow lines. Empirical estimates are shown by black points.}
    \label{fig::annmax_yearly_grad}
\end{figure}

\begin{figure}[h]
    \centering 
    \includegraphics[width=0.4\textwidth]{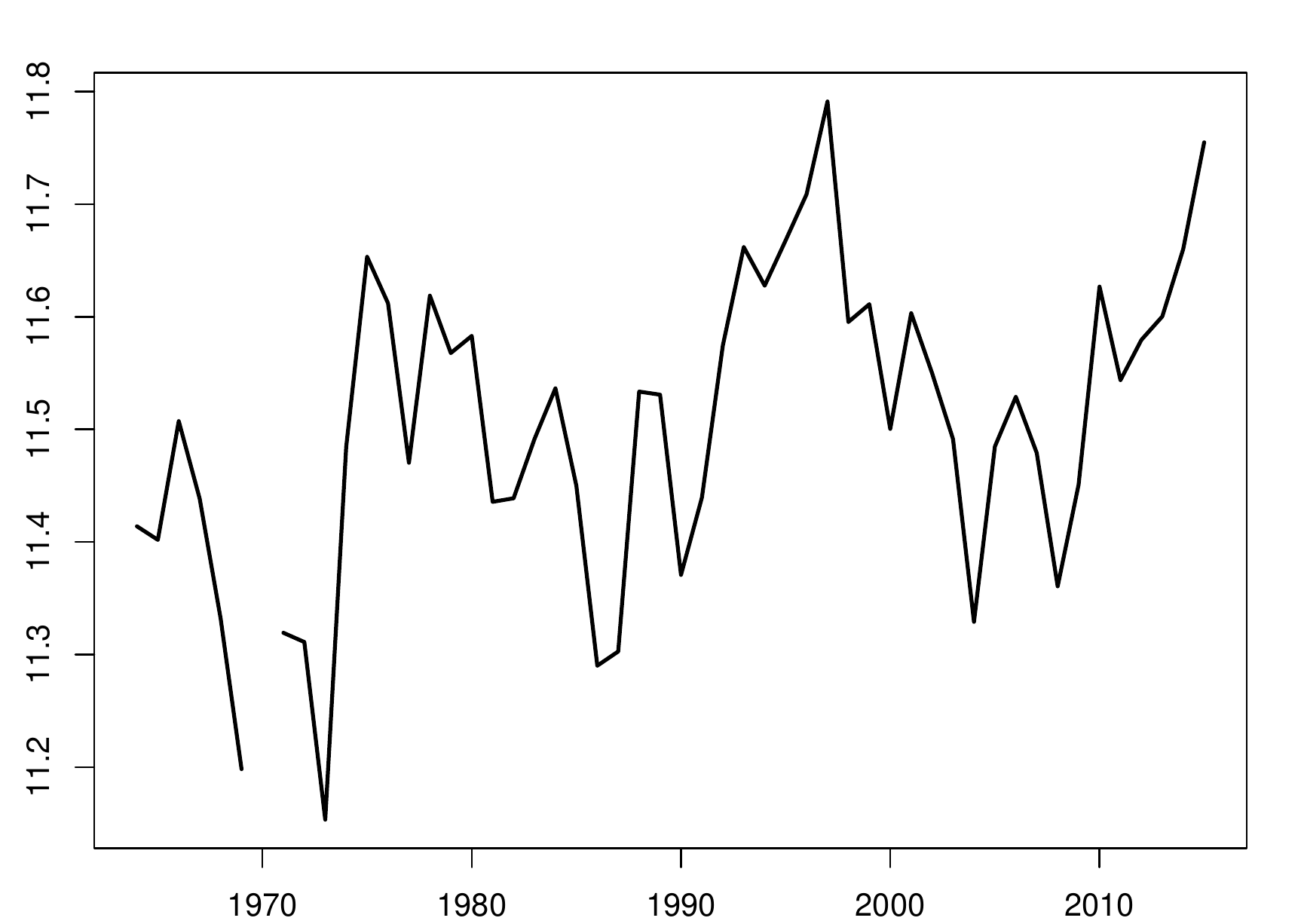}
    \includegraphics[width=0.4\textwidth]{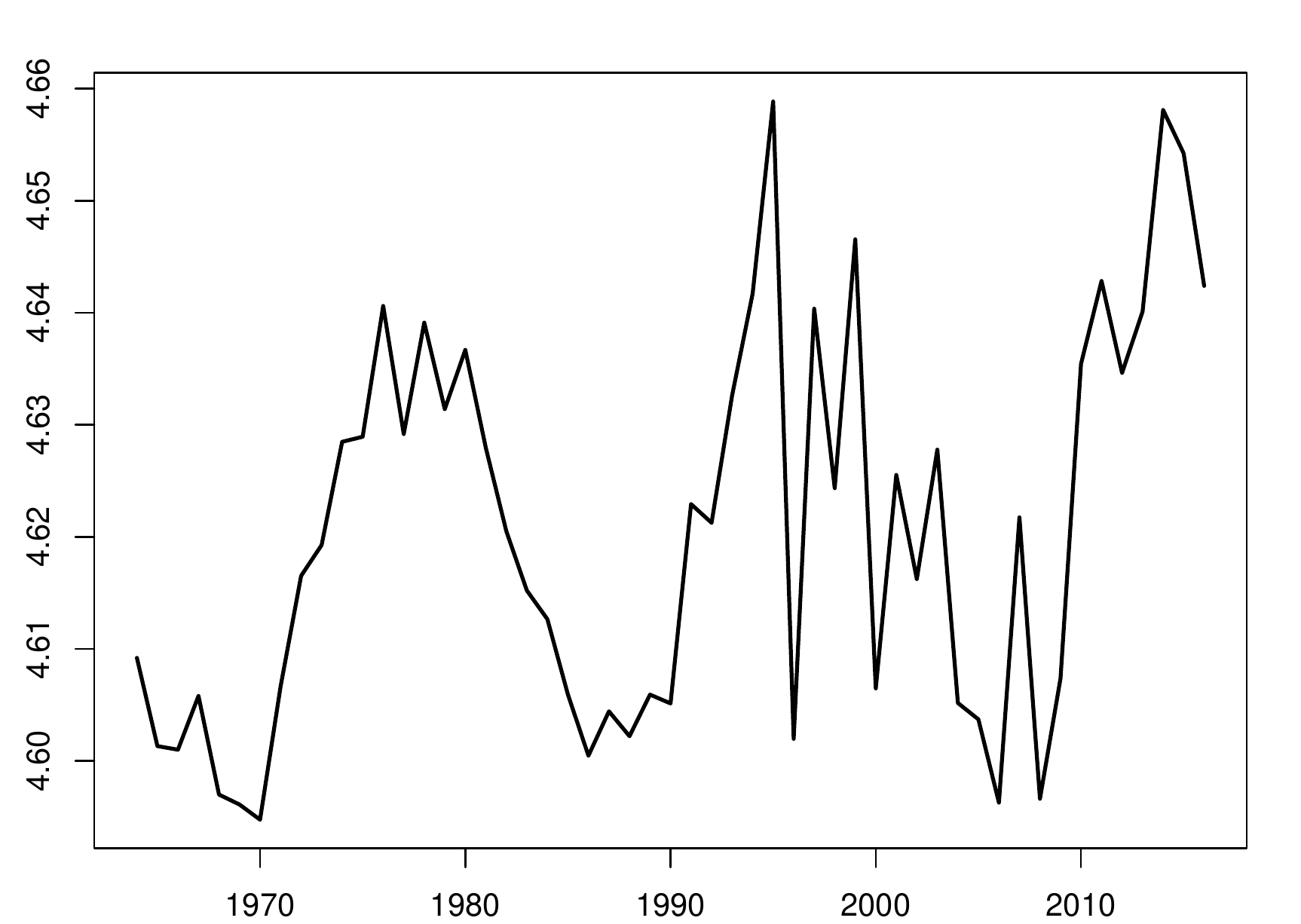} \\
    \includegraphics[width=0.4\textwidth]{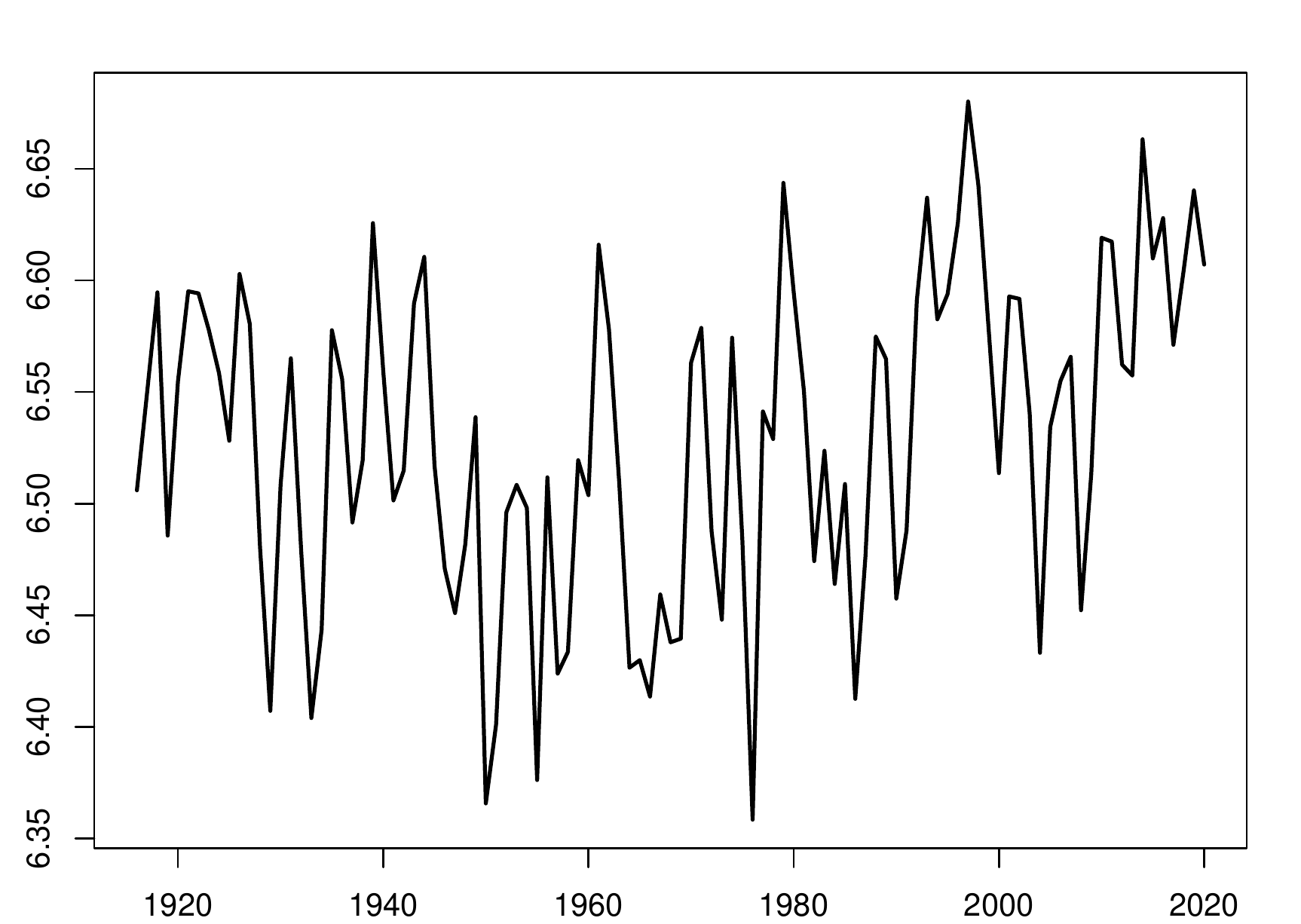}
    \includegraphics[width=0.4\textwidth]{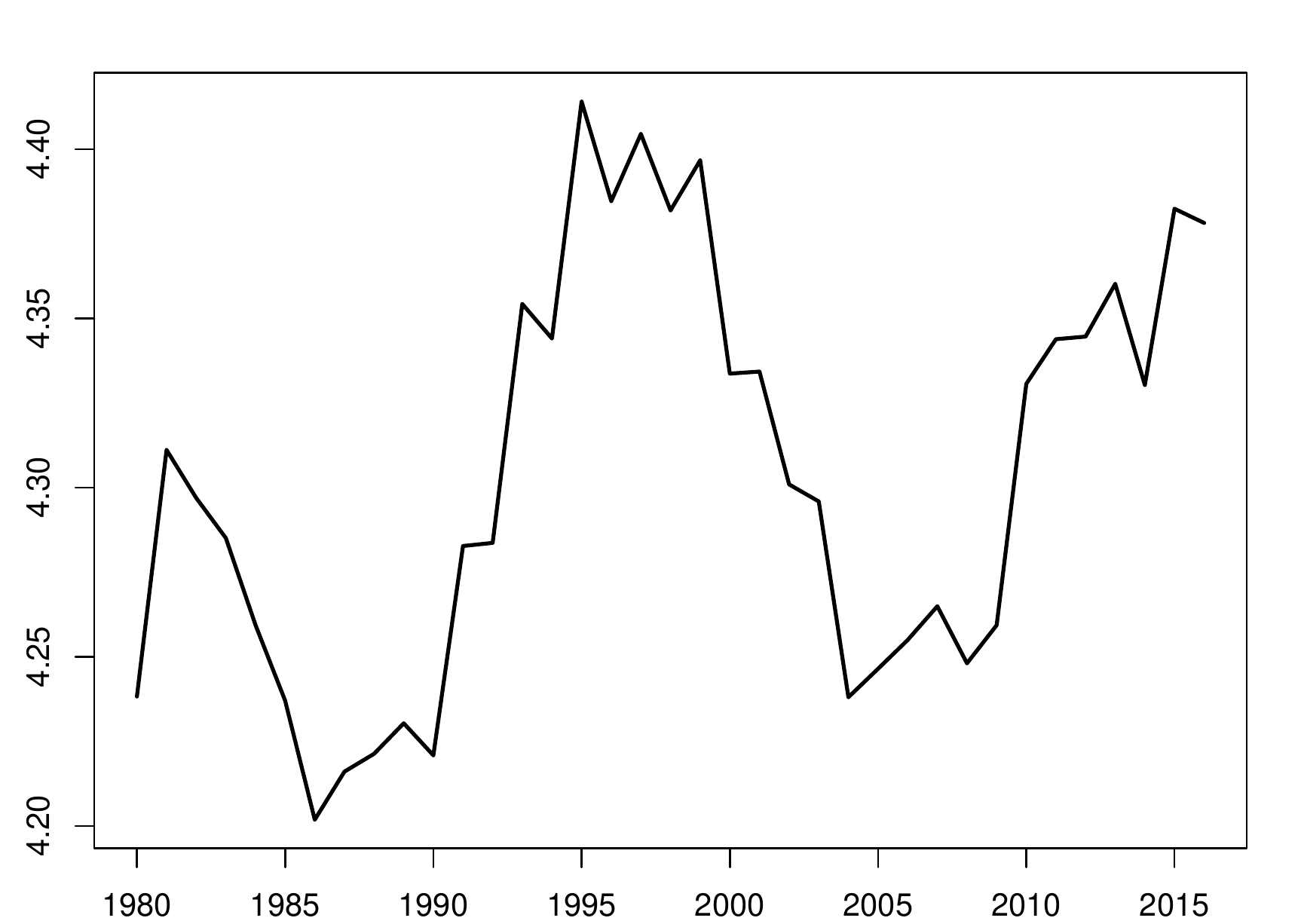}
    \caption{100 year return level estimates for Heysham (top left), Lowestoft (top right), Newlyn (bottom left) and Sheerness (bottom right) estimated using the year-specific final model over the years of observation.}
    \label{fig::annmax_yearly_overyr}
\end{figure}

\begin{figure}[h]
    \centering
    \includegraphics[width=0.5\textwidth]{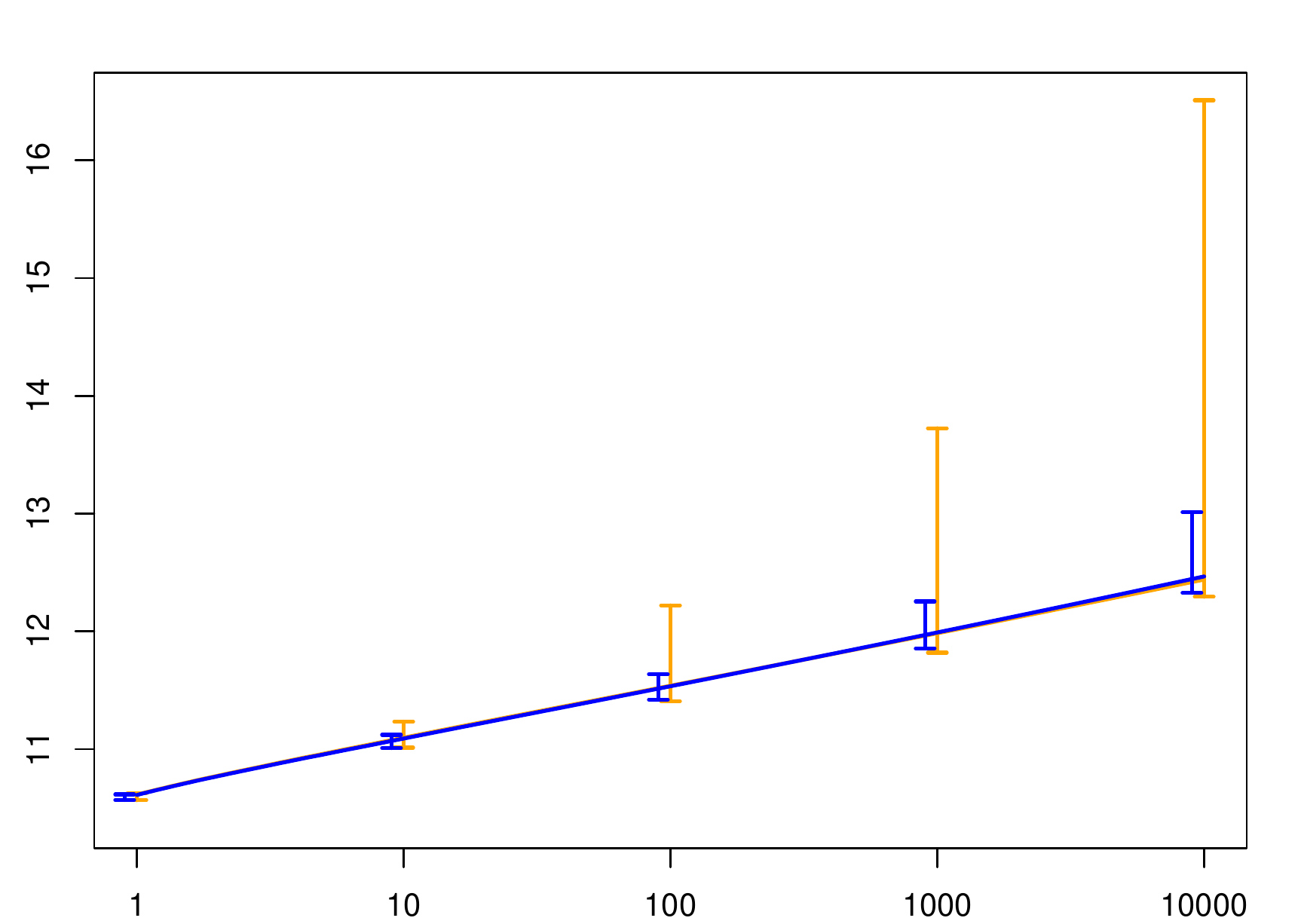}\includegraphics[width=0.5\textwidth]{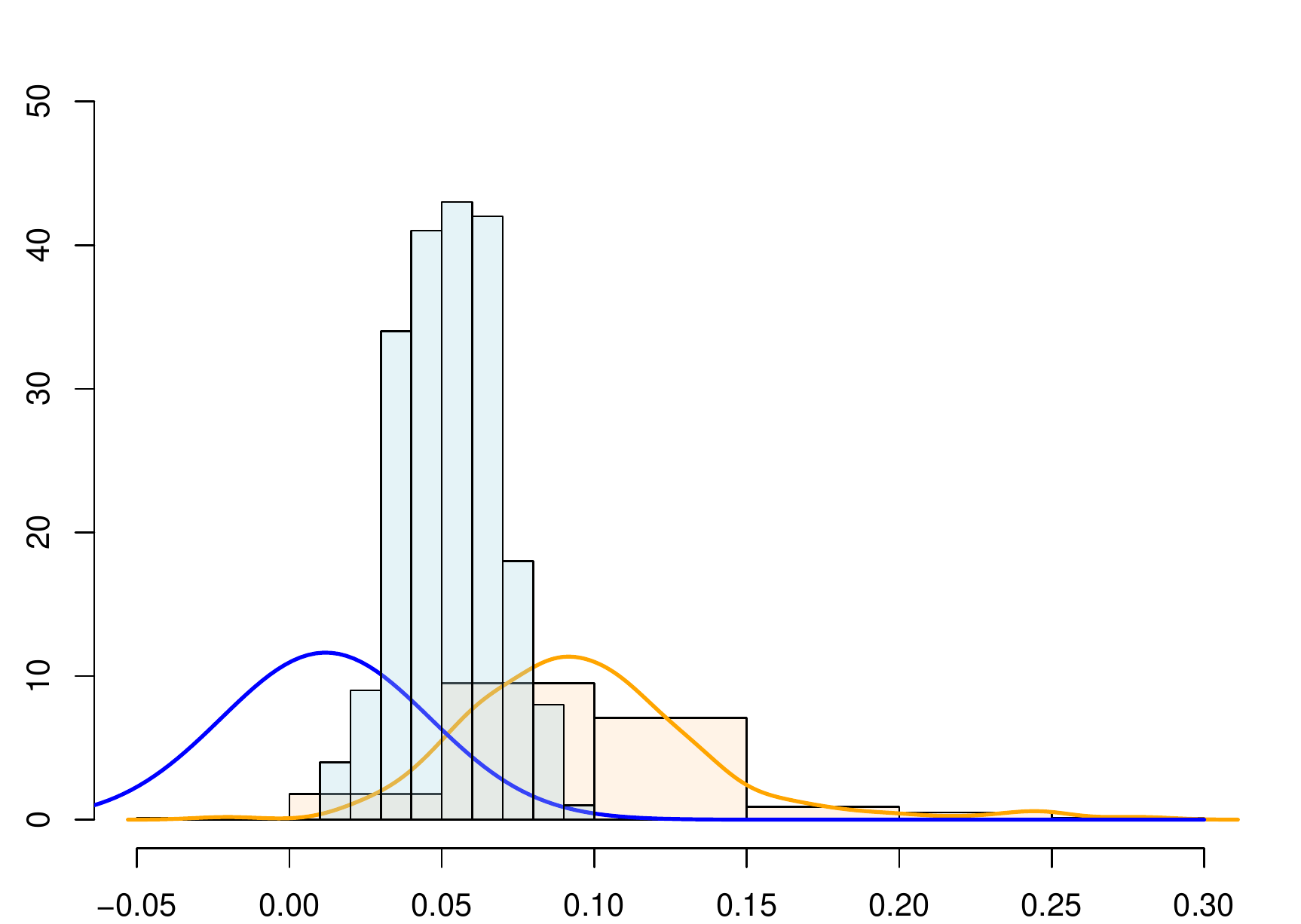}
    \includegraphics[width=0.5\textwidth]{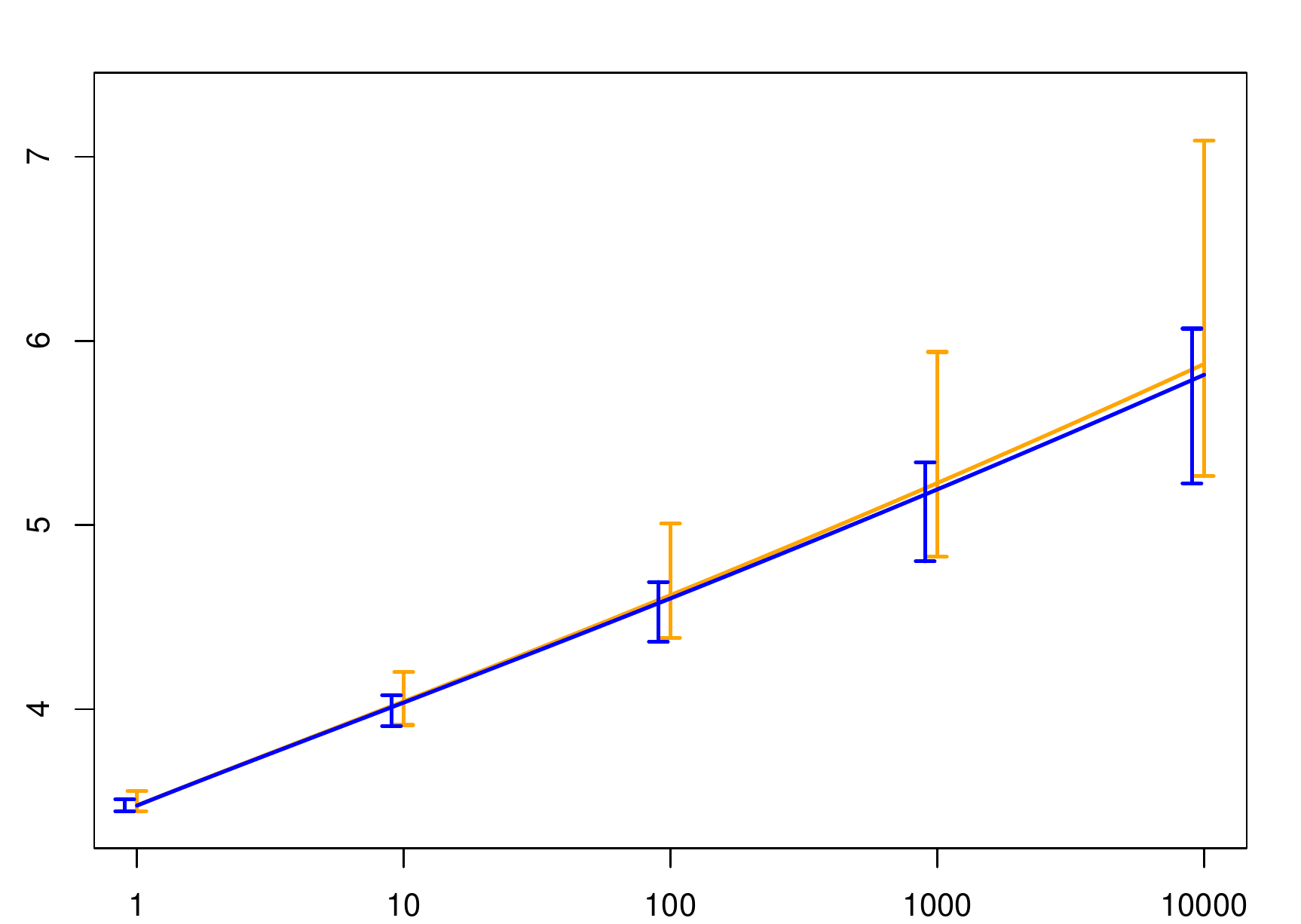}\includegraphics[width=0.5\textwidth]{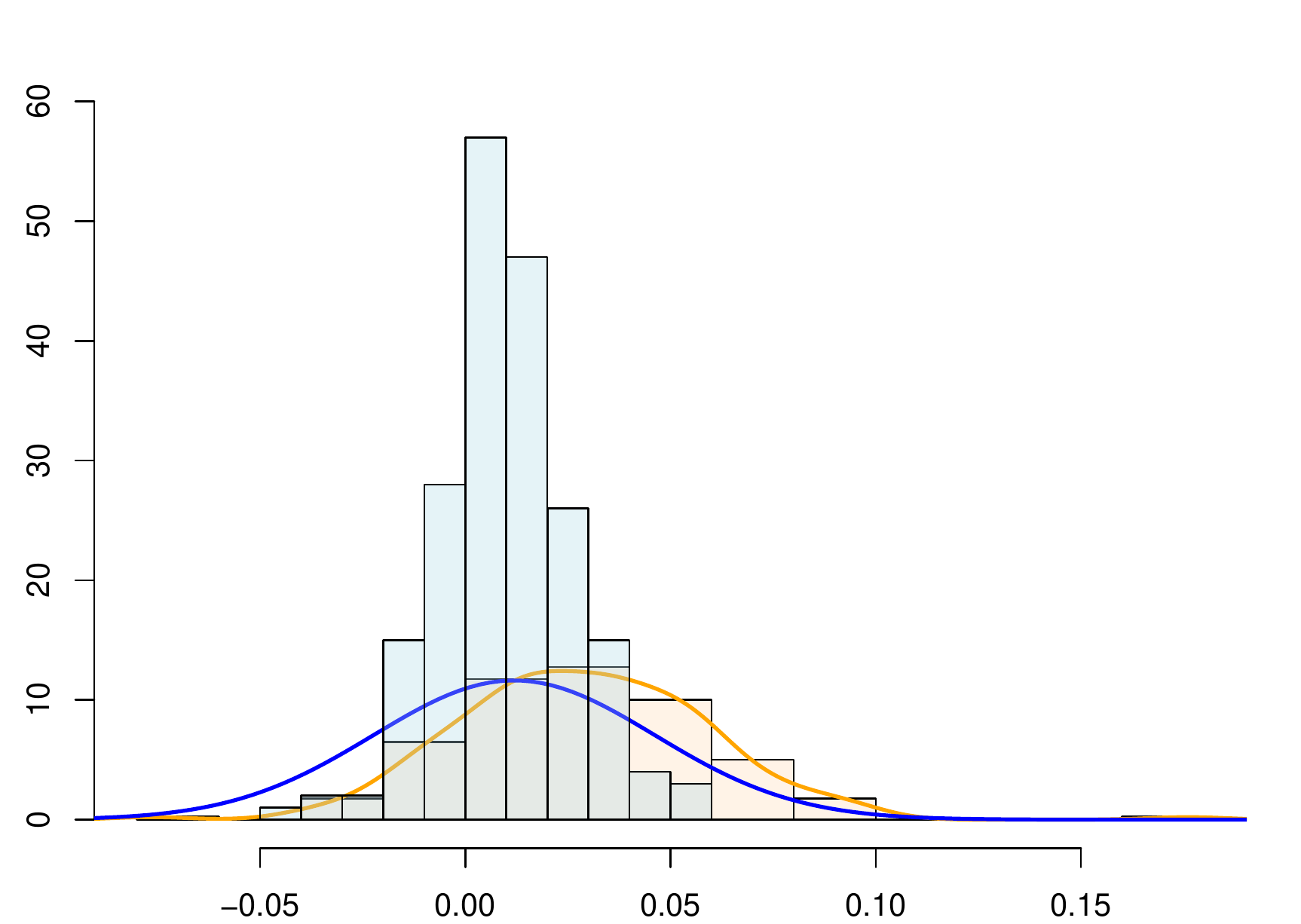}
    \includegraphics[width=0.5\textwidth]{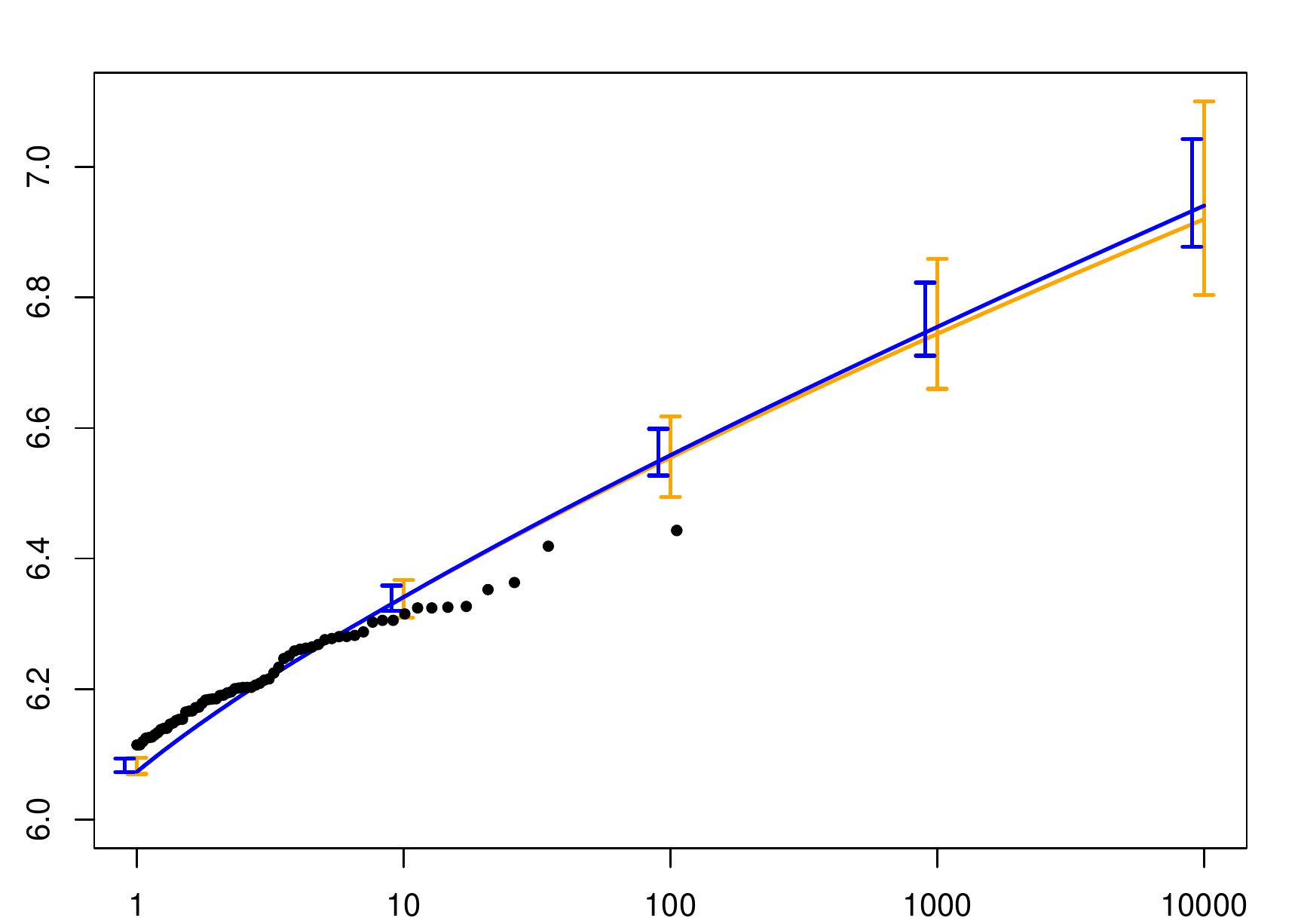}\includegraphics[width=0.5\textwidth]{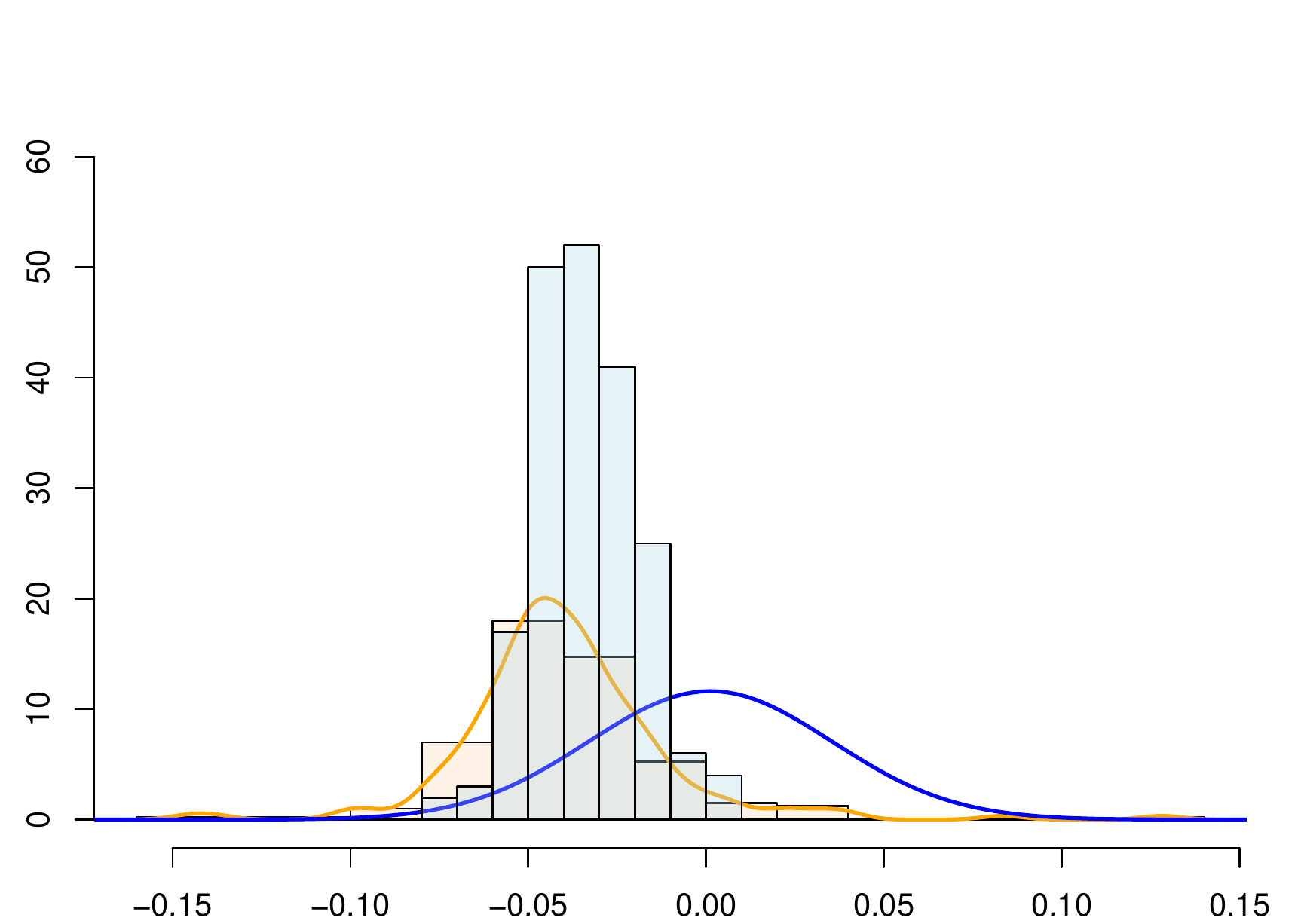}
    \caption{95\% bootstrap confidence intervals on the final (\textit{temporal dependence}) return level estimates at Heysham (top row), Lowestoft (middle row) and Newlyn (bottom row) before (orange) and after (blue) adding a prior distribution to the shape parameter (left). Empirical estimates are shown by black points. Histograms of the shape parameter estimates and their densities (right) for these two models in their corresponding colours.}
    \label{fig::all_boots}
\end{figure}

\begin{figure}[h]
    \centering
    \includegraphics[width=0.9\textwidth]{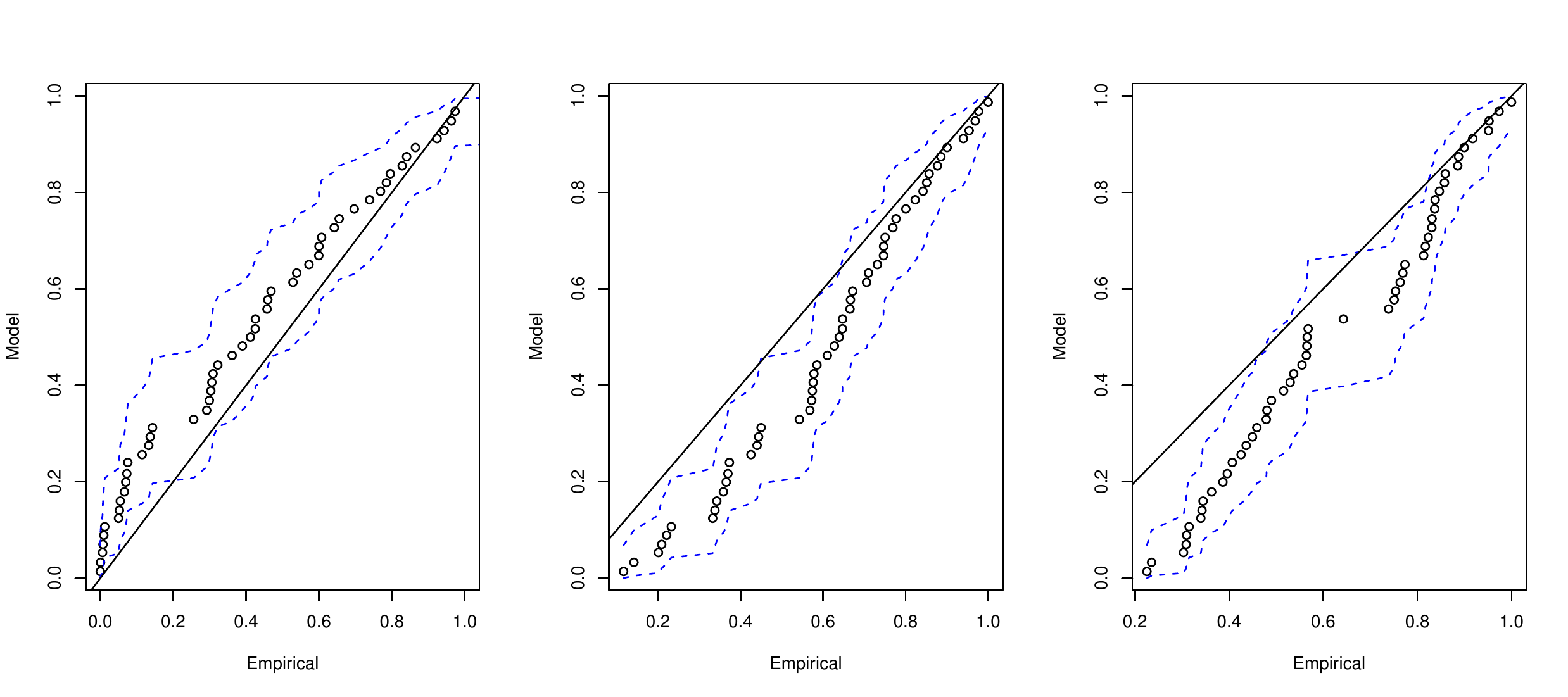}
    \includegraphics[width=0.9\textwidth]{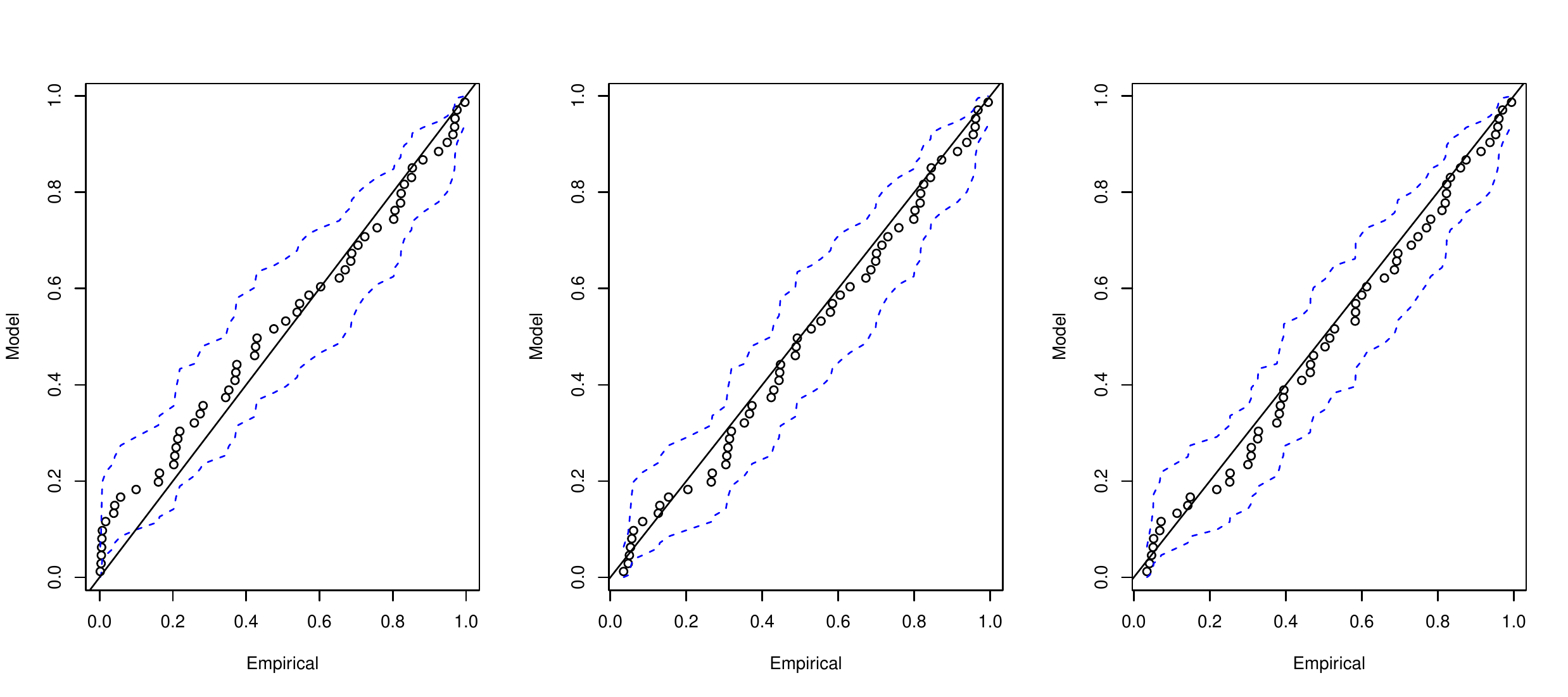}
    \includegraphics[width=0.9\textwidth]{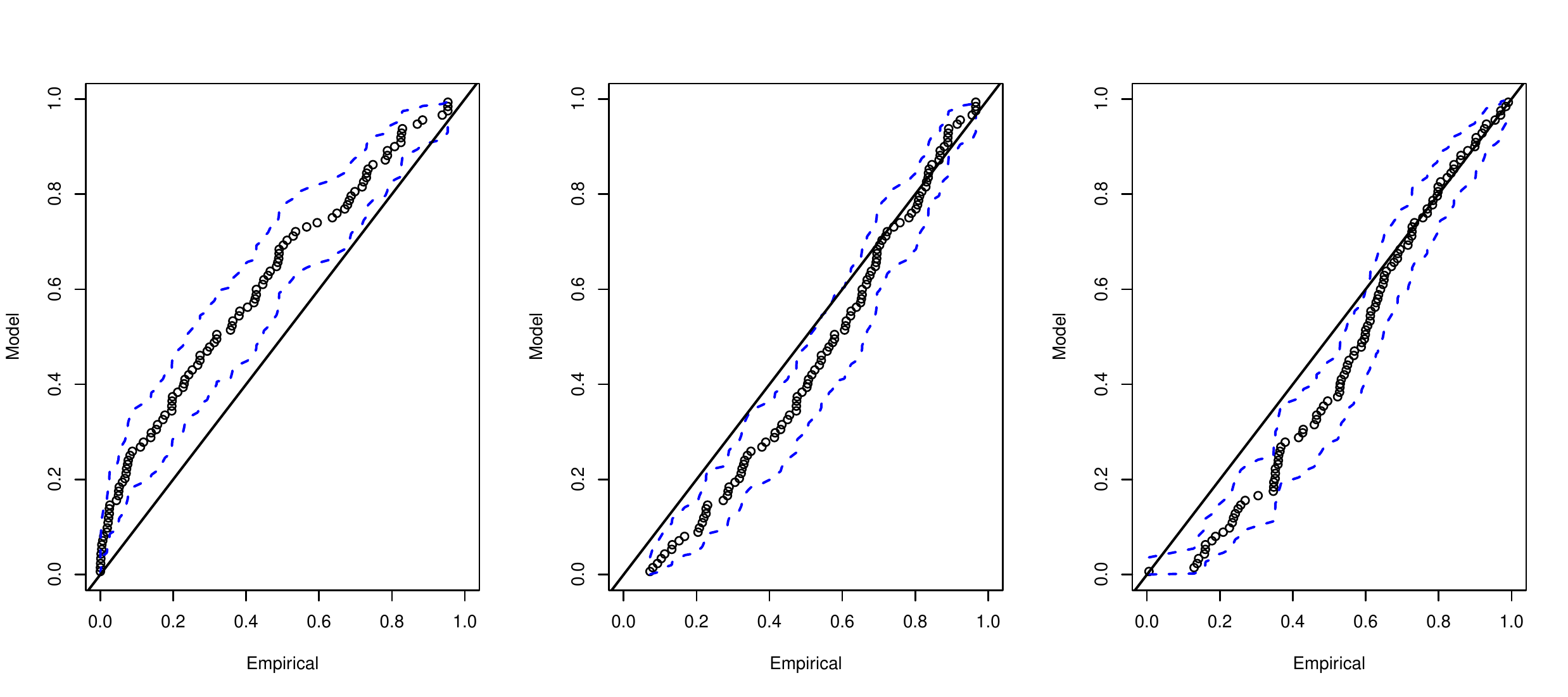}
    \caption{PP plots for the transformed annual maximum sea levels to a uniform scale at Heysham (top row), Lowestoft (middle row) and Newlyn (bottom row). These are transformed using the \textit{baseline} (left column) and final (\textit{temporal dependence}) (central column) distribution function for the annual maxima, as well as the year specific final model (right column). The black line shows the line of equality, $y=x$.}
    \label{fig::mapannmax_all}
\end{figure}

\begin{figure}[h]
    \centering
    \includegraphics[width=0.49\textwidth]{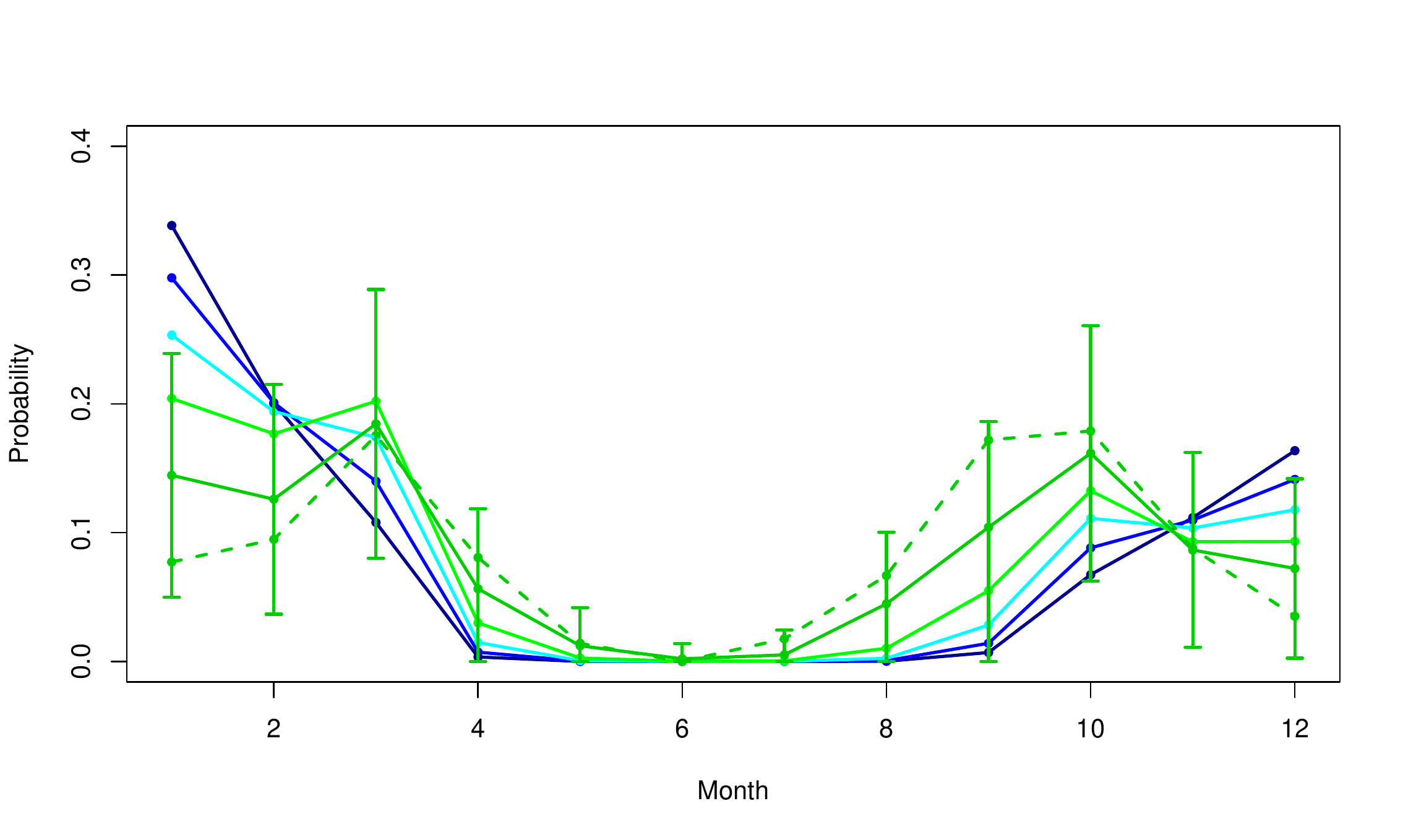}
    \includegraphics[width=0.49\textwidth]{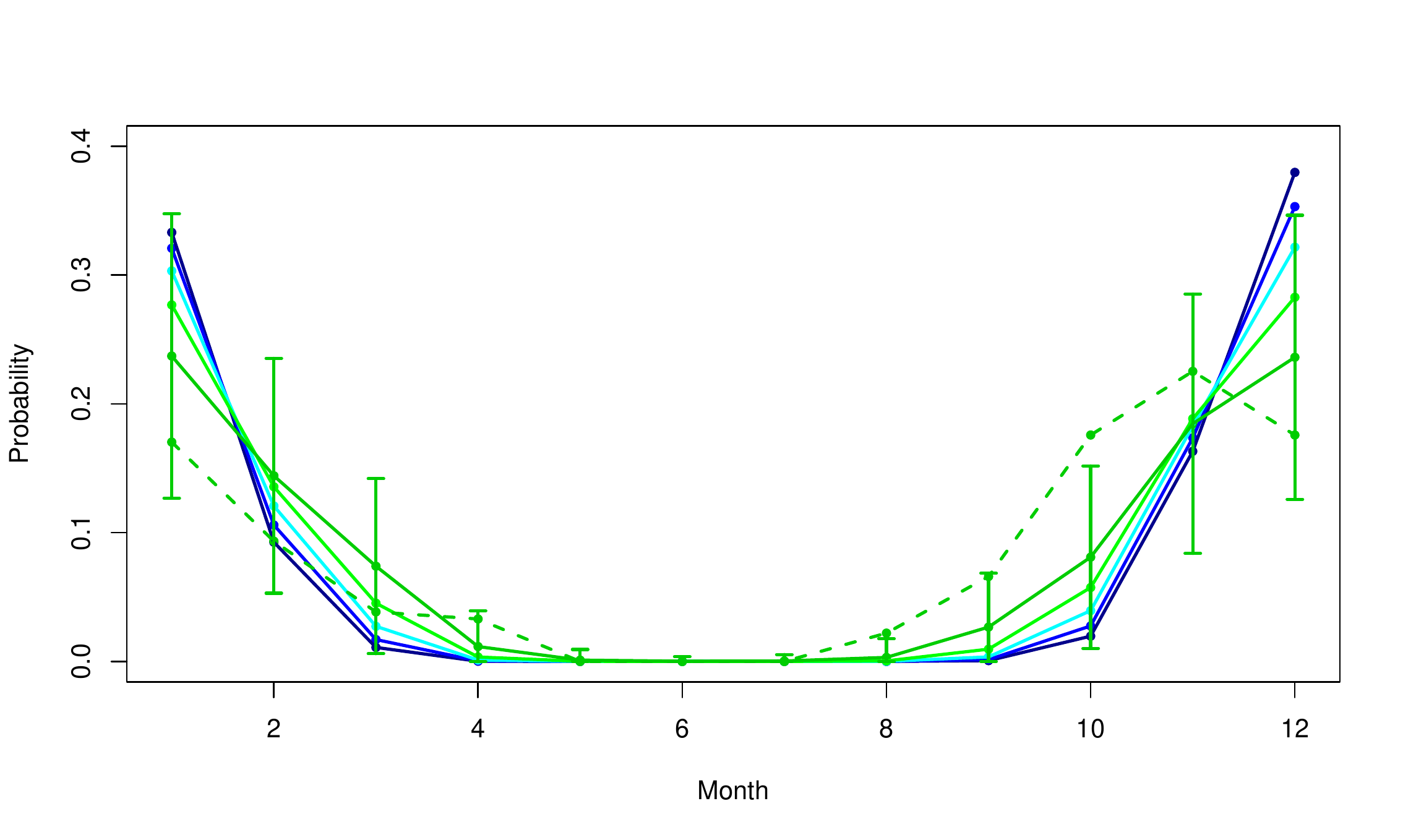} \\
    \includegraphics[width=0.49\textwidth]{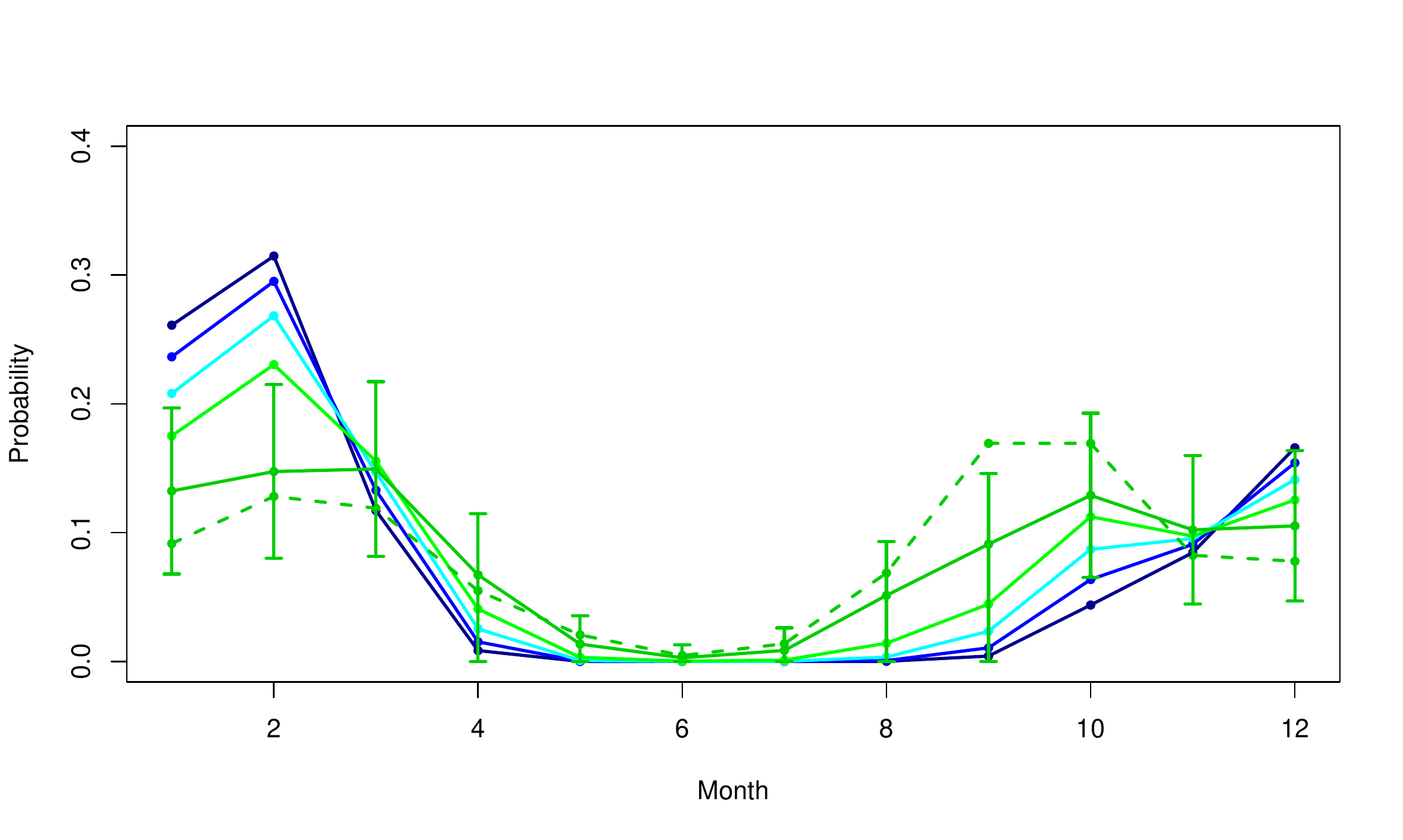}
    \caption{Estimates of $\hat P_M(j;z)$ for months $j=1-12$ at Heysham (top left), Lowestoft (top right) and Newlyn (bottom), for $p=1$ (dark green), 0.1 (green), 0.01 (cyan), 0.001 (blue), and 0.0001 (dark blue). The dashed dark green line is the empirical estimate $\tilde P_M(j;z_1)$, 95\% confidence intervals are for $\hat P_M(j;z_1)$.}
    \label{fig::Annmaxprob_HEYNEW}
\end{figure}

\clearpage
\bibliographystyleSM{apalike}
\bibliographySM{ref.bib}


\end{document}